\documentclass[%
superscriptaddress,
 amsmath,amssymb,
 aps
]{revtex4-1}
\pdfoutput=1
\usepackage{bmpsize}
\usepackage{graphicx}
\usepackage{xcolor}
\usepackage{ulem}
\usepackage{dcolumn}
\usepackage{bm}
\usepackage[mathlines]{lineno}
\usepackage{booktabs}
\usepackage{afterpage}
\usepackage{nicefrac}

\begin{document}

\title{\huge Precision Studies of QCD in the Low Energy Domain of the EIC }

\newcommand*{\LBL}{Lawrence Berkeley National Laboratory, Berkeley, CA 94720, USA} 
\newcommand*{\SBU}{Stony Brook University, Stony Brook, NY 11794, USA}
\newcommand*{\JLAB}{Thomas Jefferson National Accelerator Facility, Newport News, Virginia 23606, USA}
\newcommand*{\TEMPLE}{Temple University, Philadelphia, Pennsylvania 19122, USA}
\newcommand*{\ANL}{Argonne National Laboratory, Lemont, Illinois 60439, USA}
\newcommand*{\MIT}{Massachusetts Institute of Technology, Cambridge, Massachusetts 02139, USA}
\newcommand*{\MSU}{Michigan State University, East Lansing, 48824, Michigan, USA}
\newcommand*{\WM}{The College of William and Mary, Williamsburg, Virginia 23185, USA}
\newcommand*{\CYP} {University of Cyprus, Department of Physics, CY-1678 Nicosia; The Cyprus Institute, CY-1645 Nicosia}
\newcommand*{\DUKE}{Duke University, Durham, NC 27708, USA } 
\newcommand*{\UCLA}{Department of Physics and Astronomy, University of California, Los Angeles, USA}
\newcommand*{\UTorino}{Physics Department, University of Turin, via P. Giuria 1, I-10125 Turin, Italy}
\newcommand*{\INFNTorino}{INFN, Turin section, via P. Giuria 1, I-10125 Turin, Italy}
\newcommand*{\UCLATheory}{Mani L. Bhaumik Institute for Theoretical Physics, University of California, Los Angeles, USA}
\newcommand*{\LANL}{Los Alamos National Laboratory, Los Alamos, NM, USA} 
\newcommand*{\CNF}{Center for Nuclear Femtography, 1201 New York Avenue, Washington, DC 20005}
\newcommand*{\CFNS}{Center for Frontiers in Nuclear Science, Stony Brook University, Stony Brook, USA}
\newcommand*{\CEA}{IRFU CEA-Saclay, 91191 Gif sur Yvette, France}
\newcommand*{\ORSAY}{Université Paris-Sud 11, Orsay, France} 
\newcommand*{\FER}{INFN Ferrara, 44122 Ferrara, Italy}
\newcommand*{\KNU}{Department of Physics, Kyungpook National University, Daegu 41566, Korea}
\newcommand*{\UMD}{University of Maryland, College Park, MD 20742}
\newcommand*{\BNL}{Department of Physics, Brookhaven National Laboratory, Upton, NY 11973, USA}
\newcommand*{\Regensburg}{Institut f\"ur Theoretische Physik, Universit\"at Regensburg, D-93040 Regensburg, Germany}
\newcommand*{\Regina}{University of Regina, Regina, SK S4S~0A2 Canada}
\newcommand*{\UVA}{University of Virginia, Physics Department, 382 McCormick Rd, Charlottesville, VA 22904}
\newcommand*{\FIU}{Florida International University, Department of Physics, Miami, FL 33199, USA}
\newcommand*{\Ugent}{Ghent University, Department of Physics and Astronomy, B9000 Ghent, Belgium}
\newcommand*{\Glasgow}{ SUPA, School of Physics and Astronomy, University of Glasgow, Glasgow, G12 8QQ UK.}
\newcommand*{\IUB}{Indiana University, Bloomington, Indiana}
\newcommand*{\ODU}{Old Dominion University, Department of Physics, 4600 Elkhorn Ave. Norfolk, VA 23529}
\newcommand*{\GWU}{The George Washington University, Department of Physics, Washington, DC 20052} 
\newcommand*{\CAM}{DAMTP, University of Cambridge, UK}
\newcommand*{\APCTP}{Asia Pacific Center for Theoretical Physics, Pohang, Gyeongbuk 37673, Korea}
\newcommand*{\Zagreb}{University of Zagreb, Faculty of Science, Department of Physics, 10000 Zagreb, Croatia}
\newcommand*{\Catania}{INFN Sezione di Catania, I-95123 Catania, Italy}
\newcommand*{\Messina}{Dipartimento di Scienze Matematiche e Informatiche, Scienze Fisiche e Scienze della Terra,\\ Universit\`{a} degli Studi di Messina, I-98122 Messina, Italy}
\newcommand*{\CNRS}{Laboratoire de Physique Joliot-Curie, CNRS-IN2P3, Universit\'e Paris-Saclay}
\newcommand*{\Riken}{RIKEN Nishina Center for Accelerator-Based Science, Wako, Saitama 351-0198, Japan}
\newcommand*{\RikenBNL}{RIKEN BNL Research Center, Brookhaven National Laboratory, Upton, New York 11973-5000, USA}
\newcommand*{\UNAM}{Instituto de F\'{i}sica, Universidad Nacional Aut\'{o}noma de M\'{e}xico, Apartado Postal 20-364,\\ 01000 Ciudad de M\'{e}xico, Mexico}
\newcommand*{\BERKS}{Science Division, Penn State University Berks, Reading, Pennsylvania 19610, USA}
\newcommand*{\UPAVIA}{Dipartimento di Fisica, Universit\`{a} di Pavia, Pavia, Italy}
\newcommand*{\INFNPAVIA}{INFN, Sezione di Pavia, Pavia, Italy}
\newcommand*{\CNYANG}{C.N. Yang Institute for Theoretical Physics, Stony Brook University, Stony Brook, NY 11794, USA}
\newcommand*{\MainzU}{Institut für Physik, Institut für Kernphysik, Johannes-Gutenberg-Universität, D-55099 Mainz, Germany}
\newcommand*{\UW} {University of Washington, Department of Physics B464, Seattle, USA }
\newcommand*{\UCR} {University of California, Riverside, Department of Physics and Astronomy} 
\newcommand*{\ORNL}{Oak Ridge National Laboratory, Oak Ridge, Tennessee 37831 USA}
\newcommand*{\UKY} {University of Kentucky, College of Arts \& Science, Physics \& Astronomy}
\newcommand*{\YERPHI}{A.I. Alikhanyan National Science Laboratory, Yerevan 0036, Armenia.}
\newcommand*{\PSU} {Science Division, Penn State University Berks, Reading, Pennsylvania 19610, USA}
\newcommand*{\IPM} {School of Particles and Accelerators,
Institute for Research in Fundamental Sciences (IPM),\\ P.O.Box 19395-5531, Tehran, Iran}
\newcommand*{\UCONN}{University of Connecticut, Department of Physics 196A Auditorium Road, Storrs, CT 06269-3046 }

\newcommand*{\URegina}{Department of Physics,  University of Regina, 3737 Wascana Parkway
Regina, CANADA}
\newcommand*{\CUHK}{School of Science and Engineering, The Chinese University of Hong Kong, Shenzhen 518172, China
} 
\newcommand*{\NANJING} {Department of Physics, Nanjing Normal University, Nanjing 210023, China.}
\newcommand*{\DUQUESNE} {Duquesne University, Pittsburgh, PA 15282, USA }
\newcommand*{\JLU}{Justus Liebig University, Physikalisches Institut, Giessen, Germany }

\newcommand*{\UFER} {Università degli Studi di Ferrara, I-44122 Ferrara, Italy}

\newcommand*{\UIC} {University of Illinois at Chicago, Chicago, Illinois 60607} 

\newcommand*{\CAIRO} {Faculty of Science, Cairo University, Cairo, Egypt}

\newcommand*{\KEK}{The High Energy Accelerator Research Organization, KEK, Japan}

\newcommand*{\USNRC}{U.S. Nuclear Regulatory Commission,Triangle Science, 138 W. Hatterleigh Avenue
Hillsborough, NC 27278 }

\newcommand*{\Kurchatov}{NRC Kurchatov Institute – IHEP, Protvino, 142281, Russia}

\newcommand*{\MEPHI}{National Research Nuclear University MEPhI, Moscow, 115409, Russia
}
\newcommand*{\Mumbai}{Nuclear Physics Division, Bhabha Atomic Research Centre, Mumbai 400 085, India.}
\newcommand*{\Munster}{Institut f\"ur Theoretische Physik, Westf\"alische
Wilhelms-Universit\"at M\"unster,
Wilhelm-Klemm-Stra\ss e 9, D-48149 M\"unster, Germany}
\newcommand*{\Lanzhou}{Institute of Modern Physics, Chinese Academy of Sciences, Lanzhou 730000, China}
\newcommand*{\Pelotas}{Institute of Physics and Mathematics, Federal University
of Pelotas, \\
  Postal Code 354,  96010-900, Pelotas, RS, Brazil}

\newcommand*{\Davidson}{Department of Physics
Department of Mathematics and Computer Science
Davidson College
Box 7133
Davidson, NC 28035}

\newcommand{\ucm}{Departamento de F\'isica Te\'orica, Universidad Complutense de Madrid and IPARCOS, E-28040 Madrid, Spain}
\newcommand{\ub}{Departament de F\'isica Qu\`antica i Astrof\'isica and Institut de Ci\`encies del Cosmos, Universitat de Barcelona, E08028, Spain}

\newcommand*{\OU}{Ohio University, Department of Physics and Astronomy, Athens OH 45701, USA}

\newcommand*{\Petersburg}{Higher School of Economics, National Research University, 194100 St. Petersburg, Russia}

\newcommand*{\Gatchina}{National Research Centre Kurchatov Institute: Petersburg Nuclear Physics Institute, 188300 Gatchina, Russia}

\newcommand*{\NIT}{Department of Physics,
Dr. B.R. Ambedkar National Institute of Technology, Jalandhar
144027 India} 

\newcommand*{\Yarmouk}{Yarmouk University, Irbid, Jordan 21163.}
\newcommand*{\INFNTrieste}{INFN, Sezione di Trieste, Trieste, Italy}
\newcommand*{\VirginiaTech}{Virginia Tech University, Blacksburg, VA 24061}

\newcommand*{\SCNUIQM}{Guangdong Provincial Key Laboratory of Nuclear Science, Institute of Quantum Matter, South China Normal University, Guangzhou 510006, China}

\newcommand*{\SCNUJLQM}{Guangdong-Hong Kong Joint Laboratory of Quantum Matter, Southern Nuclear Science Computing Center, South China Normal University, Guangzhou 510006, China}

\newcommand*{\UWF}{University of West Florida, Pensacola, FL 23514, USA} 
\newcommand*{\YORK}{University of York, School of Physics, Engineering, and Technology, Heslington YO10 5DD , UK}

\newcommand*{\JINR}{Joint Institute for Nuclear Research, Bogoliubov Laboratory of Theoretical Physics. }

\newcommand*{\UPINDIA}{Dr Rammanohar Lohia Avadh University, Ayodhya-224001, U.P., INDIA}
\newcommand*{\AMU}{Adam Mickiewicz University, ul. Uniwersytetu Poznanskiego 2, 61-614 Poznan, Poland.}
\newcommand*{\UCSP}{Universidade Cidade de São Paulo, Rua Galvão Bueno 868, São Paulo, 01506-000, SP, Brazil. }
\newcommand*{\ITA}{Instituto Tecnol\'ogico de  Aeron\'autica
12.228-900, S\~ao Jos\'e dos Campos, Brazil}
\newcommand*{\UMICH}{Department of Physics, University of Michigan, Ann Arbor, Michigan 48109, USA}

\newcommand*{\CWM}{Department of Physics and Astronomy, College of William and Mary, Williamsburg, Virginia, USA}
\newcommand{\AGHUST}{AGH University of Science and Technology, FPACS, Cracow 30-059, Poland
}
\newcommand*{\DESY}{DESY, Hamburg, Germany} 
\newcommand*{\BRUNEL}{Brunel University London
Uxbridge, Middlesex
UB8 3PH,  UK}
\newcommand*{\LFTC}{Laboratory of Theoretical and Computational Physics-LFTC,
Cruzeiro do Sul University / São Paulo City University,
015060-000, São Paulo, SP, Brazil}
\newcommand*{\LAMAR}{Department of Physics Lamar University, TX, USA
}
\newcommand*{\DESUCRE}{Departamento de Física, Universidad de Sucre,
Carrera 28 No. 5-267, Barrio Puerta Roja, Sincelejo, Colombia}
\newcommand*{\SPAULO}{Laboratório de Física Teórica e Computacional, Universidade Cidade de São Paulo, Rua Galvão Bueno 868, 01506-000 São Paulo, SP, Brazil}
\newcommand*{\UMADRID}{Dpto. de F\'isica Te\'orica \& IPARCOS, Universidad Complutense de Madrid, E-28040 Madrid, Spain}
\newcommand*{\NCSU}{North Carolina State University, Raleigh, North Carolina.}
\newcommand*{\IOWASU}{Iowa State University, Iowa City, IA, USA.}
\newcommand*{\VUU}{Virginia Union University, Department of Natural Sciences}
\newcommand*{\UKANSAS}{Department of Physics and Astronomy, Malott Hall, 1251 Wescoe Hall Dr.Lawrence, KS 66045}
\newcommand*{\FUE}{ Street 90, Fifth Settlement, 11835 New Cairo, Egypt}
\newcommand*{\Khartoum}{Department of Physics, Faculty of Science, University of Khartoum, P.O. Box 321, Khartoum 11115, Sudan }
\newcommand{\Shandong}{Key Laboratory of Particle Physics and Particle Irradiation (MOE), Institute of Frontier and Interdisciplinary Science, Shandong University (QingDao), 266237, China}
\newcommand*{\CASKLTP}{CAS Key Laboratory of Theoretical Physics, Institute of Theoretical Physics, Chinese Academy of Sciences, Beijing 100190, China}
\newcommand*{\SPSUCAS}{School of Physical Sciences, University of Chinese Academy of Sciences, Beijing 100049, China}
\newcommand*{\Tsinghua}{Physics Department, Tsinghua University, 30 Shuangqing Road, Haidian District, Beijing 100084}
\newcommand*{\Guangdong}{Guangdong Provincial Key Laboratory of Nuclear Science, Institute of Quantum Matter, South China Normal University, Guangzhou 510006, China}
\newcommand*{\HongKong}{Guangdong-Hong Kong Joint Laboratory of Quantum Matter, Southern Nuclear Science Computing Center, South China Normal University.}
\newcommand*{\CAQS}{Center of Advanced Quantum Studies, Department of Physics, Beijing Normal University, Beijing 100875, China}
\newcommand*{\CHEP}{Center for High Energy Physics, Peking University, Beijing 100871, China}
\newcommand*{\MissSU}{Mississippi State University, Mississippi State, MS 39762-5167, USA}
\newcommand{\Indiana}{CEEM, Indiana University, Bloomington, IN 47408, USA}
\newcommand*{\EPTX}{CPHT, CNRS, Ecole polytechnique, I.P. Paris, 91128 Palaiseau, France}

\newcommand*{\Lehigh} {Physics Department, Lehigh University 16 Memorial Drive East Office 406, Bethlehem, PA 18015}

\newcommand*{\LMSU}{Skobeltsyn Nuclear Physics Institute and Physics Department, Lomonosov Moscow State University, Russia}

\author{V.D.~Burkert}
\affiliation{\JLAB}

\author{L.~Elouadrhiri}
\affiliation{\JLAB}

\author{A.~Afanasev}
\affiliation{\GWU}

\author{J.~Arrington}
\affiliation{\LBL}

\author{M.~Contalbrigo} 
\affiliation{\FER}

\author{W.~Cosyn}
\affiliation{\FIU}
\affiliation{\Ugent}

\author{A.~Deshpande}
\affiliation{\SBU}

\author{D.I.~Glazier}
\affiliation{\Glasgow}

\author{X.~Ji}
\affiliation{\UMD}
\affiliation{\CNF}

\author{S.~Liuti}
\affiliation{\UVA}

% ...

\maketitle
\newpage
\newpage
\vspace{1.0cm}
\large{~~~~~\underline {\bf Editorial Board}}
\vspace{0.5cm}

\begin{table}[h]
    \begin{tabular}{l l}
   \vspace{0.5cm}
    A. Afanasev      & \hspace{3cm}George Washington University \\
    \vspace{0.5cm}
     J. Arrington    & \hspace{3cm}Lawrence Berkeley National Laboratory\\
     \vspace{0.5cm}
     V. Burkert$^*$     & \hspace{3cm}Jefferson Laboratory \\
     \vspace{0.5cm}
     M. Contalbrigo  & \hspace{3cm}INFN Ferarra\\
     \vspace{0.5cm}
     W. Cosyn & \hspace{3cm}Florida International University \\ 
     \vspace{0.5cm}
     A. Deshpande    & \hspace{3cm}Stony Brook University\\
     \vspace{0.5cm}
     L. Elouadrhiri$^*$  & \hspace{3cm}Jefferson Laboratory \& Center for Nuclear Femtography\\
     \vspace{0.5cm}
     D. Glazier      & \hspace{3cm}Glasgow University\\
     \vspace{0.5cm}
     X. Ji           & \hspace{3cm}University of Maryland\\
     \vspace{0.5cm}
     S. Liuti        & \hspace{3cm}University of Virginia\\
     \vspace{0.5cm}
     Y. Oh          & \hspace{3cm}Kyungpook National University\\  
     \vspace{0.5cm}
     D. Richards     & \hspace{3cm}Jefferson Laboratory\\
     \vspace{0.5cm}
     T. Satogata     & \hspace{3cm}Jefferson Laboratory\\
     \vspace{0.5cm}
     A. Vossen       & \hspace{3cm}Duke University\\ 
    \end{tabular}
    \label{tab:my_label}
\end{table}

~$^*$ Principal Investigator

\vspace{0.1cm}
$^{**}$ Corresponding Author (burkert@jlab.org)
\newpage

\centerline{\bf\large FOREWORD}
\vspace{1.5cm}

The Electron-Ion Collider (EIC), is a powerful new facility to be built in the
the U.S. Department of Energy’s Brookhaven National
Laboratory in partnership with the Thomas Jefferson National Accelerator
Facility. Its main focus is to explore the most fundamental building blocks of the visible matter in the universe and reveal the properties of the strong force of nature.  

\bigskip

The initiative to develop this white paper followed DOE's approval of ``mission need'' (known as CD-0) in December 2019. Since then the EIC has achieved Critical Decision 1 (CD-1) approval on July 6, 2021. This milestone marks the start of the project execution phase for a next-generation nuclear physics facility, making the present initiative timely.
\bigskip

The EIC is designed to have two interaction regions that are suitable for the installation of large-scale detector systems for high priority nuclear physics experiments. The goal of the initiative leading to this white paper was to take a fresh look at the changing landscape of the science underlying the need of a complementary approach towards the overall optimization and the execution of the EIC science program, and include, where appropriate, recent scientific advancements and challenges that go beyond the original motivation for the EIC. Several of the highly rated science programs proposed for the EIC were selected, as well as recent developments that have opened up new directions in nuclear science. It also included discussions on the machine requirements and performance of detection systems for the successful and efficient execution of the EIC science program.  
\bigskip 

The organizing team held a preparatory coordination meeting on December 15–16, 2020~\cite{IR2atEIC} bringing in experts in the field to discuss the science of the EIC second interaction region, its instrumentation, and explore ways of its implementation in order to maximize the scientific impact of the EIC. The goal of this meeting was also to define the scientific program and the agenda for subsequent workshops.
\bigskip 

The first workshop took place remotely on March 17-19, 2021, and was co-hosted by Argonne National Laboratory and the CFNS. Over 400 members of the international nuclear science community 
registered as participants~\cite{IR2atEIC-1}. This first workshop highlighted the science 
that will benefit the most from a second EIC interaction region, including the science of deep inelastic exclusive and semi-inclusive processes, the physics with jets, heavy flavor production, spectroscopy of exotic hadrons, and processes with light and heavy ions. This workshop was very timely as Brookhaven National Laboratory and Jefferson Laboratory had just announced the “Call for Collaboration Proposals for Detectors to be located at the EIC” in two interaction regions. Detector 2 could complement the project detector 1 and may focus on optimizing particular science topics or addressing topics beyond the requirements defined in previous published EIC documents. It also refers to possible optimization of the second interaction region towards such aims.
\bigskip 

The second workshop~\cite{PSQatEIC} Precision Studies of QCD at EIC, co-hosted by Asia Pacific Center for Theoretical Physics (APCTP) and the CFNS, took place on July 19-23, 2021. This workshop examined the science requiring high luminosity at low to medium center of mass energies ($20~ {\rm to}~ 60$~GeV). The goal of this workshop was to motivate the study of high impact science in the context of the overall machine design, EIC operation, and detector performance, focusing on science highlights, detector concepts, and science documentation. As a result of this workshop technical working groups were formed to develop this white paper. It identifies part of the science program in the precision studies of QCD that require or greatly benefit from the high luminosity and low to medium center-of-mass energies, and it documents the scientific underpinnings in support of such a program. The objective of this document is to help define the path towards the realization of the second interaction region.

\newpage
\tableofcontents
\newpage
\section{Executive Summary}

\vspace{0.3cm}

The fundamental building blocks of ordinary matter in the universe, proton and neutron, together known as nucleons, have been discovered during the early part of the twentieth century~\cite{Rutherford:1911zz,Chadwick:1932ma}. For over half a century we have known that these nucleons are further composed of quarks and gluons. We also know that global properties of nucleons and nuclei, such as their mass and spin, and interactions are the consequences of the underlying physics of quarks and gluons, governed by the theory of strong interaction, Quantum-Chromo-Dynamics (QCD), whose fiftieth anniversary we celebrate in 2022. Yet we still do not understand how the properties of nucleons emerge from the fundamental interaction. This has resulted in the development of a new science of emergent phenomena in the nuclear medium and the 3D nuclear structure: nuclear femtography. A significant part of the science program currently at the Jefferson Laboratory 12~GeV CEBAF facility is aimed at this new science in the range where valence quarks dominate the internal structure and dynamics; the US Electron Ion Collider (EIC) in its low-to-medium center-of-mass energy is preferential for studying the region of $x_B$ from 0.01 to 0.1 where non trivial flavor and quark-anti-quark differences are expected from Chiral Symmetry Breaking.  

These capabilities will open the door to the exploration of the three-dimensional distributions in coordinate space and in momentum space of the quarks and gluons over an unprecedented kinematic range that connects to the range currently explored at lower energies in fixed-target scattering experiments.  The combined result will be an unparalleled exploration of the way in which the phenomena of nuclear physics, the mass, and the spin, and the mechanical properties emerge from the fundamental interactions of the partons, and how these properties are distributed in the confined space inside nucleons and nuclei.

The EIC in its full range of 20 to 140~GeV center-of-mass energy and featuring high luminosity operation will be a powerful facility for the exploration of the most intricate secrets of the strong interaction, and the potential discovery of phenomena not observed before. Much of the compelling science program has been described in previous documents~\cite{Accardi:2012qut, Proceedings:2020eah,AbdulKhalek:2021gbh}.      

\begin{figure}[ht!]
\centering{\includegraphics[width=1.0\columnwidth]{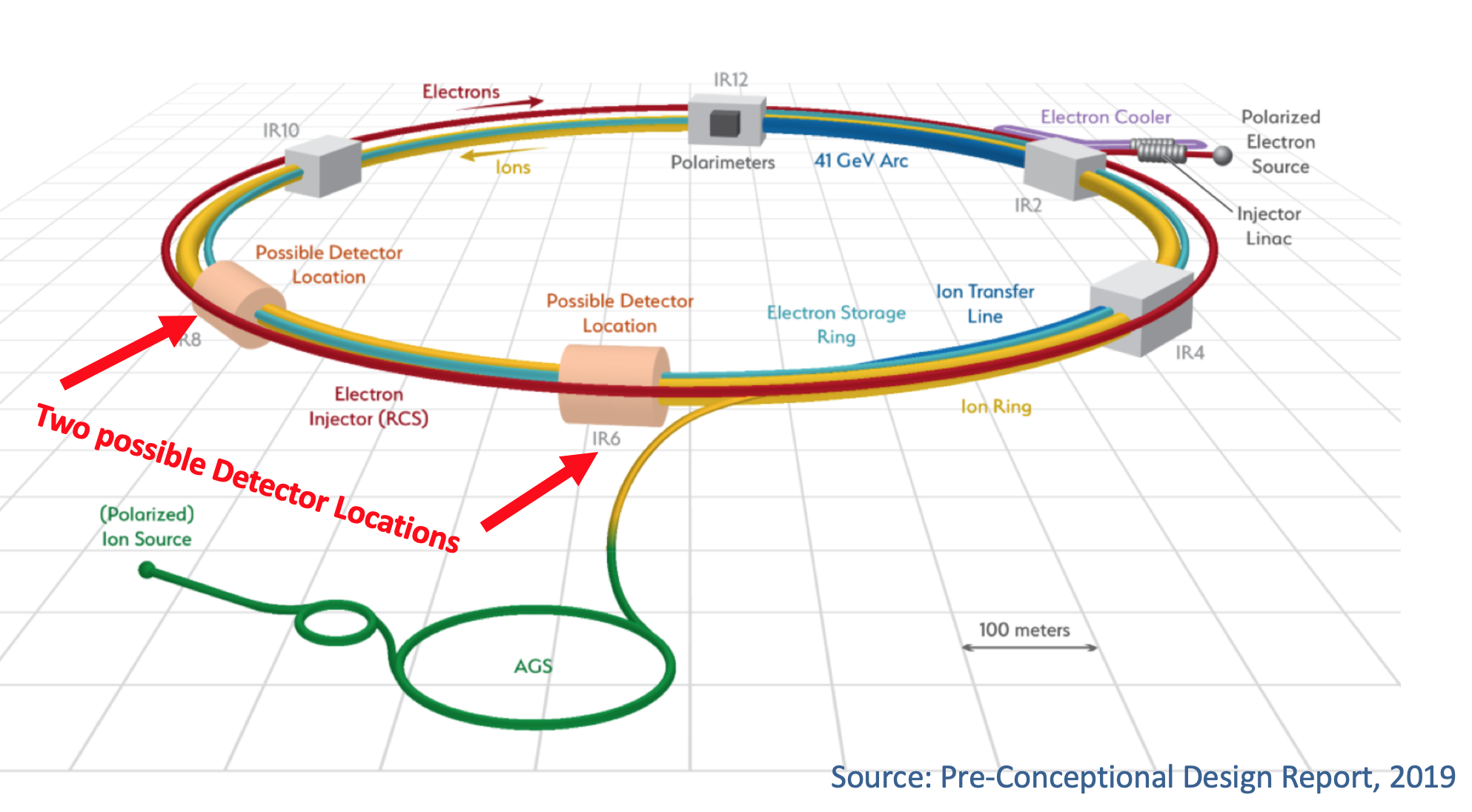}}
\caption{The EIC concept at Brookhaven National Laboratory~\cite{EIC:CDR}.  The electron and the ion beams are clearly identified. There are several beam intersection points, one at the 6 o'clock (IP6) location and at the 8 o'clock (IP8) location are suitable for the installation and operation of large scale detector systems. Interaction point IP8 may be most suitable for high-luminosity optimization at low to intermediate center-of-mass energies as well as for the installation of a secondary focus for forward processes requiring high momentum resolution. The electron beam energy ranges from 2.5 GeV to 18 GeV, while for protons the ion beam allows selected energies between 41 GeV and 275 GeV covering a collision center-mass energy from 20 GeV 
to 140 GeV.  
The ion beam is circulating counter clockwise, and the new electron ring with electrons circling clockwise. Both beams will be highly polarized with both electron and proton polarizations greater than 70\%. The EIC will benefit from two existing large detector halls in IP6 and in IR8, both fully equipped with infrastructure.}
\label{EIC-concept}
\end{figure} 
\begin{figure}[!ht]
\centering{\includegraphics[width=0.95\columnwidth]{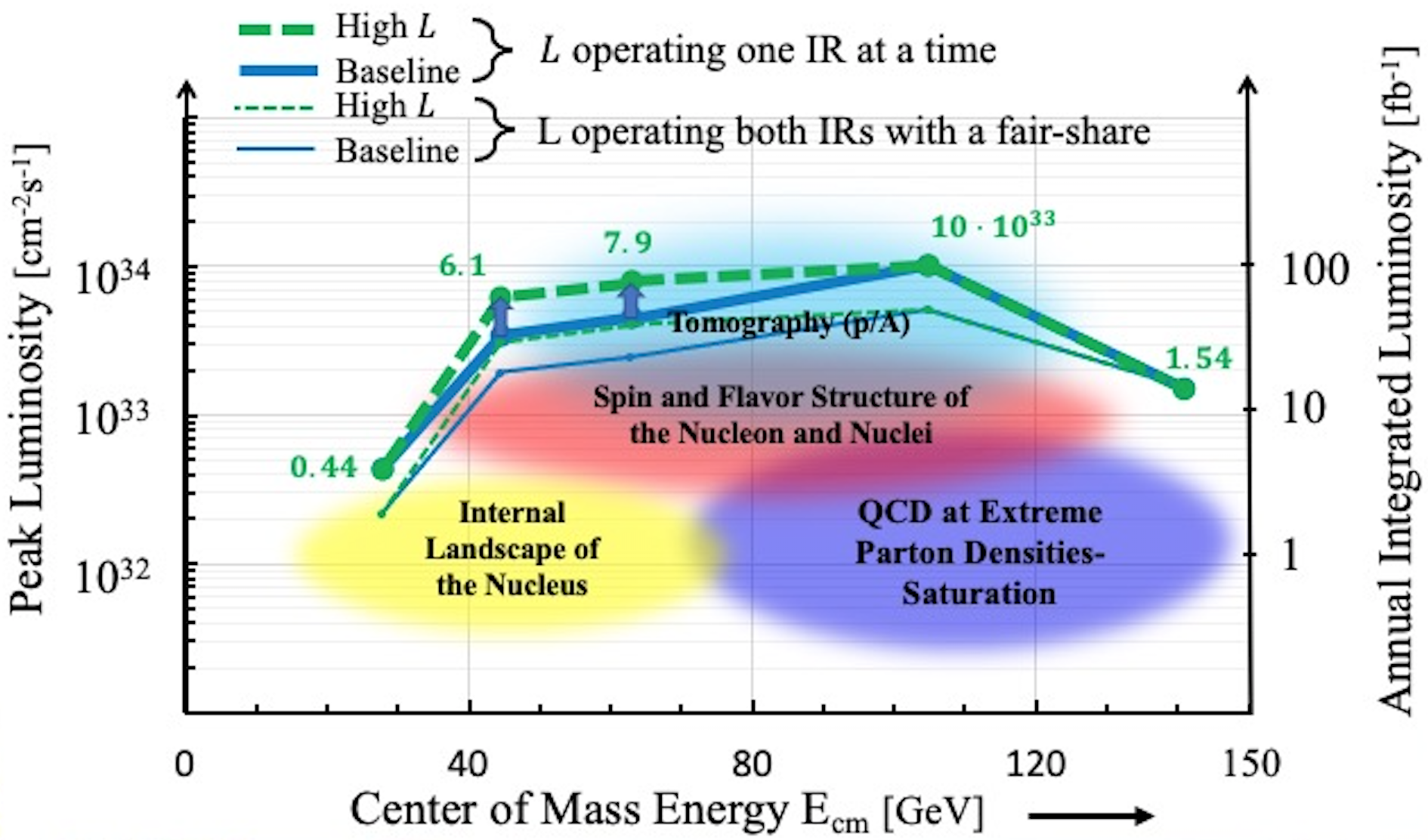}}
\caption{Estimated luminosity versus center-of-mass energies for the operation of one (thick lines) or two (thin lines) interaction regions. The blue lines show the baseline performance. The green lines show the high luminosity operation for improved beam optics and cooling. The strong drop in luminosity from the CM energy 44.7~GeV to 28.6 ~GeV is caused by increased beam-beam interactions as the proton beam energy is reduced from 100 GeV to 41 GeV while keeping the electron energy of 5 GeV. This problem is still being studied by machine experts. One option might be to keep the proton energy at 100 GeV, thus avoiding an increase in beam-beam interactions and lower the electron beam energy from 5 GeV to 2.5 GeV, resulting in 31.6 CM energy.} 
\label{EIC-lumi}
\end{figure}

The EIC project scope includes the development of an interaction region (IR) and day-one detector at IP6 and the baseline of an interaction region design for a second detector at IP8. 
A second EIC detector would be located at IP8 that will include a second focus approximately 50~m downstream of the collision point at a location with a large dispersion. Such an innovative design would enable a high-impact and highly complementary physics program to the day-one detector. The second focus thus makes it possible to move tracking detectors very close to the beam at a location where scattered particles separate from the beam envelope, thereby providing exceptional near-beam detection. This in turn creates unique possibilities for detecting all fragments from breakup of nuclei, for measuring light nuclei from coherent processes down to very low $p_T$, and greatly improves the acceptance for protons in exclusive reactions - in particular at low $x$. As such, a second detector at IP8 will significantly enhance the capabilities of the EIC for diffractive physics and open up new opportunities for physics with nuclear targets.  

With this document we highlight the science benefiting from an optimized operation at instantaneous luminosity from $0.5\times10^{34}$cm$^{-2}$s$^{-1}$ up to $1.0\times 10^{34}$~cm$^{-2}$s$^{-1}$, which is achievable in the center-of-mass range of 45 to 100~GeV, with significantly lower luminosity at 28 and 140~GeV. Furthermore, with a projected $10^7$~sec of operation (100\% equivalent) annually, the maximal integrated luminosity is 100fb$^{-1}$.   

This White Paper aims at highlighting the important benefits in the science reach of the EIC. High luminosity operation is generally desirable, as it enables producing and harvesting scientific results in a shorter time period. It becomes crucial for programs that would require many months or even 
years of operation at lower luminosity. 

We also aim at providing the justification for the development of either or both EIC detectors with characteristics that will provide support for an exciting science program at low to medium-high center-of-mass electron-ion collisions that address many of the high impact physics topics. In particular, 
the 3D-imaging of the nucleon, requiring a large amount of data in order to fill the multi-dimensional kinematic space with high statistics data, including  combinations of spin-polarized electrons and longitudinal and transverse spin-polarized protons. We also emphasize the importance of, in the future,  including positrons for processes that can be isolated through the measurement of electrical charge differences in electron and positron induced processes. Furthermore, the availability of high spin polarization for both the electron and proton beam, in the longitudinal and in the transverse spin orientation, is critically important for the measurement of the quark angular momentum distribution in the proton.      
 
\bigskip
\noindent{\bf Generalized Parton Distributions}: The discovery of the Generalized 
Parton Distributions (GPDs) and the identification 
of processes that are accessible in high energy scattering experiments, has opened
up an area of research with the promise to turn experimentally measured quantities into objects 
with 3-dimensional physical sizes at the femtometer scale. It requires precision measurements 
of exclusive processes, such as deeply virtual Compton scattering (DVCS) and deeply virtual meson production (DVMP). 
The tunable energy of the EIC combined with an instantaneous luminosity of up to 
$L=10^{34}$cm$^{-2}$s$^{-1}$ and high spin 
polarization of electrons, proton, and light nuclei, makes the EIC a formidable instrument 
to advance nuclear science from the one-dimensional imaging of the past to the 3-dimensional 
imaging of the quark and gluon structure of particles. This science is one of the cornerstones of the EIC experimental program 
and is complemented by theoretical advances as a result of precise computations on the QCD lattice
and through QCD-inspired pictures of the nucleon. To fully capitalize on these experimental and theoretical efforts 
demands operation of the EIC with high luminosity at low to medium center of mass energies. 
This will enable connecting the valence quark region, which is well probed in fixed target 
experiments, to sea quarks and gluon dominated regions at medium and small values of 
the quark longitudinal momentum fraction $x$ correlating the quarks spatial distribution with 
its momentum. The great potential of the EIC for imaging is illustrated in Fig.~\ref{fig:CFF} 
with the extraction of Compton Form Factor $\cal{H}$ covering a large $x$ range.  
\bigskip

\noindent {\bf Gravitational Form Factors}: 
Knowledge of the GPDs facilitated the development of a novel technique to employ the 
correspondence of the GPDs to the gravitational form factors (GFFs) through the moments 
of the GPDs. The GFFs are form factors of the nucleon matrix element of the energy-momentum 
tensor and are related to the mechanical properties of the proton. The Fourier transform 
over their t-dependence can be related to the distribution of forces, of mass, and of 
angular momentum. The femto-scale images obtained will provide an intuitive understanding 
of the fundamental properties of the proton, and 
how they arise from the underlying quarks and gluon degrees of freedom as described by the 
QCD theory of spin-1/2 quarks and spin-1 gluons. This is one of the most important goals in nuclear 
physics. The feasibility of this program has been demonstrated at experiments at lower energy, and expected results at the EIC have been 
simulated. 
\bigskip
 
\noindent {\bf Mechanical Properties of Particles}: In the QCD studies, it has been realized that the matrix elements, and the quark and gluon GFF,  measured through DIS momentum sum rule 
and also the source for gravitational fields of the nucleon, play important roles in understanding 
the spin and mass decomposition. The interpretation of the GFF $D(t)$ in terms of 
mechanical properties has most recently generated much interest as its relations to deeply virtual Compton scattering (DVCS) and deeply virtual meson production (DVMP) have been established. 
Moreover, the gluon GFF are directly accessible through near-threshold heavy-quarkonium production as well. 
Furthermore, the beam charge asymmetry in DVCS with a 
future positron beam will have important impact in directly accessing the $D(t)$ form factor~\cite{Accardi:2020swt}.
Figure~\ref{tangential} shows examples of estimated normal and shear force distributions 
inside the proton that will become accessible with the EIC. 
\bigskip

\noindent {\bf Nuclear Structure in Momentum Space}: As the GPDs relate to imaging in transverse Euclidean and longitudinal momentum space, 
the nucleon's 3-dimensional momentum structure may be accessed through measurements of 
transverse momentum dependent parton distribution functions employing semi-inclusive 
deep-inelastic scattering as a central part of the scientific mission of the EIC. 
This program focuses on an unprecedented investigation of the parton dynamics and correlations 
at the confinement scale and will benefit substantially by an increased luminosity at 
medium energies. 
Structure functions appearing at sub-leading twist are suppressed by a kinematic factor $1/Q$, which makes data at relatively low and medium $Q^2$ the natural domain for their measurement. 
Similarly, effects from the intrinsic transverse momentum dependence are suppressed at high $Q^2$, when most of the observed transverse momenta are generated perturbatively. As a consequence, the signal of TMDs is naturally diluted at the highest energies. 
At the same time $Q^2$ has to be high enough for the applicability of factorization theorems.

Dedicated running of the EIC at low to medium CM energy would therefore occupy kinematics where non-perturbative and subleading effects are sizeable and current knowledge allows the application of factorization to extract the relevant quantities~\cite{Grewal:2020hoc}. The strong impact of a high luminosity EIC on the 
determination of the structure function $g_T$ is demonstrated in Figure~\ref{fig:gT} 
in comparison with the existing data.
\bigskip

\noindent {\bf Exotic Mesons in Heavy Quark Spectroscopy}: The spectroscopy of excited mesons and baryons has played an essential role in the development 
of the quark model and its underlying symmetries, which led to the decoding of what was then called the 
``Particle Zoo'' of hundreds of excited states.     
Modern electro/photo-production facilities, such as those operating in Jefferson Lab, have demonstrated 
the effectiveness of photons as probes of the hadron spectrum. However the energy ranges of these 
facilities are such that most states with open or hidden heavy flavor are out of reach. Still, there is  
 significant discovery potential for photoproduction in this sector. 
Already electron scattering experiments at HERA observed low-lying charmonia, demonstrating the 
viability of charmonium spectroscopy in electro-production at high-energies but were limited by 
luminosity. Now the EIC, with orders of magnitude higher luminosity, will provide a suitable facility 
for a dedicated photoproduction spectroscopy program (by post-tagging the near $0^\circ$ scattered electron) extended to the heavy flavor sectors. In particular, 
the study of heavy-quarkonia and quarkonium-like states in photon-induced reactions while complementary 
to the spectroscopy programs employing other production modes will provide unique clues to the 
underlying non-perturbative QCD dynamics.

\bigskip
\noindent{\bf Unique science with nuclei}:  The EIC will enable deep inelastic scattering off of all nuclei with its polarized electron beam for the first time in a collider geometry. Lightest nuclei like deuteron or helium would serve as surrogates for neutrons to study flavor dependent parton distributions in kinematic regions that remain unexplored to-date. 
EIC's high luminosity and unique far-forward detection capabilities will enable detailed measurements of nuclear breakup, spectator tagging, and -- in the case of light ions -- coherent scattering reactions, far beyond what is possible in the past fixed target facilities.  Such measurements, would allow additional valuable controls over measurements and promise to understanding reaction mechanisms and to study nuclear configurations that are believed to play crucial role in the scattering process. Coherent scattering measurements in exclusive reactions enable 3D tomography of light ions in their quark-gluon degrees of freedom.
Nuclei can be used to study the influence of nuclear interactions on non-perturbative properties of the nucleon (nuclear medium modifications).  Precision measurements of the $Q^2$ dependence of the EMC effect will pin down the influence of higher twist contributions on the medium modifications of partonic distributions. The broad Bjorken-$x$ range covered by the EIC makes it an ideal machine to study the gluon EMC effect.  

\bigskip
\noindent{\bf Paper organization:}
\noindent The WP is organized in 10 sections, with section I through section V outlining 
an experimental science program. Section VI is dedicated to the 
increasing role Lattice QCD will play in supporting the high level experimental analysis, 
as well as opening up avenues of research that require information not (yet) available from  
prior experiments for the interpretation. Section VII discusses aspects of the science 
requiring special instrumentation in the far forward region of the hadron beam, and for 
the second interaction region at IP8 the option of implementing a high-resolution forward ion  
spectrometer. Radiative effects are discussed in section VIII, which all experimental analyses 
have to deal with, and may present special challenges in part of the phase space covered by 
the EIC detection system, covering nearly the full phase space available. Section IX outlines 
some of the experimental and analysis aspects that offer significant benefits from developing 
and employing artificial intelligence (AI) procedures in controlling hardware and in guiding 
analysis strategies that can be widely developed before that EIC will begin operation. 
Section X discusses the two interaction regions that can house dedicated detector systems, 
with emphasis on their complementarity in performance at different center-of-mass energies 
and optics parameters.

\newpage 
\section{GPDs - 3D Imaging and mechanical properties of the nucleon}

\label{sec:GPD}
\subsection{Introduction \& background} 
The discovery of GPDs~\cite{Muller:1994ses,Ji:1996ek,Radyushkin:1997ki} has 
opened a window on three-dimensional imaging of the nucleon, going far beyond the one dimensional longitudinal 
structure probed in deeply inelastic scattering (DIS) and the transverse structure encoded in the different 
form factors.  This discovery facilitated the development of a novel technique that employs the remarkable 
correspondence of the GFF and the second x-moments of the generalized parton distribution 
functions, and relate them to the shear stress and pressure in the proton and the distribution of orbital 
angular momentum. 
These femto-scale images (or femtography) will provide an intuitive understanding of how the fundamental 
properties of the nucleon, such as its mass and spin, arise from the underlying quark and gluon degrees of 
freedom. And then, for the first time, we will have access to the forces and pressure distributions inside 
the nucleon.
This science is one of the cornerstones of the EIC experimental program and is complemented by theoretical 
advances as a result of lattice QCD calculations and through QCD-inspired pictures of the nucleon. 
To fully capitalize on these experimental and theoretical efforts demands operation of the EIC with 
high luminosity at low to medium center of mass energies.

\begin{figure}[htb]
\centering
\begin{minipage}{0.45\textwidth}
\centering
\includegraphics[width=0.85\textwidth]{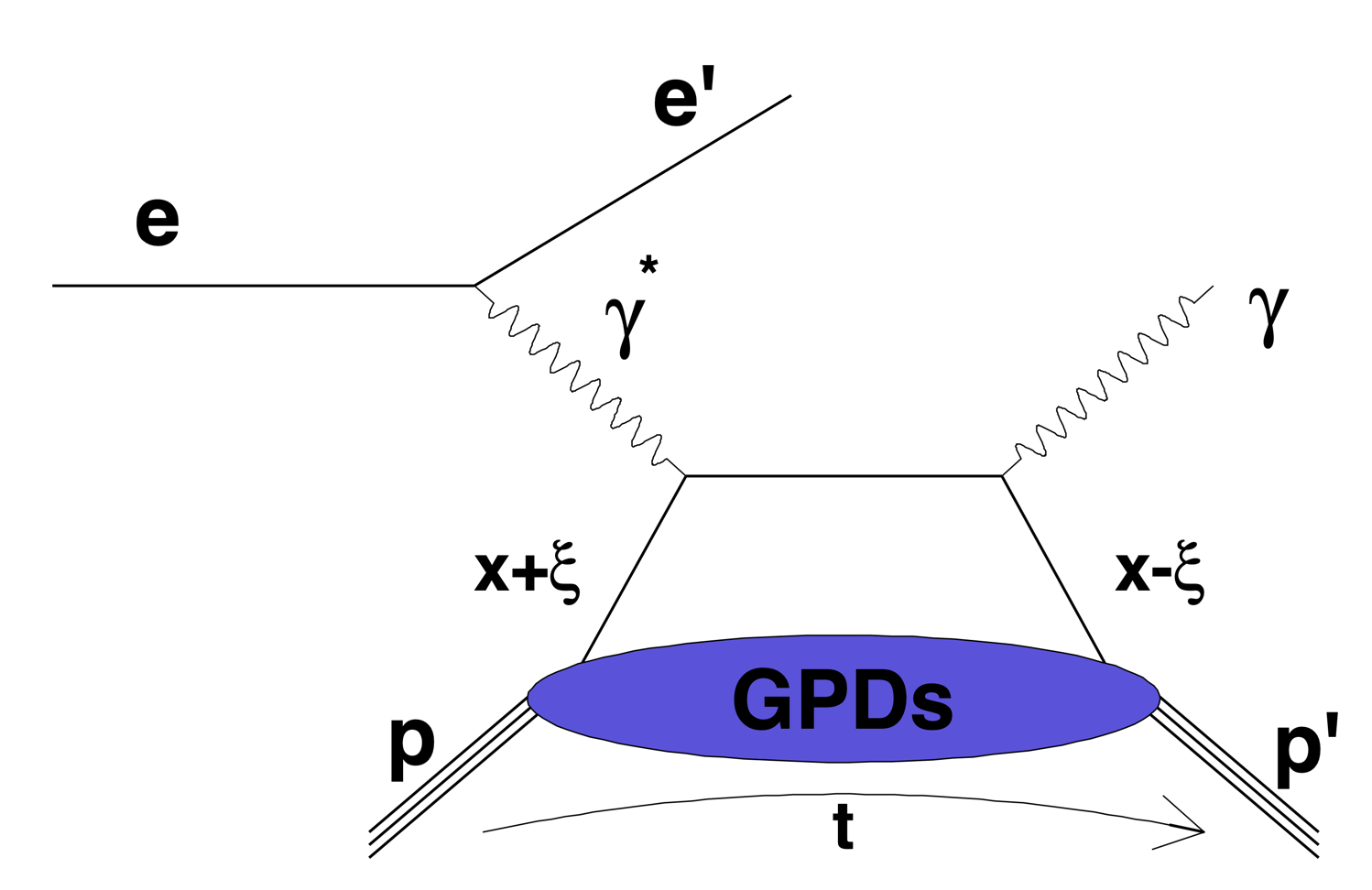}
\end{minipage}
\begin{minipage}{0.45\textwidth}
\centering
\includegraphics[width=0.85\textwidth]{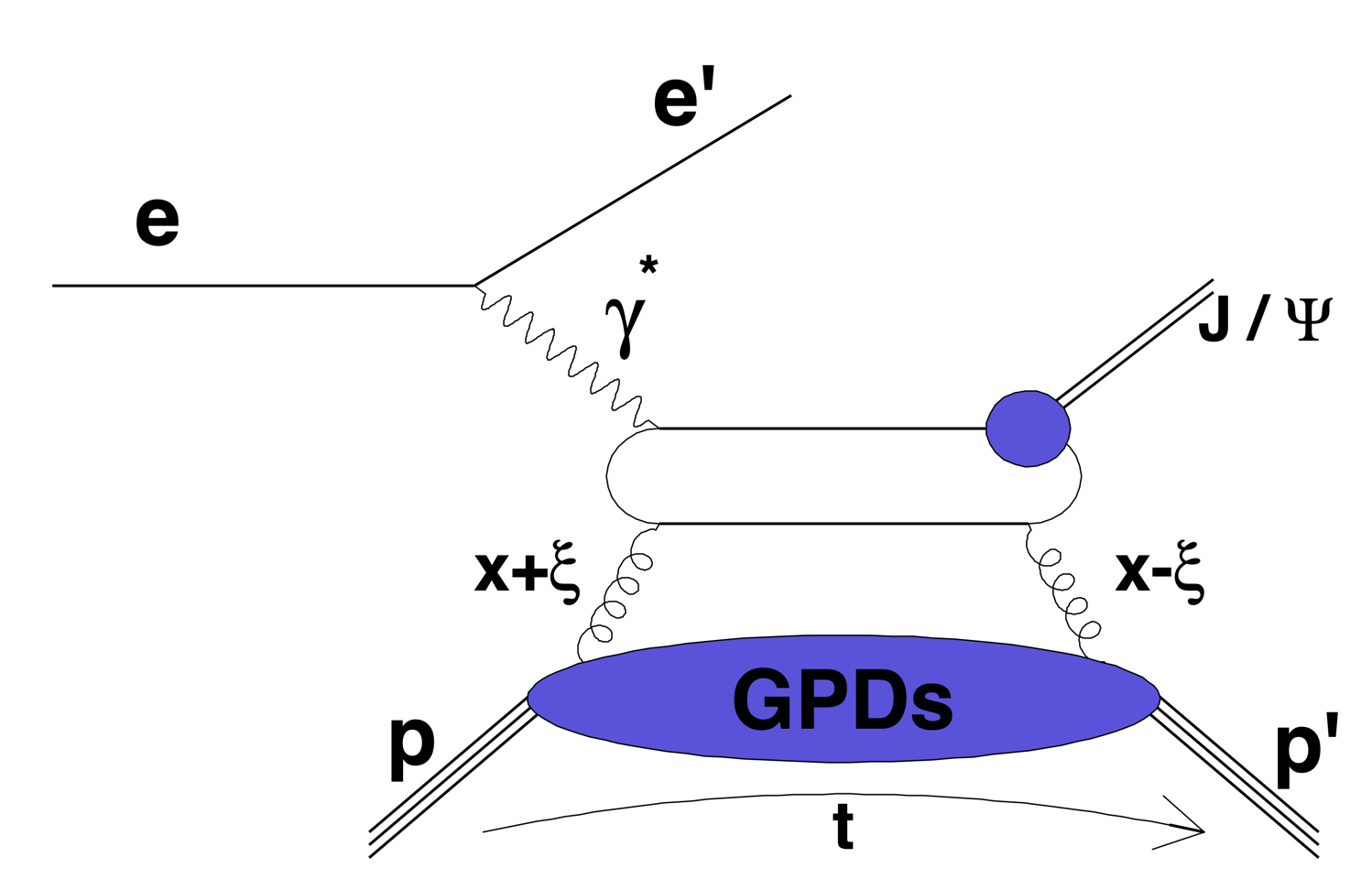}
\end{minipage}
    \caption{Deeply virtual exclusive processes in electron scattering, 
    as hard scattering events to probe the 3D quark distribution (left) and 
    gluon distribution (right).}\label{fig:handbag}
\end{figure}

The standard approach of imaging is through diffractive scattering. The deeply virtual exclusive processes allow probing entirely new structural information of the nucleon through QCD factorization (see Fig.\ref{fig:handbag}).

The golden process to study the quark GPDs is DVCS, where a virtual photon interacts with a single quark deep in the hadron, 
and the quark returns to the hadron initial ground state by emitting a high energy photon in the final state. Experimental observables in DVCS are parameterized by Compton Form Factors (CFFs)~\cite{Belitsky:2010jw}.
From the analysis of data from DESY, as well as the results of new dedicated experiments at JLab, and at CERN, early experimental constraints on CFFs have been obtained from global extraction fits~\cite{Kumericki:2011rz,Guidal:2013rya,Kumericki:2016ehc,Moutarde:2019tqa}.  However, data covering a sufficiently large kinematic range, and the different required polarization observables, have not been systematically available. 
The future EIC with high luminosity at large range in CM energies will provide comprehensive information on these hard diffractive processes, entering the precision era for GPD studies.

In what follows, after a brief review of the formalism in Section \ref{subsubDVCS}, we describe state of the art analysis methods in Section \ref{subsec:GPDanalysis}, and the study of the extraction of GFF performed at Jefferson Lab Hall B (Section \ref{sec:mechanical}). Additional processes sensitive to GPDs  complementing the main EIC focus, as well as an outlook are presented in \ref{sec:GPDoutlook}.
\subsection{Generalized Parton Distributions and Nucleon Tomography}
\label{subsec:GPDs}
GPDs, their theoretical properties, as well as phenomenological aspects related to their extraction from 
deeply virtual exclusive processes, have been the object of several review papers \cite{Ji:2013gla,Diehl:2003ny,Belitsky:2005qn,Kumericki:2016ehc,Goeke:2001tz,Ji:2004gf,Ji:2020ena} as well as of reports supporting the design of the upcoming EIC \cite{Accardi:2012qut,AbdulKhalek:2021gbh}. The main properties of GPDs are outlined below while reminding the reader that many open questions concerning constraints on GPD models, such as the application of positivity bounds \cite{Pobylitsa:2002iu,Pire:1998nw}, dispersion relations \cite{Teryaev:2005uj,Anikin:2007yh,Diehl:2007jb,Goldstein:2009ks,Pasquini:2014vua},
flavor dependence \cite{Kriesten:2021sqc}, NLO perturbative evolution, as well as the separation of twist-2 and twist-3 contributions in the deeply virtual exclusive cross sections, are still intensely debated. The ultimate answer to many of these questions will be found in the outcome of carefully designed experiments at the EIC. It is therefore mandatory to define analysis frameworks to extract GPDs from data. Various approaches, listed in Section \ref{subsec:GPDanalysis}, have been developed which represent a new step towards realizing the goal of nucleon tomographic imaging.    

\subsubsection{Deeply Virtual Exclusive Processes, GPDs and Compton Form Factors}
\label{subsubDVCS}
\begin{figure}[ht!]
  \centering
    \includegraphics[width=0.5\columnwidth]{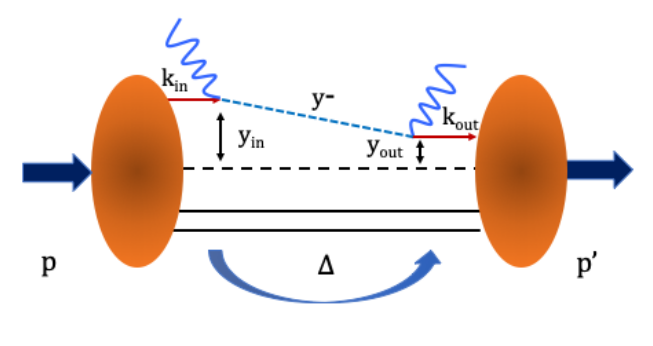}
\begin{align}
  \begin{cases}
  b_{\perp} = \dfrac{y_{in_{\perp}}+y_{\,out_{\perp}}}{2} \\  
  \Delta_{_{3/4/22}} = k_{in} - k_{out} = p - p\prime \ \;\;\;   
  \end{cases}
  \begin{cases}   y = {y_{in}-y_{out}}\\ k = \dfrac{k_{in}+k_{out}}{2} 
\end{cases} \nonumber
\end{align}      
    \caption{Correlation function for the GPDs defined in Eq.(\ref{eq-gpd}), highlighting both momentum and Fourier conjugate spatial coordinates.}
    \label{fig:correlator}
\end{figure}

The non-perturbative part of the handbag diagram in Fig.~\ref{fig:handbag}(left) is parameterized by GPDs
\begin{eqnarray}
\frac{P^+}{2\pi}\int\text{d}y^-\,\text{e}^{ixP^+y^-}\langle p'|\bar{\psi}_q(0)\gamma^+(1+ \gamma^5)\psi_q(y)|p \rangle
 & = & \bar{U}(p',\Lambda')\left[{H^q(x,\xi,t)}\gamma^+ + {E^q(x,\xi,t)}i\sigma^{+\nu}\frac{\Delta_\nu}{2M}\right. \nonumber\\
 & + & \left.{\widetilde{H}^q(x,\xi,t)}\gamma^+\gamma^5 + {\widetilde{E}^q(x,\xi,t)}\gamma^5\frac{\Delta^+}{2M}\right]U(p,\Lambda)\
\label{eq-gpd}
\end{eqnarray}
\noindent where the index $q$ refers to the quark flavor; $P=\frac{1}{2}\left(p+p'\right)$ is the average proton 4-momentum, while $\Delta= p'-p$ is the 4-momentum transfer to the proton,
$t=\Delta^2$. The Fourier transform is performed along the light-cone (LC) with $y^+=\vec{y}_\perp=0$ (Fig.\ref{fig:correlator}). 

The active quark carries light cone momentum fractions $x+\xi$ and $x-\xi$, respectively, in the initial and final states, so that the average quark LC momentum is, 
$k^+=xP^+$ and the LC momentum difference is, $\Delta^+= p'^+-p^+ =-2\xi P^+$.

\noindent Ordinary parton distribution functions (PDFs) can be recovered from GPDs at $\xi=0, t=0$ as,
\begin{equation}
\frac{1}{4\pi}\int\text{d}y^-\,\text{e}^{ixp^+y^-}\langle p|\bar{\psi}_q(0)\gamma^+\psi_q(y)|p \rangle =  H^q(x,\xi=0,t=0) = q(x)\
\end{equation}
\noindent and similarly $\widetilde{H}^q(x,\xi=0,t=0) = \Delta q(x)$. Furthermore, like ordinary parton distributions, all of the expressions considered here depend on the hard scale for the scattering process, $Q^2$, which is omitted in the expressions for ease of presentation.  
\noindent Because of Lorentz covariance, the nth Mellin moment of a GPD is a polynomial in $\xi$ of order (n+1)~\cite{Ji:1998pc}. Because of parity and time reversal invariance, these polynomials are even for the GPDs of spin-1/2 targets such as the proton. The coefficients of each power of $\xi$ are functions of $t$, which constitute 
generalized form factors. For n=0 in particular, the moments are independent of $\xi$ and give the familiar elastic form factors. In section~\ref{sec:mechanical} we will use the 2nd Mellin moments of GPD $H$ and GPD $E$ when discussing the GFF of the proton.

\begin{subequations}
\begin{eqnarray}
\int_{-1}^1\text{d}x\,H^q(x,\xi,t) = F_1^q(t) , \qquad &  & \qquad \int_{-1}^1\text{d}x\,E^q(x,\xi,t) = F_2^q(t)  \\
\int_{-1}^1\text{d}x\,\widetilde{H}^q(x,\xi,t) = G_A^q(t) ,  \qquad & & \qquad  \int_{-1}^1\text{d}x\,\widetilde{E}^q(x,\xi,t) = G_P^q(t)  
\end{eqnarray}
\end{subequations}

\begin{figure}[h!]
 \centering
    \includegraphics[width=0.4\columnwidth]{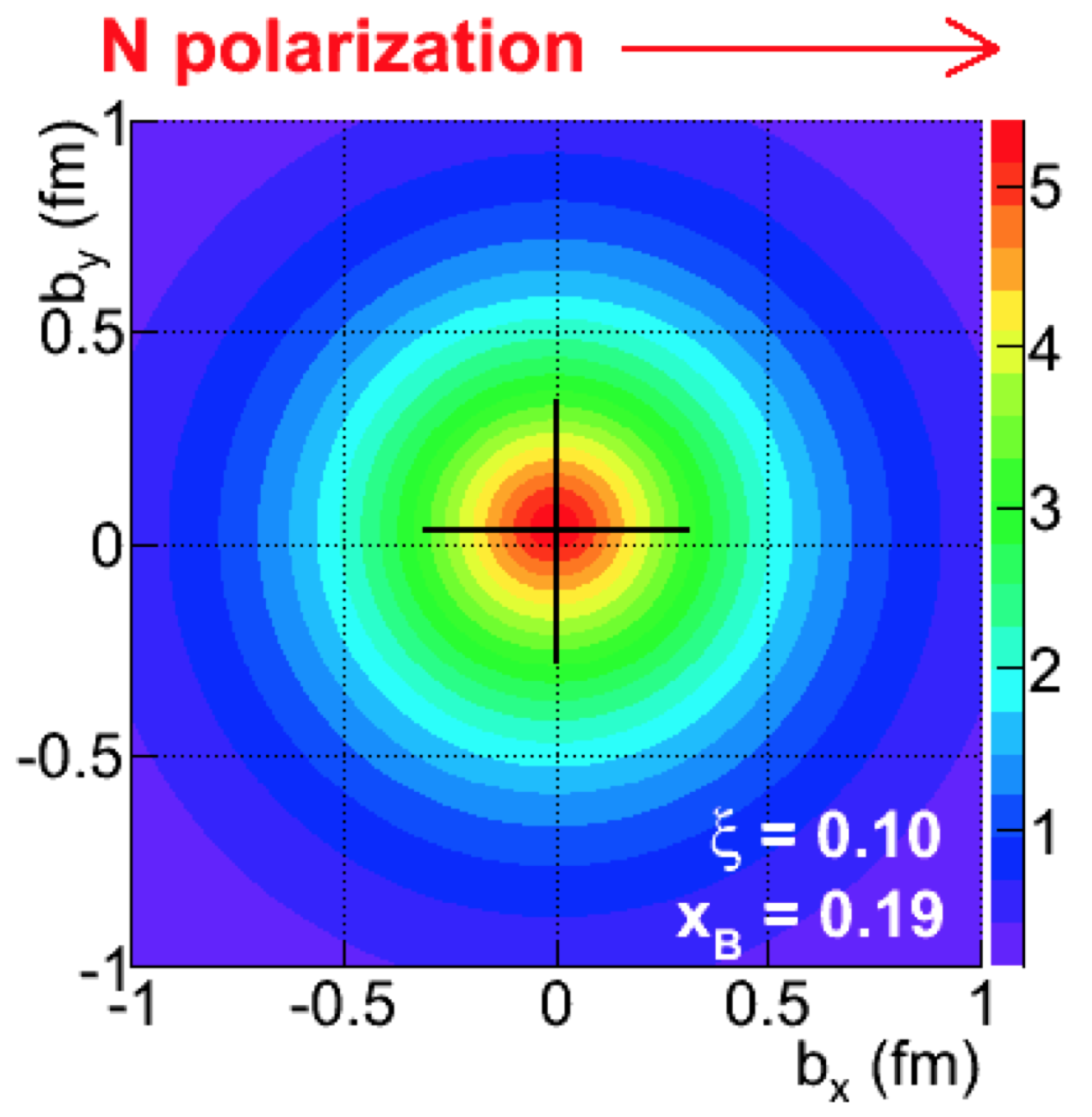}
\hspace{1.0cm}    \includegraphics[width=0.4\columnwidth]{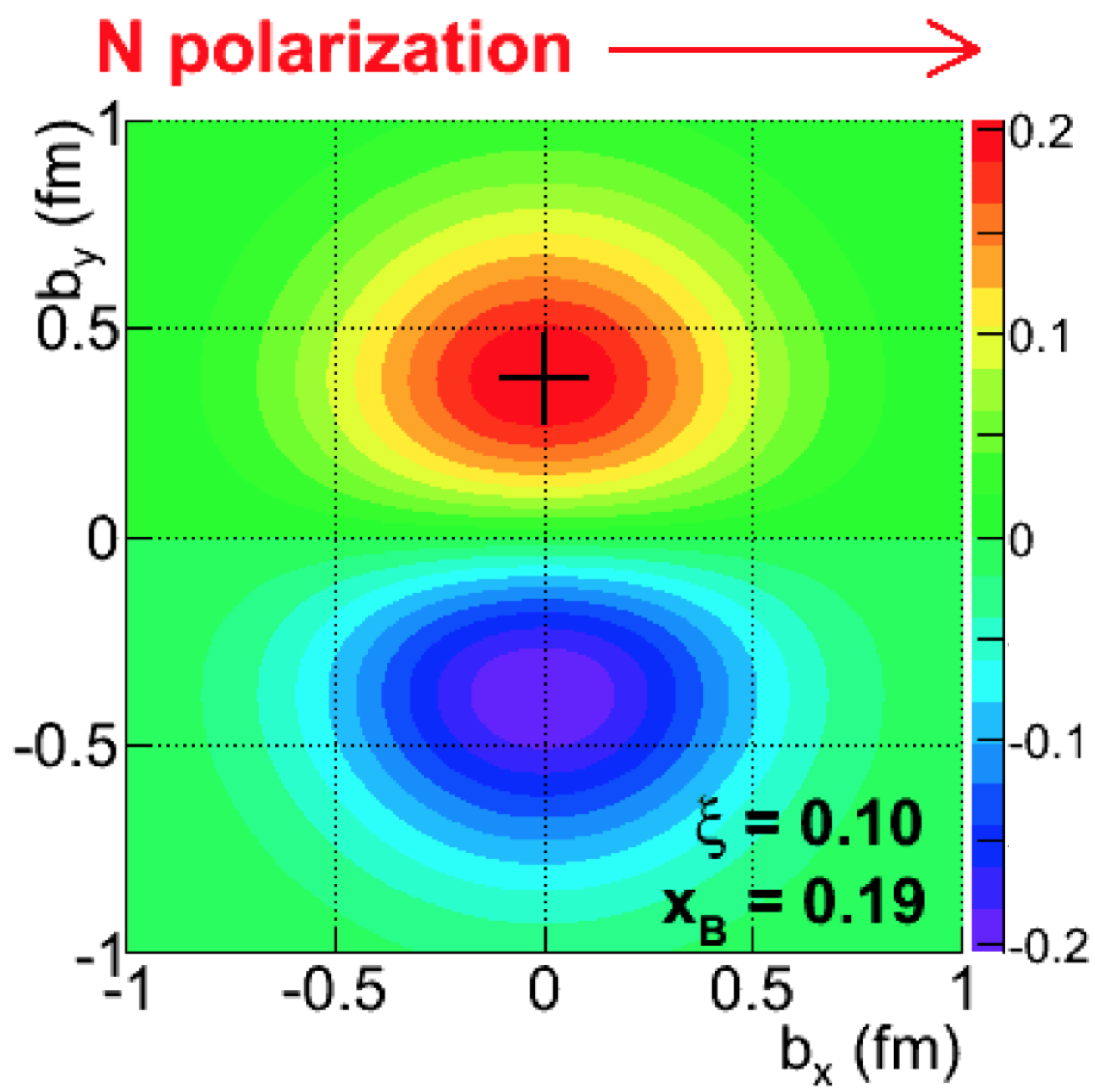}
    \caption{Left: Image of the 2-dimensional distribution of $\cal{H} + \cal{E}$  in the valence region for a 
    spin-polarized proton with the polarization axis parallel to $b_x$. The polarization causes a vertical 
    shift of the center. Right: Same as on the left, but showing the distribution of GPD $\cal{E}$ separately, with the effect of the polarization more dramatically seen as a clear spatial separation of electrical charges, i.e. u- and d-quarks in $b_y$ space, generating a flavor-dipole. Note that the color codes on the left and right panels have different scales to account for the much smaller amplitude of the $\cal{E}$ CFF.}
    \label{fig:E-H}
\end{figure}

\noindent GPDs also encode information on the joint distributions of partons as functions of both the longitudinal momentum fraction $x$ 
and the transverse impact parameter $\vec{b}_\perp$. For a nucleon polarized along the transverse $X$ direction they are given by~\cite{Guo:2021aik},
\begin{equation}
q^{In}_X(x,{\bf b}_\perp)=  \int \frac{\text{d}^2{\bf \Delta}_\perp}{(2\pi)^2}\exp [i  {\bf b}_\perp \cdot {\bf \Delta}_\perp] \left[H_q(x,0,-\Delta^2) +i\frac{\Delta_y}{2M}\left(H_q(x,0,-\Delta^2) + E_q(x,0,-\Delta^2)\right)\right] 
\end{equation}
\noindent Figure~\ref{fig:E-H} shows one of the projected results for the 2-dimensional images of the CFF $\mathcal{E(\xi,\it t)}$ and $\mathcal{H(\xi,\it t)}$ Fourier transformed into impact parameter space $(b_x, b_y)$. 
The image was extracted from simulated CLAS12 measurements of different polarization asymmetries and cross sections with the proton 
transversely polarized. 

\begin{figure}[htb!]
\centering

\includegraphics[width=12.cm]{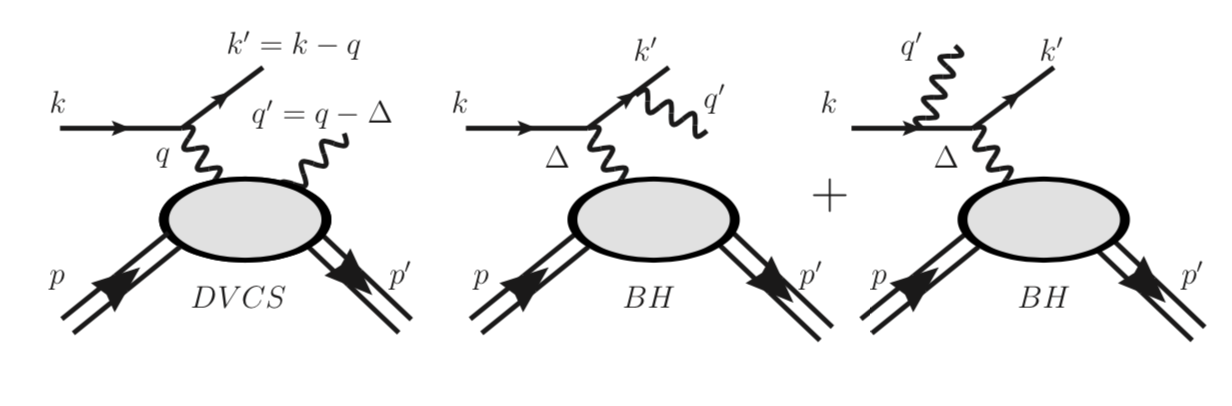}
\caption{Exclusive photon electroproduction through DVCS (left) and BH processes (middle and right).}
\label{fig:dvcs}
\end{figure}

\noindent In the following we focus on the DVCS process shown in Fig.\ref{fig:dvcs} (left). DVCS can be considered the prototype for all deeply virtual exclusive scattering (DVES) experiments and as such it has been the most studied process.
The DVCS matrix elements are accessed through exclusive photoproduction, 
\[ ep \rightarrow e'p'\gamma \]
where the final photon is produced at the proton vertex. 
A competing background process given by the Bethe-Heitler (BH) reaction is also present, where the photon 
is emitted from the electron and the matrix elements measure the proton elastic form factors, Fig.\ref{fig:dvcs} (right). 
The cross section is a function of four independent kinematic variables besides the electron-proton center-of-mass energy  
$\sqrt{s}$, the scale $Q^2$, the skewness $\xi$,  related to Bjorken x$_{B}$ as $\xi \approx x_{B}/(2- x_{B})$, $t$, 
and the angle between the lepton and hadron planes, $\phi$. 

The CFFs are complex quantities which at leading order in perturbative QCD, are defined through the convolution integral,
\begin{eqnarray} 
\label{eq:GPD-CFF}
{\mathcal F}(\xi,t; Q^2)   =   
\int_{-1}^{1} \text{d}x \left[ \frac{1}{\xi-x-i\epsilon} \pm  \frac{1}{\xi+x-i\epsilon} \right] F(x,\xi,t; Q^2)  
\end{eqnarray},
where $\mathcal{F} = \mathcal{H}, \mathcal{E}, \widetilde{\mathcal{H}}, \widetilde{\mathcal{E}}$, and $\pm$ 
indicates helicity independent (-) or helicity dependent (+) GPDs. 

\begin{figure}[htbp]
\centering
\includegraphics[width=0.9\columnwidth]{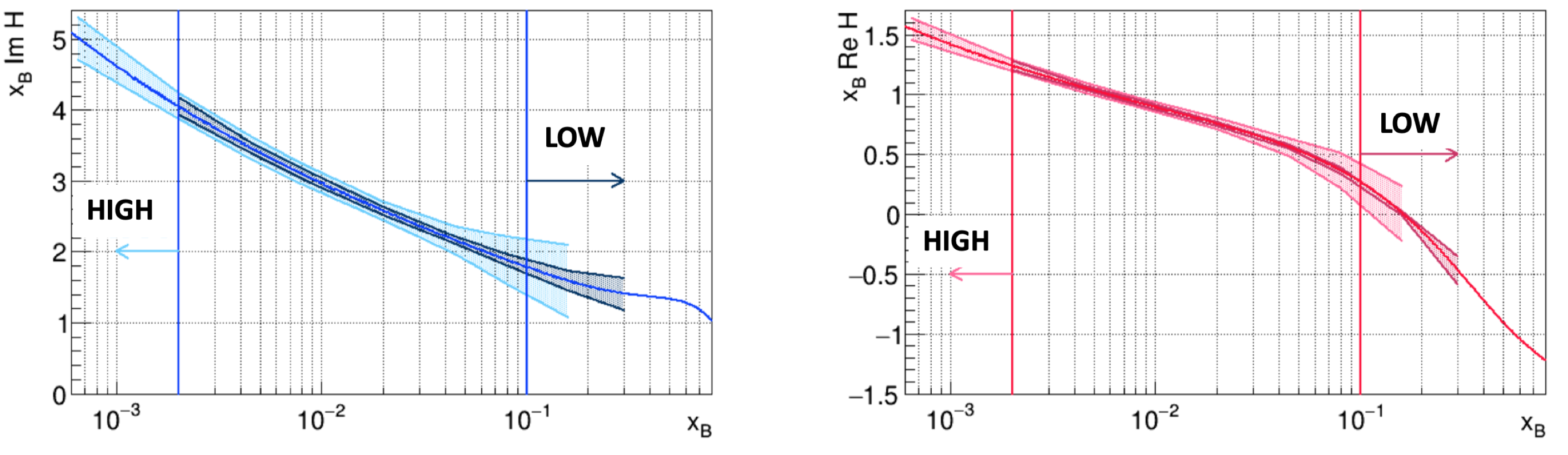}
\caption{Compton form factors Im$\cal{H}$ and Re$\cal{H}$ extracted at local $x_B$ values from simulated DVCS events at different CM beam energies, $\sqrt{s} = 31.6$~GeV (LOW) and $\sqrt{s}\ge 100$~GeV (HIGH). The dark shaded bands represent the reach and the uncertainties at the lower CM-energy. The lighter shaded bands represent the higher CM-energy. The $x_B$ regions labeled LOW can only be covered at the low CM-energy with reasonable uncertainties. The $x_B$ region labeled HIGH can only be reached with the high CM-energy. The widths of the bands indicate the estimated uncertainties due to overall reconstruction effects, statistics and systematic uncertainties. For each of the two CM-energies a combined integrated luminosity of 200~fb$^{-1}$ equally split between longitudinally polarized and transversely polarized proton runs is assumed. At $x_B > 0.1$ smaller uncertainties can be achieved at the low CM-energy, which provides overlapping $x_B$ kinematics with the JLab 12 GeV experiments (not shown). The region $x_B < 2\times 10^{-3}$ can only be reached at the high CM-energy. Note, that the CFF $\cal{E}$ and $\widetilde{\cal{H}}$ are determined simultaneously. Here we have used same integrated luminosity for the two CM energies. The results are statistics limited and may be scaled for different assumptions. Regarding the luminosity assumptions at the low CM energy see comments in the caption of Fig.~\ref{EIC-lumi}.} 
\label{fig:CFF}
\end{figure}

Figure~\ref{fig:CFF} displays estimates of $ x_{B} \text{Re}{\mathcal H}$ and $x_{B} \text{Im}{\mathcal H}$ at fixed value of $t$. 

It is however important to keep in mind that a study of various processes is necessary to access GPDs in a controllable way. Firstly, the 
time-like counterpart of DVCS, named time-like Compton scattering (TCS) \cite{Berger:2001xd,Pire:2011st}, accessed through the nearly forward photoproduction of a lepton pair $\gamma N \to \gamma^* N^\prime$ is crucial to test the universality and the analytical properties (in $Q^2$) of the factorized scattering amplitude \cite{Mueller:2012sma}. Deeply virtual meson production (DVMP) amplitude has also been proven to factorize but 
current data seem to delay the onset of the scaling regime, which makes the study of the process ($\gamma^* N \to {\cal{M}} N'$) an important
laboratory for the study of next to leading twist processes. Secondly, a new class of factorized amplitudes has emerged \cite{Qiu:2022bpq} where the 
hard scattering process is a $2 \to 3$ process. The case of the process $\gamma N \to \gamma \gamma N'$ with a large invariant mass of the 
diphoton \cite{Pedrak:2020mfm,Grocholski:2022rqj} and a quasi-real or virtual initial photon is particularly interesting since it probes 
the charge-conjugation odd part of the quark GPDs in contradistinction with the DVCS/TCS probe. Other processes where a meson-meson 
\cite{Ivanov:2002jj} or photon-meson pair (with a large invariant mass) is produced have been studied \cite{Boussarie:2016qop,Duplancic:2018bum}; when a transversely polarized $\rho$ meson enters the final state, they
should give access to the eluding transversely quark GPDs.

The electroweak production of a single charmed meson has also been proposed \cite{Pire:2021dad} to access in a new way these transversely quark GPDs. Reconstructing the final state $D$ or $D^*$ meson is however an experimental challenge.

All these new reactions have quite small cross-sections and would greatly benefit from a high luminosity option in the low energy range of the EIC. More detailed
feasibility studies need to be performed but first order of magnitude estimates show that they need a quite large coverage of photon detection which seems in line with current detector designs.

\subsubsection{Analysis methods}
\label{subsec:GPDanalysis}
\noindent GPDs are projections of Wigner distributions that give access to the unknown mechanical properties of the nucleon involving both space and momentum correlations. Among these are the quark and gluon angular momentum, along with spin directed $qgq$ interactions \cite{Diehl:2003ny,Belitsky:2005qn,Kumericki:2016ehc,Goeke:2001tz,Ji:2004gf,Ji:2020ena}. An accurate knowledge of GPDs would unveil an unprecedented amount of information on nucleon structure and on the working of the strong interactions. Nevertheless, after two decades of experimental and phenomenological efforts, it has been, so far, impossible to extract these important quantities directly from experiment. 
The problem lies at the core of their connection with observables: the cleanest probe to observe GPDs is from the matrix elements for deeply virtual Compton scattering (DVCS) (Fig.\ref{fig:dvcs}, and Sec.\ref{subsec:GPDs}). In a nutshell, GPDs are multi-variable functions depending on the kinematic set of variables, $x, \xi, t, Q^2$ (see eq.[\ref{eq-gpd})], which enter the DVCS cross section in the form of convolutions with complex kernels, calculable in perturbative QCD, known as Compton Form Factors (CFFs). Furthermore, because GPDs are defined at the amplitude level, they appear in bilinear forms, in all observables, including various types of asymmetries. An additional consequence is that all four GPDs, $H$, $E$, $\widetilde{H}$, $\widetilde{E}$, enter simultaneously any given beam/target spin configuration. It is therefore necessary to consider simultaneously a large array of different observables in order to extract the contribution of each individual GPD, even before addressing the issues of their flavor composition, and of the sensitivity of observables to quark/antiquark components (for a detailed analysis of the DVCS cross section we refer the reader to \cite{Braun:2014sta,Kriesten:2019jep,Guo:2021gru}).

For high precision femtography, which is required to obtain proton structure images, the hadron physics community has been developing sophisticated analyses. The success of Machine Learning (ML) methodologies in modeling complex phenomena make this a prime choice for GPD extraction. 

\noindent Three main frameworks using ML are currently being pursued aimed at the extraction of GPD from data, which differ in the techniques, methodologies, and in the types of constraints derived from theory. In this respect, it is has become clear that the use of lattice QCD results will be indispensable in GPD analyses \cite{Lin:2017snn,Constantinou:2020hdm} and efforts in this direction are under way. 

The Zagreb group \cite{Cuic:2020iwt,Kumericki:2011rz,Kumericki:2013br} addresses the extraction of CFF from experimental data on various DVCS observables for different beam and target polarizations based on a neural network  (NN) architecture, or a multilayer perceptron. The recent analysis in Ref.\cite{Cuic:2020iwt} introduces variable network configurations depending on whether the model is for an unflavored or flavored quark. The use of theoretical constraints is explored, in this case given by the assumption that the CFFs obey a dispersion relation \cite{Anikin:2007yh,Diehl:2007jb,Goldstein:2009ks}. Results of the fit highlight the existence of hidden correlations among CFFs arising from different harmonics in $\phi$ appearing in the cross section formulation of Refs.\cite{Belitsky:2002tf,Belitsky:2010jw}. Comparisons with previous, unconstrained results, and with a
standard least-squares model fit to the same data show large uncertainties and often an inversion of the trend of data as a function of $\xi$ and $t$.

The PARTONS group addresses two different stages of the analysis,  namely, the extraction of CFF from data \cite{Moutarde:2019tqa,Moutarde:2018kwr}, and, most recently, the determination of GPDs \cite{Dutrieux:2021wll}.  
CFFs are extracted in Refs.\cite{Moutarde:2019tqa,Moutarde:2018kwr} from global fits of all available DVCS data using a standard NN augmented by a genetic algorithm. This work's purpose is to help benchmarking the group's future NN based analyses. The GPD effort is centered around the concept of ``shadow GPDs" \cite{Dutrieux:2021nlz}, which broadly define the set of all local minima generated by regression analysis using given functional parametrizations. Shadow GPDs propose a practical pathway to solve the inverse problem of extracting GPDs from CFFs. The practicality of the concept still remains to be demonstrated.  

More recently, the UVA group developed an analysis initially focused on the DVCS cross section \cite{Grigsby:2020auv}. The framework devised in Ref.\cite{Grigsby:2020auv} serves as a first step towards the broader scope of developing a complete analysis for the extraction of CFFs and GPDs from experimental data. Industry standard ML techniques are used to fit a cross section model based on currently available DVCS experimental data, allowing for  efficient and accurate  predictions interpolating between experimental data points across a wide kinematic range. Estimating model uncertainty allows one to make informed decisions about predictions well outside of the region defined by data, extrapolating to  unexplored kinematic regimes. While the results of this analysis show that, for instance, the network can effectively generalize in $t$, even in regions with no data, the study also points out  several of the practical challenges of fitting the sparse NN with significant experimental uncertainty, as defined by current DVCS data availability. Another important aspect of this study is the handling of the uncertainties from experimental data which is ubiquitous to physics analyses but less commonly considered in building ML models.

Standard least-squares based model fits are also currently being performed at this stage to provide a baseline for new more exploratory approaches. The result of one of these studies are presented in Fig.\ref{fig:CFF} and in Section\ref{sec:mechanical}. The latter are equivalent to local fits where CFFs are independently determined from measurements between different kinematic bins. In a more recent development, the free coefficients of a given CFF parameterization are matched to experimental data and the kinematic bins are no longer treated independently, allowing for interpolation between measurements of the same observable on neighboring kinematic bins. This method also affords to extrapolate outside the experimental data, paving the way for impact studies. However, a systematic uncertainty is introduced by the functional choice of a parameterization, which could potentially impact the predictivity of the approach. Furthermore, while ML based approaches provide solutions to overcome the occurrence of local minima, standard fits are not flexible in this respect.  This approach can be most useful in the earlier phase of an experimental program when insufficient data are available, preventing use of more flexible alternatives. 

All of the studies mentioned above are not only beneficial to the physics community but provide an interesting overlay of objectives for the physics, applied math, computer science and data science communities. A future investment of resources to bring together all communities will allow for a precise extraction of the 3D structure of the nucleon by using a wide range of new methodologies: from including  the simulation uncertainties directly in the training procedure, to developing unsupervised (or weakly supervised) procedures, improving the calibration of simulations, developing new inference techniques to improve the efficiency in using simulations, and many more ongoing developments.

In the next section we describe a CFF extraction method based on dispersion relations \cite{Anikin:2007yh,Diehl:2007jb}.
The foremost advantage of this approach is that it reduces the number of unknown parameters to be extracted,
by calculating the Real part of the amplitude from the corresponding Imaginary part plus a subtraction constant. 
The key observation here is that the same subtraction constant (with a flipped sign) enters in the dispersion relations 
for the CFFs $\mathcal H$ and $\mathcal E$, while the subtraction constants for CFFs $\tilde{\mathcal {H}}$ and $\tilde{\mathcal {E}}$ vanish.
These global fits require to be performed with analytical parameterizations of the CFFs dependences, 
since one needs to extrapolate beyond the available data to perform the full dispersion integral. Furthermore, it is known that dispersion relations are affected by a kinematic, $t$-dependent threshold dependence which partially hampers a direct connection to GPDs and affects the extraction of the subtraction term \cite{Goldstein:2009ks}. Although the precision of present data does not allow for a full evaluation of these systematic uncertainties, a dedicated study will be possible in the wider kinematic range of the EIC.

\subsection{{\it D}-term form factor, and mechanical properties of the nucleon - beyond tomography}
\label{sec:mechanical}
\noindent In section~\ref{subsec:GPDs} tomographic spatial imaging was discussed through access to GPDs employing the DVCS process. This section discusses how to obtain information about gravitational/mechanical properties of the proton. Mechanical properties that relate to gravitational coupling, such as the internal mass distributions, the quark pressure, and the angular momentum distribution inside the proton, are largely unknown. These properties are encoded in the proton's matrix element of the Energy Momentum Tensor (EMT)~\cite{Kobzarev:1962wt,Pagels:1966zza} and are expressed through the GFF~\cite{Ji:1996ek}.
\begin{eqnarray}
 \langle p_2|\hat{T}^{q,g}_{\mu\nu}|p_1\rangle \!=\! \bar{u}(p_2)\!\left [A^{q,g}(t)\frac{P_\mu P_\nu}{M} + B^{q,g}(t)\frac{i(P_\mu\sigma_{\mu\rho}+P_\nu\sigma_{\mu\rho})\Delta^\rho}{2M} +   D^{q,g}(t)\frac{\Delta_\mu\Delta_\nu - g_{\mu\nu}\Delta^2}{4M}+ M\bar{c}^{q,g}(t)g_{\mu\nu}\right]\! u(p_1)
 \label{EMT}
\end{eqnarray}
The form factors $A^{q,g}(t)$, $B^{q,g}(t)$, $\bar{c}^{q,g}(t)$, $D^{q,g}(t)$ encode information on the distributions of energy density, angular momentum, and internal forces in the interior of the proton as described in detail in Sec.~\ref{Sec:mechanical-explained}.
By virtue of energy-momentum conservation, the terms $\bar{c}^{q,g}(t)$ contribute 
to both the quark and to the gluon part with same magnitude but with opposite signs, so that $\sum_q\bar{c}^q(t) + \bar{c}^g(t)= 0$. Experimental information on the gluon contribution may come from trace anomaly measurements in $J/\Psi$ production at threshold, or possibly with the help from LQCD.  

The superscripts $q,g$ indicate that the breakdown is valid for both quarks $q$ and gluons $g$. Most of the discussion in this section is related to the quark contributions, and we will omit the reference to the gluon part for the remainder of this subsection. 
The GFFs of quarks and gluons also depend on the renormalization scale $\mu^2$ (associated with the hard scale $Q^2$ of the process) that we omit in the formalism for simplicity. The total GFFs, $A(t)=\sum_q A^q(t)+A^g(t)$ and analog for $B(t)$ and $D(t)$, are renormalization scale independent.

The GFF are the entry into the mechanical and other properties of the protons. However, there is not a practical, direct way to measure these form factors as it would require measurements employing the graviton-proton 
interaction, a highly impractical proposition due to the extreme weakness of the gravitational interaction~\cite{Kobzarev:1962wt,Pagels:1966zza}. More recent theoretical development showed that the GFFs may be 
indirectly probed in deeply virtual Compton scattering (DVCS)~\cite{Ji:1996nm}. DVCS allows probing  
the proton's quark structure expressed in the GPDs, as the basis for the exploration of its mechanical 
or gravitational properties~\cite{Polyakov:2002yz}. 

The handbag diagram for the DVCS amplitude~\ref{fig:handbag} contains contributions  from non-local operators with collinear twist 2, 3, and 4, where the latter two can be neglected at large $Q^2$. These operators can be expanded through the operator product 
expansion in terms of local operators with an infinite tower of $J^{PC}$ quantum numbers. This includes operators with the quantum numbers of the graviton, so information about how the target would 
interact with a graviton is encoded within this tower. 
The GPDs $H^q$ and $E^q$ are mapped to the GFF $D^q(t)$, $A^q(t)$, and $J^q(t)=\frac12\,A^q(t)+\frac12\,B^q(t)$ in the Ji sum rule ~\cite{Ji:1996nm}, involving  the second Mellin moment of the GPD $H^q$ and $E^q$ as
\begin{eqnarray}
\int \mathrm{d}x \, x[H^q(x, \xi, t) +  E^q(x, \xi, t)]  =  2J^q(t), \label{gpd-E} \\
\int \mathrm{d}x \, x H^q(x, \xi, t)  =  A^q(t) + \xi^2 D^q(t). \label{gpd-H}
\label{mellin}
\end{eqnarray}
 
In the following we focus on the term $D^q(t)$ that encodes information about mechanical properties, see  Sec.~\ref{Sec:mechanical-explained}.

This new direction of nucleon structure research has recently resulted in the first estimate of the pressure distribution inside the proton based on experimental data~\cite{Burkert:2018bqq}, employing CLAS DVCS-BH 
beam-spin asymmetry data~\cite{CLAS:2007clm} and differential cross sections \cite{CLAS:2015uuo}, and constraints from parameterized data covering the full phase space.  

\begin{figure}[htb!]
\centering{
\includegraphics[width=0.95\columnwidth]{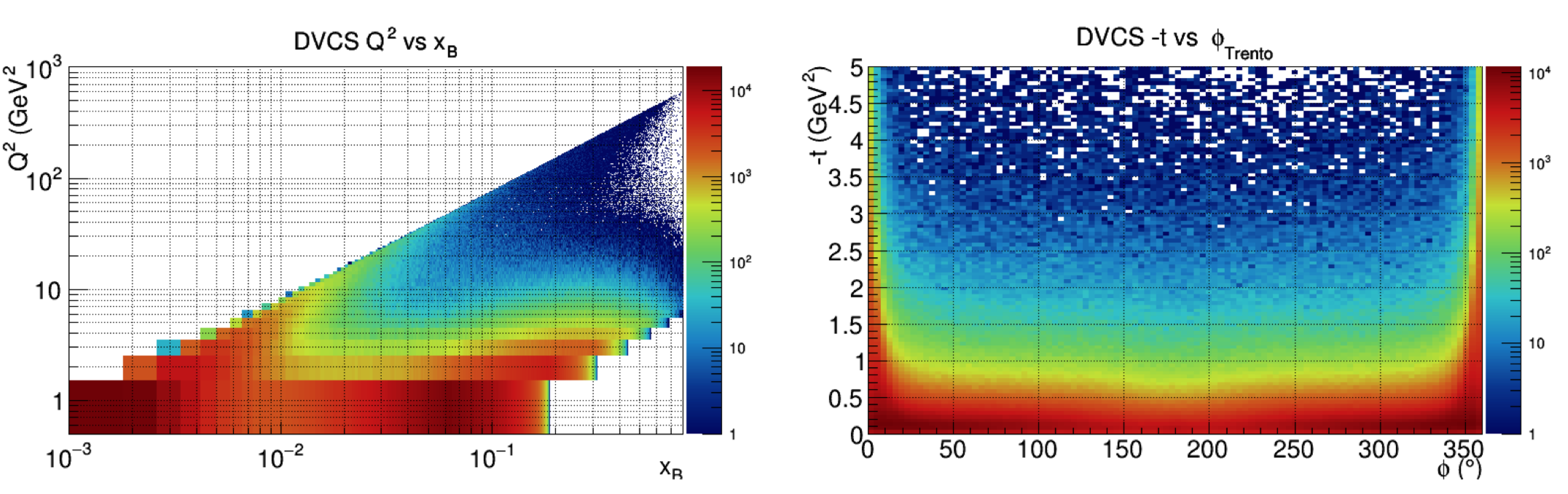}}
\caption{Accessible ranges in $x_B$ vs $Q^2$ (left), and $t$ vs azimuth angle $\phi$ (right) for the DVCS process 
at a center-of-mass energy $\sqrt{s}=28$~GeV. The color code indicates the number of events per pixel for a given luminosity.} 
\label{fig:DVCS_kinematics}
\end{figure}

With the EIC as a high luminosity machine and a large energy reach these properties can be accessed covering a large range in  $x_B$, $Q^2$ and $-t$ in the exclusive DVCS process. As shown in Figure~\ref{fig:DVCS_kinematics}
the lower EIC CM energy range of $3\times 10^{-3}<x_B<0.1$ will cover the valence quark and sea-quark domains, while at the high CM energies the gluon contributions will be accessible at $10^{-4}<x_B<10^{-2}$. 

Ideally, one would determine the integrals in Eqs.(\ref{gpd-E}) and (\ref{gpd-H}) by measuring GPD $H$ and $E$ in the entire $x$ and $\xi$ space and in a large range of $t$. 
For the DVCS experiments, such an approach is impractical as the GPDs are not directly 
accessible in the full $x,\xi$-space, but only at the constrained kinematics $x = \pm \xi$. The GPDs also do not 
directly appear in the experimental observables. Instead, GPDs appear inside the Compton Form Factors defined in Eqn. (\ref{CFF}) that depend only on the two variables $\xi$, $t$. 

\noindent where one has traded the real function of 3 parameters $H(x, \xi, t)$ with the complex 
functions of 2 parameters  $\text{Re} {\mathcal H}(\xi,t)$ and $\text{Im}{\mathcal H}(\xi,t)$ 
that can be related more directly to experimentally accessible observables.
The CFF appear in experimental cross sections and in polarization observables. 
CFF $\mathcal {H}(\xi, t)$ as well as $\mathcal{E}(\xi, t)$ are thus accessible through a 
careful analysis of differential cross sections and the responses to spin polarization of the electron and the proton beam.   

As discussed in section \ref{subsec:GPDanalysis}, the extraction of the Im${\mathcal H(\xi,t)}$ and Re$\mathcal{H}(\xi, t)$ CFF has been pursued by employing global parameterizations for the $\xi$ and $t$ dependencies~\cite{Burkert:2018bqq} and using machine learning (ML) and artificial neural networks approaches~\cite{Moutarde:2019tqa,Kumericki:2016ehc,Grigsby:2020auv}

In order to determine the $D^q(t)$ form factor we can employ a subtracted fixed-$t$ dispersion relation that
relates the real and imaginary parts of the CFF $\mathcal{H}$ to a subtraction term $\Delta^q(t)$ whose determination requires additional experimental information. 
The dispersion relation  and its relationship to the subtraction term $\Delta^q(t)$ is given as 
\begin{eqnarray}
 {\rm Re}{\cal H}^q(\xi,t) = \Delta^q(t) +\frac{ 1}{\pi} {\cal P} \int_0^1 \text{d}x \ \left[\frac{1}{\xi-x}  -\frac{1}{\xi+x}\right] {\rm Im}{\cal H}^q(x,t), \label{DR}
\end{eqnarray}
where $\mathcal{P}$ is the principal value of the Cauchy integral, for simplicity written without threshold effects~\cite{Anikin:2007yh,Goldstein:2009ks}.

The subtraction term $\Delta^q(t)$ was shown to be related to the D-term~\cite{Anikin:2007yh,Diehl:2007jb} through the series of Gegenbauer polynomials. When only the first term in the series is retained and we assume $D^u(t)\approx D^q(t)$ based on large-$N_c$ predictions \cite{Goeke:2001tz} and neglect strange and heavier quark contributions which at JLab energies is a good approximation (recall that in DVCS the contributions of different quark flavors enter weighted by squares of the fractional quark charge factors), then we obtain:
\begin{eqnarray}
D^Q(t) = \sum_q D^q(t) \approx \frac{18}{25}\sum_q e_q^2\Delta^q(t) \label{delta}
\end{eqnarray}
This truncation of the Gegenbauer polynomials causes a model-dependence as the higher order terms can not be isolated with DVCS measurements alone, and must currently be computed in models. The chiral Quark Soliton Model~\cite{Kivel:2000fg} predicts a 30\% contribution due to the next term in the Gegenbauer expansion. Computations of the next to leading term may in future become possible  from LQCD (see also section~\ref{subsubsec:GPDs} for more detailed discussion on LQCD contributions to GPDs and 3D imaging). 
  
It is important to remark that the different terms in the Gegenbauer expansion of $\Delta^q(t)$ have different renormalization scale dependencies. The broader $Q^2$-coverage at EIC may therefore provide the leverage to discriminate between the different terms and help to isolate the leading term related to $D^q(t)$. In the limit of the renormalization scale going to infinity, all higher Gegenbauer terms vanish and asymptotically $\Delta^q(t)\to 5\,D^q(t)$ \cite{Goeke:2001tz}.
We note that in the limit renormalization scale going to infinity it is $\sum_qD^q(t)\to D(t)\,N_f/(N_f+4C_F)$ and $D^g(t)\to D(t)\,4C_F/(N_f+4C_F)$ where $D(t)$ is the total GFF, $N_f$ is the number of flavors and $C_F = (N_c^2-1)/(2N_c)$ \cite{Polyakov:2018zvc}.

\subsection{Backward hard exclusive reactions and probing TDAs with high luminosity EIC}
\label{subsec:backward}


A natural and promising extension of the EIC experimental program for hard exclusive processes is the study of hard exclusive electroproduction and photoproduction reactions in the near-backward region~\cite{Gayoso:2021rzj}. 
These measurements will allow further exploration of  hardronic structure in terms of baryon-to-meson and baryon-to-photon Transition Distribution Amplitudes~\cite{Pire:2021hbl} which extend both the concepts of Generalized Parton Distributions (GPDs) and baryon Distribution Amplitudes (DAs).  

\begin{figure}[h]

\includegraphics[width=0.30\textwidth]{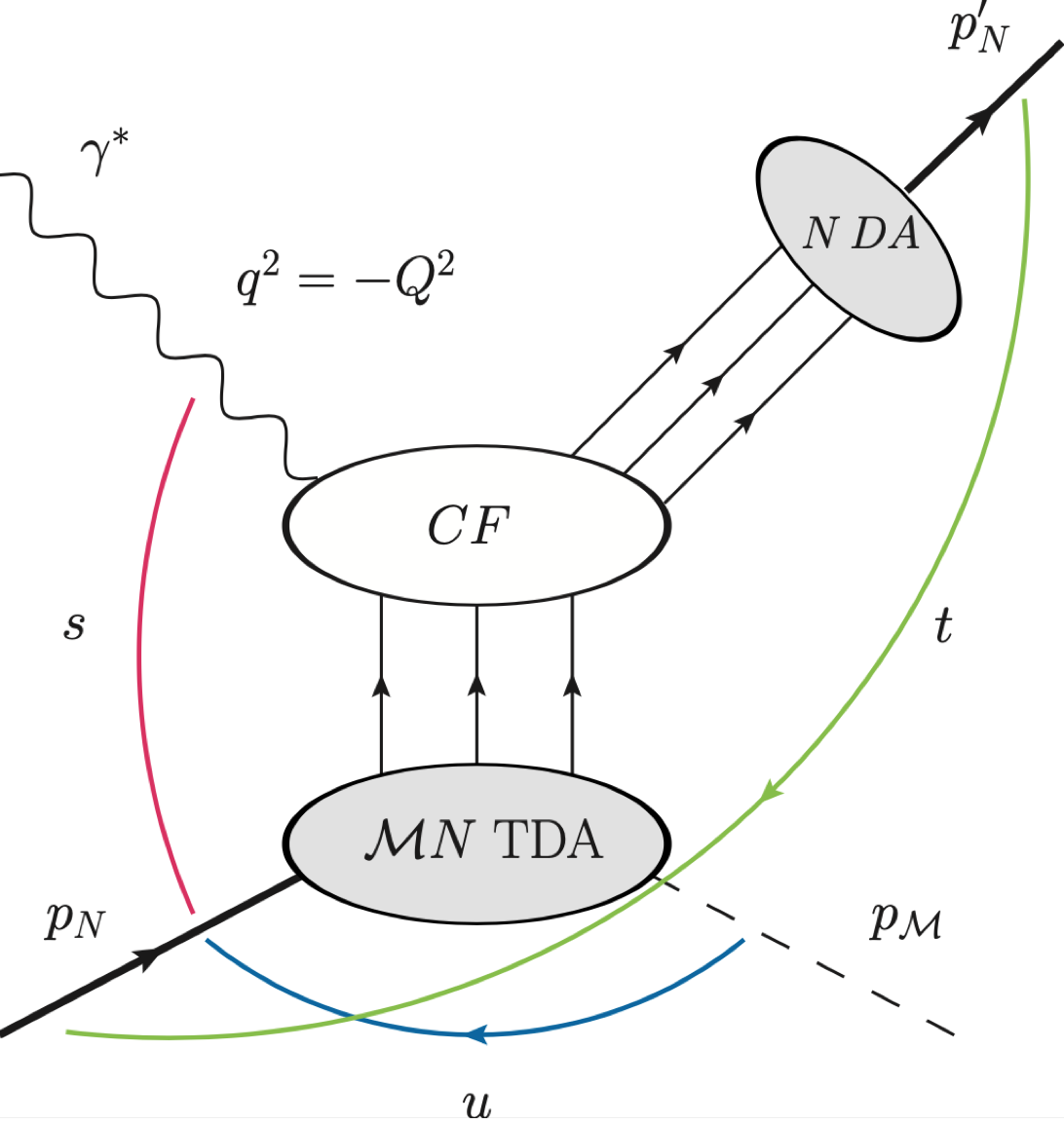}  
\hspace{1.5cm}\includegraphics[width=0.37\textwidth]{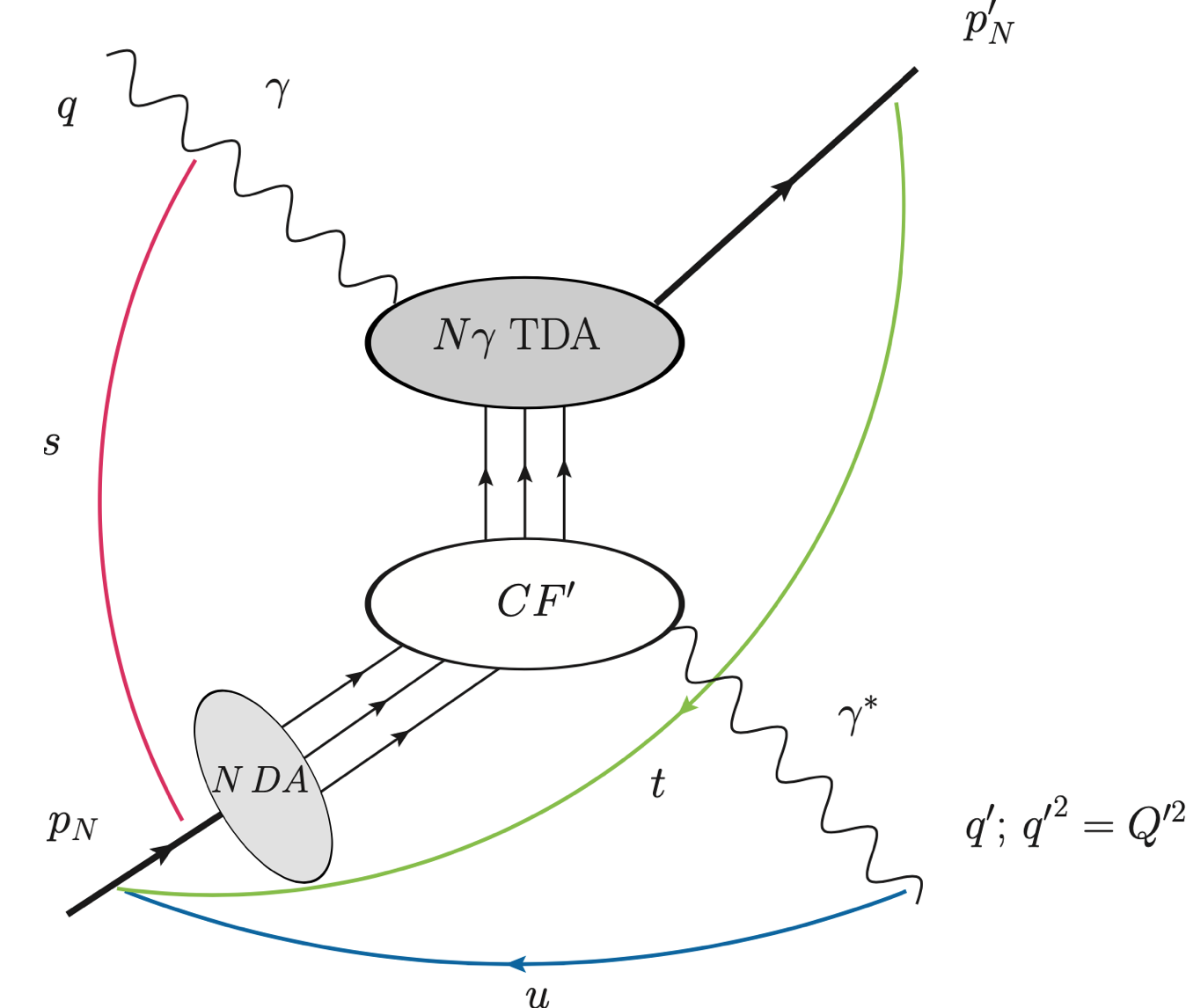}
    \caption{ Left: Collinear factorization mechanism  for hard exclusive electroproduction of mesons  ($\gamma^* N \to N' {\cal M}$) in the  near-backward  kinematic regime (large $Q^2$, $W^2$; fixed $x_{B}$; $|u|\sim 0$). Right: Collinear factorization  of TCS ($\gamma N \to \gamma^* N' $)  in the  near-backward  kinematic regime (large $Q'^2$, $W^2$; fixed $\tau \equiv \frac{Q'^2}{2 p_N \cdot q}$;   $|u|\sim 0$);
  ${\cal M} N$ ($N\gamma$) TDA stands for the transition distribution amplitudes from a nucleon-to-a-meson (photon-to-a-nucleon); $N$ DA stands for the nucleon distribution amplitude;
      $CF$ and $CF'$ denote the corresponding hard subprocess amplitudes (coefficient functions).  }
\label{Fig_TDAfact}
\end{figure}

Baryon-to-meson (and baryon-to-photon) TDAs arise within the collinear factorization framework for hard exclusive reactions in a kinematic regime that is complementary to the usual near-forward kinematic in which a familiar GPD-based description applies for hard exclusive meson 
electroproduction reactions and DVCS. Technically, TDAs are defined as 
transition matrix elements between a baryon and a meson (or a photon) states
of the same non-local three-quark operator on the light-cone occurring in the 
definition of baryon DAs. In Fig.~\ref{Fig_TDAfact} we sketch the collinear factorisation reaction mechanism involving TDAs (and nucleon DAs) for hard 
exclusive near-backward electroproduction of a meson off a nucleon target
$\gamma^* N \to N' {\cal M}$ \cite{Lansberg:2011aa} and of hard exclusive near-backward photoproduction of a lepton pair off a nucleon target
(backward Timelike Compton Scattering (TCS))
$\gamma N \to \gamma^* N' \to \ell^+ \ell^- N'$ \cite{Pire:2022fbi}.

The physical contents of baryon-to-meson and baryon-to-photon TDAs is conceptually similar to that of GPDs and baryon DAs. Since the non-local QCD operator defining TDAs carries the 
quantum numbers of a baryon it provides access to the momentum distribution of baryonic number inside hadrons. It also enables the study of non-minimal Fock components of hadronic light-front wave functions. Similarly to GPDs, by switching to the impact parameter space, one can address the distribution of the baryonic charge inside hadrons in the transverse plane. This also enables to study the mesonic and electromagnetic clouds surrounding hadrons and provides new tools for ``femtophotography'' of hadrons. 
Testing the validity of the collinear factorized
description in terms of TDAs for hard backward  electroproduction and photoproduction reactions requires a 
detailed experimental analysis. 
The very first experimental indications of the relevance of the TDA-based description for hard electroproduction of backward  mesons
off nucleons 
were recently obtained at JLab in the studies of 
backward pseudoscalar meson  electroproduction 
$$ep \to e'n \pi^+$$ by the CLAS collaboration and in Hall A ~\cite{CLAS:2017rgp,CLAS:2020yqf}, 
and  backward vector meson electroproduction $$ep \to e'p' \omega$$ by Hall C \cite{JeffersonLabFp:2019gpp}.
This latter analysis enabled checking one of the crucial predictions of the TDA-based formalism, the 
dominance of the transverse cross section  $\sigma_T$. 
A dedicated study of backward neutral pseudoscalar meson production 
with a complete Rosenbluth separation of the cross section to challenge $\sigma_T \gg \sigma_L$
condition is currently prepared by Hall C \cite{Li:2020nsk}. 

The hard exclusive backward reactions to be studied with the EIC include the hard exclusive
backward electroproduction of light pseudoscalar unflavored $\pi$, $\eta$, and strange mesons $K$ 
and vector $\rho$, $\omega$, $\phi$ mesons as well as backward DVCS.
Another option can be the study of hard exclusive backward photoproduction of lepton pairs 
(backward TCS) and of heavy quarkonium. 
The peculiar EIC kinematics, as compared to fixed target experiments, allows, in principle, a thorough analysis of the backward region
pertinent to TDA studies.
Higher $Q^2$ providing a larger lever arm to test the characteristic scaling behavior would be accessible in a domain of moderate
$\gamma^*N$ energies, {\it i.e.} rather small values of the usual $y$ variable and not too small values of $x_B$. It worth mentioning that since TDA-related cross sections are usually small the high luminosity is definitely needed to scan a sufficiently wide $Q^2$ range. This will allow the new domain of backward hard exclusive reactions physics to be further explored.

The detection of $u$-channel
exclusive electroproduction: $$e+p \to e'+p'+\pi^0$$ seems easily feasible thanks to the $4 \pi$ coverage of EIC detector package.
A preliminary study documented in \cite{AbdulKhalek:2021gbh} shows the feasibility of detecting exclusive $\pi^0$ production at $u\sim u_0$. The
scattered electrons are well within the standard detection specification. The two photons (from decaying $\pi^0$) project a
ring pattern at the zero degree calorimeter (tagging detector along the incidence proton beam) close to the effective
acceptance, while recoiled proton enters forward EM calorimeter at high pseudorapidity. The detector optimization and efficiency for detecting these process is currently undergoing.

Also a rough vector meson dominance model based estimates of backward TCS cross section for the EIC kinematical conditions 
presented in \cite{Pire:2022fbi} suggest a considerable number of events within the high luminosity regime to study 
photon-to-nucleon TDAs. 

More phenomenological prospective studies and further theoretical development are needed to establish a sound experimental program focusing on TDAs for EIC.
\subsection{Outlook - Beyond the EIC initial complement}
\label{sec:GPDoutlook}
Spin polarized electron and proton beams lead to single-spin dependent cross sections that are proportional to the imaginary part of the DVCS-BH interference amplitude. Double-spin dependent cross sections provide an access to the real part of the  interference amplitude but suffer from strong to dominant contributions of the BH amplitude which makes difficult and  inaccurate the experimental determination of the real part from this observable. An indisputable and precise determination of this quantity is required to unravel the mechanical properties of the nucleon.

\begin{figure}[ht!] 
\includegraphics[width=0.45\columnwidth]{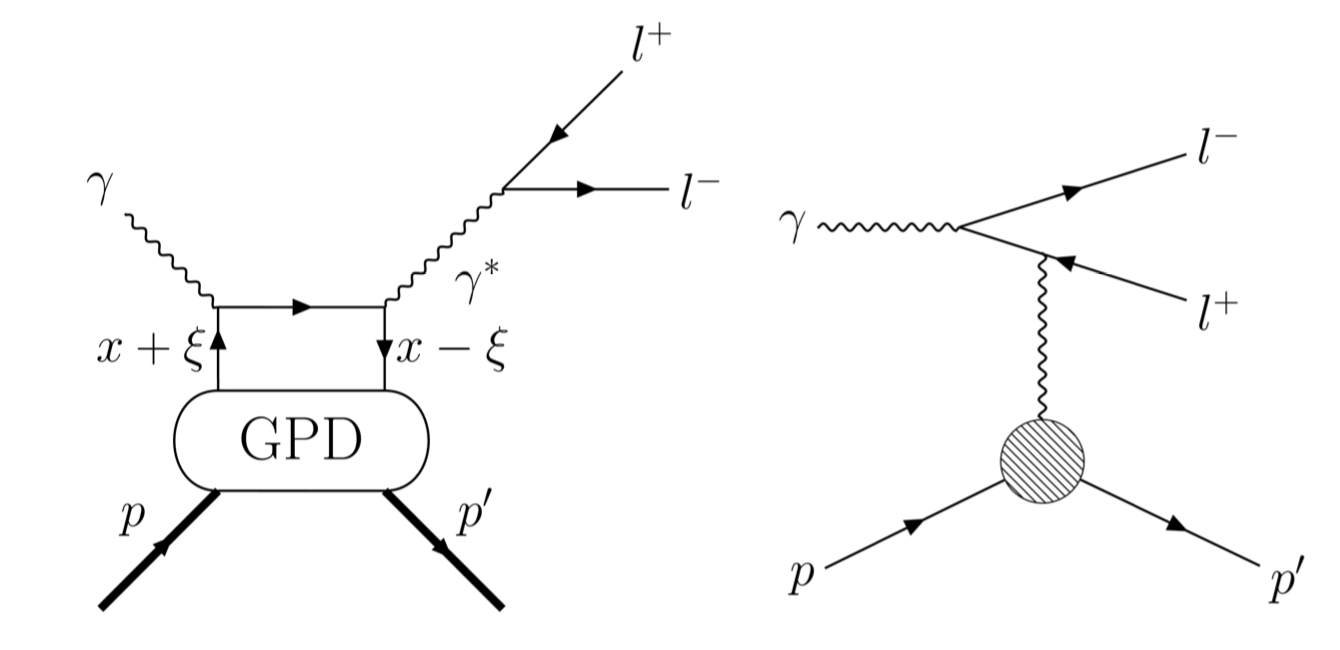}
\includegraphics[width=0.35\columnwidth]{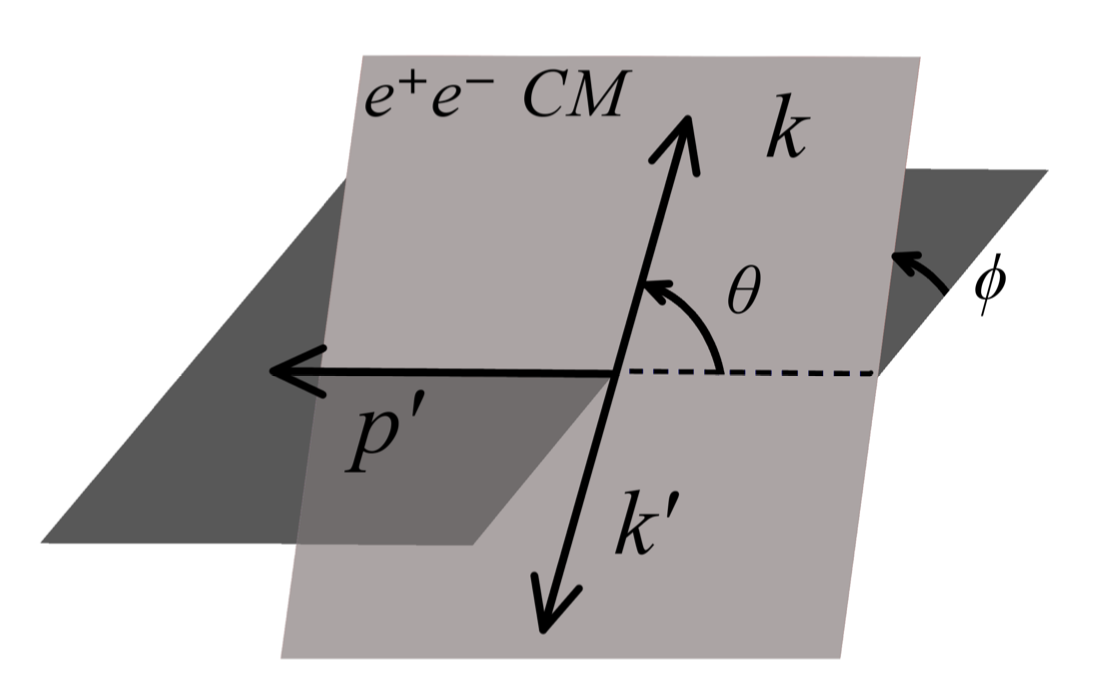}
\caption{\footnotesize Left: Handbag diagram of the TCS process. Middle: Diagram of the BH processes. Right: Relevant angles for the TCS kinematics in CMS to isolate the Re$\mathcal{H}$  contribution in the interference term.}
\label{TCS}
\end{figure}

Accessing the real part of interference amplitude is significantly more challenging than the imaginary part. It appears in the unpolarized cross sections for which either the BH contribution is dominant, or all three terms (pure BH, pure DVCS, and DVCS-BH interference amplitudes) are comparable. The DVCS and interference terms can be separated in the unpolarized cross-sections by exploiting their dependencies on the incident beam energy, a generalized Rosenbluth separation. This is an elaborated experimental  procedure, which needs some theoretical hypothesis to finally extract an ambiguous physics  content~\cite{Defurne:2017paw,Kriesten:2020apm,JeffersonLabHallA:2022pnx}. Time-like Compton scattering (TCS), $\gamma p \to l^+ l^- p$ is another process which can, in principle, provide direct but luminosity challenging access to the Re$\mathcal{H}(\xi,t)$ in a back-to-back configuration~\cite{CLAS:2021lky} as displayed in Fig.~\ref{TCS}. TCS requires zero-degree electron scattering, generating $l^+l^-$ pairs in quasi-real photo-production over a continuous mass range above resonance production. 
The feasibility of measuring TCS, and its strong sensitivity to the D-term, has already been established at CLAS12~\cite{CLAS:2021lky}. 

A more convenient access to the real part of the interference amplitude is obtained from the comparison between  unpolarized electron and positron beams~\cite{Voutier:2014kea}. Indeed, at leading twist, the electron-positron unpolarized DVCS cross section difference is a pure interference signal, linearly dependent on the real part of the DVCS-BH interference term. As such it provides the cleanest access to this crucial observable, without the need for additional theoretical assumptions in the CFFs extraction procedure~\cite{CLAS:2021gwi}. Implementation of a positron source, both polarized and  unpolarized~\cite{PEPPo:2016saj}, at the EIC would thus significantly enhance its capabilities in the high impact 3D imaging science program, with respect, for instance, to the extraction of the CFF  Re$\mathcal{H}(\xi,t)$ and of the gravitational form factor $D^q(t)$.

\section{Mass and spin of the nucleon}
\label{sec:mass}

\noindent
The most fundamental physical properties of the nucleons as well as other hadrons are their masses and spins. Understanding how they arise from the QCD theory of light spin-1/2 quarks and massless spin-1 gluons is one of the most important goals in nuclear physics~\cite{NASreport}. The experimental study  
of the proton spin structure began in the 1980's and has continuously driven the field of hadronic physics for the last thirty years~\cite{EuropeanMuon:1989yki}. Despite much effort, a complete picture of the proton spin structure is still missing~\cite{Ji:2020ena}. The origins of the proton mass have mostly been a theoretical interest in QCD-motivated models or effective approaches such as chiral perturbation theory, and its understanding in the QCD-based framework and related experimental tests have gained attentions only recently~\cite{Ji:2021mtz}. 

Gaining insight into the emergence of hadron mass from the experimental results on the pion/kaon electromagnetic form factors and PDFs analyzed within the Continuum Schwinger Method (CSM) represents an important aspect of efforts in experiments of the 12 GeV era at JLab \cite{Roberts:2021nhw} and those foreseen at the EIC in the US \cite{Arrington:2021biu} and at the EiCC in China \cite{Anderle:2021wcy}. A successful description of the electroexcitation amplitudes of the $\Delta(1232)3/2^+$, $N(1440)1/2^+$, and $\Delta(1600)3/2^+$ resonances of different structure \cite{Mokeev:2022xfo} has been achieved within the CSM \cite{Segovia:2015hra,Segovia:2014aza} employing the same momentum-dependent dressed quark mass evaluated from the QCD Lagrangian \cite{Roberts:2020hiw} and supported by the experimental results on the structure of the pion/kaon and the ground state nucleon. This success has demonstrated a promising opportunity to address challenging and still open problems in the Standard Model on the emergence of hadron mass by confronting the predictions from QCD-rooted approaches on a broad array of different hadron structure observables with the results from experiments with electromagnetic probes already available and those foreseen from intermediate energy facilities at the luminosity frontier.
 
In the QCD studies, it has been realized that the matrix elements/form factors of the quark and gluon energy momentum tensor (EMT),  measured through DIS momentum sum rule and also the source for gravitational fields of the nucleon, play important roles in spin and mass~\cite{Ji:1994av,Ji:1996ek}. 
Moreover, the interpretation of the GFF $C(Q^2)$
in terms of mechanical properties has generated much interest~\cite{Polyakov:2018zvc}. 
Experimentally, the form factors of EMT can be accessed through the second-order moments of quark and gluon GPDs which can be probed through DVCS and DVMP as discussed in the early sections~\cite{Ji:1996ek}. EIC is particularly important for probing the GPDs of gluons which are a crucial part of the nucleon~\cite{Accardi:2012qut}. It has been suggested recently that the gluon EMT form factors might be directly accessible through near-threshold heavy-quarkonium production~\cite{Guo:2021ibg}. 

\subsection{Nucleon mass}
\label{subsec:nucl-mass}

Unlike non-relativistic systems in which the masses mostly arise from the fundamental constituents, masses of relativistic systems arise predominantly through interactions. Indeed, without the strong interactions, three current quarks making up the nucleon weigh about $\sim 10$ MeV (at $\mu_{\rm \overline{MS}}\sim 2$ GeV), presumably from electroweak symmetry breaking, which is about 1\% of the bound state mass~\cite{PDG2020}. Schematically, we can write 
the nucleon mass in terms of quark masses and the strong interaction scale $\Lambda_{\rm QCD}$, 
\begin{equation}
     M_N = \sum_i \alpha_i m_i + \eta \Lambda_{\rm QCD} \ , 
\end{equation}
where $\alpha_i$ and $\eta$ are dimensionless coefficients determined from the strong interaction dynamics. Note that $\Lambda_{\rm QCD}$ is a free
parameter of QCD, which in principle can take any value, and therefore, the nucleon mass can be 10 TeV or 100 MeV, independent of the details of strong interaction physics. One cannot hope, therefore, to explain from QCD itself why the nucleon mass is 940 MeV, not any other value, without invoking more fundamental theories such as grand unifications which may explain why $\Lambda_{\rm QCD}$ takes the value that we measured~\cite{Georgi:1974sy}. 

In the nucleon models, $\Lambda_{\rm QCD}$ scale has generally been replaced with some parameters with more direct physical interpretations. For instance, in the models emphasizing chiral symmetry breaking, $\Lambda_{\rm QCD}$ is superseded by the chiral symmetry breaking scale and the constituent quark and/or gluon masses~\cite{Manohar:1983md}. On the other hand, in the models such as the MIT bags which stress the color confinement, $\Lambda_{\rm QCD}$ has been associated with the energy density of the false vacuum inside a bag~\cite{Chodos:1974je}. In the instanton liquid models, $\Lambda_{\rm QCD}$ is reflected through typical instanton size and density~\cite{Schafer:1996wv}. Unfortunately, the effective degrees of freedom in models cannot be studied directly in experiments, and therefore the pictures cannot be directly verified without additional assumptions.
In lattice QCD calculations, $\Lambda_{\rm QCD}$ is tied with lattice spacing $a$ which is an ultraviolet momentum cut-off
and the strong coupling associated with 
the cut-off. As we shall discuss below, a model-independent way to introduce this scale might be through the gluonic composite scalar field which breaks the scale symmetry, a Higgs-like scale-generation mechanism~\cite{Ji:2021qgo}. 

So then what are the meaningful questions one can ask about the nucleon mass, and can they be answered through experiments at EIC? The most discussions so far in the literature are about mass distributions into different dynamical sources and about spatial distributions inside the nucleon. For example, what will be the proton mass if all quark masses where zero? This question has been studied in chiral perturbation theory in 1980's~\cite{Gasser:1982ap}. 
Through Lorentz symmetry relation, it has been found that the quark and gluon kinetic energy contributions to the nucleon mass can be studied through deep-inelastic scattering~\cite{Ji:1994av}. Moreover, 
it has been suggested that the trace anomaly contribution to the nucleon mass can be measured directly as well~\cite{Kharzeev:1995ij}. All of these studies are based on understandings of the energy sources in the strong interaction Hamiltonian, $H_{\rm QCD}$. Experimental measurements and theoretical calculations of these mass contributions constitute important tests on an important aspect of our understandings of the nucleon mass. 

The spatial distributions of mass/energy densities are an important concept in gravitational theories as they are sources of gravitational potentials. In the limit when the quantum mechanical fluctuations can be neglected or the mass is considered heavy, the proton can have a fixed center-of-mass position with spatial profiles of mass and other densities. Studies of these profiles can be done through the GFF as one has learned about the spatial distributions of the electric charges and currents~\cite{Polyakov:2018zvc}. Moreover, 
the trace anomaly contribution is related to the scalar form factor which maps out the dynamical ``bag constant''~\cite{Ji:2021qgo}.  

\subsubsection{Masses in dynamical energy sources}

A complete picture of the mass distributions into different sources starts from the QCD Hamiltonian~\cite{Ji:1994av}. 
In relativistic theories, the Hamiltonian is a spatial integral of (00)-component of the second-order EMT $T^{\mu\nu}$. Despite that field theories are full of UV divergences, the full EMT is conserved 
and hence finite. This second-rank tensor can be uniquely decomposed into a trace term proportional 
to the metric tensor $g^{\mu\nu}$ and a traceless term $\bar T^{\mu\nu}$. They are separately finite due to Lorentz symmetry. Thus the QCD Hamiltonian contains two
finite pieces, the scalar and (second-order) tensor terms, 
\begin{equation}
      H = H_S+H_T \ . 
\end{equation}
A general feature of the Lorentz-symmetric QFT in (3+1)D is that the $H_S$ contributes 1/4 of a bound state mass, 
and the tensor term $H_T$ contributes
3/4~\cite{Ji:1994av}, namely
\begin{equation}
    E_{S,T} = \langle P| H_{S,T}|P \rangle;~~~~~ E_T = 3 E_S = \frac{3}{4}M \ , 
\end{equation}
where the expectation value is taken in a static hadron (nucleon) state $|\vec{P}=0\rangle$. Again, this is independent of any other specifics of an  underlying theory. 

A further decomposition of the tensor part of the Hamiltonian (energy) can be done through quark and gluon contributions, 
\begin{equation}
    E_T= E_{Tq}(\mu) + E_{Tg}(\mu) \ . 
\end{equation}
These energy sources can be probed through the matrix elements of the corresponding parts in the EMT in terms
of the momentum fractions of the parton distributions, $E_{Tq,g}(\mu) = (3/4) M_N \langle x\rangle_{q,g}(\mu)$, where the quark and gluon $\langle x\rangle_{q,g}(\mu)$ can be obtained 
from the phenomenological PDFs~\cite{Ji:1994av}. Therefore, a major part of the proton mass can be understood in terms of quark and gluon kinetic energy contributions, although the latter separation depends on scheme and scale as indicated by argument $\mu$. 

\begin{figure}[t]
 \includegraphics[height=5.5cm]{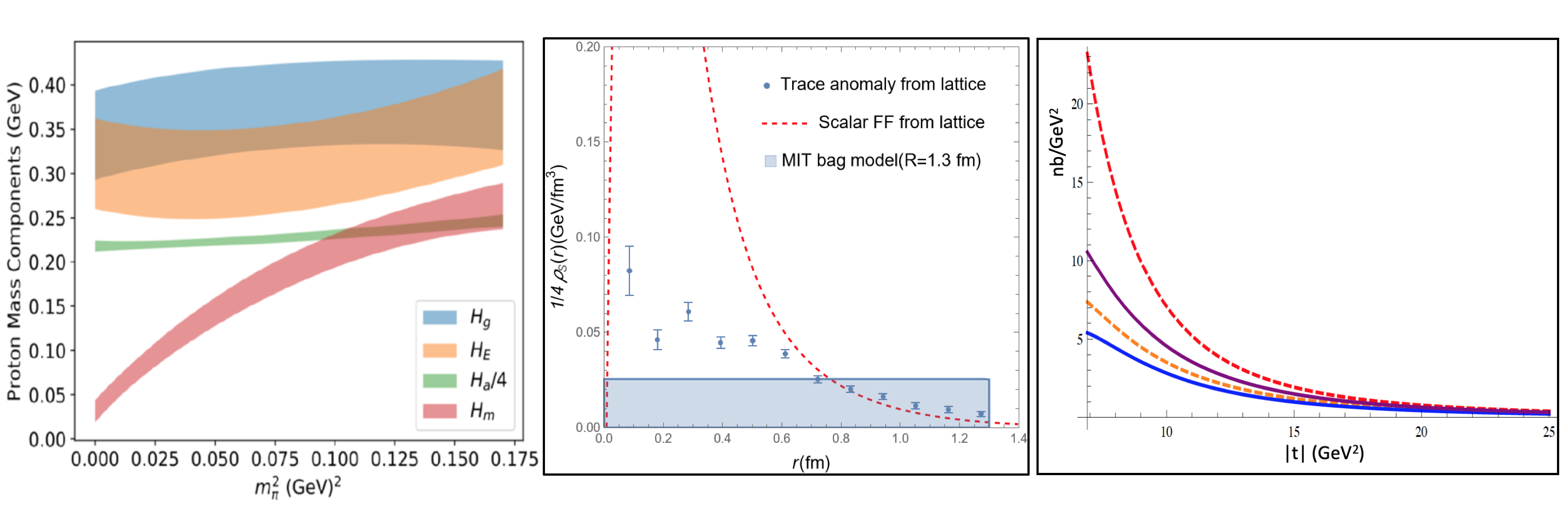}
\caption {Left: the proton mass decomposition, calculated from lattice QCD, into different sources, including the quark mass ($\sl{H_m}$), quark and gluon kinetic and potential energy ($\sl{H_g},\sl{H_E}$), and quantum anomalous energy contributions ($\sl{H_a}$)~\cite{Yang:2018nqn,Alexandrou:2017oeh}. Middle: the scalar density 
distribution in space which can be constructed from the GFF ~\cite{He:2021bof,LHPC:2007blg,Shanahan:2018pib}. Right: Differential cross section $d\sigma/dt$ 
in units of nb/GeV$^{2}$ for exclusive threshold $J/\Psi$ production at EIC as a function of $|t|$ at $W=4.4$~GeV, $Q^2=64$~GeV$^2$. The dashed curves are for $D^g=0$ and the solid curves are for nonzero $D^g$ (from LQCD). The split between the two solid curves, or two dashed curves is caused by the variation in the gluon scalar matrix element $0<b<1$~\cite{Boussarie:2020vmu}.}
\label{fig:mass}
\end{figure}

The scalar energy that contributes to the 1/4 of the proton mass comes from the following matrix element, 
\begin{equation}
    E_S = \frac{1}{8M}\langle P|(1+\gamma_m)m\bar\psi\psi + \frac{\beta(g)}{2g} F^2|P\rangle \ ,  
\end{equation} 
where $\gamma_m$ and $\beta$ are perturbative anomalous dimension and (appropriately normalized) QCD beta function, respectively. The operator is  twist-four in high-energy scattering and its matrix element is difficult to measure directly. However, the up and down quark mass contribution has been historically related to the so-called $\pi$-N $\sigma$-term which can be extracted from experimental data~\cite{Gasser:1990ce}. The strange quark mass contribution is related the baryon-octet mass spectrum through chiral perturbation theory~\cite{Gasser:1984gg}. A lattice QCD calculation of various contributions to the proton mass is shown on the left panel in Fig. 
\ref{fig:mass}~\cite{Yang:2018nqn,Alexandrou:2017oeh}.

The most interesting and surprising is the contribution of the gluon trace-anomaly term $F^2$, which sets the scale for other
contributions. To understand the physics
of this contribution, one can consider the composite scalar field $\phi\sim F^2$ which has a vacuum expectation value through the gluon condensate. Inside the nucleon, however, the $\phi$ field is not the same. In fact, $\phi$ gets a contribution through 
its static response to the valence quarks inside the nucleon, with physics similar to the MIT bag model constant $B$, shown as the dots and  shaded area on the mid-panel in Fig. \ref{fig:mass}. This response can also be calculated dynamically as the exchange of 
a series of $0^{++}$ scalar particles. If this is dominated by a single scalar particle like the $\sigma$ meson, the mechanism of mass generation is then identical to the Higgs mechanism. 

It has been suggested that this matrix element can be measured through the threshold heavy-quarkonium production of photon or electron
on scattering on the proton target~\cite{Kharzeev:1995ij,Wang:2019mza}. However, due to large differences between the initial and final nucleon momenta, the interpretation has initially been suggested in the vector dominance model (VDM).
A better phenomenological description might be through AdS/CFT models~\cite{Hatta:2018ina,Mamo:2022eui}. At EIC, one may consider deeply-virtual $J/\Psi$ production to directly measure gluon matrix elements. In the large $Q^2$ and skewness-$\xi$ limit, the twist-2 gluon GFF and twist-4 $F^2$ matrix (enhanced by $1/\alpha_s$) elements may dominate. Shown on the right panel in Fig. \ref{fig:mass} is the sensitivity of the cross section on the anomaly matrix element~\cite{Boussarie:2020vmu}. 

An indirect approach to access the scalar matrix element is to use the momentum-current conservation, $\partial_\mu 
T^{\mu\nu}=0$, from which the form factors
of the tensor part is related to that of the scalar part. 
The GFF were defined in equation~\ref{EMT}, which is reproduced here for reference: 
\begin{eqnarray}
 \langle p_2|\hat{T}^{q,g}_{\mu\nu}|p_1\rangle \!=\! \bar{u}(p_2)\!\left [A^{q,g}(t)\frac{P_\mu P_\nu}{M} + B^{q,g}(t)\frac{i(P_\mu\sigma_{\mu\rho}+P_\nu\sigma_{\mu\rho})\Delta^\rho}{2M} +   D^{q,g}(t)\frac{\Delta_\mu\Delta_\nu - g_{\mu\nu}\Delta^2}{4M}+ M\bar{c}^{q,g}(t)g_{\mu\nu}\right]\! u(p_1) \nonumber
\end{eqnarray}

One of the combinations yields the (twist-four) scalar form factor 
~\cite{Ji:2021mtz} 
\begin{equation}
    G_s(t)= MA\left(t\right) +B(t)\frac{t}{4M}
-D(t)\frac{3t}{4M} \ ,
\end{equation}
which contains only the twist-two contributions from the tensor part due to the conservation law. Thus, to get the contribution of the trace anomaly term, either in experiments or from lattice QCD simulations, one needs to measure the form factors $A$, $B$ and $D$ from combined quark and gluon contributions. 

The Fourier transformation of the $G_s(t)$ 
from lattice QCD~\cite{LHPC:2007blg,Shanahan:2018pib} is shown as the dotted line in the middle panel on Fig. \ref{fig:mass}. Shown also as dots in the same panel is the anomaly contribution from lattice QCD~\cite{He:2021bof}. 

\subsubsection{Mass radius and ``confining" scalar density}

The energy density profile in space requires study of the elastic form factors of the EMT as in the case of electric charge distribution. The relevant mass/energy ($T^{00}$) form factor in the Breit frame is
\begin{equation}
    G_m(t)= MA\left(t\right) +B(t)\frac{t}{4M}
-D(t)\frac{t}{4M} \ . 
\end{equation}
As discussed extensively in the literature, when a particle has a finite mass, the spatial resolution 
of a coordinate-space distribution is limited by its Compton wavelength. In the case of the nucleon, this is about 0.2 fm. Since the nucleon charge diameter is around 1.7 fm, one can talk about an approximate coordinate-space profile. Thus, one can define the spatial distribution of energy as the Fourier transformation of the mass form factor~\cite{Polyakov:2018zvc}
\begin{equation}
     \rho_m(r) = \int \frac{ {\text d}^3 \bf q}{(2\pi)^3} e^{i \bf {q}{\cdot }{\bf r}} G_m(t)   \ . 
\end{equation}
The alternative is 
to interpret the nucleon form factors in the infinite momentum frame, which yield a 2D profile~\cite{Freese:2021mzg}. 

From the spatial energy distribution, one can define the Sachs-type mass radius as
\begin{equation}
\langle r^2 \rangle_{m}
     = 6 \left.\frac{{\text d}G_{m}(t)/M}{{\text d}t}\right|_{t=0} = \left.6\frac{\text{d}A(t)}{{\text d}t} \right|_{t=0} - 3 \frac{D(0)}{2M^2} \ . 
\end{equation}
The recent data from $J/\psi$ production at threshold has motivated extracting the proton's mass radius using either VDM or AdS/CFT type interpretation~\cite{Kharzeev:2021qkd,Mamo:2021krl}.
A QCD factorization study indicates
that a connection with the gluon contribution can be established, while the quark contribution can
be obtained through a similar form factor. Both contributions 
have been computed on the lattice QCD~\cite{LHPC:2007blg,Shanahan:2018pib}, from which one can 
extract the mass radius as 0.74~fm ~\cite{Ji:2021mtz}. 

Another interesting quantity is the scalar density, 
\begin{equation}
     \rho_s(r) = \int \frac{{\text d^3} {\bf q}}{(2\pi)^3} e^{i{\bf q}{\cdot }{ \bf r}} G_s(t)\ , 
\end{equation}
defining a scalar field distribution inside the nucleon. $G_s(t)$ can either be deduced directly from the trace part of the EMT or indirectly through the form factors of the twist-2 tensor, as discussed above. This scalar field is the analogue of the MIT bag constant $B$, 
which is a constant inside the nucleon but zero outside, and may be  
considered as a confining scalar field. A plot of a LQCD calculation of the scalar density~\cite{Shanahan:2018pib} is shown in the middle  panel of Fig.~\ref{fig:mass}.

One can define the scalar or confining radius as , 
\begin{equation}
    \langle r^2 \rangle_{s}
     = 6 \left.\frac{{\text d}G_{s}(t)/M}{dt}\right|_{t=0} =
    6\frac{{\text d} A(t)}{{\text d} t} - 9\frac{D(0)}{2M^2} \ , 
\end{equation}
which can be compared with the bag radius.
The difference between the confining and
mass radii is 
\begin{equation}
    \langle r^2\rangle_s - \langle r^2\rangle_m
    = - 6 \frac{D(0)}{2M^2} \ . 
\end{equation}
Therefore, a consistent physical picture that the confining radius is larger than the mass radius requires the $D$-term 
$D(0)<0$~\cite{Ji:2021mtz}. 
 
\subsection{Nucleon Spin Structure}

The spin structure of the nucleon has been one of the most important driving forces in hadronic physics research in the last thirty years. Non-relativistic quark models have simple predictions about the spin structure, which have been shown  incorrect through dedicated deep-inelastic scattering studies~\cite{EuropeanMuon:1989yki}. On the other hand, this is not unexpected because QCD quarks probed by high-energy scattering are different from the constituent quarks used in the simple quark models, and a connection between them is difficult to establish.  

\begin{figure}[t]
   \includegraphics[height=6.5cm]{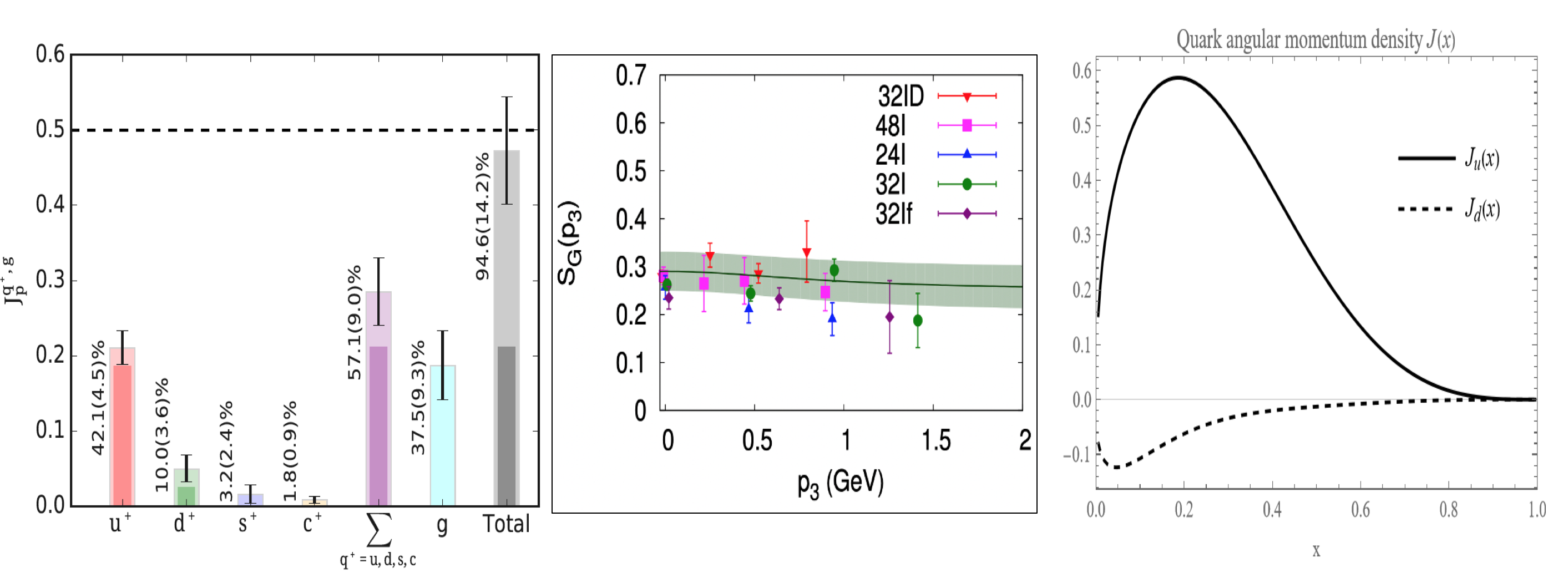}
\caption{Proton spin structure calculated from lattice QCD. (Left panel) the covariant spin decomposition~\cite{Alexandrou:2020sml}. (Middle panel) the gluon helicity contribution $\Delta G$ calculated from large momentum effective theory~\cite{Yang:2016plb}, $\rm p_3$ is the absolute value of the 3-momentum $\rm {\bf p}(0,0,p_3)$. (Right panel) Integrated quark transverse angular momentum density versus quark momentum fraction $j_q(x)$ of the proton from LQCD, which can be measured through twist-2 GPD $E(x)$.}
\label{fig:spin}
\end{figure}

\subsubsection{Longitudinal-Spin Sum Rules}

The most common approach to study the proton
spin is to understand the longitudinal polarization in the infinite momentum frame in which the quasi-free 
quarks and gluons are probed in high-energy scattering~\cite{Jaffe:1989jz}. In particular, quark and gluon helicity contributions can
be measured through summing over parton helicities $\Delta \Sigma =\int dx \sum_i \Delta q^+(x) $
and $\Delta G= \int dx \Delta g(x)$ which appear in the leading-twist scattering observables, where $+$ indicates summing over quarks and antiquarks. 
The EIC planned at BNL will make an important study of $\Delta G$ through $Q^2$ evolution and two-jet production~\cite{Accardi:2012qut}. A 
complete spin sum rule also requires measurement of the 
partonic orbital contributions $l_{q,g}= \int dx l_{q,g}(x)dx$, where $l_{q,g}(x)$ are orbitial angular momentum carried by quarks and gluons with momentum fraction $x$~\cite{Hagler:1998kg}, such that
\begin{equation}
    \frac{1}{2}\Delta \Sigma + \Delta G
     + l_q + l_g = \hbar /2 \ . 
\end{equation}
This spin sum rule was derived from QCD angular momentum operator by Jaffe and Manohar~\cite{Jaffe:1989jz}. 
Since the proton helicity does not grow as the
momentum of the proton, it is a twist-three quantity 
in high-energy scattering. 
Thus, a measurement of partonic $l_q(x)$ and $l_g(x)$ 
requires experimental data on 
twist-three generalized parton distributions~\cite{Hatta:2011ku,Ji:2012ba,Hatta:2012cs}, which will 
be challenging at EIC~\cite{Guo:2022cgq,Bhattacharya:2022vvo}. 

Therefore, it appears that 
the longitudinal spin 
structure is not simple to measure and interpret in the IMF. This, however, is not the case if instead considering a gauge-invariant sum rule~\cite{Ji:1996ek}, 
\begin{equation}
    \frac{1}{2}\Delta \Sigma + L_q + J_g = \hbar/2 \ , 
\end{equation}
which are not based on partons, where $L_q$ and $J_g$ are related to the GFF 
through $J_g = (A_g(0)+B_g(0))/2$, $J_q = \Delta \Sigma/2 + L_q = (A_g(0)+B_g(0))$. This sum rule is frame-independent, and does not have a simple partonic interpretation 
when going to the IMF. On the other hand, $J_q$ and $J_g$ can
be extracted from twist-2 GPDs, 
\begin{equation}
    J_{q, g} = \frac{1}{2}
    \int {\text d}x x (E_{q,g}(x,\xi,t=0) + H_{q,g}(x, \xi, t=0)) \ . 
\end{equation}
In the IMF, the twist-2 $L_q$ contains both the 
twist-three parton orbital angular momentum $l_q$ and a contribution from potential 
orbital angular momentum. This connection between twist-2 and twist-3 observables is a reflection of Lorentz symmetry, through which, one can construct the frame-independent longitudinal spin sum rule by measuring the twist-two GPDs~\cite{Ji:1997pf}.

Lattice QCD calculations of the angular momentum structure of the nucleon have been investigated by a number of groups (see a review in~\cite{Ji:2020ena}). In particular, the frame-independent longitudinal spin sum rule has been explored with gauge invariant
operators on the lattice. Shown on the left panel in Fig. \ref{fig:spin} is a calculation
of the spin sum rule by the ETMC collaboration~\cite{Alexandrou:2020sml}.
A more recent result from the $\chi$QCD collaboration can be found in~\cite{Wang:2021vqy}.
The gluon helicity contribution $\Delta G$ has been extracted from polarized RHIC experiments and calculated in the large momentum effective field theory~\cite{Yang:2016plb}, shown on 
the middle panel in the same figure.

\subsubsection{Transverse-Spin Sum Rules}

The spin structure of a transverse polarized proton has been less studied both theoretically and experimentally. However,
it is not widely known that 
the transverse spin in the IMF is simpler to understand than the longitudinal one~\cite{Ji:2012vj}. This is due to that the transverse angular momentum $J_\perp$
grows with the momentum of nucleon, 
\begin{equation}
       J_\perp \sim \gamma \to \infty 
\end{equation}
where $\gamma$ is the Lorentz boost factor~\cite{Ji:2020hii}. $J_\perp$ is then a leading-twist quantity and has a simple twist-2 partonic interpretation. 

Introducing the parton's transverse angular momentum 
distribution $j_q(x)$ for quarks and $j_g(x)$ for gluon, one has 
\begin{equation}
    j_{q,g}(x) = \frac{1}{2} x
    \Big(E_{q,g}(x,t=0) + \{q,g\}(x)\Big) \ . 
\end{equation}
Physically, $j_{q,g}(x)$ is the transverse angular momentum
density of the quarks and gluons when the partons 
carry the longitudinal momentum fraction $x$~\cite{Ji:2012vj}. These densities
represent the total angular momentum contributions which cannot be separated into spin and orbital ones, as the former is sub-leading for the transverse 
polarization. Using the above, one has
the simple twist-2 partonic sum rule for transverse spin
\begin{equation}
    \int^1_0 {\text d}x \left(\sum_q j_q(x)+j_g(x)\right) = \hbar/2
    \end{equation}
which is the analogy of the well-known momentum sum rule. 
Physically, experimental measurements of $E_{q,g}(x,t)$ are
best performed with transversely polarized targets with leading-twist observables. An example of $j_{u,d}(x)$ is shown on the right panel of Fig. \ref{fig:spin}, which 
is obtained from lattice calculation of $E_q(x)$ and phenomenological $q(x)$. 

There is another transverse spin sum rule at the twist-3 level, which is the rotated version of the Jaffe-Manohar
sum rule for longitudinal spin~\cite{Guo:2021aik},
\begin{equation}
    \frac{1}{2}\Delta \Sigma_T + \Delta G_T
     + l_{qT} + l_{gT} = \hbar /2 \ . 
\end{equation}
The numerical values of these quantities are the same
as the ones without the $T$ subscript. However,
they are integrated from twist-3 parton densities, e.g., 
 $\Delta\Sigma_T = \sum_q \int {\text d}x~ (\Delta q^+(x) + g_2^q(x))$, where $g_2$ is a well-known transverse-spin
 distribution which
integrates to zero, and similarly for others. Like
the Jaffe-Manohar sum rule, the twist-3 parton
densities pose great challenges to measure experimentally. 

\subsection{D-term and strong forces in the interior of the nucleon}
\label{Sec:mechanical-explained}
\noindent 
The gravitational form factors 
$A^{q,g}(t)$, $B^{q,g}(t)$, $\bar{c}^{q,g}(t)$, $D^{q,g}(t)$ defined 
in Eq.~(\ref{EMT}) contain information on the spatial distributions 
of the energy density, angular momentum, and internal forces. The 
interpretation in the Breit frame, where $P^\mu=\frac12(p'+p)^\mu=(E,0,0,0)$ 
and $\Delta^\mu =(p'-p)^\mu=(0,\vec{\Delta})$, is done by introducing the 
static EMT by means of a 3D Fourier transform as \cite{Polyakov:2002yz}
\begin{eqnarray}
    T_{\mu\nu}(\vec{r}) = 
    \int\frac{d^3\Delta}{2E(2\pi)^3}\,e^{-i\vec{\Delta}\cdot\vec{r}}
    \langle p_2|\hat{T}_{\mu\nu}|p_1\rangle\, . 
 \label{Eq:static-EMT}
\end{eqnarray}
The interpretation can be performed also in frames other than
Breit frame \cite{Lorce:2018egm} or in terms of 2D densities 
\cite{Lorce:2018egm,Freese:2021czn,Freese:2021mzg} with 
Abel transformations allowing one to switch back and forth
between the 2D and 3D interpretations \cite{Panteleeva:2021iip}. The 
consideration of 2D densities for a nucleon state boosted to the 
infinite momentum frame is of particular advantage as then the 
transverse center of mass of the nucleon is well-defined 
\cite{Burkardt:2000za}. In other frames and in the 3D case, 
this is not possible impeding the 3D spatial EMT distributions 
from being exact probabilistic parton densities.
The reservations are similar to the interpretation of the electric
form factor $G_E(t)$ in terms of a 3D electrostatic charge distribution
and the definition of a charge radius (which, despite all caveats, gives 
us an idea of the proton size).
The 3D formalism is nevertheless mathematically 
rigorous \cite{Polyakov:2018zvc} and the 3D interpretation is valid 
from a phase-space point of view \cite{Lorce:2020onh} becoming 
exact for the nucleon in the limit of a large number of colors 
$N_c$ \cite{Goeke:2007fp,Polyakov:2018zvc,Lorce:2022cle}.

In Eq.~(\ref{Eq:static-EMT}) we quote the total static EMT, 
$T_{\mu\nu}={T}^q_{\mu\nu}+{T}^g_{\mu\nu}$, but one can also define 
the separate quark and gluon static EMTs \cite{Polyakov:2002yz}.
The meaning of the different components of the static EMT is intuitively
clear with $T_{00}(\vec{r})$ denoting the energy density which yields the 
nucleon mass when integrated over space, and $T_{0k}(\vec{r})$ being 
related to the spatial distribution of the angular momentum which 
upon integration over space yields the nuclen spin $\frac12$.
The distributions of energy density and angular momentum are unknown,
but in both cases we at least know very well their integrals, namely
the total nucleon mass and total spin $\frac12$. 

The arguably most interesting components of the static EMT are 
$T_{ij}(\vec{r})$, for two reasons. First, they describe the stress 
tensor and the distribution of internal forces \cite{Polyakov:2002yz}
and are related to the $D$-term, a property on the same footing as 
mass, spin and other fundamental characteristics of the proton
\cite{Polyakov:1999gs} which was completely unknown until recently. It is worth pointing out that a free non-interacting fermion has a mass and spin but no $D$-term \cite{Hudson:2017oul} which hence emerges as a particle property generated by the dynamics and the interactions in a theory.
Second, in order to access the quark and gluon distributions of energy 
density and angular the knowledge of all GFFs is needed which are encoded 
in GPDs via Eqs.~(\ref{gpd-E},~\ref{gpd-H}) which in turn are encoded in 
the Compton form factors in Eq.~(\ref{CFF}), the actual observables in DVCS. 
In comparison to that, information on the GFF $D^q(t)$ can be inferred
much more directly from measurements of the Compton form factors via
the fixed-$t$ dispersion relation in Eq.~(\ref{DR}). 
\subsubsection{Stress tensor}
The key to investigating the mechanical properties of the proton is
the stress tensor $T_{ij}(\vec{r})$ which is symmetric and can be 
decomposed in terms of a traceless part and a trace as
\begin{equation}
T^{ij}(\vec{r}) = \biggl(e_r^i\,e_r^j-\frac13\,\delta^{ij}\biggr)\,s(r)
               + \delta^{ij}\,p(r)\,\label{Eq:Tij-p-s}
\end{equation}
with $s(r)$ known as the distribution of shear forces and $p(r)$ 
known as the distribution of pressure forces while $e_r^i$ are the 
components of the radial unit vector $\vec{e}_r = \vec{r}/|\vec{r}|$. 
The distributions 
$s(r)$ and $p(r)$ are not independent of each other but related
by the differential equation $\frac23\,s'(r)+\frac2r\,s(r)+p'(r)=0$
which originates from energy-momentum conservation
$\nabla^iT^{ij}(\vec{r}) = 0$. 
At this point it is worth stressing that the distributions of energy
density and angular momentum can be equally well discussed in the 2D
interpretation. But pressure, i.e.\ force acting on a surface element,
is intrinsically a 3D concept. (One can introduce the notion of a
2D pressure \cite{Lorce:2018egm,Freese:2021czn,Freese:2021mzg},
but in that case one looses the connection to the familiar meaning of 
pressure in physics and in the daily life.)

If the form factor $D(t)$ is known, the distributions $s(r)$ and $p(r)$ can
be determined via the relations \cite{Polyakov:2018zvc}
\begin{eqnarray} 
s(r) &=& -\frac{1}{2M}r\frac{d}{dr}\frac{1}{r}\frac{d}{dr}\widetilde{D}(r) \, , \\
p(r) &=& \frac{1}{6M}\frac{1}{r^2}\frac{d}{dr}r^2\frac{d}{dr}\widetilde{D}(r) \, ,\\
\nonumber \\ \nonumber 
{\rm with} ~~~ \widetilde{D}(r) &=& \int\frac{d^3{\bf \Delta}}{(2\pi)^3}\exp^{-i{\bf \Delta r}} D(-{\bf \Delta}^2) \,.
\end{eqnarray} 
If the separate $D^q(t)$ and $D^g(t)$ form factors are known, one can 
analogously define ``partial'' quark and gluon shear forces $s^q(r)$ and 
$s^g(r)$. Also ``partial'' pressures $p^q(r)$ and $p^g(r)$ can be defined, 
but for that besides respectively $D^q(t)$ and $D^g(t)$ one needs also the 
form factor $\bar{c}^q(t)=-\bar{c}^g(t)$ which is responsible for the 
``reshuffling'' of forces between the gluon and quark subsystems inside 
the proton \cite{Polyakov:2018exb}. The instanton vacuum model predicts 
$\bar{c}^q(t)$ to be very small \cite{Polyakov:2018exb} which would allow
one to define partial quark pressures $p^q(r)$ in terms of $D^q(t)$ alone. 
The form factor $\bar{c}^q(t)$ is difficult to access experimentally
but it can be computed in lattice QCD.

An equivalent, compact way to express the relation of $s^q(r)$ and
$p^q(r)$ and the form factor $D^q(t)$ is given by (for gluons analogously)
\begin{eqnarray} 
D^q(t) &=& 4M \int{d^3{\bf r}} {\frac{j_2(r\sqrt{-t})}{t}} s^q(r) \label{sr} \\ \ 
D^q(t) &=& 12M \int{d^3{\bf r}}{\frac {j_0(r\sqrt{-t})} {2t}} p^q(r). \label{pr} \
\end{eqnarray} 
where $M$ is the proton mass, $j_0$ and $j_2$ are spherical 
Bessel functions of zeroth and second order, respectively. 
Taking the limit $t\to 0$ in Eqs.~(\ref{sr},~\ref{pr}) one obtains 
two equivalent expressions for the $D$-term $D=D(0)$ given by
\begin{eqnarray} 
D = - \frac{4}{15}\,M \int{d^3r}\; r^2 s(r) 
  =    M \int{d^3r}\;r^2 p(r), \label{Eq:D-term} \,.
\end{eqnarray} 
The derivation of (\ref{Eq:D-term}) requires the use of the von 
Laue condition $\int_0^\infty dr\,r^2p(r)=0$ \cite{von-Laue:1911},
a necessary but not sufficient condition for stability which follows
from energy-momentum conservation.

The stress tensor $T^{ij}(\vec{r})$ is a $3\times3$ matrix which can
be diagonalized. One eigenvalue is the normal force per unit area given 
by $p_n(r)=\frac23\,s(r)+p(r)$ with the pertinent eigenvector $\vec{e}_r$
while the other two eigenvalues are degenerate in spin-0 and 
spin-$\frac12$ cases, with the degeneracy lifted only for higher spins,
are referred to as tangential forces per unit area and are given by
$p_t(r)=-\,\frac13\,s(r)+p(r)$ whose eigenvectors can be chosen to be unit 
vectors in $\vartheta$- and $\varphi$-directions in spherical 
coordinates \cite{Polyakov:2018zvc}.

\subsubsection{Mechanical stability - connection to neutron stars}
The normal force makes appearance if we consider the force 
$F^i=T^{ij}dS^j=[\frac23\,s(r)+p(r)]\,dS\,e_r^i$ within the proton 
acting on an area element $dS^j = dS\,e_r^j$. Mechanical stability 
requires this force to be directed towards the outside, otherwise
the system would implode. This implies that the normal force per 
unit area must be positive definite~\cite{Perevalova:2016dln}.
\begin{equation}
\frac23\,s(r)+p(r)>0\;.\label{Eq:normal-force-positivity}
\end{equation}
At this point it is instructive to notice that this is exactly the 
condition which is imposed when calculating the radius of a neutron
star. Neutron stars are basically macroscopic hadronic systems 
(``giant nuclei'') in which gravity and general relativity effects 
cannot be neglected.
Based on a chosen model for the equation of state of nuclear matter,
one solves the Tolman-Oppenheimer-Volkoff equation which yields the 
radial pressure inside the neutron star as function of the distance 
$r$ from the center of the neutron star. In our notation, the radial 
pressure corresponds to $\frac23\,s(r)+p(r)$.
The solution of the Tolman-Oppenheimer-Volkoff equation yields a 
radial pressure which is positive in the center and decreases 
monotonically until it drops to zero at some $r=R_\ast$ and
would become negative for $r>R_\ast$. This would correspond to 
a mechanical instability and is avoided by defining the
point $r=R_\ast$ to be the radius of the neutron star, 
see for instance \cite{Prakash:2000jr}. 
In this way, within the neutron star the mechanical stability 
condition (\ref{Eq:normal-force-positivity}) is always valid, 
and the point where the normal force per unit area drops to 
zero coincides with the ``edge'' of the system. 

The proton has of course no sharp ``edge'' being ``surrounded'' 
by a ``pion cloud'' due to which the normal force does not drop 
literally to zero but exhibits a Yukawa-tail-type suppression at 
large $r$ which becomes proportional 
to $\frac{1}{r^6}$ in the chiral limit \cite{Goeke:2007fp}.
In the less realistic but nevertheless very instructive 
and inspiring bag model, cf.\ Sec.~\ref{subsec:nucl-mass}, 
one does have an ``edge'', namely at the bag boundary, 
where the normal force drops to zero \cite{Neubelt:2019sou}. 
However, in contrast to the neutron star one does not determine the 
``edge'' of the bag model in this way. Rather the normal force drops 
``automatically'' to zero at the bag radius which reflects the fact 
that from the very beginning the bag model was thoughtfully constructed 
as a simple but mechanically stable model of hadrons \cite{Chodos:1974je}.
\subsubsection{Charge and mechanical radius of proton and of neutron}
The normal force per unit area $\frac23\,s(r)+p(r)$ is an ideal quantity to define the 
size of the system, thanks to positivity in Eq.~(\ref{Eq:normal-force-positivity}) 
guaranteed by mechanical stability. Notice that a quantity like electric charge
distribution can be used to define an electric charge radius for
the positively charged proton which is a meaningful proxy
for the ``proton size''. However, for an electrically neutral hadron
this is not possible. One can still define an electric mean square charge
radius $r_{\rm ch}^2 = 6\,G'_E(0)$ in terms of the derivative of the electric 
form factor $G_E(t)$ at $t=0$. But for the neutron $r_{\rm ch}^2<0$ which
gives insights about the distribution of the electric charge inside the 
neutron, but does not tell us anything about its size. This is ultimately 
due to the neutron's charge distribution not being positive definite.

The positive-definite normal force per unite area $\frac23\,s(r)+p(r)$, 
Eq.~(\ref{Eq:normal-force-positivity}), allows us to define the  
{\it mechanical radius} as follows \cite{Polyakov:2018guq,Polyakov:2018zvc}
\begin{equation}
\label{Eq:mech-r}
r_{\rm mech}^2 
= \frac{\int d^3r\,r^2\,\biggl(\frac23\,s(r)+p(r)\biggr)}
       {\int d^3r\,\biggl(\frac23\,s(r)+p(r)\biggr)} 
= \frac{6\, D(0)}{\int_{-\infty}^0dt\,D(t)}\,.
\end{equation}
Interestingly, this is an ``anti-derivative'' of a form factor 
as opposed to the electric mean square charge radius defined in
terms of the derivative of the electric form factor at $t=0$.
With this definition the proton and neutron have the same
radius (up to small isospin violating effects). Another
advantage is that the (isovector component of the) electric 
mean square charge radius diverges in the chiral limit 
which makes it an inadequate proxy for the proton size
in the chiral limit, while the mechanical radius 
in Eq.~(\ref{Eq:mech-r}) remains finite in
the chiral limit \cite{Polyakov:2018zvc}.
The mechanical radius of the proton is predicted to be somewhat smaller 
than its charge radius in soliton models at the physical value of the 
pion mass \cite{Goeke:2007fp,Cebulla:2007ei}. 
In quark models both radii become equal when one takes the 
non-relativistic limit \cite{Neubelt:2019sou,Lorce:2022cle}.

An immediate consequence of the positive-definite nature of the
normal force per unit area $\frac23\,s(r)+p(r)$ in 
Eq.~(\ref{Eq:normal-force-positivity}), is that the $D$-term 
$D=D(0)$ is negative \cite{Perevalova:2016dln}. This has been
confirmed in model and lattice QCD calculations, see e.g.\
\cite{Goeke:2007fp,Cebulla:2007ei,Kim:2012ts,Jung:2014jja,Neubelt:2019sou,Perevalova:2016dln,Neubelt:2019sou,Lorce:2022cle} and the
review \cite{Polyakov:2018zvc}. The behaviour of the 
EMT spatial distributions at large-$r$ is dictated by the
behavior of the GFFs at small $t$ which can be studied in 
chiral perturbation theory \cite{Alharazin:2020yjv}.
This allows one to derive a model-independent bound 
formulated in terms of a low-energy constant.
According to this bound the $D$-term of the nucleon is 
negative and $D\le -0.20\pm 0.02$ \cite{Gegelia:2021wnj}.
\subsubsection{D-term and long range forces}
Among the open questions in theory is the issue of how to define the
$D$-term in the presence of long-range forces such as the electromagnetic interaction. It was shown
in a classical model that the $D(t)$ of the proton diverges for $t\to0$ 
like $1/\sqrt{-t}$ when QED effects are included \cite{Varma:2020crx}. 
The form factor $D(t)$ 
exhibits a divergence for $t\to0$ due to QED effects also for charged
pions \cite{Kubis:1999db}. Similar behavior was observed for $D(t)$ of 
the electron in 1-loop QED calculations \cite{Metz:2021lqv}. 
Also for the H-atom, a bound state of the electromagnetic interaction, does one find conflicting results \cite{Ji:2021mfb,Ji:2022exr}. These findings are not 
entirely surprising as the presence of a massless state (the photon) 
in a theory may have profound consequences. Notice that $D(t)$ is the
only GFF which exhibits a divergence for $t\to 0$ when QED effects are included.
Also this is not surprising given the relation of $D(t)$ to the forces acting in a system. 
The behavior of $D(t)\propto 1/\sqrt{-t}$ at small-$t$ 
is relevant only in the unmeasurable region of very small
$|t|<10^{-3}\rm GeV^2$ such that this is of no practical
concern for experiments \cite{Varma:2020crx}.
However, a satisfactory theoretical definition of the $D$-term 
may require not only the inclusion of electromagnetic forces but also gravitational forces which, 
no matter how weak, are present in every system and are also 
long-range forces \cite{Hudson:2017xug}.
Notice that despite the divergence of $D(t)$ due to QED effects, the accompanying prefactor 
$(\Delta_\mu\Delta_\nu - g_{\mu\nu}\Delta^2)$ ensures that the matrix element 
$\langle p_2|\hat{T}^{q,g}_{\mu\nu}|p_1\rangle$ is overall well-behaving in the $t\to0$ forward limit.

\begin{figure}[ht!] 
\includegraphics[width=0.4\columnwidth]{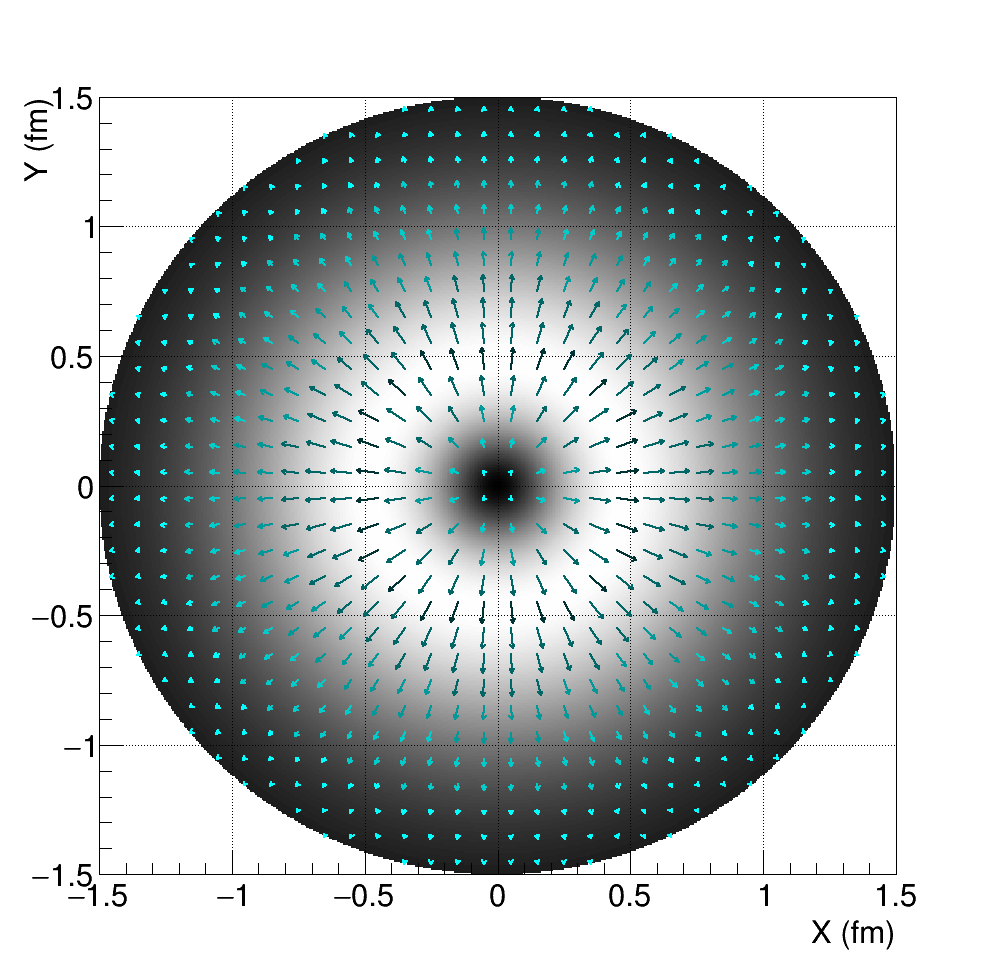}
\includegraphics[width=0.4\columnwidth]{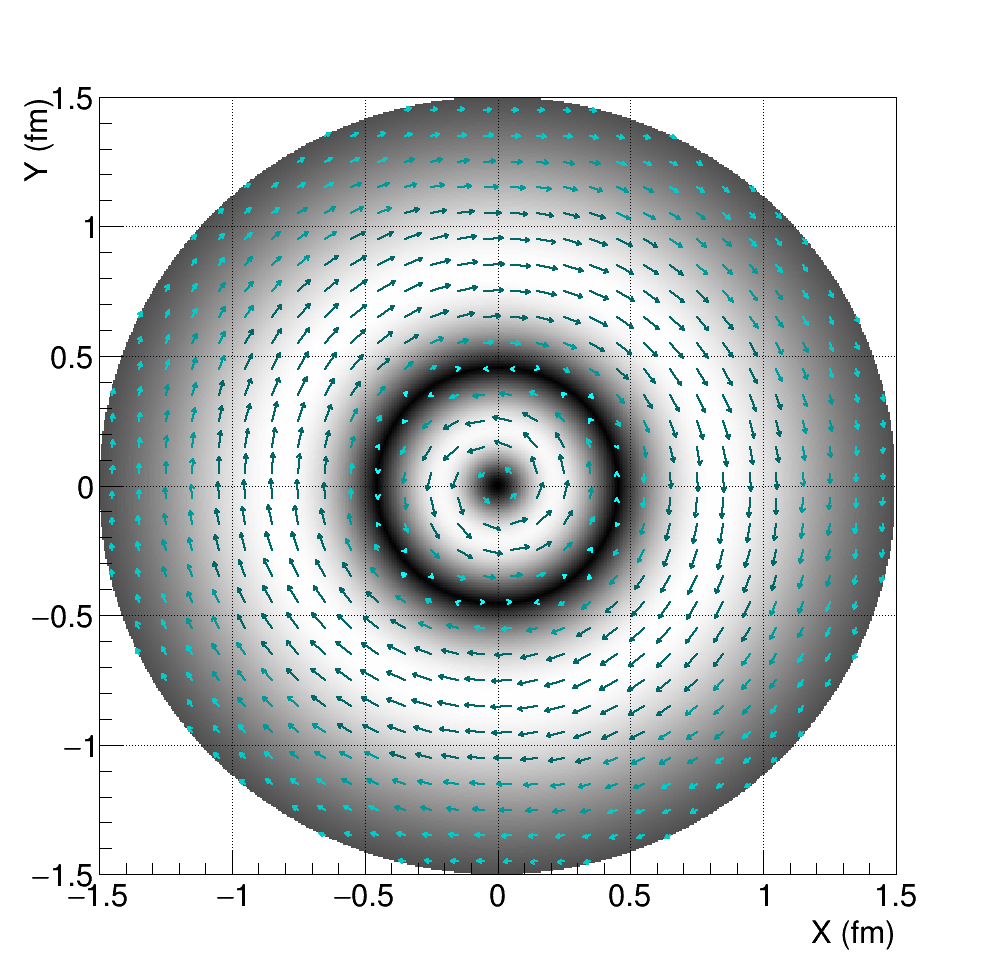}
\caption{\footnotesize Left: Spatial distribution of radial force, which has a positive sign everywhere. Right: Distribution of tangential force, which exhibits a node near a distance $r \approx  0.45$fm from the center, where it also reverses sign as indicated by the direction of the arrows. The lines represent the magnitude of force acting along the orientation of the surface. Note that pressure acts equally on both sides of a hypothetical pressure gauge immersed in the system. A positive magnitude of pressure means that an element of the proton is being pushed on from both direction,. i.e. it is being "squeezed", while a negative magnitude means it is being pulled on from both directions, i.e. it is being "stretched". \cite{Lorce:2018egm,Freese:2021qtb}.}
\label{tangential}
\end{figure} 

The first experimental information from Jefferson Lab experiments allows
one to present  first visualization of the pressure inside the proton.
Using expression for $D^q(t)$ in (\ref{delta}) and the parameterization 
of $\Delta(t)$ in~\cite{Burkert:2021ith} the Fourier transforms 
(\ref{sr}) and (\ref{pr}) can be inverted to determine respectively 
$s^q(r)$ which is also referred to pressure anisotropy, and 
$p^q(r)$ which is also referred to as the isotropic pressure.

Figure~\ref{tangential} shows an example of a tangential pressure distribution inside the proton using parameterizations of $\mathcal{H}(\xi,t)$ and $\Delta(t)$. We stress that these results have been obtained with paramterizations of the kinematic observables $\xi$ and $t$ extrapolated into unmeasured physical territory. The extension of these measurements to higher energies, including into the EIC kinematics domain and the availability of transversely polarized protons, will enable experiments with strong sensitivity to the CFF $\mathcal{E}(\xi, t)$ and $\mathcal{H}(\xi, t)$ and unprecedented kinematic coverage.  


\section{Accessing the Momentum Dependent Structure of the nucleon in Semi-Inclusive Deep Inelastic Scattering}
\label{sec:tmd}
\subsection{Overview}
\begin{figure}
    \centering
    \includegraphics[width=0.95\textwidth]{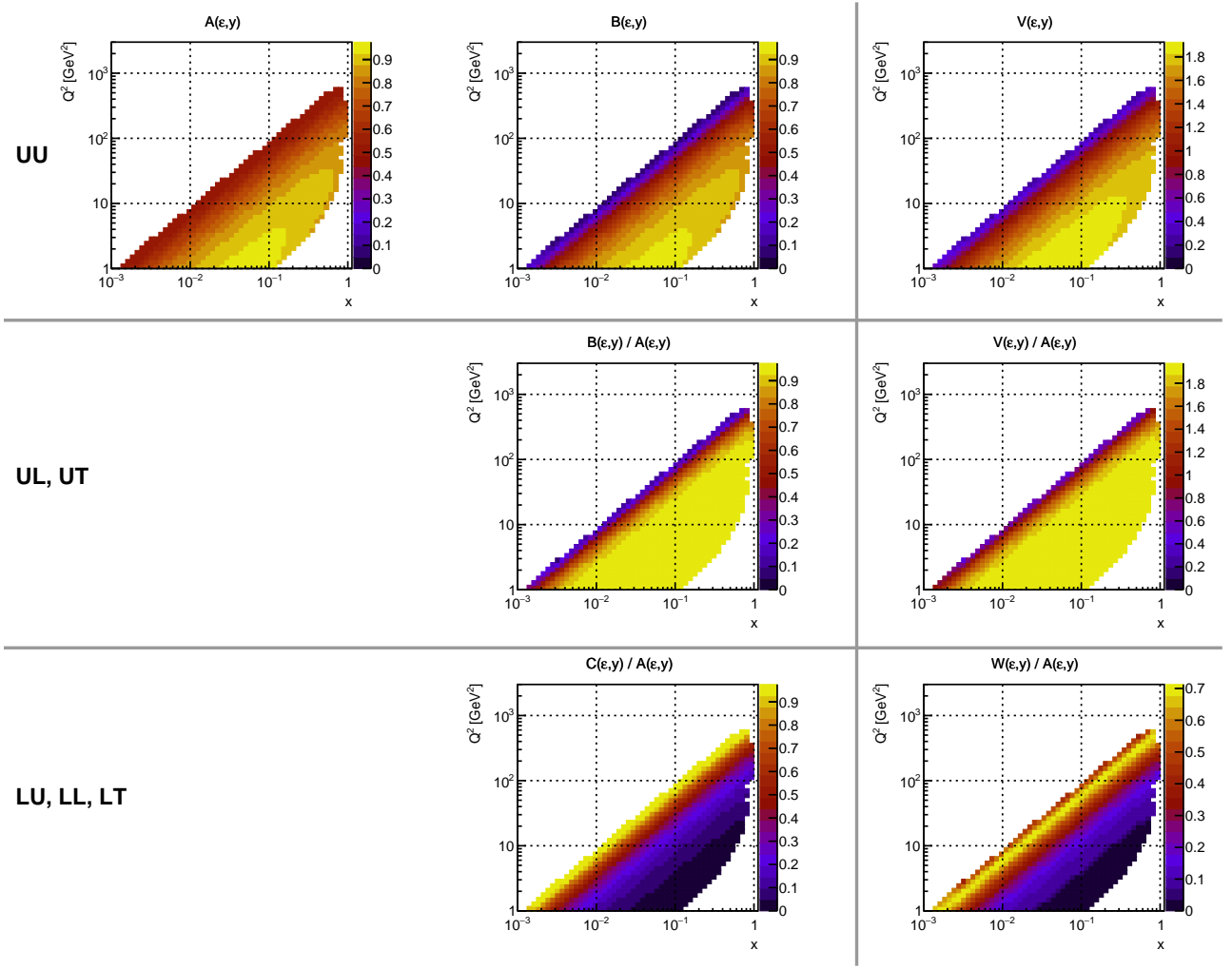}
    \caption{Relative kinematic factors entering beam and target spin asymmetries and polarization independent cross-section for $5\times 41$ beam energy. These so-called depolarization factors are dependent on $y$ and $\epsilon=\nicefrac{(1-y-1/4\gamma^2 y^2)}{(1-y+1/2y^2+1/4\gamma^2y^2)}$ where $\gamma=\nicefrac{2Mx}{Q}$~\cite{Bacchetta:2006tn}. The nomenclature using $A,B,C,V,W$ is taken from~\cite{Gliske:2014wba}.
    They can be approximated by $A\approx (1-y+\frac{1}{2} y^2)$, $B\approx (1-y)$, $C\approx y(1-\frac{1}{2} y)$, $V\approx(2-y)\sqrt{1-y}$ and $W\approx y\sqrt{1-y}$.
    The rows indicate the different beam and target polarization combinations while the first two columns relate to twist-2 quantities and the third column to twist-3 quantities. 
    For the spin independent cross-section, the factor $A$ impacts the transverse momentum independent part, $B$ the asymmetry relating to the Boer-Mulders $h_1^\perp$ function and $V$ the asymmetry relating to the twist-3 FF $D_{1T}$. For the target-spin asymmetries $UL$ and $UT$, the factor $B/A$ impacts the extraction of the transversity, pretzelosity $h_{1T}^\perp$ and worm-gear $h_{1L}^\perp$ asymmetries, whereas $V/A$ impacts the extraction of $h_L$ from $UL$ asymmetries.
    While the factors described so-far become small only for large $y$, the factors entering asymmetries with respect to the beam helicity, $LU$, $LL$ and $LT$ shown in the third row become small for medium and small $y$. Here the $C/A$ factor enters the extraction of the wormgear (LT) and helicity dependent FFs, whereas $W/A$ enters the extraction of the twist-3 PDFs $g_T$ and $e$. As illustrated by the figures, beam helicity dependent asymmetries are significantly suppressed at low values of $y$. This restricts the minimal $Q^2$ value that can be accessed and limits the statistical precision of the measurement.
    \label{fig:depolFactors5_41}}
\end{figure}

\begin{figure}
    \centering
    \includegraphics[width=0.95\textwidth]{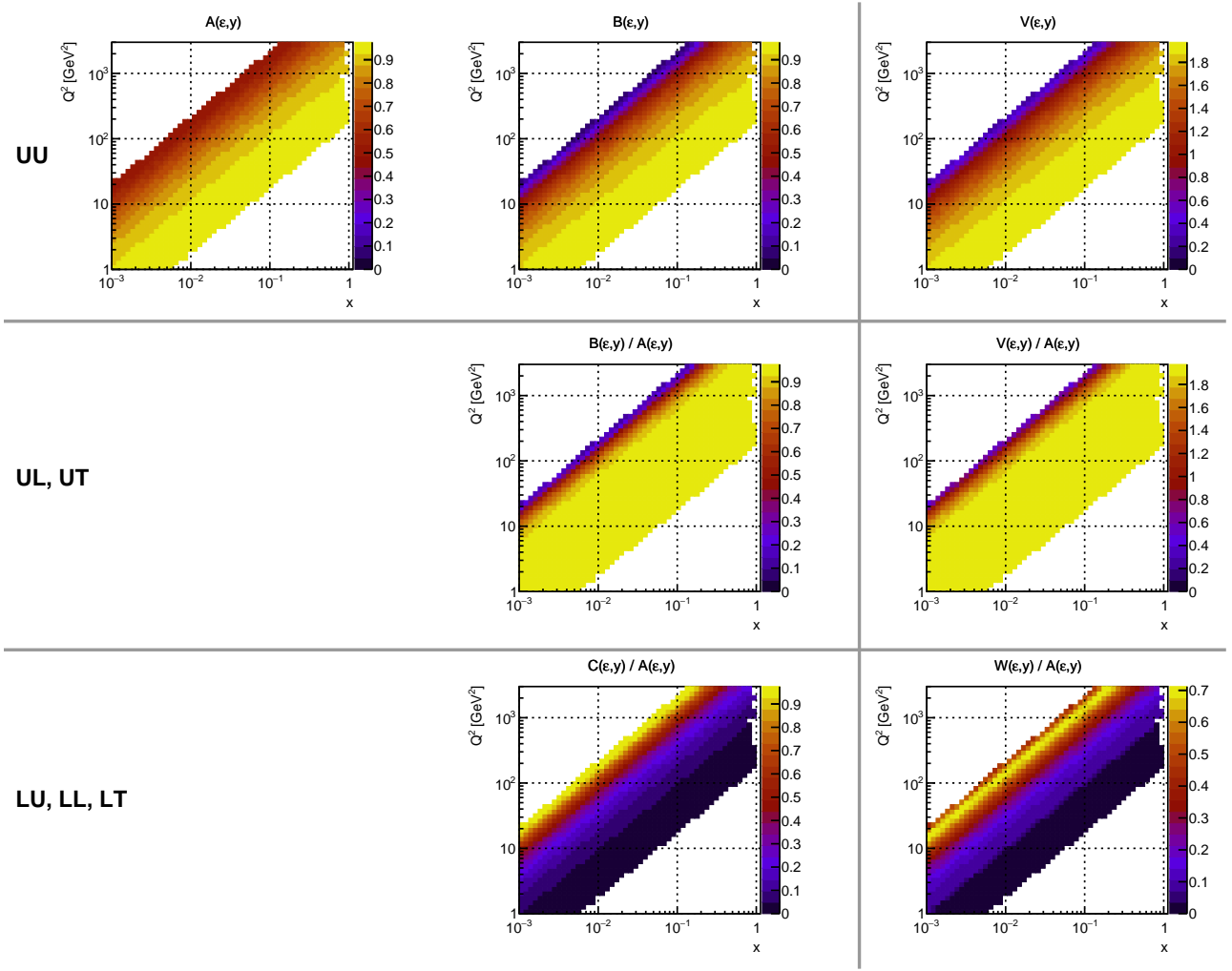}
    \caption{Like Fig.~\ref{fig:depolFactors5_41} but for $18 \times 275$ beam energies. Due to the higher $\sqrt{s}$, the accessible $Q^2$ range for TMDs extracted from beam-helicity dependent asymmetries is higher. At large $\sqrt{s}$ a large fraction of the data is at low $y$, making these measurements even more challenging.} 
    \label{fig:depolFactors18_275}
\end{figure}

\begin{figure}
    \centering
    \includegraphics[width=0.95\textwidth]{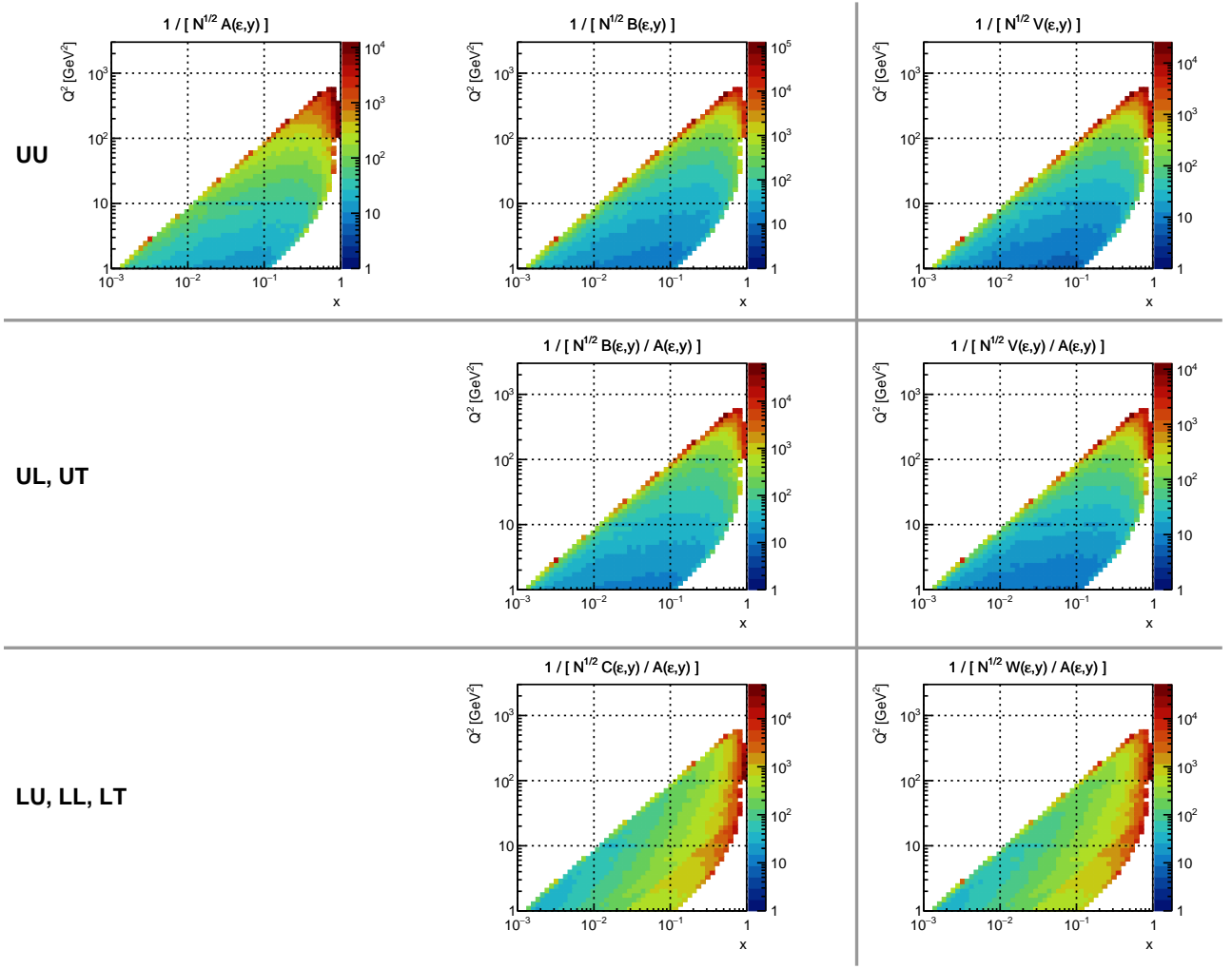}
    \caption{Quantity $(\sqrt{N_i}/ d)^{-1}$ for the $5\times 41$ configuration, where $N_i$ is the normalized count rate in a bin and $d$ the depolarization factor. The quantity is proportional to the relative statistical uncertainty in the respective bin with a proportionality factor of $N^{-1}_\textrm{total}$. This illustrates the relative statistical uncertainties one can reach for TMDs dependent on different polarization factors. 
    \label{fig:depolStatFactors5_41}}
    \end{figure}

\begin{figure}
    \centering
    \includegraphics[width=0.95\textwidth]{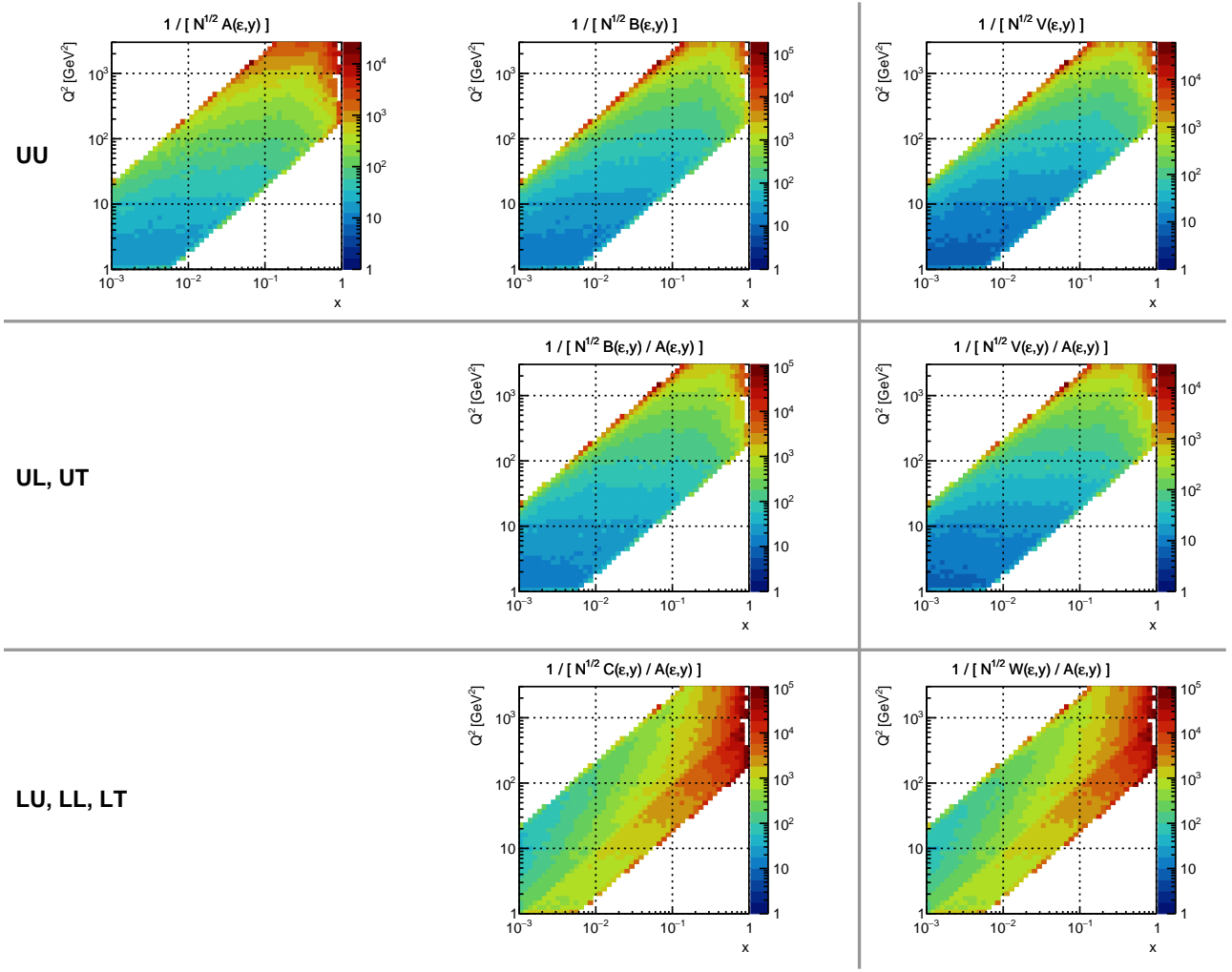}
    \caption{Same as Fig. ~\ref{fig:depolFactors5_41}, but for the $18 \times 275$ beam energy configuration. The relative impact of the depolarization factor on asymmetries dependent on the electron beam helicity is increased due to the phase space distribution of the data. \label{fig:depolStatFactors18_275}}
    \end{figure}

\begin{figure}[ht!]
    \centering
    \includegraphics[width=0.49\textwidth]{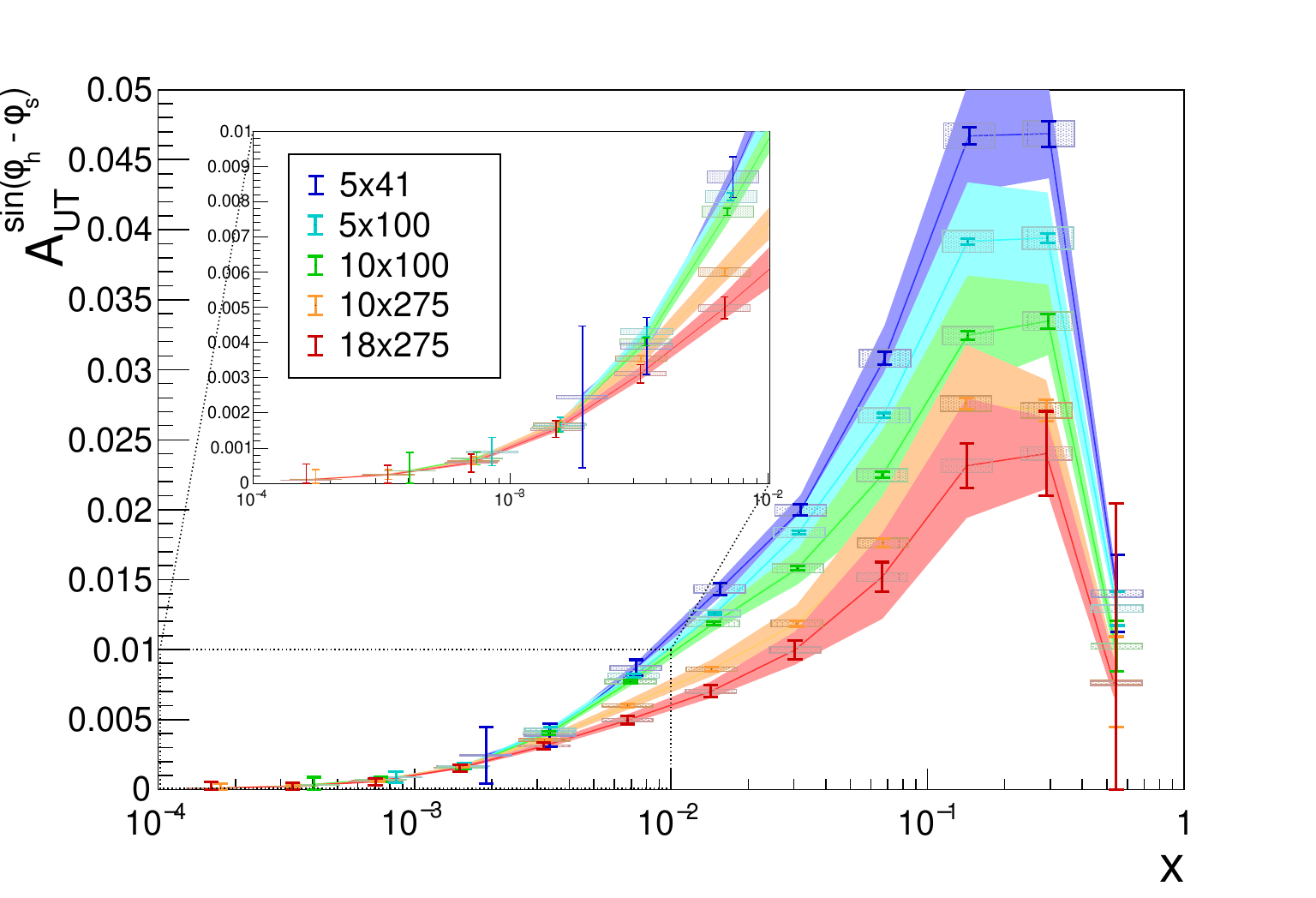}
    \includegraphics[width=0.49\textwidth]{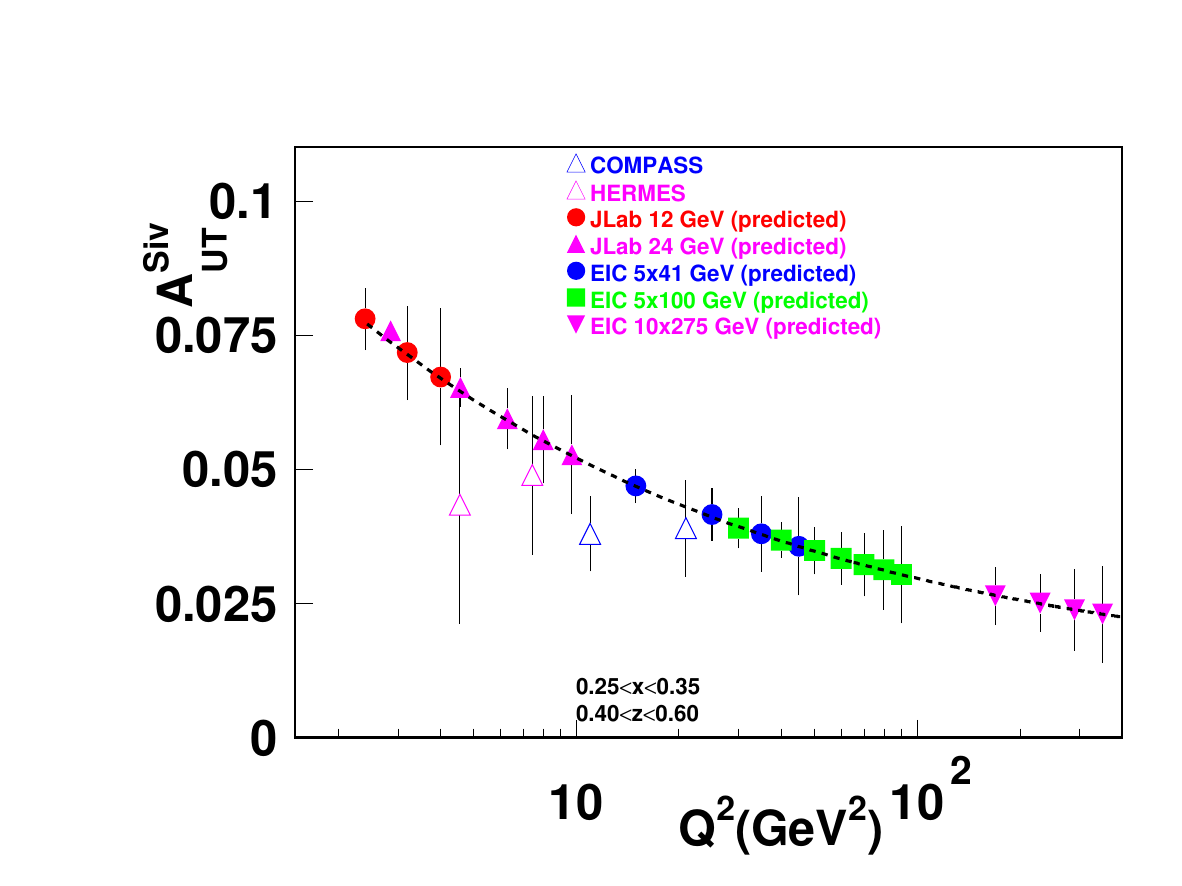}
    \caption{Left: Projected Sivers asymmetry for various EIC run settings. (Example for ATHENA pseudodata), 2\% point-to-point systematic uncertainties assumed. Right: projected Sivers asymmetries for 100 days of data taking at each CM setting with the baseline luminosity vs. $Q^2$ for $0.25 < x < 0.35 $ and $0.4<z< 0.6$ at the luminosity optimized EIC, JLab12 and the proposed JLab24. For the JLab projections, the acceptance of the CLAS detector is used. The proposed SoLID experiment will be able to run at higher luminosity values and is expected to improve on these projections~\cite{Chen:2014psa,SolidUpdateCDR}.
    The drop of the amplitude with $Q^2$ is evident. At the same time the projected uncertainties rise, as the valence quark region is harder to access  at high $Q^2$. A constraint of $y>0.05$ is used for this figure.}
    \label{fig:SiversVsE}
\end{figure}

\begin{figure}[ht]
    \centering
    \includegraphics[trim=0.5cm 10cm 0.5cm 10cm ,clip,width=0.7\textwidth]{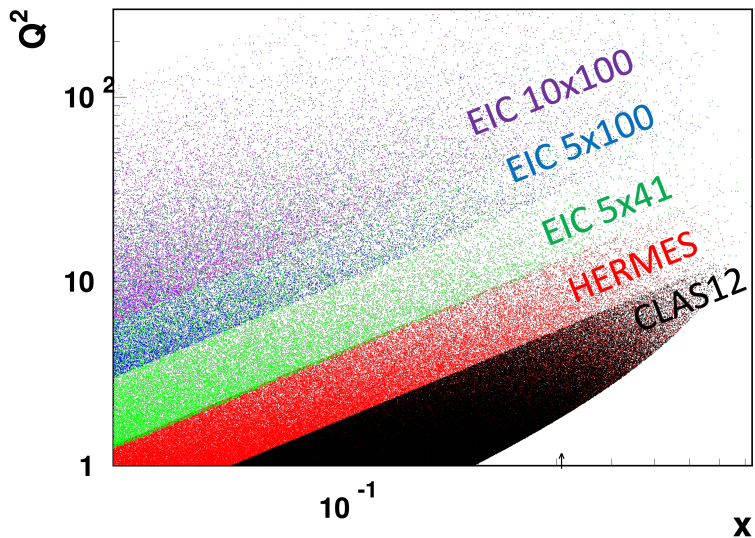}
    \caption{Estimated coverage of JLab12, HERMES and EIC data for different energy configurations. The need to deliver high luminosity for the low and medium energy configurations to fill in the phase space between fixed target experiments and the higher EIC options is obvious. The data are constrained to $y>0.05$.}
    \label{fig:PhaseSpace}
\end{figure}

\begin{figure}[ht]
    \centering
    \includegraphics[width=0.95\textwidth]{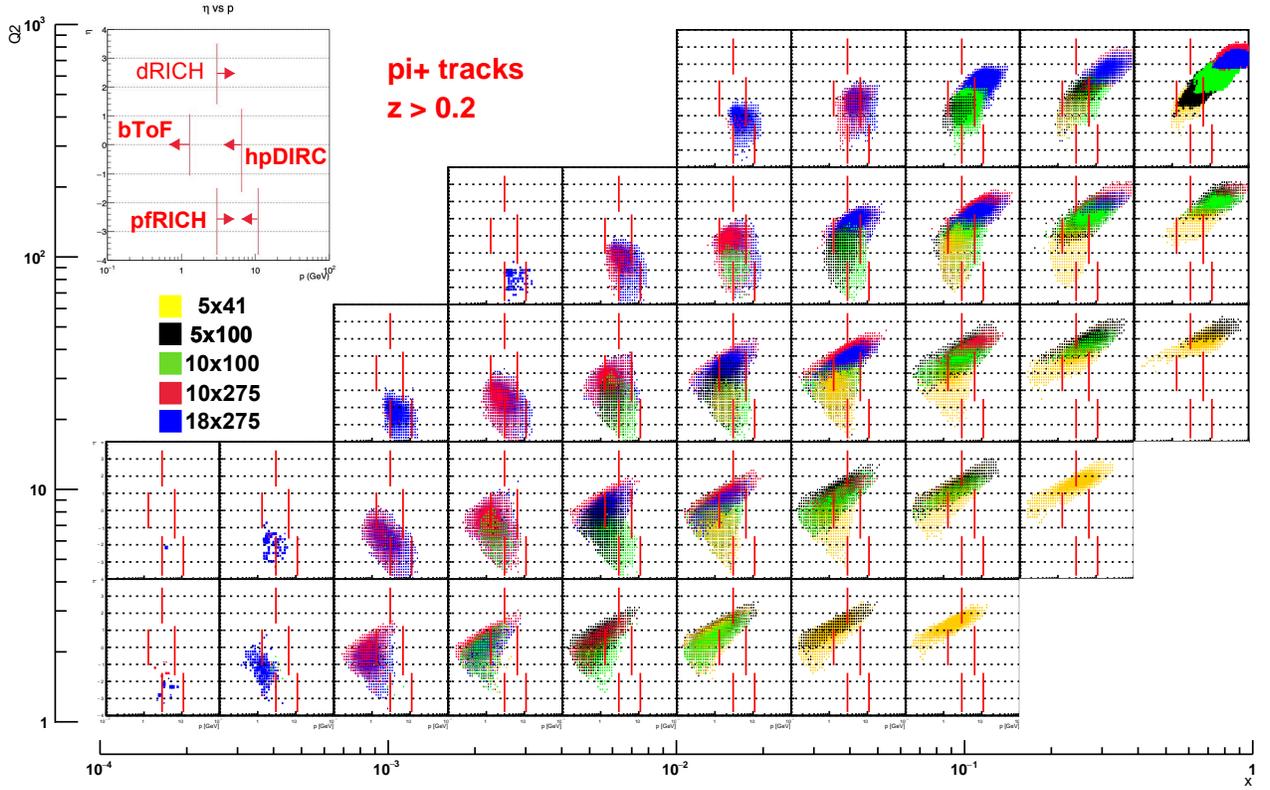}
    \caption{Acceptance of an exemplary EIC detector (here: ATHENA) in laboratory frame $\eta / p$ for various energy configurations and $x,Q^2$ regions. PID limits exemplary for the ATHENA proposal are indicated with red lines. At the highest energies a significant fraction of high $z$ particles is outside the PID range. The horizontal axes are momenta from 0.1 to 100~GeV, and the vertical axes are pseudo-rapidity from -4 to +4.  }
    \label{fig:PIDcoverage}
\end{figure}

Accessing the spin dependent and spin averaged nucleon structure encoded in Transverse Momentum Dependent parton distribution functions (TMD PDFs, or simply TMDs) as well as subleading twist parton distribution functions (twist3 PDFs) in semi-inclusive deep-inelastic scattering~\cite{Anselmino:2020vlp} is a central part of the scientific mission of the EIC~\cite{AbdulKhalek:2021gbh}. This program focuses on an unprecedented investigation of the parton dynamics and correlations at the confinement scale and will benefit substantially by an increased luminosity at medium energies for the following reasons. 
\begin{itemize}
    \item 
Structure functions appearing at sub-leading twist are suppressed by a kinematic factor $1/Q$, which makes data at relatively low and medium $Q^2$ the natural domain for their measurement. 
Similarly, effects from the intrinsic transverse momentum dependence are suppressed at high $Q^2$, when most of the observed transverse momenta are generated perturbatively. As a consequence, the signal of TMDs is naturally diluted at the highest energies. 
However, at the same time $Q^2$ has to be high enough for the applicability of factorization theorems, which makes most fixed target data already challenging. Running the EIC at low- to medium-CM energies might therefore occupy a sweet spot at which non-perturbative and subleading effects are sizeable and current knowledge allows the application of factorization to extract the relevant quantities~\cite{Grewal:2020hoc}. 
The Sivers asymmetry, related to one of the most intriguing parton dynamics which will be discussed below, is shown in Fig.~\ref{fig:SiversVsE} for different EIC energy options, illustrating the rapid fall of the expected TMD signal as higher and higher $Q^2$ is accessed. 

\item At fixed $Q^2$ but lower $\sqrt{s}$, the fractional energy transfer of the virtual photon $y$ is higher, which is helpful for the extraction of TMDs due to the more advantageous kinematic factors for asymmetries sensitive to the helicity of the electron beam and the higher resolution of the reconstruction of kinematic variables as will be described further below. 

The kinematic factor of relevance here, is commonly known as the depolarization factor. It exhibits a strong $y$ dependence and is small for electron beam helicity dependent asymmetries in phase space with low $y$~\cite{Bacchetta:2006tn}. Following the nomenclature from Ref.~\cite{Gliske:2014wba}, we use the symbols $A,B,C,V,W$ for the different depolarization factors. They can be approximated by $A\approx (1-y+\frac{1}{2} y^2)$, $B\approx (1-y)$, $C\approx y(1-\frac{1}{2} y)$, $V\approx(2-y)\sqrt{1-y}$ and $W\approx y\sqrt{1-y}$~\cite{Bacchetta:2006tn,Gliske:2014wba}. For the spin independent cross-section, the factor $A$ impacts the transverse momentum independent part, $B$ the asymmetry relating to the Boer-Mulders $h_1^\perp$ function and $V$ the asymmetry relating to the twist-3 FF $D_{1T}$. For the target-spin asymmetries $UL$ and $UT$, the factor $B/A$ impacts the extraction of the transversity, pretzelosity $h_{1T}^\perp$ and worm-gear $h_{1L}^\perp$ asymmetries, whereas $V/A$ impacts the extraction of $h_L$ from $UL$ asymmetries.
While the factors involved in the unpolarized cross-section and asymmetries with unpolarized electron beam ($A,B,V$ as well as $B/A$ and $V/A$) become small only for large $y$, the factors entering asymmetries with respect to the beam helicity, $LU$, $LL$ and $LT$ become small for medium and small $y$. Here the $C/A$ factor enters the extraction of the wormgear (LT) and helicity dependent FFs, whereas $W/A$ enters the extraction of the twist-3 PDFs $g_T$ and $e$. 


This is demonstrated in Figures~\ref{fig:depolFactors5_41} and~\ref{fig:depolFactors18_275} which show the magnitude of the  depolarization factors for the relevant target and beam spin asymmetries as well as the polarization independent cross-section vs $x$ and $Q^2$. 
As illustrated by the figures, beam helicity dependent asymmetries are significantly suppressed at low values of $y$. This restricts the minimal $Q^2$ value that can be accessed and limits the statistical precision of the measurement.

Figure~\ref{fig:depolFactors5_41} shows the factor for the $5 \times 41$ beam energy combination and Fig.~\ref{fig:depolFactors18_275} for the $18 \times 275$ combination. 
As discussed, the combinations $C/A$ and $W/A$ are suppressed at low $y$ which has a significant impact at larger $\sqrt{s}$. The factor $C/A$ appears in front of the wormgear PDF in the $A_{LT}$ asymmetry and the factor $W/A$ in front of $e$ and $g_T$ twist-3 asymmetries in $A_{LU}$ and $A_{LT}$ asymmetries, where the subscript indicate beam and target polarizations as customary. Figures~\ref{fig:depolStatFactors5_41} and~\ref{fig:depolStatFactors18_275} show the impact on the depolarization factors on the expected statistical uncertainties vs. $x$ and $Q^2$. 

Furthermore, at low $y$ the reconstruction of the relevant kinematics in the Breit-frame suffers from low resolution. These issues have been shown to be significantly improved using the hadronic final state as input to ML/AI methods~\cite{Pecar:2022vuo} or translating the kinematics into the lab-frame~\cite{Gao:2022bzi}. However, even with these improvements, larger $y$ still offers advantages in the resolution that can be reached.

\item To map out the structure of the nucleon encoded in TMDs and twist3 PDFs, high precision, multi-dimensional measurements are needed, which requires very high statistics. For our understanding of the evolution and proper domain of these objects, it is essential to cover an extended kinematic phase space region connecting the future collider to the ongoing fixed-target precision measurements, {\it e.g.} by the JLab experiments. Figure~\ref{fig:PhaseSpace} shows the estimated phase space covered by the existing JLab12 program compared to the lowest and highest EIC energy options. 
\item Finally, intermediate energies have an advantage for a SIDIS program, as its foremost detector requirements are excellent tracking and particle identification. The most significant signals are expected for particles that carry a large momentum fraction $z$ of the fragmenting quark, as these particles are most closely connected to the original quark properties. As illustrated in Fig.~\ref{fig:PIDcoverage}, at intermediate EIC energies, all particles that are detected at mid-rapidity are within the momentum acceptance range of the reference detectors. This is not necessarily true for the highest energies, when particle identification within the typical EIC detector dimensions becomes challenging.
\end{itemize}

The remainder of this section is organized as follows. Section~\ref{sec:twist3} will discuss the physics case for twist-3 observables, Sec.~\ref{subsec:tmd} will give a short overview of the TMD framework and impact studies for unpolarized and Sivers TMD, which were identified as golden channels in the Yellow Report. This section will also briefly discuss TMDs in medium. Finally, Sec.~\ref{sec:jets} will introduce the case for jet physics at intermediate energies and  high luminosity.
Radiative corrections might complicate the picture, as the impact on cross-sections and asymmetries can be sizable, depending on the kinematic regime. The interplay between radiative corrections and TMD extraction is still very much under investigation with recent studies~\cite{Akushevich:2019mbz,Liu:2021jfp} showing potential significant effects on the angular reconstruction for TMDs in certain parts of the phase space. However, as those studies are still in their initial stages, these effects are not considered for the studies shown in this section.


\subsection{Accessing Quark-Gluon Correlations at sub-leading Twist}
\label{sec:twist3}
The interest for contributions that are suppressed by factors of $(M/Q)^{t-2}$ has recently grown with the possibility to access them in low-energy experiments, such as HERMES and CLAS. Moderate $Q^2$ values at EIC will offer unique opportunities for precision analyses of higher-twist distribution functions. Such PDFs are often associated to multi-parton correlations as, to some extent, the operator that defines such objects is made of quarks and gluon fields. Such operators are almost unexplored by phenomenology~\cite{Efremov:2002ut,Accardi:2009au,Sato:2016tuz,Aschenauer:2015ndk,Boglione:2015zyc,Anselmino:2020vlp}.  As argued below, the physics of twist-$3$ distributions is broader than the already important quark-gluon-quark interaction, whose third Mellin moments receive an interpretation in terms of forces~\cite{Burkardt:2008ps}.

\begin{figure}
    \centering
    \includegraphics[width=1.0\textwidth]{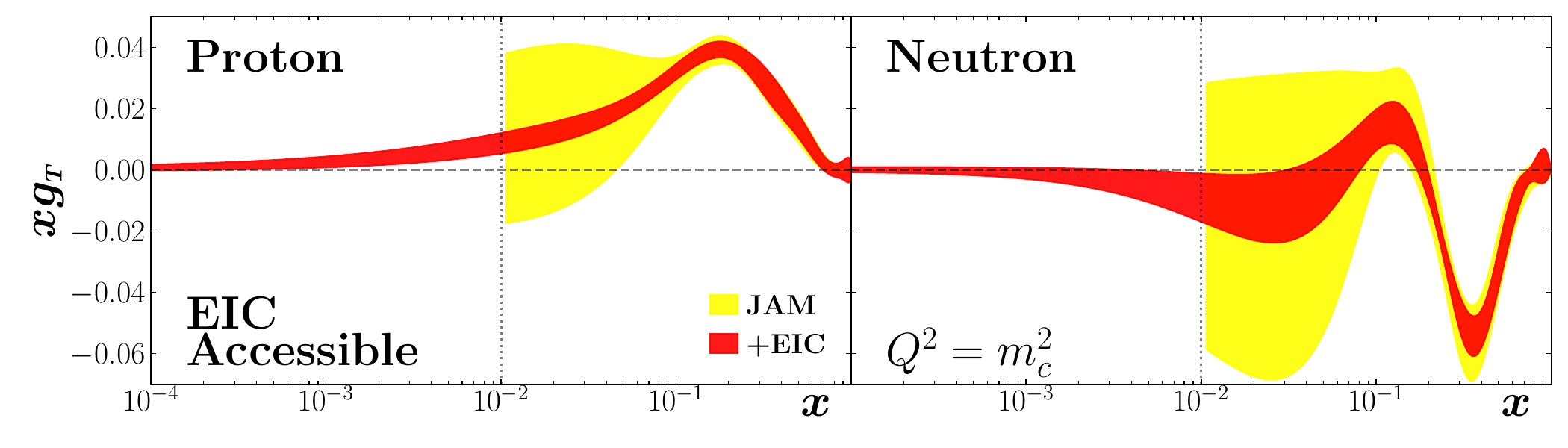}
    \caption{Impact of EIC data with high luminosity at low/medium energies on $g_T$ extraction. The improvement at high x is moderate (but not zero) due to pre-existing data. This extraction uses data at $18\times 275$, $10 \times 100$, $5 \times 100$ and $5 \times 41$, assuming an integrated luminosity of $10$fb$^{-1}$ at 18x275 and the other energies scaled according to their relative instantaneous luminosities. Figure produced for the Yellow Report~\cite{AbdulKhalek:2021gbh}.
\label{fig:gT}}
   
\end{figure}

A well-known example of higher-twist objects is the twist-$3$ contribution to the axial-vector matrix element, $g_T$. The latter can be expressed in terms of a leading-twist distribution through the  Wandzura-Wilczek relation, and a genuine twist-$3$ contribution. Data have shown that the genuine term is not necessarily small~\cite{Accardi:2009au,Sato:2016tuz}. In the Yellow Report for the EIC, the access to $g_T$ through double-spin asymmetry $A_{LT}$ in inclusive DIS has been proposed as the golden channel towards the study of multi-parton correlations. It was shown that the impact on the uncertainty, based on the previous JAM analyses, is expected to be significant. Figure~\ref{fig:gT} shows the impact of the EIC data with high luminosity at low and medium energies on $g_T$ extraction.   

The scalar PDF, $e(x)$, is preeminent in that it relates to diverse aspects of non-perturbative dynamics, such as the scalar charge of the nucleons and an explicit quark-mass term, in addition to the quark-gluon correlations. The scalar charge is particularly interesting in view of the mass decomposition of the proton as it constitutes a unique avenue towards the phenomenological extraction of the scalar condensate~\cite{Efremov:2002qh}.  While there exist semi-phenomenological approaches to the determination of the pion-nucleon sigma-term, {\it e.g.}~\cite{Alarcon:2011zs,Hoferichter:2016ocj}, the twist-$3$ $e(x)$ can provide a determination that is minimally biased by the underlying theoretical assumptions. Some model dependence is, based on our current understanding, inevitable, since the extraction of the sigma requires knowledge of $e(x)$ in particular down to $x=0$, which is not experimentally accessible. 
The access to the scalar PDF through longitudinal beam-spin asymmetries in (dihadron) SIDIS~\cite{Bacchetta:2003vn} was proposed as a silver channel in the Yellow Report. Up to date, the scalar PDF has been accessed at JLab, in CLAS~\cite{CLAS:2020igs} and CLAS12~\cite{Hayward:2021psm}, for low values of $Q^2$ and $x$ ranging from $0.1-0.5$, leading to the first point-by-point phenomenological extraction~\cite{Courtoy:2022kca}.
While the parameterization of $e(x)$ is still a work in progress, the impact from the EIC was shown to be significant thanks to the broad kinematical reach. The $x$ range will be extended towards small-$x$ values, in the region relevant for the evaluation of the sum rules -- such as the relation to the scalar charge. The $Q^2$ range, spanning a broad window of mid-$Q^2$ values, will allow analyses that account for QCD evolution effects on each contribution. EIC thus represents a unique opportunity to expand the curent exploratory studies towards global QCD analyses of the rich phenomenology of higher-twist distribution functions.

\begin{figure}[ht]
\centering
\includegraphics[width=0.9\textwidth]{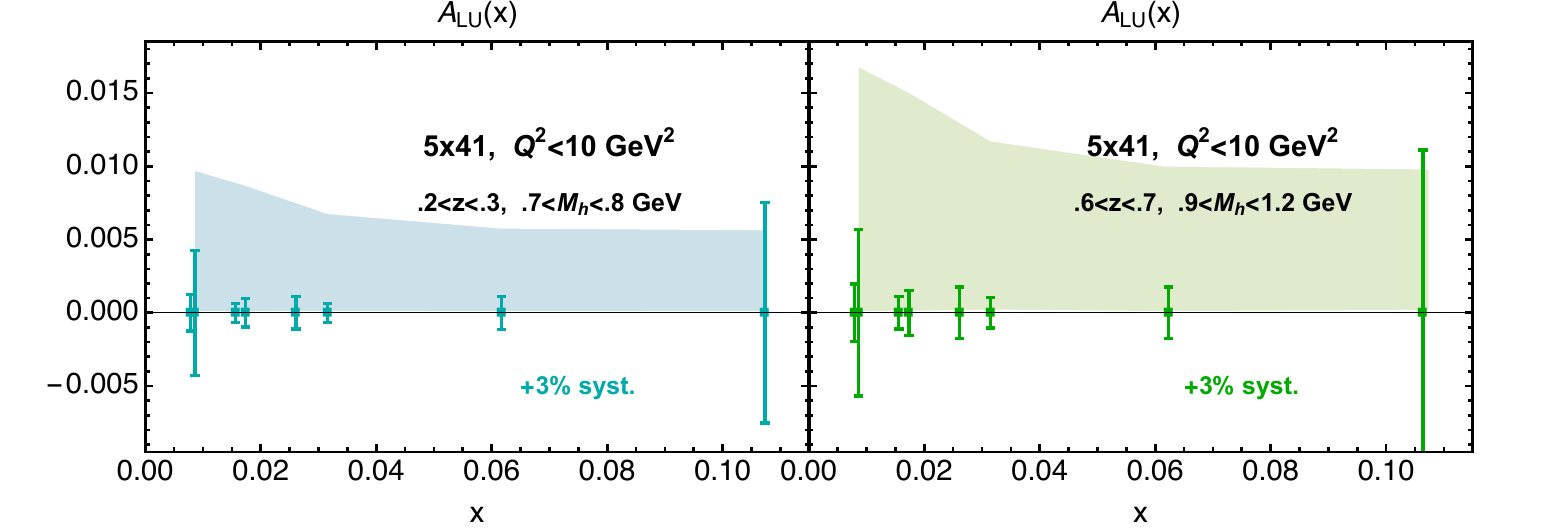}
\includegraphics[width=0.9\textwidth]{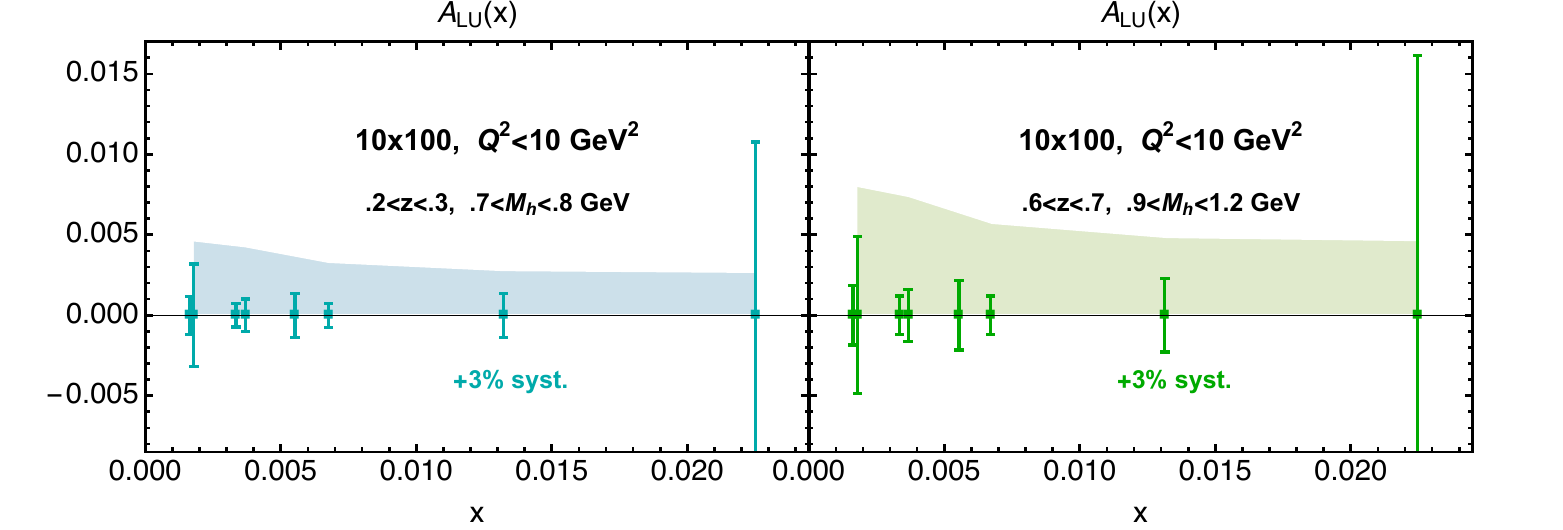}
\caption{Beam Spin Asymmetry in semi-inclusive di-hadron production. Predictions corresponding to $Q^2=1$ GeV$^2$ based on the di-hadron fragmentation functions of Ref.~\cite{Radici:2015mwa}, low-energy models for the twist-$3$ PDF $e(x)$ (see text) and MSTW08 for the unpolarized PDF at LO. Figure is  taken from the Yellow Report~\cite{AbdulKhalek:2021gbh}. The twist-3 fragmentation is neglected. The upper and lower panel show two different energy configuration ; the left (blue) and right (green) plots correspond, respectively, to the fragmentation kinematics of ($0.2<z<0.3$,\, $0.7<M_h<0.8$ GeV) and ($0.6<z<0.7$,\, $0.9<M_h<1.2$ GeV). The bands give the envelope of the model projections discussed in the text folded with the uncertainty of the interference fragmentation function. The projected statistical uncertainties are plotted at zero and correspond to 10~fb$^{-1}$ at each CM setting. This illustrates that the data at lower $\sqrt{s}$ will have a larger impact on constraining $e(x)$. Furthermore, the $Q^2< 10$ GeV$^2$ data, where the signal is still expected sizable, is restricted to low $x$ for large $\sqrt{s}$, where in turn $e(x)$ is expected to be small.}
\label{fig:ex}
\end{figure}

In Fig.~\ref{fig:ex} the theoretical predictions are shown for the contribution of $e^a(x)$ to the beam spin asymmetry in semi-inclusive di-hadron production in
the collinear framework for two different center of mass energies, showing larger projected asymmetries for lower energies as expected. This asymmetry receives a contribution not only from $e^a(x)$ but also from a term involving a twist-3 di-hadron fragmentation function together with $f_1^a(x)$~\cite{Bacchetta:2003vn}. The latter has not been considered here~\cite{Courtoy:2022kca}. 
The uncertainties in Fig.~\ref{fig:ex} come from the envelope of the uncertainties on the interference fragmentation function~\cite{Radici:2015mwa} and two models for $e^a(x)$, the light-front constituent quark model~\cite{Pasquini:2018oyz} and model of the mass-term contribution to $e^a(x)$ with an assumed constituent quark mass of $~300$ MeV and the unpolarized PDF from MSTW08LO. All PDFs and fragmentation functions are taken at $Q^2=1$ GeV$^2$ and the projected uncertainties for the EIC are shown only for $Q^2$ values smaller than $10$ GeV$^2$.

As the leading twist analysis addressed further below, all higher-twist analyses will rely on the possibility to separate the contributions of the various flavors from different observables, and mostly from different targets. In particular, deuteron and $^3$He nuclei will provide effective neutron targets to complement the proton data. 

The phenomenological efforts can be paired with the progress made from the lattice~\cite{Lin:2017snn,Constantinou:2020hdm}. Moments of higher-twist distributions have been determined on the lattice~\cite{Bhattacharya:2021uob}, frameworks for quasi-PDFs are being studied as well~\cite{Braun:2021gvv}.

Beyond the collinear twist-$3$ mentioned above, there is a plethora of higher-twist TMDs that could be studied at the EIC. Moreover, the second IR will grant us the opportunity to explore the relations between twist-$3$ collinear PDFs and twist-$2$ TMDs, the understanding of which is key for the interpretation of low-energy dynamics.

\subsection{Measurements of TMDs}
\label{subsec:tmd}
The lepton-hadron semi-inclusive deep inelastic scattering (SIDIS) at the EIC will provide excellent opportunities to probe the confined motion of quarks and gluons inside the colliding hadron, which are encoded in the transverse momentum dependent parton distribution functions (TMD PDFs, or simply, TMDs).  With the scattered lepton and an observed hadron (or jet) with sensitivity to transverse momentum in the final-state, SIDIS provides not only a hard scale $Q \gg \Lambda_{\rm QCD} $ from the virtuality of the exchanged virtual photon to localize an active quark or gluon inside the colliding hadron, but also a natural ``soft'' scale from the momentum imbalance between the observed lepton and hadron in the final-state, which is sensitive to the transverse momentum of the active quark or gluon.  

With the one-photon approximation, the ``soft'' scale is the transverse momentum of the observed hadron in the photon-hadron (or the Breit) frame, ${\bf P}_{h_T} \gtrsim \Lambda_{\rm QCD}$. When $Q\gg |{\bf P}_{h_T}|$, the unpolarized SIDIS cross section can be factorized as \cite{Bacchetta:2006tn},
\begin{equation}
\frac{d\sigma^{\rm SIDIS}}{dx_B dQ^2 d^2{\bf P}_{h_T}}
\propto x \sum_i e_i^2 \int d^2{\bf p}_T\, d^2{\bf k}_T\, 
\delta^{(2)}({\bf p}_T - {\bf k}_T - {\bf P}_{h_T}/z) \,
\omega_i({\bf p}_T,{\bf k}_T) f_{i}(x,p_T^2) D_{h/i}(z,k_T^2)
\equiv {\cal C}\left[\omega f D\right]\, ,
\label{e.sidis}
\end{equation}
which provides the direct access to the TMD PDFs, $f_{i}(x,p_T^2)$ of flavor $i$ and transverse momentum $p_T^2\equiv {\bf p}_T^2$, and TMD fragmentation functions (FFs), $D_{h/i}(x,k_T^2)$ for a parton of flavor $i$ and transverse momentum $k_T^2\equiv {\bf k}_T^2$, to evolve into the observed hadron $h$ of transverse momentum $P_{h_T}$ in this photon-hadron frame. In Eq.~(\ref{e.sidis}), the 
$\omega_i({\bf p}_T,{\bf k}_T)$ is a known function depending on the kinematics, the type of TMDs and corresponding angles between the parton transverse momenta.  

With many more TMDs than PDFs, it will be possible to learn much more on QCD dynamics that holds the quarks and gluons together to form the bound hadron, despite being harder to extract and separate these TMDs from experimental data. On the other hand, with a good detector able to cover the angle distribution between two well-defined planes, the leptonic plane determined by the colliding and scattered leptons, and the hadronic plane defined by the colliding and observed hadrons, SIDIS measurements at the EIC will allow the extraction of various TMDs by evaluating independent angular modulations of the angle distribution between the two planes as well as the distribution between the hadron spin vector and one of the planes.

\subsubsection{Impact on the understanding of TMD factorization and applicability to fixed target data}
\begin{figure}[ht]
\begin{center}
\includegraphics[width=1.0\textwidth]{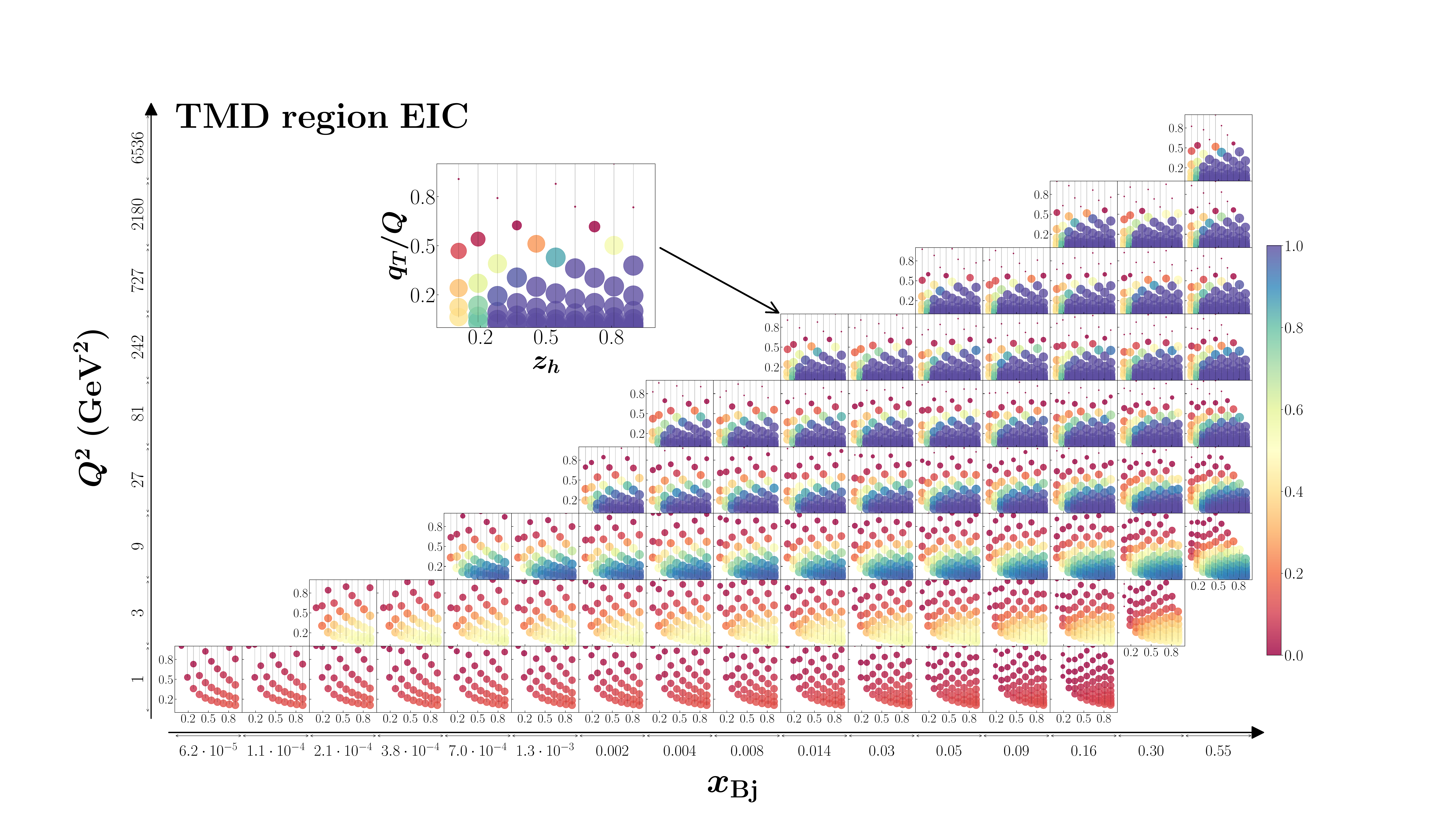}
\caption{\label{fig:affinity} 
TMD affinity for EIC kinematics. Bin centers are located in the points corresponding to the bin averaged values of $x_b$ and $Q^2$, and in each of these bins various values of $z_h$ and $q_T/Q$ can be measured. In each bin of fixed $z_h$ and $q_T/Q$, the affinity is indicated by a dot with size proportional to the corresponding affinity value. The affinity is color coded according to the scheme on the right of the panels: red (and smaller) symbols correspond to low TMD affinity, while dark blue (and larger) symbols correspond to high TMD affinity. The plot is from Ref.~\cite{Boglione:2022gpv}}

\end{center}
\end{figure}

 The TMD factorization formula Eq~\ref{e.sidis}  receives corrections which enter in terms of powers of $\delta\sim P_{hT}/z/Q$. Identifying the domain of applicability of TMD factorization is not trivial \cite{Boglione:2016bph}. In recent analyses, usually the choice $\delta<0.25$ is adopted, at least for high $Q$ \cite{Bacchetta:2017gcc,Scimemi:2017etj,Scimemi:2019cmh,Bacchetta:2019sam}. These restrictions reduce the significance of a large amount of existing measurements, in particular a majority of data from existing fixed target experiments. 
Figure~\ref{fig:affinity} illustrates this issue by showing the results of Ref.~\cite{Boglione:2022gpv} where the regions of pion production in SIDIS at the EIC are studied using results of Ref.~\cite{Boglione:2019nwk}.  The so-called affinity to TMD factorization region (i.e. the probability that the data can be described by TMD factorization) is calculated for each bin of the EIC measurements. The affinity represents the probability of the bin to belong to TMD factorization region and spans from 0\% to 100\%, indicated by color and symbol size in the figure. One can see from Fig.~\ref{fig:affinity} that only at relatively high $z$ and $P_{hT}$ (and relatively large $x$ and $Q^2$) corrections to the TMD factorization description are expected to be negligible. 
 The reach of the EIC data into other regions, will be important for the study the connections to other types of factorization, for instance the collinear factorization or the region accessed by fixed target experiments, where sizable corrections to the current TMD formalism are expected.  Comparing this figure with the reach of the different energy option shown in Fig.~\ref{fig:PIDcoverage}, it can be seen that intermediate beam energy option such as $10\times 100$ GeV$^2$ operate largely in a region where TMD factorization holds, but also contain phase space in the transition region towards other QCD regimes. The flexibility to go from one regime of factorization to the other will be a crucial ingredient in our understanding of QCD, and in the interpretation of the vast amount of fixed target data, which has a low TMD affinity.

\begin{figure}[ht]
\begin{center}
\includegraphics[width=0.48\textwidth]{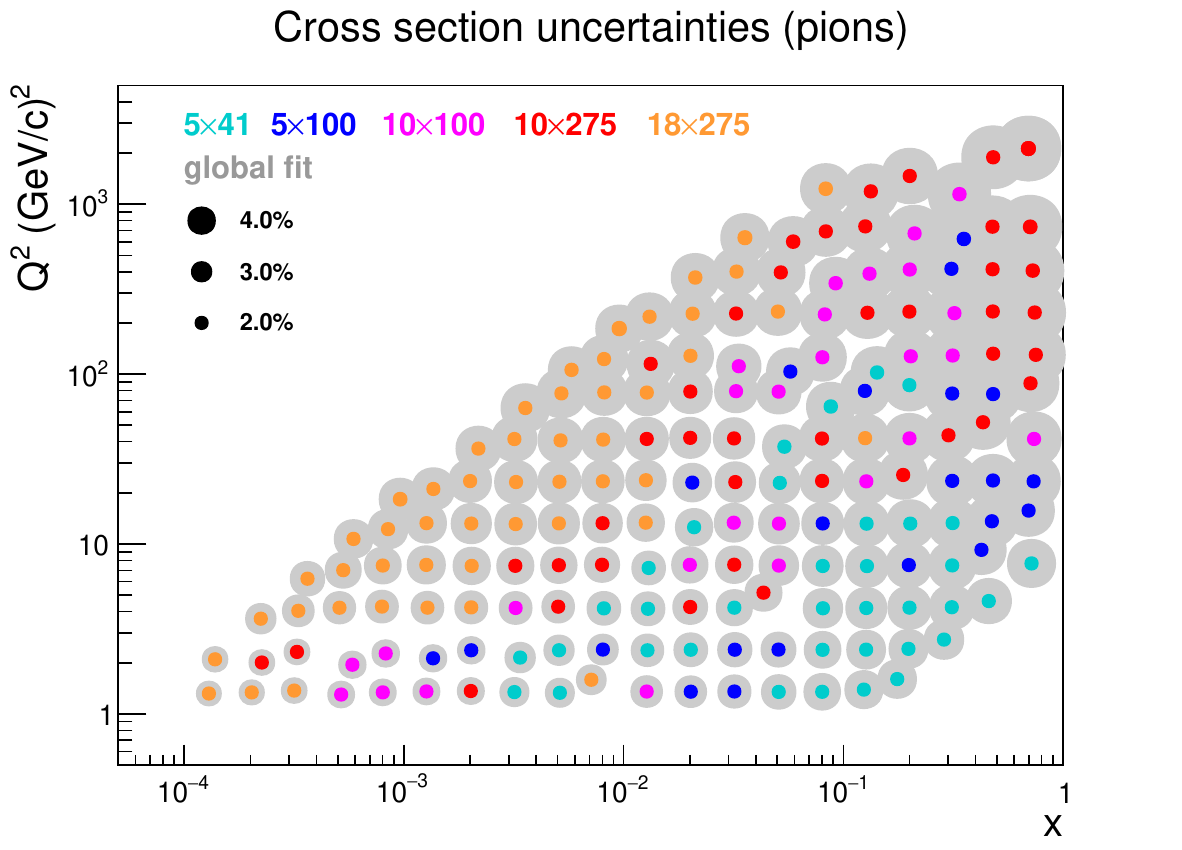}
\includegraphics[width=0.4\textwidth]{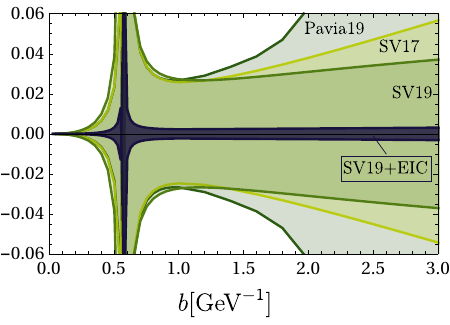}
\caption{\label{fig:tmdImpact}Left: Impact on unpolarized TMD measurements integrated within $0<q_T/Q<1.0, z>0.2$ , figure from Athena Proposal. Fit based on {\texttt{P17}}. The color-code shows the datasets with the highest impact at a given $x,Q^2$ point. The assumed systematic uncertainty of 2\% point-to-point is dominating. However, the extraction of a specific point in $b$ is sensitive to the collected statistics as shown in the right plot.
Right: Impact of the EIC data on the extraction of the CS kernel as function of $b$ (GeV$^{-1}$) at $\mu=2$~GeV using \texttt{SV19} as a baseline compared to several other global extractions not using EIC data. Figure from the Yellow Report~\cite{AbdulKhalek:2021gbh}.}
\end{center}
\end{figure}
\subsubsection{Impact on TMD PDF extraction}

The theoretical description of TMDs has been extensively studied in coordinate space labeled by $b$ as  the conjugate variable of transverse momentum. In the large $b$ region (small $q_T\approx p_T/z$), TMDs  are non-perturbative and encode intrinsic properties of hadrons while in  the small $b$, TMDs  are dominated by QCD radiation which is calculable in perturbative QCD. In the latter, TMDs can be connected with their corresponding collinear counterparts such as PDFs and fragmentation functions offering a new venue to constrain collinear distributions using TMD observables. While the experimental data is sensitive to all regions in coordinate space, as discussed above, the relative contribution of each region to the physical observables depends on the kinematics of the final state particles accessible at a given collision energy. Because of this, different collision energies from low to high at high luminosity are needed at the EIC in order to systemically probe TMDs at different regions of coordinate space. In the sections below, we concentrate on the impact on the unpolarized TMD PDFs as well as the Sivers TMD PDF as exemplary cases that would profit from increased precision at moderate energies. 

\subsubsection{The impact study on the unpolarized TMDs}
\label{sec:unpolarized}
The unpolarized TMD distributions and fragmentation functions have been extracted in Refs.~\cite{Scimemi:2017etj,Bacchetta:2017gcc,Scimemi:2019cmh,Bacchetta:2019sam,Bacchetta:2022awv} (\texttt{SV17}, \texttt{PV17}, \texttt{SV19}, \texttt{PV19}, \texttt{MAPTMD22}) with high perturbative accuracy up to NNLO and up to N$^3$LL of TMD logarithmic resummation. The data used in these global analyses includes Drell-Yan and SIDIS processes measured at fixed target experiments  \cite{ZEUS:1995acw,H1:1996muf,Asaturyan:2011mq,HERMES:2012uyd,COMPASS:2013bfs,COMPASS:2017mvk,Ito:1980ev,Moreno:1990sf,E772:1994cpf} at relatively low energies, and the collider measurements at higher energy scales \cite{PHENIX:2018dwt,CDF:1999bpw,CDF:2012brb,D0:1999jba,D0:2007lmg,D0:2010dbl,ATLAS:2014alx,ATLAS:2015iiu,CMS:2011wyd,CMS:2016mwa,LHCb:2015okr,LHCb:2015mad,LHCb:2016fbk}. The span in the resolution scale $Q$ and in observed transverse momentum $q_T$ allows for an extraction of the non-perturbative Collins-Soper kernel (CS-kernel) and the unpolarized TMDs. These extractions demonstrate an agreement between the theory and the experimental measurements. 

The extremely precise LHC measurements at $Q\simeq M_Z$ provide very stringent constraints on the CS-kernel and TMDs in the region of small values of $b$. However, the uncertainty of extractions grows in the region of $b>1$ GeV$^{-1}$ due to the lack of the precise low-$q_T$ data.  The large $b$ region is important for the understanding of the non-perturbative nature of TMDs and the primordial shapes TMDs and CS-kernel. 
In particular for the $Q$ range accessed by intermediate energies, $Q \geq 5-10$ GeV, TMDs are only very poorly constrained.
Low and intermediate energies at the EIC will naturally provide precision data in this kinematic regime as shown below. Predictions from various groups are different in this region, see Ref.~\cite{Vladimirov:2020umg}, and also disagree with the lattice measurements \cite{LatticeParton:2020uhz,Shanahan:2020zxr,Schlemmer:2021aij}. This disagreement is problematic since it points to a limited understanding of the TMD evolution encoded in the CS-kernel, which dictates the evolution properties of all TMDs and describes properties of the QCD vacuum~\cite{Vladimirov:2020umg}. 
The measurements from the EIC  will fill in the gap between the low-energy and high-energy experiments, and will pin down these functions at higher values of $b$ corresponding to lower values of $k_T$. Ultimately, it will help to unravel the 3D nucleon structure in a very wide kinematic region. 

\begin{figure}[b]
\begin{center}
\includegraphics[width=0.95\textwidth]{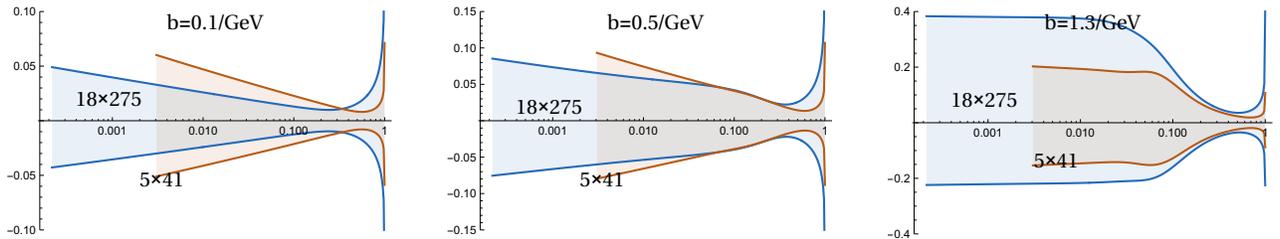}
\caption{\label{fig:uTMDimpact} Comparison of relative uncertainty bands for unpolarized u-quark TMD PDFs at different values of $b$ as a function of $x$. Lighter blue band is the impact of $18\times 275$ data, light brown band is the impact of $5\times 41$ EIC pseudo data. The dataset used for these projections is the same as used for the Yellow Report~\cite{AbdulKhalek:2021gbh}. In particular all energy options use the same integrated luminosity.}
\end{center}
\end{figure}

The unpolarized structure function is the leading contribution to the differential SIDIS cross-section and also serves as the weight for polarized asymmetries. As discussed above, mapping the unpolarized TMD over the full phase space is a also necessary to probe TMD evolution effects which partially cancel in the extraction of spin asymmetries. Therefore, the knowledge of unpolarized TMDs is of paramount importance for the whole momentum tomography program.

To demonstrate the impact, in particular of medium- and low energy data, we consider the \texttt{PV17} and \texttt{SV19}-fits. Figure~\ref{fig:tmdImpact}, left shows the relative impact of the different energy options on the extraction of the \texttt{PV17} based TMD fit. It is evident, that low and medium energies dominate over a wide range of phase space, in particular at intermediate $x-Q^2$. This is even more impressive considering that the impact plot is based on the baseline luminosities.

The estimation of the impact on the nonperturbative parts of the CS-kernel and unpolarized TMDs has been done using the \texttt{SV19}-fit as the baseline. The analysis was performed with the inclusion of EIC pseudo-data (in $5\times 41$, $5\times 100$, $10 \times 100$, $18\times 100$ and $18\times 275$ beam-energy configurations). The pseudo-data, generated by {\sc pythia} \cite{Sjostrand:2006za}, includes expected statistical and estimated systematic uncertainties, for a hand-book detector design with moderate particle identification capability. The estimate for the improvement in the uncertainties for the extraction of the unpolarized TMDs is shown in the right panel in  Fig.~\ref{fig:tmdImpact} exemplary for $f_{1T}^u$. In general, the main impact in the unpolarized sector occurs for the CS-kernel, whose uncertainty reduces by a factor of $\sim 10$. This is only possible  with precise and homogeneous coverage of the $(Q,x,z)$ domain, which can efficiently de-correlate the effects of soft gluon evolution and internal transverse motion. 

Fig. \ref{fig:uTMDimpact} shows the
impact of the same integrated luminosity with the highest, $18\times 275 $, energy configuration and the lowest, $5\times 45$ energy configuration on the extraction of the unpolarized u-quark TMD PDFs at different values of $b$ as a function of $x$.  As expected, the lower energy data has a significant impact to constrain the PDF in the valence quark region for all $b$ and over the majority of the $x$ range at higher values of $b$. 
 This is thanks to the sensitivity to smaller values of $p_T$. Notice that the high energy option has little impact in the valence region, as large $x$ values can only be accessed at large $Q^2$. The combination of low and high energy measurements will have the most homogeneous coverage of the kinematics required for the studies of TMDs.

\subsubsection{The impact study on the Sivers functions}
\label{sec:sivers}
\begin{figure}[htb]
    \centering
    \includegraphics[width=0.45\textwidth]{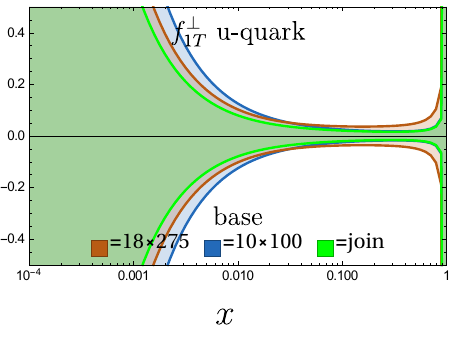}\includegraphics[width=0.45\textwidth]{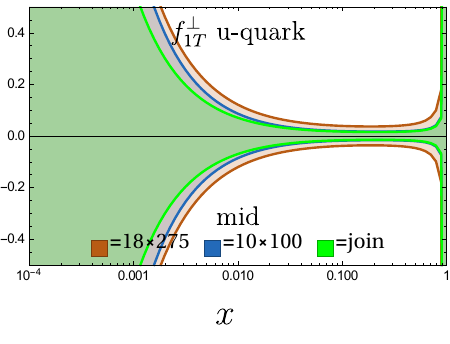}
     \caption{Expected impact on the u-quark  Sivers functions as a function $x$ as obtained from semi-inclusive pion and kaon EIC pseudo-data for $10\times 100$, $18\times 275$) beam-energy configurations and the combined impact. Fit uses pseudodata from the EIC reference detector described in the Yellow Report~\cite{AbdulKhalek:2021gbh} and \texttt{SV19} fit. Left: impact of equal time data taking with the base configuration, right: impact of proposed luminosity increase at low and mid energies.}
    \label{fig:siversu}
\end{figure}

The non-vanishing Sivers asymmetry triggered a lot of interest in the physics community and many groups have performed  extractions of the Sivers functions from the available experimental data~\cite{Anselmino:2005ea,Collins:2005ie,Vogelsang:2005cs, Anselmino:2008sga,Bacchetta:2011gx,Sun:2013hua,Echevarria:2014xaa,Boglione:2018dqd,Luo:2020hki,Cammarota:2020qcw,Bacchetta:2020gko,Echevarria:2020hpy,Bury:2020vhj,Bury:2021sue}. However, currently the global pool of Sivers asymmetry measurements offers a relatively small number of data points that could be consistently analysed using the TMD factorization approach. The future measurements by the EIC will provide a significant amount of new data in a wide and unexplored kinematic region, and thus have a decisive impact in the determination of the Sivers functions. 

To determine the impact of EIC measurements on the Sivers function, the pseudo-data generated by {\sc Pythia-6} \cite{Sjostrand:2006za} was used with a successive reweighing by a phenomenological model for the Sivers and unpolarized structure functions from Ref.~\cite{Anselmino:2008sga}. The pseudo-data for $\pi^\pm$ and $K^\pm$ production in $e+p$ and $e+^3He$ collisions at the highest ($18\times 275$) and the lowest ($5\times 41$) beam-energy configurations were analyzed. The resulting pseudo-data set is about two orders of magnitude larger in comparison with the current data. Performing the fit of the new pseudo-data with the initial set of Sivers functions taken from the global analysis made in Refs.~\cite{Bury:2020vhj,Bury:2021sue} and based on the current SIDIS \cite{HERMES:2009lmz,COMPASS:2012dmt,COMPASS:2016led,COMPASS:2008isr,JeffersonLabHallA:2011ayy} and Drell-Yan \cite{COMPASS:2017jbv,STAR:2015vmv} measurements, a substantial reduction of uncertainties is obtained. The uncertainty bands are reduced by an order of magnitudes for all flavors.

Fig.~\ref{fig:siversu} shows the impact on the uncertainty of the u-quark Sivers function at  $b=0$ GeV$^{-1}$ as a function of $x$. The distribution of impact between $5\times 41$ and $18\times 275$ beam-energy configurations is similar to the unpolarized case. Namely, $5\times 41$ configuration constrains mainly the large-$x$ region, while $18\times 275$ configuration constrains the low-$x$ region. The combined set of the data gives the most homogeneous error reduction. In turn, it reduces significantly uncertainties of the integral characteristics. For example, the integral over Qiu-Sterman function has about $3\%$ uncertainty (in the combined case) versus $6\%$ (for $18\times 275$ case) or $12\%$ (for $5\times 41$ case).
Figure~\ref{fig:SiversVsE} shows the projected experimental uncertainties compared to projections based on the extraction in Ref.~\cite{Bacchetta:2020gko} for more energy options and vs $Q^2$. Intermediate energies are most advantageous, since the expected asymmetries are large while still enough statistics for a multi-dimensional analysis are collected. This is in particular evident when plotting the asymmetries vs $Q^2$ where the drop of the expected asymmetries at high $Q^2$ can be observed as well as the drop of statistics expected from the EIC in the valence region at high $Q^2$.
\subsubsection{TMDs in nuclei}
\label{sec:tmd_nuclei}

QCD multiple scattering in the nuclear medium has been demonstrated to be responsible for the difference between TMDs in bound and free nucleons within a generalized high-twist factorization formalism~\cite{Liang:2008vz} and the dipole model~\cite{Mueller:2016gko,Mueller:2016xoc}. In these models, the scale of the power corrections which modify the relevant distribution for the process is proportional at leading order to $\alpha_s(Q)$, which becomes small at large $Q$, see for instance \cite{Qiu:2003vd,Qiu:2004da}. Thus while the EIC will be capable of performing $e-A$ collisions for a wide range of nuclear targets, a low center of mass energy is optimal for probing nuclear medium modifications to TMDs.

From a phenomenological standpoint, nuclear modifications to collinear PDFs have been performed in Refs.~\cite{Eskola:1998df,deFlorian:2003qf,Hirai:2007sx,Eskola:2007my,Schienbein:2009kk,AtashbarTehrani:2012xh,Khanpour:2016pph,Eskola:2016oht,Walt:2019slu,Kovarik:2015cma,AbdulKhalek:2019mzd,AbdulKhalek:2020yuc} and for the collinear fragmentation function in Ref.~\cite{Sassot:2009sh,Zurita:2021kli}. In these global analyses, the medium modifications to the distributions enter into the non-perturbative parameterizations. In the TMD description, the QCD multiple scattering naturally leads to a broadening of the transverse momentum distributions. Recently, the first extraction of the unpolarized nuclear modified TMDs have been performed in Ref.~\cite{Alrashed:2021csd}. The authors of this paper performed a global analysis at NLO+NNLL to the world set of experimental data from hadron multiplicity production ratio at HERMES~\cite{HERMES:2007plz}, Drell-Yan reactions at Fermilab~\cite{Alde:1990im, NuSea:1999egr} and RHIC~\cite{Leung:2018tql}, as well as $\gamma^*/Z$ production at the LHC~\cite{CMS:2015zlj,ATLAS:2015mwq}. In analogy to the work that has been done in the past, this analysis took the medium modifications to enter into the non-perturbative parameterization of the collinear distributions as well as the parameterization for the non-perturbative Sudakov factor, which controls the broadening of the transverse momentum distribution. Despite the success of work in \cite{Alrashed:2021csd} in describing the world set of experimental data, there are currently few data points which can be used in order to constrain the TMD FFs. While the HERMES measurement of the hadron multiplicity ratio probed a relatively wide kinematic region, the stringent kinematic cuts applied to ensure the data are within the proper TMD region vastly reduces the total number of useful experimental points. Since Semi-Inclusive DIS is sensitive to both the TMD PDFs as well as the TMD FFs, experimental measurements within the broad kinematical reach of EIC at small and medium $Q$ represents the optimal process for probing nuclear modifications to TMDs. 

\subsection{Jet Hadronization Studies}
\label{sec:jets}

Jets are collimated sprays of particles, which are observed in collider experiments. They exhibit a close connection to energetic quarks and gluons that can be produced in hard-scattering processes at the EIC~\cite{Hinderer:2015hra,Abelof:2016pby,Boughezal:2018azh,Borsa:2020ulb,Arratia:2019vju,Page:2019gbf}. Besides event-wide jet measurements, significant progress has been made in recent years to better understand jet substructure observables, see Refs.~\cite{Larkoski:2017jix,Kogler:2018hem,Marzani:2019hun} for recent reviews. Jet substructure observables can be constructed to be Infrared and Collinear Safe making them less sensitive to experimental resolution effects. Nevertheless, hadronization corrections can be sizable for these observables. For several jet substructure observables it is possible to connect the relevant hadronization correction to universal functions. The scaling of these functions can be predicted from first principles which can be tested experimentally by studying jets at different energies and by varying parameters of specific observables. EIC jets at different center of mass energies have different quark/gluon fractions and a different quark flavor decomposition. Therefore, the measurement of jets at high luminosity and low center of mass energies can provide important complementary information to better disentangle the flavor decomposition of the hadronization corrections of jets and also to study their correlation with different initial state PDFs. Several jet observables in the literature have been studied which are particularly sensitive to the quark flavor and quark/gluon differences. Examples include jet angularities~\cite{Lee:2006nr,Ellis:2010rwa,Aschenauer:2019uex,Caletti:2021oor}, the jet charge~\cite{Waalewijn:2012sv,Kang:2020fka}, angles between jet axes~\cite{Cal:2019gxa}, groomed jet substructure~\cite{Hoang:2019ceu}, flavor correlations~\cite{Chien:2021yol}, energy-energy correlators~\cite{Dixon:2019uzg,Chen:2020adz,Li:2021zcf}, jets at threshold~\cite{Dasgupta:2007wa,Arratia:2020ssx}, and T-odd jets~\cite{Liu:2021ewb,Lai:2022aly}. The EIC provides a clean environment with a minimal background contamination from the underlying event/multi-parton interactions making it an ideal place to study low-energy aspects of jets. In addition, the measurements of jets for multiple jet radii at different energies may help to explore in detail the connection of hadron and jet cross sections. Recently, it was demonstrated that inclusive hadron cross sections can be obtained from inclusive jet calculations by taking the limit of a vanishing jet radius~\cite{Neill:2020bwv,Neill:2020tzl}.

An important aspect of jet observables is their sensitivity to TMD PDFs and FFs. For example, lepton-jet cross sections in the laboratory frame~\cite{Liu:2018trl,Kang:2021ffh,Hatta:2021jcd} and the Breit frame~\cite{Gutierrez-Reyes:2018qez,Gutierrez-Reyes:2019vbx,Gutierrez-Reyes:2019msa} give access to (spin-dependent) quark TMD PDFs where the final state radiation can be calculated perturbatively. Similarly, di-jet production can be used to study gluon TMD PDFs~\cite{Kang:2018vgn,delCastillo:2020omr}. Moreover, the transverse momentum of hadrons inside the jet relative to the jet axis can provide access to TMD FFs, which is independent of initial state TMD PDFs~\cite{Arratia:2020nxw}. Here the choice of the jet axis is important and different physics can be probed~\cite{Neill:2016vbi}. Especially, due to the separation of initial and final state TMD PDFs and FFs, jet observables can provide important complementary information to semi-inclusive deep inelastic scattering. All of these observables and the information content they provide benefit greatly from measurements over a wide kinematic range. In particular, high luminosity at the EIC will allow for a unique quark flavor decomposition.

A measurement that is in particular luminosity hungry, is the detection of diffractive di-jet events. This observable is sensitive to the elusive Generalized TMDs (GTMDs)~\cite{Hatta:2016dxp,Hatta:2019ixj} of gluons. Lower collision energies provide constraints for the moderate $x$-range of the gluon distribution, while higher energies are sensitive to the small-$x$ gluon distribution. If, as typically assumed, the gluon spin (helicity and orbital angular momentum) is sizable at moderate $x$, it is critical to have very high luminosity at lower/intermediate collision energies at the EIC.

\section{Exotic meson spectroscopy}
\label{sec:exotic}
\subsection{Motivations for an exotic spectroscopy program at the EIC}

Modern electro/photoproduction facilities, such as those operating in Jefferson Lab, have demonstrated the effectiveness of photons as probes of the hadron spectrum. However the energy ranges of these facilities are such that most states with open or hidden heavy flavor are out of reach. This is unfortunate as there remains significant discovery potential for photoproduction in this sector. Already electron scattering experiments at HERA \cite{ZEUS:2002wfj,H1:2005dtp} observed low-lying charmonia, demonstrating the viability of charmonium spectroscopy in electroproduction at high-energies but were limited by luminosity. Now the proposed EIC, with high luminosity, will provide a suitable facility for a dedicated photoproduction spectroscopy program extended to the heavy flavor sectors. In particular, the study of heavy-quarkonia and quarkonium-like states in photon-induced reactions will not only be complementary to the spectroscopy programs employing other production modes but may give unique clues to the underlying non-perturbative QCD dynamics. 

\begin{figure}
  \centering
    \includegraphics[width=\textwidth]{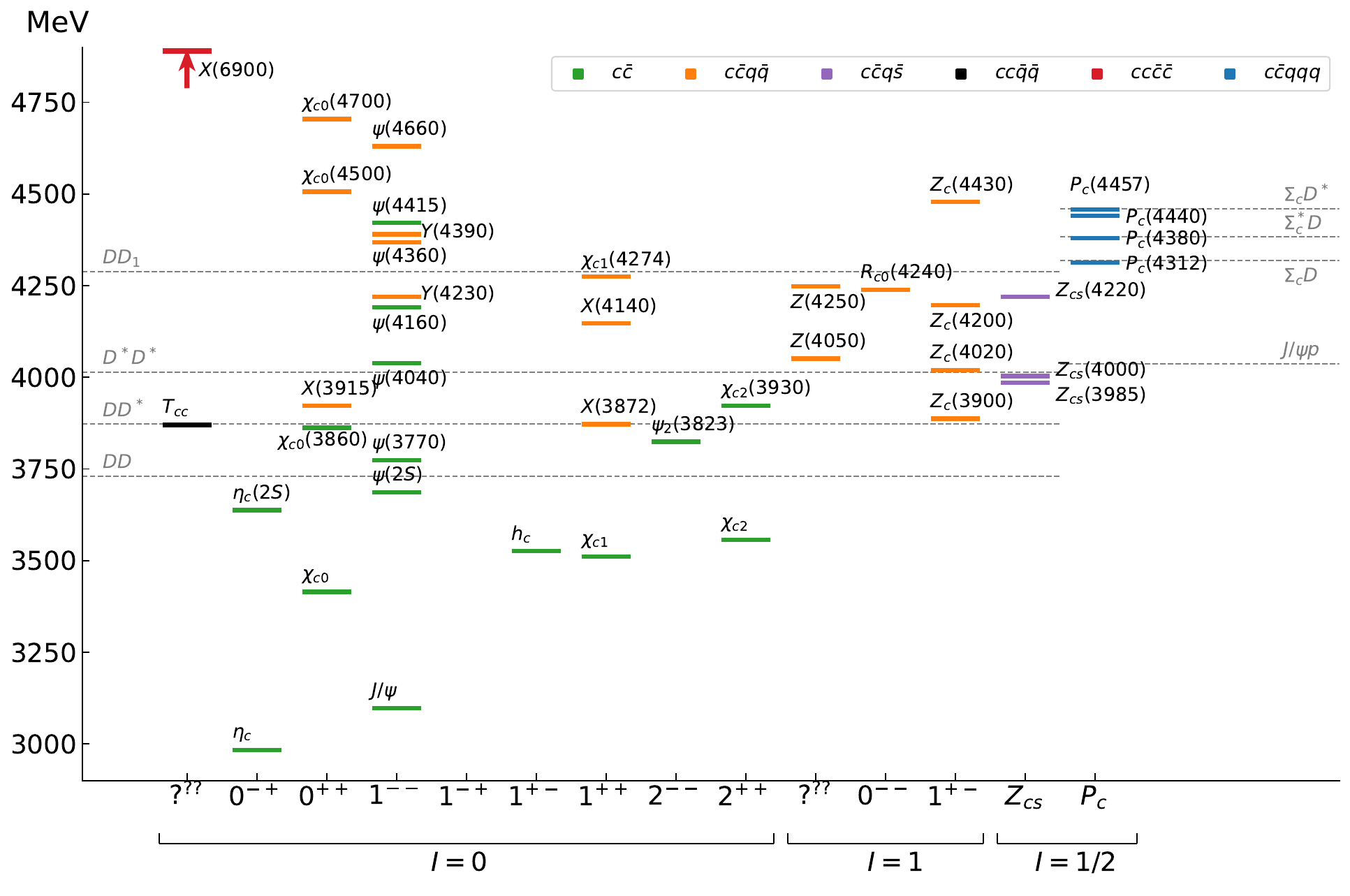}
    \caption{Experimentally measured charmonia, XYZ and pentaquark spectra from \cite{jpaccollaboration2021novel}. A '?' refers to unknown spin or parity.}
    \label{spect_jpac}
  \hfill
 \end{figure}

One of the most striking features of quarkonium spectra is the wealth of observed experimental signals which seem to indicate an exotic QCD structure beyond conventional $Q\bar{Q}$ mesons. Starting with the observation of the narrow $\chi_{c1}(3872)$ in the $J/\Psi \, \pi^+\pi^-$ invariant mass spectrum by the BELLE Collaboration in 2003 \cite{Belle:2003nnu}, these states, collectively denoted the $XYZ$'s, now number in the dozens. The dramatic change in landscape from 2003 up to 2021 is illustrated in figure \ref{spect_jpac} where new states beyond quark model charmonium are highlighted. These states exhibit properties which are not consistent with expectations of conventional QCD bound states, for example : large isospin violation in the case of the $\chi_{c1}(3872)$; iso-vector quarkonium-like character for the $Z$'s; supernumeracy of the vector $Y$ states. We refer to reviews such as \cite{Brambilla:2019esw,OlsenRevModPhys.90.015003} for more detailed discussion. The underlying dynamics governing their nature is not unambiguously known. The experimental signals of these states, usually in the form of sharp peaks in invariant mass spectra or broader enhancements that are required to describe distributions in a more complex amplitude analysis, allow multiple interpretations of their structure, e.g. multi-quark states, hadron-hadron molecules, kinematic cusps or triangle singularities. Disentangling these possibilities is one of the foremost missions of exotic spectroscopy and would further our understanding of the non-pertubative nature of QCD in heavy sectors. 

One challenge in this endeavor is that, with few exceptions, the $XYZ$ signals have only been observed in single production modes, usually $e^+e^-$ annihilation or $B$ meson decays. 
Observation of any of these states at the EIC through photoproduction would thus provide independent and complementary verification of their existence. Further, an ubiquitous feature of $XYZ$ signals is their proximity to open thresholds and the presence of additional particles in the reconstructed final state. This complicates the interpretation of experimental peaks as complicated kinematic topologies involving nearby open channels may modify or mimic a resonant signal.
Here photoproduction provides a unique opportunity to produce $XYZ$ in isolated final states, thus alleviating the role of kinematic singularities. In this way a null result may be equally important towards uncovering the spectrum of genuine bound-states. Additionally the polarized electron and proton beam setups enable the determination of spin-parity assignments of states for which these are not yet known. 
The EIC would also have real discovery potential for exotic heavy flavor mesons.

 A dedicated spectroscopy effort can make meaningful contributions to several aspects of non-exotic quarkonium physics. Theoretical understanding of photoproduction processes conventionally rely on Regge theory and exchange phenomenology which have been tested extensively in the light sector \cite{Nys:2018vck}. Measurement of quarkonium photoproduction cross-sections serves as a testing ground of scattering phenomenology in heavy sectors where perturbative QCD inputs may also be used. In particular the microscopic structure of $\gamma Q\bar{Q}$ interaction and assumptions such as Vector Meson Dominance (VMD) may be tested \cite{Mamo:2021tzd,vmdXu2021}. 
 
Beyond the charmonium sector, the energy reach of the EIC will also allow the study of near-threshold bottomonium photoproduction which may be sensitive to the trace anomaly contribution to the nucleon mass and would be complementary to ongoing studies of $J/\Psi$ photoproduction studies currently underway at Jefferson Lab \cite{GlueX:2019mkq,Meziani:2016lhg}. 
 Further, this mass range is predicted to also exhibit a rich landscape of pentaquark-like structures 
\cite{P_bPhysRevD.101.074010,Paryev:2020jkp}
 the as yet unobserved hidden-bottom partners of the $P_c$ signals observed in the $J/\Psi p$ mass spectra in $\Lambda_c$ decays.
 
\subsubsection{Photoproduction with the EIC}

Given the many physics opportunities around photoproduction of heavy quarkonia, new measurements at the EIC will be essential for understanding both exotic and conventional quarkonium spectra. Photoproduction provides a flexible production mode, able to produce the full spectrum of hadrons of any quantum number. This gives such measurements significant discovery potential and allows mapping out of patterns within the observed spectrum.
The trade-off however is that the cross sections for photoproducing heavy mesons are small, only up to $\mathcal{O}(1$ nb), meaning a dedicated spectroscopy program will require high luminosity at sufficiently large centre-of-mass energies to make a meaningful contribution. The proposed EIC, maintaining high luminosity at its lower centre-of-mass energies, would be well placed to meet these conditions. In particular, even with the lower centre-of-mass settings of 29 and 45 $(GeV/c^{2})$ there is sufficient energy to directly produce many exotic states of interest in the charmonium sector without the constraints in bounds from parent masses which occur in decay processes. Kinematic generation of peaks through final state interactions, such as triangle diagrams, will also be suppressed over the entire $W$ range.

When combined with complete measurement of the final state, the polarized electron and proton beams offer means for detailed partial wave analysis to disentangle overlapping states, deduce the quantum numbers of resonant states and study production mechanisms. This is of particular importance for many of the excited $XYZ$ states which have intrinsically greater decay widths and contribute to more complicated final states. The use of partial-wave analysis through polarized photoproduction set-ups for exotic searches is currently being pursued in the light-quark sector at the GlueX experiment and much of the expertise will be readily applicable to the EIC setup. This includes the possibility to measure polarized cross-sections, spin density matrix elements, and asymmetries.

The variable beam setups of the EIC allow exploration of Primakoff production of axial vector charmonium \cite{Albaladejo:2020tzt} and simultaneous measurement of charged charmonium-like isospin multiplets with deuteron beams. Additionally, the electroproduction mode of the EIC allows measurement of $Q^2$ dependence and photocouplings, a detailed study of which may be a reliable probe of the microscopic nature of exotic hadrons \cite{Kawamura:2013wfa,Anikin:2004vc}. Electroproduction studies are of particular importance for the $\chi_{c1}(3872)$ and the closely related $\tilde{X}(3872)$ candidate claimed in muoproduction by the COMPASS experiment in the $J/\Psi \pi^+\pi^-$ mass spectrum \cite{COMPASS:2017wql}. Although this new state closely resembles the $\chi_{c1}(3873)$ in mass and width, its dipion mass distribution was suggestive of a scalar wave instead of the usual $\rho J/\Psi$ decay mode of the $\chi_{c1}(3872)$, implying a different C-parity. Further this state was observed in production with an additional pion in the final state but not in exclusive production, raising further questions as to the nature of the muoproduced peak. Detailed study of the $J/\Psi \pi \pi$ mass spectra in virtual photoproduction would help to understand the COMPASS result.

\subsubsection{States of interest}

The first goal of an exotic spectroscopy program will be to identify the production of the most
established states, $\chi_{c1}(3872)$, Y(4260) and  $Z_{c}(3900)$. The decay of these states to a $J/\Psi$ and pions will provide a clean and well studied final state and we discuss in Section (\ref{sec:spectro_detection}) the prospects for measuring this with the EIC. After that there are many open questions in XYZ physics, particularly with respect to the nature of peaks in invariant mass distributions which we hope to address. Here we consider a few examples with decays which should be readily measurable and make rate estimates for these in Section (\ref{spec_Estimates}).

A recent publication from LHCb show structure in the $J/\Psi K^{+}$ mass spectra which they can reproduce with the addition of two new resonances with strangeness and hidden charm, $Z_{cs}(4000)$ and $Z_{cs}(4220)$ \cite{LHCb:2021uow} with widths around 100-200 MeV. A similar, narrower state, the $Z_{cs}(3985)$, has also been seen in $K^{+}D\bar{D}^{*}$ by BESIII \cite{BESIII:2020qkh}.

The X(6900) or $T_{c\bar{c}c\bar{c}}(6900)$ tetraquark candidate has been seen from its decay to 2$J/\Psi$ \cite{LHCb:2020bwg}. Analogue Z states have been seen in the b-quark sector by Belle, with the $\Upsilon$ or $h_{b}$ mesons in combination with a charged pion \cite{ZbPhysRevLett.108.122001}. Production of these states are also well within EIC centre-of-mass energies. In addition, spectroscopy at the EIC will be able to search in a variety of other final states replacing pions for other mesons such as vectors.
We can  also look for charm quarks via reconstructing D mesons
the most accessible decay mode of which will be $K^{-}\pi^{+}$ with a branching ratio of 
around 4\%, while the decay of XYZ into final states with D mesons is likely to be quite high. 
As seen later XYZ decay products populate the detector region relatively uniformly giving good 
potential for reconstructing events including pairs of D mesons. This would be particularly useful for investigating the molecular picture of these states.

\subsection{Estimates for the EIC}

\subsubsection{JPAC Photoproduction Amplitudes}

In order to estimate the feasibility of quasi-real photoproduction for states of interest at EIC energies we followed the approach of a recent JPAC Collaboration study in \cite{Albaladejo:2020tzt}. Here, general principles are used to construct exclusive photoproduction amplitudes of the charmonium states of interest on the per-helicity-amplitude basis. In this way, full kinematic dependence is retained and the production may be propagated along decay chains to reconstructed final states. 

In general the amplitude of producing a meson, $\mathcal{Q}$ via the exchange of a particle, $\mathcal{E}$ with spin $j$ take the form:
    \begin{equation}
        \langle \lambda_\mathcal{Q} \, \lambda_{N^\prime} |T_\mathcal{E}|\lambda_\gamma, \lambda_{N}\rangle 
        = \mathcal{T}^\mu_{\lambda_\gamma \, \lambda_\mathcal{Q}}\; \mathcal{P}_{\mu\nu}^{(\mathcal{E})} \; \mathcal{B}^\mu_{\lambda_N \, \lambda_{N^\prime}} 
    \end{equation}

\noindent where $\mathcal{T}$ and $\mathcal{B}$ are Lorentz tensors of rank-$j$ and given by effective interaction Lagrangians which provide an economical way to satisfy kinematic dependencies and discrete symmetries of the reaction. Such methods have been widely used to motivate searches for exotic hadrons through photoproduction \cite{Wang:2015jsa,Cao:2020cfx,Winney:2019edt,Wang:2019krd,Karliner:2017qje,HillerBlin:2016odx,Wang:2015lwa,Galata:2011bi,Lin:2013mka}
The form of the exchange propagator, $\mathcal{P}$, provides means to consider production in two kinematic regions of interest: near-threshold and at high-energies, where production is expected to proceed through exchanges of definite-spin and Reggeized particles respectively. The center-of-mass range available at the EIC provides wide coverage in energy, thus for first estimates we used a simple linear interpolation between the low- and high-energy models provided in \cite{Albaladejo:2020tzt}.

\subsubsection{Electroproduction}

We generalized the aforementioned (real) photoproduction to consider exclusive electroproduction with low-$Q^2$ quasi-real virtual photons via a factorized model whereby the amplitude for producing a virtual photon beam is followed by the t-channel photoproduction of the meson. The produced meson subsequently decays to specific final states which can be measured in the EIC detector:

\begin{equation}
    \frac{d^{4}\sigma}  {ds \, dQ^{2\,} dt \,d\phi} =  \Gamma(s,Q^{2},E_{e}) \; \frac{d^{2}\sigma_{\gamma *+p \rightarrow V+p} (s,Q^{2})}  {dt \, d\phi}
    \label{eqn:Spectroscopy_diffCross}
\end{equation}

$\Gamma(s,Q^{2},E_{e})$ is the virtual photon flux and $\frac{d^{2}\sigma_{\gamma *+p \rightarrow V+p} (s,Q^{2})}  {dt d\phi}$ is the two-body photoproduction cross section calculated from the model of \cite{Albaladejo:2020tzt}, modified by an additional $Q^{2}$ dependence taken from \cite{H1:1999pji}. Eqn. (\ref{eqn:Spectroscopy_diffCross}) was integrated numerically to give the total cross section for determining event rates. Note, the virtual photon flux integration leads to a factor of around 0.2 for the case of  $\chi_{c1}(3872)$ production relative to real photoproduction for the 5x41 GeV beams.

\subsubsection{Other Models}

To estimate how reliable our production rates may be we compared to other approaches that have been published recently. 

In \cite{Yang:2021jof} a semi-inclusive production mechanism for hadron molecules was investigated. Here the molecular constituents were first photoproduced via Pythia, and then allowed to interact together in given $X$ and $Z$ states. Cross sections for semi-inclusive production were given for the highest proposed EIC centre-of-mass energy for $\chi_{c1}(3872)$, $Z_c$ and $Z_{cs}$ and are compared to our estimates for exclusive production in Table \ref{tab:spec_models}. While the estimates for $\chi_{c1}(3872)$ are an order of magnitude lower than this work, the $Z_c$ cross section is an order of magnitude higher. 
We note that the calculations of \cite{Yang:2021jof} should be valid for larger $Q^{2}$, in the central region (large $p_{T}$), those from \cite{Albaladejo:2020tzt} should be valid at $Q^2 < 1 (GeV/c^{2})^2$, and peak in the peripheral region (small $p_{T}$), where we expect the bulk of events to be produced. 

Using the same method, Ref.~\cite{Shi:2022ipx} estimates the semi-inclusive production rates of more exotic hadrons, and finds that copious $P_{cs}$ pentaquarks and $\Lambda_c\bar \Lambda_c$ dibaryons can be produced at EIC. It is also promising to search for double-charm tetraquarks at EIC. In addition, Ref.[2208.02639] also suggests that the possible 24 GeV upgrade of CEBAF [proper ref.] can play an important role in the search of hidden-charm tetraquarks and pentaquarks.

A very similar approach to the current work is taken in \cite{Xie:2021sik}, where the models of  \cite{Albaladejo:2020tzt} were coupled to a virtual photon produced from electron-proton scattering interactions. Their results are compared to ours in table \ref{tab:spec_models}, where our estimates are  just over a factor 2 lower for the low energy setting and more comparable for the high energy setting. The differences are likely due to our interpolation of low and high models, or handling of phase space and virtual photon flux factors when performing the integration. The threshold at $Q^2>0.01~(GeV/c^2)^2$ is applied in this comparison but not in our later results where we integrate the full allowable $Q^{2}$ range.

In general we can expect integrated cross sections for electroproduction of up to order 1 nb for production of mesons with charm quarks.

\begin{table}[h]
   \centering
        \caption{Model Comparisons. Note, in the Lanzhou calculations cuts are applied to $Q^{2}$ and W, as indicated in the column title with units in GeV. The same cuts are applied to our calculation when comparing to Lanzhou, but not to the comparisons with Yang. The cut on $W >20~GeV/c^2$ has a very large effect on the calculated electroproduction cross sections as the photoproduction cross section for X and Z of \cite{Albaladejo:2020tzt} falls rapidly.}
        \label{tab:spec_models}
        \begin{tabular}{c|c|c|c|c|c|c|}
            \cline{2-7}
             & \multicolumn{2}{|c|}{3.5x20 $Q^{2}>0.01$;$W<16$} & \multicolumn{2}{|c|}{18x275 $Q^{2}>0.01$; $20<W<60$} & \multicolumn{2}{|c|}{18x275 $Q^{2}>0$}\\
            \cline{2-7}
             & JPAC & Lanzhou\cite{Xie:2021sik} & JPAC & Lanzhou\cite{Xie:2021sik} & JPAC & Yang\cite{Yang:2021jof} \\
            \hline
            \multicolumn{1}{|c|}{$\chi_{c1}(3872)$} 
            & \multicolumn{1}{|c|}{0.47 nb}  & \multicolumn{1}{|c|}{1.2 nb}
            & \multicolumn{1}{|c|}{0.00014 nb}  & \multicolumn{1}{|c|}{0.00021 nb}
            & \multicolumn{1}{|c|}{3.5 nb}  & \multicolumn{1}{|c|}{0.216-0.914 nb}\\
            \hline
            \multicolumn{1}{|c|}{$Y(4260)$} 
            & \multicolumn{1}{|c|}{0.06 nb}  & \multicolumn{1}{|c|}{0.2 nb}
            & \multicolumn{1}{|c|}{1.5 nb}  & \multicolumn{1}{|c|}{2.0 nb}
            & \multicolumn{1}{|c|}{14 nb}  & \multicolumn{1}{|c|}{-}\\
            \hline
            \multicolumn{1}{|c|}{$Z_{c}^{+}(3900)$} 
            & \multicolumn{1}{|c|}{0.06 nb}  & \multicolumn{1}{|c|}{0.16 nb}
            & \multicolumn{1}{|c|}{0.00018 nb}  & \multicolumn{1}{|c|}{0.00048 nb}
            & \multicolumn{1}{|c|}{0.41 nb}  & \multicolumn{1}{|c|}{3.8-14 nb}\\
            \hline
         \end{tabular}
\end{table}

\subsubsection{Estimates}\label{spec_Estimates}
 
In table \ref{tab:spec_estimates} we give estimates for the production of a variety of exotic states with the EIC. These are based on the models and parameters detailed in \cite{Albaladejo:2020tzt}, with the addition of the $Z_{cs}(4000)$ production using kaon exchange; and the modification of the X(6900) model to use a higher branching ratio to $\Psi\omega$ of 3\%,  which was previously taken as 1\%.
These estimates assume a luminosity of $10^{34}$ cm$^{-2}$s$^{-1}$. The additional branching ratios, used to calculate events per day, of $J/\Psi \rightarrow e^{+}e^{-}$ was taken as 6\% and $\Upsilon(2S) \rightarrow e^{+}e^{-}$ as 1.98\%.

Current measurements of X and Y states contain up to order 10 thousand and 1 thousand events respectively. This is similar to the daily production rate of our estimates. So with an overall detector acceptance of order 10 \% the EIC would be able to make significant contributions to our understanding of these states.

We note that a previous investigation of charged final states in electroproduction at an electron-ion collider \cite{PhysRevC.100.024620} through a Regge exchange mechanism found similar production rates for the $Z_{c}(4430)$, approximately a factor 2 lower than our estimates for the $Z_{c}(3900)$. They also conclude that the final state rapidity depends on the beam energy, at lower center of mass energies production shifts toward mid-rapidity, where the final state may be reconstructed in a central detector.

\begin{table}
    \centering
       \caption{Summary of results for production of some states of interest at the EIC electron and proton beam momentum $5\times100 (GeV/c)$ (for electron x proton). Columns show : the meson name;  our estimate of the total cross section; production rate per day, assuming a luminosity of $6.1\times10^{33}$ cm$^{-2}$s$^{-1}$; the decay branch to a particular measurable final state; its ratio; the rate per day of the meson decaying to the given final state. }
       \label{tab:spec_estimates}
        \begin{tabular}{|c|c|c|c|c|c|c|}
        \hline
        Meson    & Cross Section (nb) & Production rate (per day) & Decay Branch    &    Branch Ratio (\%)   & Events (per day) \\
        \hline
        $\chi_{c1}(3872)$         & 2.3       & 2.0 M     & $J/\Psi\; \pi^{+}\pi^{-}$   & 5   & 6.1 k       \\
        \hline
        $Y(4260)$                 & 2.3       & 2.0 M     & $J/\Psi\; \pi^{+}\pi^{-}$   & 1   & 1.2 k       \\
        \hline
        $Z_{c}(3900)$          & 0.3      & 0.26 M    & $J/\Psi\; \pi^{+}$       &  10  & 1.6 k      \\
        \hline
        $X(6900)$            & 0.015      & 0.013 M    & $J/\Psi\; J/\Psi$       &  100 & 46     \\
        \hline
        $Z_{cs}(4000)$            & 0.23      & 0.20 M    & $J/\Psi\; K^{+}$       &   10  & 1.2 k      \\
        \hline
        $Z_{b}(10610)$           & 0.04      & 0.034 M    & $\Upsilon(2S)\;\pi^{+}$   &   3.6   & 24      \\
        \hline
        
\end{tabular}
\end{table}

\subsubsection{Detection of final states}\label{sec:spectro_detection}

Meson photoproduction at the EIC will require a detector with full hermicity. Quasi-real photoproduction results in the scattered electron being very close to the incident beam line. t-channel production provides very little transverse momentum for the recoiling baryon, which will likewise be scattered within a degree or so of the beam. On the other hand the meson itself will be produced relatively centrally at the lower centre-of-mass settings making for excellent detection of its decay products.

\begin{figure} 
\hspace{-0.2cm}\includegraphics[width=0.85\columnwidth]{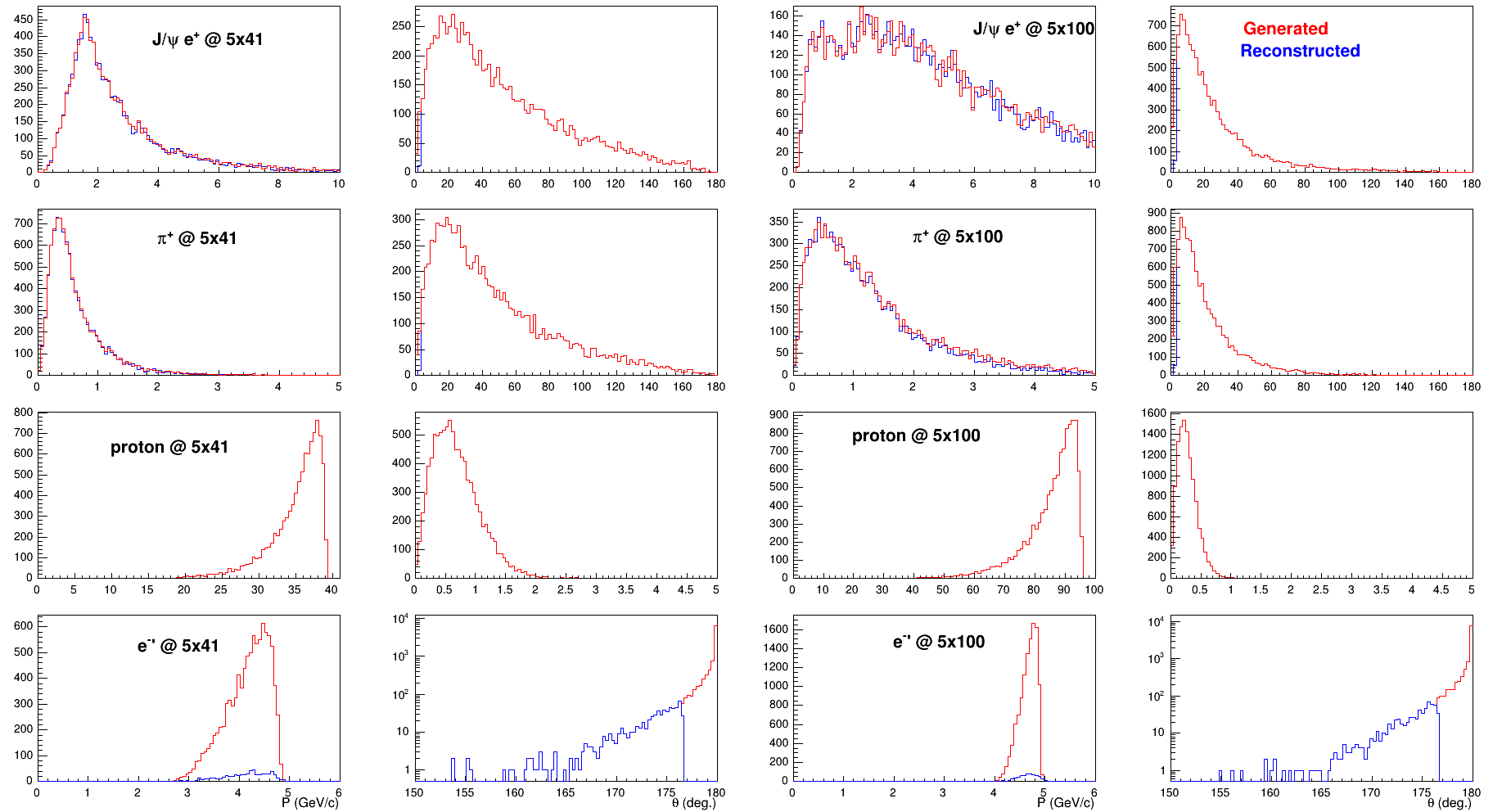}
\caption{Momentum and angle distributions for X production. Left(right) columns are for beam configuration 5x41(5x100). Rows, from top to bottom, show $J/\Psi$ decay $e^{+}$; X decay $\pi^{+}$; the scattered proton; and the scattered electron. Red lines show the true generated distributions while blue are the detected particles as expected with the EIC Yellow Report matrix detector.}
\label{fig:spect_particle_momentum}
\end{figure}

The individual particle momentum distributions for the 5x100 centre-of-mass setting are shown in Fig. \ref{fig:spect_particle_momentum}. Also shown are the distributions expected when reconstructed with the EIC Yellow Report matrix detector via the eic-smear package \cite{AbdulKhalek:2021gbh}.
It is clear the meson decay products are almost entirely directed at the high acceptance central detector region. Protons pass to the far-forward detector region, while there is some electron detection in the backward electron region.

For final states including a $J/\Psi$, which are mostly under consideration here, excellent electron/pion separation will allow a clean tag of $J/\Psi$ events through its narrow width in the $e^{+}e^{-}$ invariant mass. Coupled with a very high detection efficiency this should allow for full identification of the meson decay products and provide a means for peak hunting in many final states including a final $J/\Psi$.

Supplementing the meson detection with far-forward and far-backward detector systems will enhance the spectroscopy program by allowing measurements of the full production process, that is measurement of the reaction variables W, from the $e^{-}$ and $t$ from the recoil baryon. Detecting the scattered electron also allows determination of the longitudinal and transverse polarisation components of the virtual photon, providing further information on the production processes through access to the meson spin density matrix elements. This can be done with the backward detector around 5-10\%  of the time when the electron beam momentum is lowest (5 GeV), due to the transverse kick to the electron from the Lorentz boost due to the more energetic proton. A dedicated far-backward electron detector such as the proposed low-$Q^{2}$ tagger could increase the electron detection rate significantly. Detection of both the electron and baryon can also allow for superior background rejection for exclusive event reconstruction.

\subsection{Outlook}

We have briefly examined the case for producing exotic mesons through quasi-real photoproduction at the EIC. Although it is difficult to make strong statements on what we might expect this is exactly due to the uncertainty around the nature and structure of the new states seen at other labs. We have shown that if real exotic states exist then many of these should have sufficiently high cross sections to be measurable. The low centre-of-mass configurations are particularly suited to mesons produced through fixed spin exchanges of light mesons, which have a high cross section close to threshold. Coupled with a high luminosity this would provide a very high production rate, while the kinematics and hermetic detector systems are ideal for reconstructing the mesons we wish to study and allow us to exploit the EIC's discovery potential in exotic heavy flavor spectroscopy.  


\section{Science highlights of light and heavy nuclei}%
 \label{sec:nuclei}

\subsection{Introduction}
\begin{figure} [htb]
        \centering
        \includegraphics[width=0.3\columnwidth]{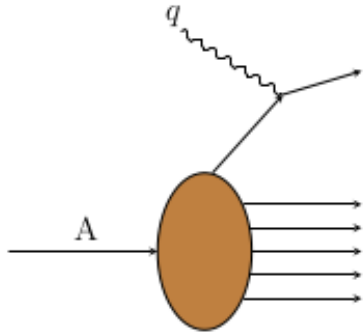}
        \caption{Schematic diagram of a nuclear breakup process.  The virtual photon $q$ interacts with a constituent 
        of the nucleus $A$ and particles originating from the breakup of the nucleus can be detected in the far 
        forward region of the EIC detector.}
        \label{fig:breakup}
    \end{figure}

Lepton-induced high-energy scattering with nuclei will be measured at fixed target facilities such as Jefferson Lab 12~GeV.  
These facilities have a rich experimental program that will yield interesting results for years to come.  To complement these 
programs, the EIC will be the first high-energy facility that has the ability to collide electrons and nuclei, which means 
it comes with unique capabilities:
\begin{itemize}
    \item The EIC has a wide kinematic range in $Q^2$ and Bjorken $x$, enabling high-energy nuclear measurements in unexplored kinematics. 
    \item The EIC can have beams of polarized light ions ($^3$He, deuteron, etc.~\cite{Bai:2011hd}), enabling studies of the
    polarized nuclear (neutron) structure, the polarized EMC effect, and nuclear spin-orbit phenomena.  The deuteron, being 
    spin-1, offers possibilities of spin studies beyond that of the nucleon.
    \item Measurements on nuclei inherently have to deal with nuclear effects such as the Fermi motion, nuclear binding and 
    correlation effects, and possible non-nucleonic components of the nuclear wave function~\cite{Frankfurt:1988nt}. 
    In inclusive measurements these nuclear effects form one of the dominant sources of systematic uncertainties. 
    With its extensive far-forward detection apparatus in both interactions regions, detecting particles originating from 
    the breakup of the nuclear target (nuclear target fragmentation region) is possible and can help to eliminate or control 
    these nuclear effects. (See Fig.~\ref{fig:breakup} for a schematic diagram.)  As a consequence, these more exclusive 
    measurements will push the capabilities of the EIC as a precision machine for high-energy nuclear physics.
\end{itemize}

   Measuring nuclear breakup reactions at the EIC has several advantages.  In collider kinematics nuclear fragments are 
   still moving forward with a certain fraction of the initial beam momentum and in non-coherent scattering  they have
   a different rigidity from the beam particles.  This makes their detection more straightforward than in fixed target
   experiments where they typically have low momenta in the laboratory frame (10s of MeV/$c$).  The detection of these 
   fragments enables additional control over the initial nuclear state in the high-energy scattering event.  It can be
   used to probe effective targets, for instance, free neutron structure  in tagged spectator 
   DIS~\cite{Frankfurt:1981mk,Sargsian:2005rm}, and pion and kaon structure in the Sullivan process~\cite{Arrington:2021biu}. 
   Nuclear breakup measurements also determine which nuclear configurations (densities, virtualities, initial 
   nucleon momentum) play a role in the process, important for instance in a multivariate disentanglement of nuclear 
   medium modification effects such as the EMC effect.  A special case of detecting fragments is coherent nuclear 
   scattering in hard exclusive reactions, where the initial nucleus receives a momentum kick but stays intact (no breakup). 
   Measurements of these coherent reactions allow us to perform tomography of light nuclei in quark and gluon degrees 
   of freedom as for the nucleon (Sec.~\ref{sec:GPD}) and to study coherent nuclear effects in these systems.

For all these reactions, having high event rates is of high importance (multidimensional cross sections measured with 
sufficient precision, probing rare nuclear configurations).  To obtain these high event rates one needs both high luminosity 
for a wide kinematic range and high acceptance for the detection of final-state particles.  In both interaction regions, the 
EIC will have a dedicated set of far-forward detectors that enable the detection of nuclear fragments with high acceptance. 
Due to the intricate engineering challenges (magnets, beam pipe, crossing angle of the beam), each interaction region will
have some holes in the acceptance.  Having these holes in different regions of the kinematic phase space would enforce the 
complementarity between the two interaction regions.  Having a secondary focus  would also increase acceptance of detected 
fragments down to lower $p_T$ values.  This is especially important for coherent scattering of light nuclei, where the $p_T$
values are much lower than for the free proton. (see Section~\ref{sec:forward}.)

In the remainder of the section we offer a brief overview of nuclear reactions that can be studied at the EIC and the physics
motivation behind them. These can all benefit from the complementarity offered by having a second IR.  We discuss these 
according to the nature of the measurements, starting with inclusive measurement, then semi-inclusive and tagged reactions, 
and we conclude with a discussion on exclusive nuclear channels and charm-flavored hypernuclei.

\subsection{Inclusive measurements}

    EIC can measure inclusive DIS on a wide range of nuclei, from the lightest to heaviest nuclei, and in a wide range of 
    Bjorken $x$ and $Q^2$.  This can shed light on the dynamics of nuclear modifications of partonic distribution functions: 
    shadowing and anti-shadowing at low values of $x$ and the so-called EMC effect at high $x$.  These high-$x$ measurements 
    benefit from lower center of mass energies and, with the $Q^2$ range that can be explored at the EIC, the $Q^2$ dependence 
    of the EMC effect could be further explored.  This would enable the disentanglement of leading and higher-twist effects 
    in the medium modifications. QCD evolution applied to the wide $Q^2$-range offers a way of getting access to the gluon 
    EMC effect at high $x$.  In addition, for polarized light nuclei the polarized EMC effect~\cite{Cloet:2005rt} could be 
    further explored, which is so far an unknown quantity that will be explored in an upcoming JLab experiment~\cite{JLAB_polEMC}.

\subsection{Semi-inclusive and tagged spectator measurements}

    The use of semi-inclusive reactions on nuclei for nuclear TMD studies was highlighted earlier in Section~\ref{sec:tmd_nuclei}.  
    Here, we focus on so-called tagged spectator measurements, where one or more nuclear fragments from the nuclear breakup are 
    detected. This helps, as previously outlined, to control the nuclear configurations playing a role in the hard scattering 
    processes. One example is the use of deuteron or $^3$He as effective neutron targets by tagging one (resp. two) spectator protons~\cite{Frankfurt:1981mk,Sargsian:2005rm,Kondratyuk:1983kq,DelDotto:2017jub,Cosyn:2019hem,Cosyn:2020kwu,Friscic:2021oti,Jentsch:2021qdp}. 
    These neutron data are an essential ingredient in the quark flavor separation of the partonic distribution functions. 
    In the tagged spectator reactions, an effective free neutron target can be probed by performing a so-called on-shell 
    extrapolation of the measured cross sections or asymmetries~\cite{Sargsian:2005rm,Jentsch:2021qdp}.  The presence of 
    polarized light ion beams enables the extraction of polarized neutron structure in this 
    manner~\cite{Kondratyuk:1983kq,Cosyn:2019hem,DelDotto:2017jub,Friscic:2021oti}.  
    
    Measuring tagged spectator reactions at larger nucleon momenta (several 100 MeV relative to the ion rest frame) is of 
    interest to several outstanding questions in nuclear physics and how these are interconnected.  What is the QCD nature 
    of the short-range, hard core part of the nucleon-nucleon force~\cite{Boeglin:2015cha,Miller:2015tjf,Tu:2020ymk}?  How 
    do nuclear medium modifications of partonic properties manifest themselves and what nuclear configurations play a role 
    in these~\cite{Hen:2016kwk}?  In these kinematics, however, the influence of final-state interactions between products
    of the hard scattering and the spectator(s) and between the spectators has to be accounted for~\cite{Cosyn:2017ekf,Strikman:2017koc} 
    in order to disentangle them from the QCD phenomenon of interest.  These final-state interactions are moreover little 
    explored in high-energy scattering and are an interesting topic that can teach us about the space-time evolution of
    hadronization dynamics.

    While technically not a nuclear process, the Sullivan process $e +p \to e' + X+ (N \;\text{or}\; Y)$ share characteristics 
    with the previously discussed processes.  The physics interest of the Sullivan process lies in the extraction of pion and kaon structure~\cite{Aguilar:2019teb,Arrington:2021biu}.  The pion being the pseudo-Goldstone boson of dynamical chiral symmetry 
    breaking, this can shed light on the mechanism of emergent hadronic mass (EHM) within QCD.  For the kaon, the presence of 
    the heavier strange quark opens up the study of the interplay between EHM and the Higgs mechanism.  In the Sullivan process, 
    a nucleon or hyperon is tagged in the far-forward region at low four momentum transfer squared $-t$. In this manner, the 
    process is dominated by meson exchange in the $t$-channel and, by extrapolating the observables to the on-shell pole of the 
    exchanged meson, one can extract pion (nucleon tagging) or kaon (hyperon tagging) structure.  Compared with the earlier HERA 
    extractions, the high luminosity and wide kinematic range of the EIC would result in an order of magnitude decrease of 
    statistical errors on the extracted pion PDFs.  These measurements require high luminosity ($ > 10^{33}$ cm$^{-2}$ sec$^{-1}$) 
    in order to compensate for the few times $10^{-3}$ fraction of the proton wave function related to the pion (kaon) pole. 
    Additionally, for kaon structure lower center of mass energies are preferable so that sufficient $\Lambda$ decays happen 
    in the far forward region, see Sec.~\ref{sec:forward}. 

    Nuclear properties beyond that of the mean-field shell model can be studied using $A(e,e'NN)$ two-nucleon knockout reactions.  
    These can especially shed light on the nature of the nuclear short-range correlations (SRCs) and their potential relation
    to nucleon medium modifications~\cite{Hen:2016kwk}.  The EIC will enable measurements of these processes up to $Q^2$ values 
    a factor of 3-4 higher than has been achieved so far in fixed target setups~\cite{Hauenstein:2021zql}.  In these two-nucleon 
    knockout reactions in selected kinematics, one \textit{leading} nucleon originates from the interaction with the photon, while 
    the other is the \textit{recoil} partner that originated from the SRC-pair.  As with the previous discussed processes, the 
    detection of recoil nucleons happens in the far forward detector apparatus, due to the boost in the collider lab frame 
    relative to the ion rest frame. Additionally, detection of nuclear fragments ($A-2$), and/or veto its breakup, could be 
    possible improving control over the reaction mechanism in these reactions~\cite{Patsyuk:2021fju}.  
    
    Measurements of 
    single-nucleon knockout reactions in mean-field kinematics are possible at EIC up to $Q^2 \approx 20~\text{GeV}^2$~\cite{Hauenstein:2021zql}.  These would help to constrain the onset of the nuclear color transparency phenomenon~\cite{Dutta:2012ii}, which has not been observed for proton knockout up to $Q^2= 14~\text{GeV}^2$~\cite{HallC:2020ijh}.  Color transparency could also be explored in other kinematics and reaction mechanisms.  One example that was recently explored is meson electroproduction on nuclei in backward kinematics~\cite{Huber:2022wns}, see also Sec.~\ref{subsec:backward}.
    
    Concerning the detection capabilities of the EIC for these 2N knockout reactions, for the leading nucleon the detection 
    region depends on the ion beam energy. With 41~GeV/A beams, the majority of the leading nucleons is detected in the central 
    detector, while for 110~GeV/A it is detected in the far-forward region, see Fig.~3 of Ref.~\cite{Hauenstein:2021zql}. 
    Moreover, at 110~GeV/A  higher acceptance for recoil nucleons is also achieved.  For leading neutrons, however, with 
    110 GeV the neutrons are outside the angular coverage of the ZDC, and these channels have to be measured at the lower 
    ion beam energy.

\subsection{Exclusive measurements}

Hard exclusive reactions on light nuclei can be measured in both the coherent and incoherent (nuclear breakup) 
channels~\cite{Dupre:2015jha}.  The coherent channel, similarly to the case of the nucleon discussed in Secs.~\ref{sec:GPD} 
and \ref{sec:mass}, would give access to 3D tomography of light nuclei in quark and gluon degrees of freedom and 
the extraction of mechanical properties of light nuclei.  It could also potentially shed light on the size of 
non-nucleonic components of the nuclear wave function.  The incoherent channel, on the other hand, can be used to 
study medium modifications of nucleon tomography~\cite{Fucini:2019xlc,Fucini:2020lxi} and to probe neutron 3D 
structure~\cite{Rinaldi:2012ft}.  Three of the lightest nuclei (d,$^3$He,$^4$He) have the interesting feature 
that they have different spin and binding energies~\cite{Cano:2003ju,Guzey:2003jh,Scopetta:2004kj,Cosyn:2018rdm,Fucini:2018gso}.  
$^4$He being spin-0 has the advantage that it has only one leading twist GPD in the chiral even sector.  $^3$He 
is a spin-1/2 nucleus, meaning that hard exclusive observables can be similarly defined to those of the free nucleon. 
Lastly, the spin-1 deuteron has a richer structure of GPDs beyond that of the nucleon (associated with its tensor 
polarization modes), meaning that new spin-orbit phenomena can be studied.  In terms of binding energy, the deuteron 
is very loosely bound, while $^4$He is very tightly bound and $^3$He falling somewhat in between.  This gives us 
access to different degrees of nuclear effects that can be studied in these systems.  Additionally, the availability 
of high-precision \textit{ab initio} nuclear wave functions for these light nuclei results in a high degree of 
theoretical control in calculations.  The challenges of detecting these exclusive reactions are covered in more 
detail in Sec.~\ref{sec:forward}.  There, the influence of a secondary focus on the lower limit of the measurable 
$t$-range for the exclusive channel especially deserves highlighting.

\subsection{Charm-flavored hypernuclei}

Hypernuclear physics has been one of the crucial tools for studying the interactions between nucleons and strange hyperons. 
Most experimental studies on hypernuclei have been focused on $\Lambda$ hypernuclei and many precise measurements have 
been performed as reviewed in Ref.~\cite{Feliciello:2015dua}.
Recently, these efforts are extended to hypernuclei with multi-strangeness such as $\Xi$ hypernuclei.

\begin{figure}[t]
\centering
         \includegraphics[width=0.8\textwidth]{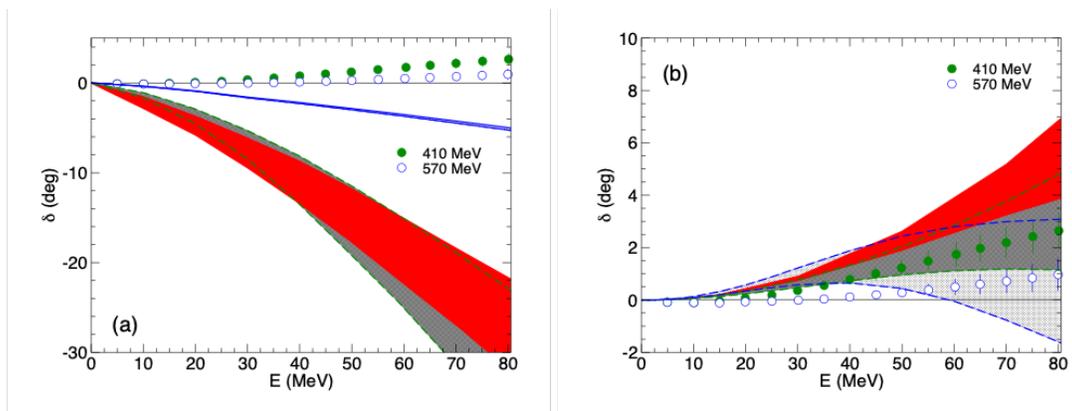}	
	\caption{Predictions for the $\Lambda_c N$ $^3D_1$ phase shift. 
	(a) Results from covariant $\chi$EFT taken from Ref.~\cite{Song:2020isu}. 
	(b) Results based on the $\Lambda_c N$ potential from Ref.~\cite{Haidenbauer:2017dua}. 
	Red (black), green (dark grey), and blue (light grey) bands correspond to $m_\pi = 138$, 410, and 570~MeV, 
	respectively. The width of the bands represent cutoff variations/uncertainties. 
	Lattice results of the HAL QCD Collaboration corresponding to $m_\pi = 410$~MeV (filled circles) and 570~MeV (open circles) 
	are taken from Ref.~\cite{Miyamoto:2019mfk}. 
	The figure is from Ref.~\cite{Haidenbauer:2021tlk}.
	}
	\label{fig:rigidity}
\end{figure}

Recently, there have been interests in charm hypernuclei of which the existence was predicted almost 45 years 
ago~\cite{Tyapkin:75,Dover:1977jw} right after the discovery of the charm quark.
As the strange hypernuclei structure heavily depends on the $\Lambda$-nucleon interactions, the stability of charm 
hypernuclei depends on the $\Lambda_c$-nucleon interactions.
Following the seminal works of 1980s, there have been many theoretical model calculations on various states of 
$\Lambda_c$ hypernuclei. 
The calculated spectra of charm hypernuclei are found to be sensitive to the $\Lambda_c$-nucleon interactions.
(See, for example, Refs.~\cite{Krein:2017usp,Wu:2020nin} for a review.) 
As there is no empirical information on the $\Lambda_c N$ interactions, various ideas were adopted for modeling the 
potential between $\Lambda_c$ and the nucleon.
In recent calculations, lattice simulation results were used to model this potential. 
However, depending on the approach to the physics point from the unphysical quark masses used in lattice calculations, 
the extrapolated potentials lead to very different results for the $\Lambda N$ interactions~\cite{Haidenbauer:2021tlk}.
Figure~\ref{fig:rigidity} shows different predictions for the $\Lambda_c N$ $^3D_1$ phase shift extrapolated from 
the same lattice calculations but with different extrapolation methods. It shows that the results are completely 
different depending on the extrapolation approaches.
Therefore, experimental measurements on charm hypernuclei are strongly required to shed light on our understanding 
of the $\Lambda_c N$ interactions.

Experimentally, earlier efforts to find charm hypernuclei started right after the seminal work of Ref.~\cite{Tyapkin:75} 
and a few positive reports on the existence of charm hypernuclei (called supernuclei at that time) were 
claimed~\cite{Lyukov:1989nn}.
However, no serious follow-up research was reported and, in practice, there is no experimental information on 
charm hypernuclei.
The experimental investigations in this topic would be possible at future hadron beam facilities~\cite{PANDA:2021ozp}.
The experimental instrumentation of the EIC allows for precise measurements and would offer a chance to 
study charm hypernuclei.
So far $\Lambda$ hypernuclei have been studied extensively with high intensity meson beams as well
as electron beams.
Electro-production of Lambda hypernuclei was studied with the
$^A Z(e,e'K^+)^A_\Lambda(Z-1)$ reaction and similar reaction,
$^A Z(e,e' D^-) ^A_{\Lambda_c^+}Z$ will produce charm hypernuclei by 
converting a neutron to $\Lambda_c^+$ and $D^-$.
Through observation of produced $D^-$ and scattered electron, the missing mass
technique can be applied to the spectroscopic study of charm
hypernuclei.
Therefore, studying charm hypernuclei with electron-ion collider would open a new way to study heavy-flavored 
nuclei with the future hadron beam facilities. 
This investigation can also be extended to the bottom sector~\cite{Feliciello:2012zz}, which is simpler than 
the charm sector as there is no Coulomb interaction between $\Lambda_b$ and nucleons. 
Therefore, comparing the properties of bottom hypernuclei and strange hypernuclei would give a clear clue on 
the mass dependence of the strong interactions.
The designed energy range of EIC would allow further investigations.


\section{Precision studies of Lattice QCD in the EIC era}
\label{sec:lattice}

Lattice QCD enables the first-principles solution of QCD in the strong-coupling regime, and thereby facilitates calculations that can both guide the analysis of key physics quantities to be determined at the EIC, and provide complementary calculations that can further the physics potential of the EIC.  The calculation of the internal structure of the nucleon, pion and other hadrons in terms of the fundamental quarks and gluons of QCD has been a key effort of lattice calculations since the inception of lattice QCD.  Notably, there have been the first-principles calculation of the electromagnetic form factors, and of the low moments of the unpolarized and polarized parton distribution functions and of the generalized form factors.  Similarly, the low-lying spectrum of QCD has been a benchmark calculation that now including the electroweak splittings.  Nevertheless, the formulation of lattice QCD in Euclidean space imposes important restrictions.  Firstly, time-dependent quantities, and in particular those related to matrix elements of operators separated along the light cone, could not be calculated, thereby precluding the computation of quantities, such as the $x$-dependent parton distribution functions. Further, scattering amplitudes, and thereby information about resonances in QCD, eluded direct computation. In both the fields of three-dimensional imaging and spectroscopy key theoretical advances have circumvented these restrictions and transformed our ability to address key questions of QCD in the strong-coupling regime.

\subsection{Three-dimensional Imaging of the Nucleon}
\label{3D-imaging}
The electromagnetic form factors, and the generalized form factors corresponding to the moments with respect to $x$ of the GPDs, can be expressed as the matrix elements of time-independent, local operators amenable to computation in lattice QCD on a Euclidean grid.  In particular, there has been a progression of calculations of the lowest moments of the isovector generalized form factors~\cite{Hagler:2003jd,LHPC:2007blg,LHPC:2010jcs} that have already provided important insight into three-dimensional imaging of the nucleon, notably in discerning the role of orbital angular momentum.  

The realization that $x$-dependent distributions including the one-dimensional parton distribution functions and the quark distribution amplitudes, and the three-dimensional GPDs could be computed from the matrix elements of operators at Euclidean separations, with its genesis in Large-Momentum Effective Theory (LaMET)~\cite{Ji:2013dva}, or quasi-PDF approach, has spurred a renewal in the first-principles calculation of hadronic and nuclear structure.  For the isovector distributions, the basic matrix elements are those of spatially separated quark and anti-quark fields, joined by a Wilson line so as to ensure gauge invariance; an alternative approach to relating the resulting lattice matrix elements to the familiar PDFs is the pseudo-PDF framework~\cite{Radyushkin:2017cyf}. While both the quasi- and pseudo-PDFs methods share the same matrix elements, the former matches the lattice data to the light-cone PDFs using a large momentum expansion, while the latter is based on a short distance expansion. A further framework that encompasses both the quasi-PDF and pseudo-PDF approaches is that of the so-called ``Good Lattice Cross Sections" method that admits spatially separated gauge-invariant operators thereby simplifying the lattice renormalization at the expense of computational cost~\cite{Ma:2014jla}.  Characteristic of any of these approaches is the need to attain high spatial momentum on the lattice in order to obtain a controlled description of the $x$-dependent PDF. For the most easily accessible isovector nucleon PDFs, there are now several calculations at the physical light- and strange-quark masses. Recent reviews can be found in Refs.~\cite{Lin:2017snn,Cichy:2018mum,Detmold:2019ghl,Ji:2020ect,Lin:2020ijm,Constantinou:2020pek,Constantinou:2020hdm,Cichy:2021lih,Constantinou:2022yye}.

Each of the approaches introduced above admits the calculation of the GPDs, and both the incoming and outgoing hadrons now have to be boosted to high but distinct spatial momenta to introduce a non-zero momentum transfer $-t$.

\subsubsection{Parton distribution functions}   

The direct calculation of distribution functions is not possible in lattice QCD as the latter is formulated with a Euclidean metric, while the former have a light-cone nature. The last decade has been instrumental in attaining the $x$-dependence of PDFs through a number of approaches, such as the hadronic tensor~\cite{Liu:1993cv}, auxiliary quark field~\cite{Detmold:2005gg,Braun:2007wv}, the quasi-PDFs~\cite{Ji:2013dva}, pseudo-PDFs~\cite{Radyushkin:2016hsy}, current-current correlators~\cite{Ma:2014jla}, and with an   OPE~\cite{Chambers:2017dov}. The most intensively-studied methods are the quasi- and pseudo-PDFs, which rely on calculation of matrix elements of non-local operators that are coupled to hadronic states that carry non-zero momentum. The non-local operators contain a straight Wilson line with a varying length in the same spatial direction as the momentum boost. Naturally, the corresponding matrix elements are defined in coordinate space, and can be transformed to the desired momentum space, $x$, with a Fourier transform. A factorization process relates the quasi and pseudo distributions to the light-cone PDFs, with the matching kernel calculated in perturbation theory. Both methods have been used for lattice calculations using ensembles of gauge configurations at physical quark masses~\cite{Lin:2017ani,Alexandrou:2018pbm,Chen:2018xof,Alexandrou:2018eet,Liu:2018hxv,Lin:2018pvv,Joo:2020spy,Bhat:2020ktg,Lin:2020fsj}. Such studies correspond to different lattice discretizations (actions) and parameters and a comparison may reveal systematic effects related to the employed methodology, discretization and volume effects.

\begin{figure}[h!]
\begin{center}
\includegraphics[scale=0.945]{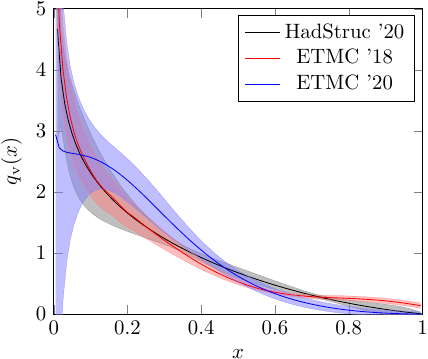}
\hspace*{0.15cm}
\includegraphics[width=0.45\textwidth]{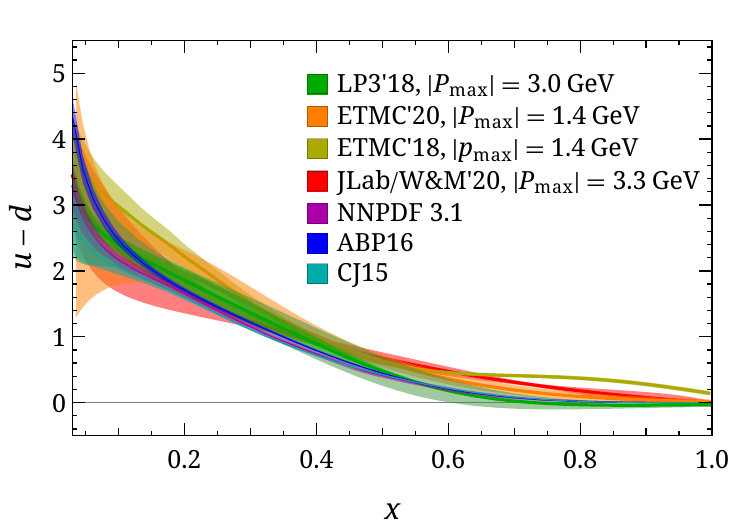}
\includegraphics[width=0.4\textwidth]{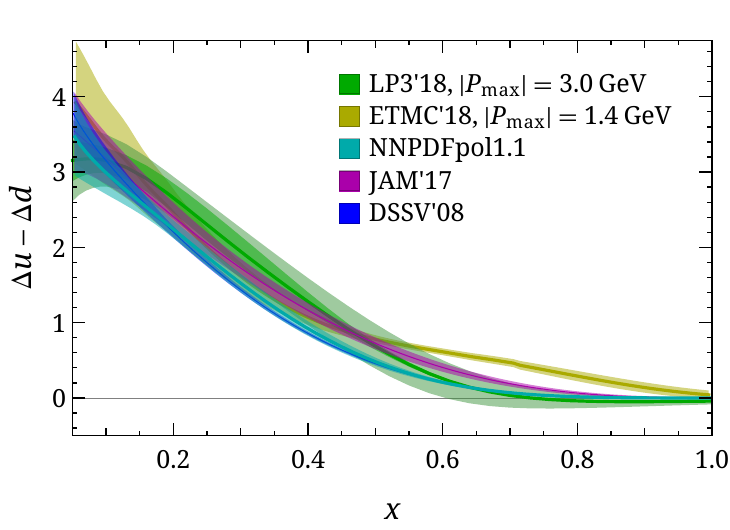}
\includegraphics[width=0.4\textwidth]{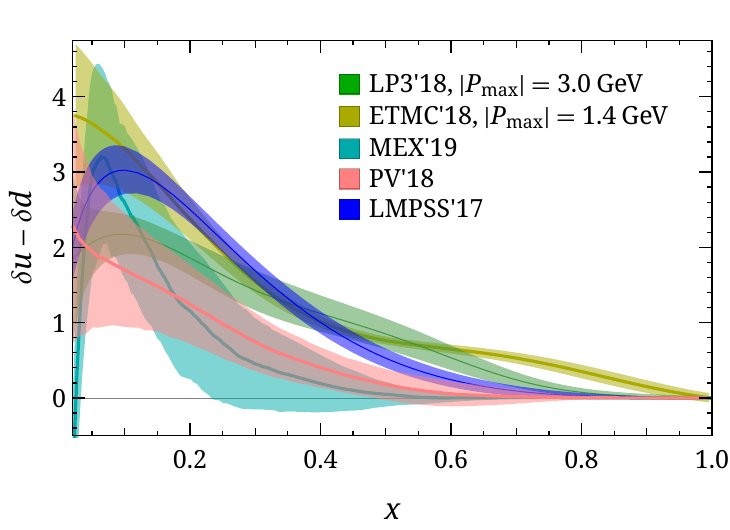}
\caption{Upper left: A selection of lattice-QCD results on the unpolarized PDF using the quasi-PDFs method~\cite{Alexandrou:2019lfo} (red band) and pseudo-ITDs from Ref.~\cite{Joo:2020spy} (gray band) and Ref.~\cite{Bhat:2020ktg} (blue band).
A comparison of unpolarized isovector nucleon PDFs from lattice QCD (upper right), helicity (lower left) and transversity (lower right) at or near the physical pion mass~\cite{Joo:2020spy,Chen:2018xof,Lin:2018pvv,Liu:2018hxv,Alexandrou:2018pbm,Alexandrou:2018eet,Alexandrou:2019lfo,Bhat:2020ktg} with global fits. Plots taken from Ref.~\cite{Constantinou:2022yye}.
All results are given in the $\overline{\text{MS}}$ scheme at a renormalization scale of 2~GeV.}
\label{fig:nucleon_QPFG_pPDF}
\end{center}
\end{figure}

In Fig.~\ref{fig:nucleon_QPFG_pPDF} we show results for the unpolarized isovector valence PDF for the proton. The results indicated by HadStruc'20~\cite{Joo:2020spy} and  ETMC '20~\cite{Bhat:2020ktg} have been obtained using the pseudo-PDFs method, while ETMC'18~\cite{Alexandrou:2018pbm} uses the quasi-PDFs approach. The results are very encouraging, exhibiting agreement for a wide range of values for $x$. The small tension at large $x$ is due to systematic effects such as higher-twist contamination and the ill-defined inverse problem in the reconstruction of the $x$ dependence of the PDFs.  In fact, Refs.~\cite{Alexandrou:2018pbm,Bhat:2020ktg} analyze the same raw data, and they differ in the analysis (quasi-PDFs versus pseudo-PDFs). This corroborates that the large-$x$ region has contamination from the aforementioned systematic effects. 
A similar tension is also present in the comparison of the lattice data, e.g, of Ref.~\cite{Joo:2020spy} with the global analyses of experimental data sets shown in the right panel of Fig.~\ref{fig:nucleon_QPFG_pPDF}. 
When predicting spin-dependent PDFs, lattice calculations may already provide comparable predictions to phenomenological global analyses. 
The lower panel of Fig.~\ref{fig:nucleon_QPFG_pPDF} summarizes the lattice predictions for helicity and transversity nucleon isovector PDFs at physical pion mass~\cite{Alexandrou:2018pbm,Lin:2018pvv,Alexandrou:2018eet,Liu:2018hxv}. 
The helicity lattice results are compared to two phenomenological fits, NNPDFpol1.1~\cite{Nocera:2014gqa} and JAM17~\cite{Ethier:2017zbq}, exhibiting nice agreement.
The lattice results for the transversity PDFs have better nominal precision than the global analyses by PV18 and LMPSS17~\cite{Lin:2017stx}.
The success in extracting the $x$ dependence of PDFs is a significant achievement for lattice QCD, and has the potential to help constrain PDFs in kinematic regions where experimental data are not available. The synergy of lattice QCD results and global analysis is currently under study and some results can be found in Refs.~\cite{Cichy:2019ebf,Bringewatt:2020ixn,DelDebbio:2020rgv}.

\subsubsection{Generalized parton distributions}
\label{subsubsec:GPDs}
\begin{figure}[!ht]
    \centering

\includegraphics[width=0.4\textwidth]{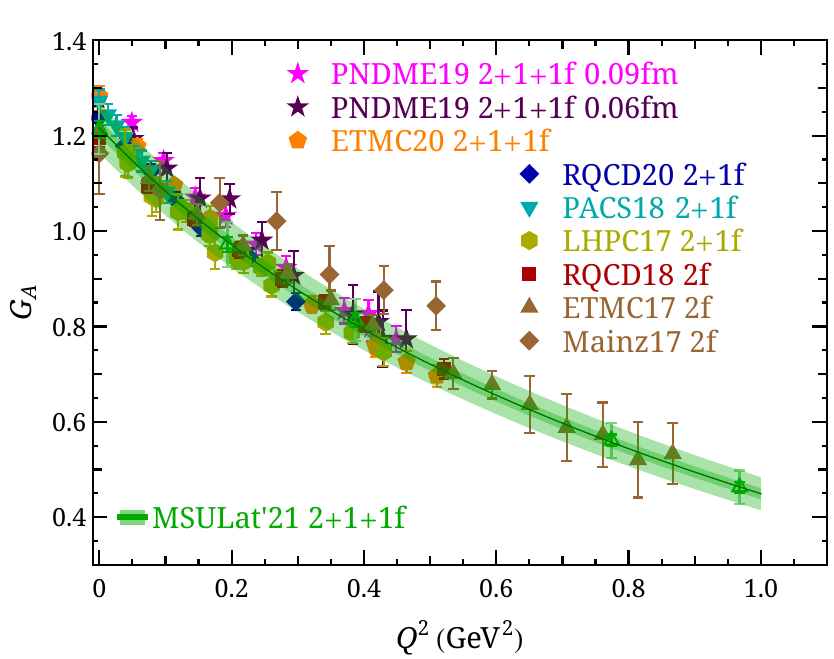}
\includegraphics[width=0.45\textwidth]{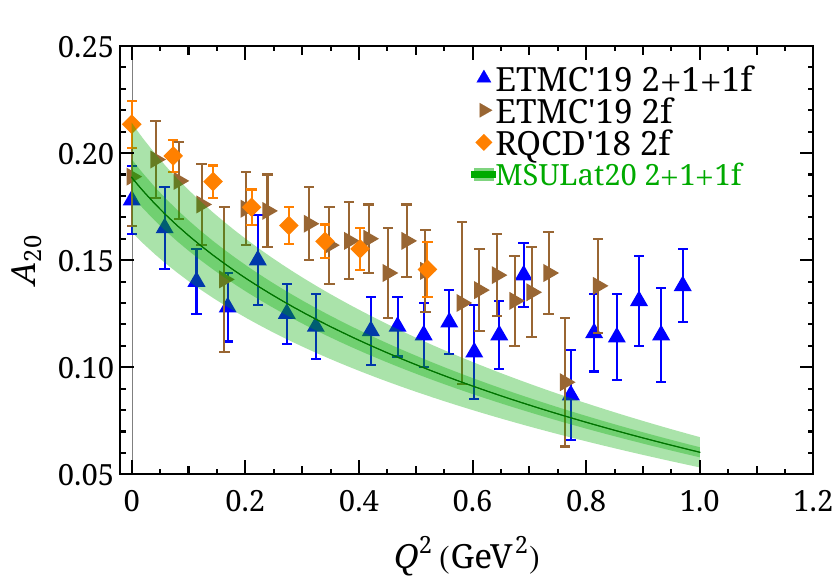}
    \caption{Summary of lattice calculations of $G_A(-t)$ (left) and $A_{20}(-t)$ (right) using ensembles at or near physical quark masses. The label of $G_A$ results correspond to: ETMC '20~\cite{Alexandrou:2020okk}, RQCD '20~\cite{RQCD:2019jai}, PNDME '19~\cite{Jang:2019vkm}, PACS '18~\cite{Shintani:2018ozy}, RQCD '18~\cite{Bali:2018qus}, ETMC '17~\cite{Alexandrou:2017hac}, LHPC '17~\cite{Hasan:2017wwt} and MSULat'21~\cite{Lin:2021brq} . 
    The corresponding results for $A_{20}$ are: ETMC '19 ~\cite{Alexandrou:2019ali}\footnote{Larger-volume results are plotted for the ETMC'19 2f calculation.}, RQCD '18~\cite{Bali:2018zgl}, and MSULat'20~\cite{Lin:2020rxa}.}
\label{fig:latt_moments_GPDs}
\end{figure}

Information on GPDs from lattice QCD is mostly extracted from their Mellin moments, that is the form factors (FFs) and generalized form factors (GFFs). This line of research has been very successful within lattice QCD, and several results for the form factors using ensembles with physical quark masses appeared in the last five years. Furthermore, the flavor decomposition for both the vector and axial form has been performed, giving the individual up, down, strange and charm contributions to these quantities~\cite{Sufian:2016pex,Sufian:2017osl,Alexandrou:2018zdf,Alexandrou:2018sjm,Alexandrou:2019olr,Alexandrou:2021wzv}. A summary of state-of-the-art calculations can be found in Ref.~\cite{Constantinou:2020hdm}. In the left panel of Fig.~\ref{fig:latt_moments_GPDs} we show results on the axial form factor at physical quark masses from various lattice groups employing different lattice discretization and analysis methods. Its forward limit is the axial charge, $g_A\equiv G_A(0)$, which is a benchmark quantity for lattice QCD, and is related to the intrinsic spin carried by the quarks in the proton. As can be seen, the results are in very good agreement, despite the fact that not all sources of systematic uncertainties have been fully quantified. The level of agreement indicates that remaining systematic effects are small. Further, $g_A$ is found to be in agreement with the world average of experimental data~\cite{Markisch:2018ndu}. This is a breakthrough for lattice QCD calculations, as they demonstrate that agreement with experiment is achieved once systematic uncertainties are eliminated.

More recently, lattice results on the GFFs associated with the sub-leading Mellin moments of GPDs (one-derivative operators) became available at the physical pion mass. In the right panel of Fig.~\ref{fig:latt_moments_GPDs} we show results on $A_{20}$, which appears in the decomposition of the energy momentum tensor. Its forward limit is the quark momentum fraction, $\langle x \rangle \equiv A_{20}$, which enters the spin decomposition~\cite{Ji:1996ek}. Extracting GFFs is more challenging for a number of reasons. First, the introduction of covariant derivatives increases the gauge noise, as well as the uncertainties due to cutoff effects. Second, in general the number of GFFs increases, requiring independent matrix elements to disentangle the GFFs. Third, beyond the NNNLO Mellin moments, there is unavoidable mixing under renormalization. The introduction of matrix elements with greater than three covariant derivatives introduces power-divergent mixing with matrix elements with few derivatives, thereby precluding the calculation of the higher Mellin moments. Consequently, there are limitations in mapping the three-dimensional structure of the nucleon from the FFs and GFFs. 

\begin{figure}[!ht]
\centering
    \includegraphics[scale=0.475]{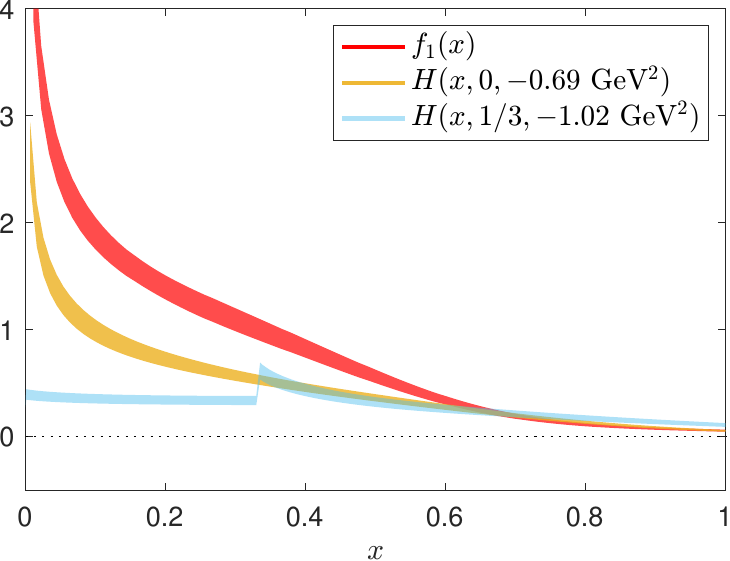}\hspace*{0.25cm}
    \includegraphics[scale=0.475]{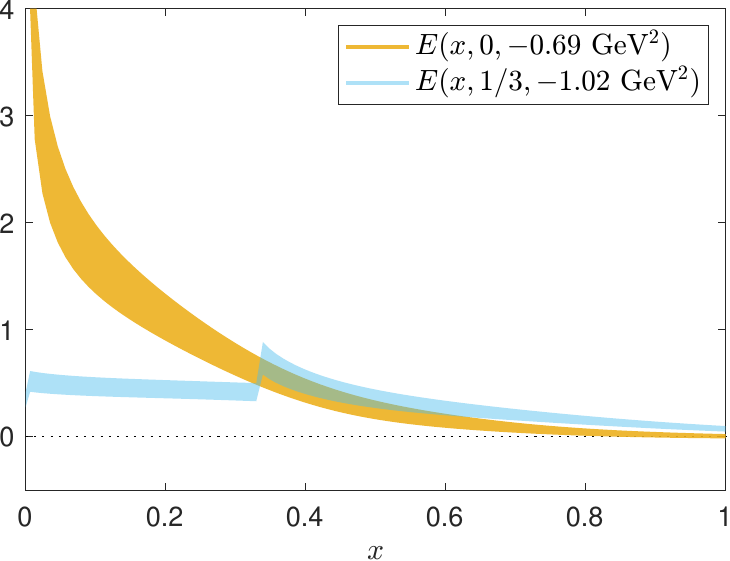}\\ \includegraphics[scale=0.475]{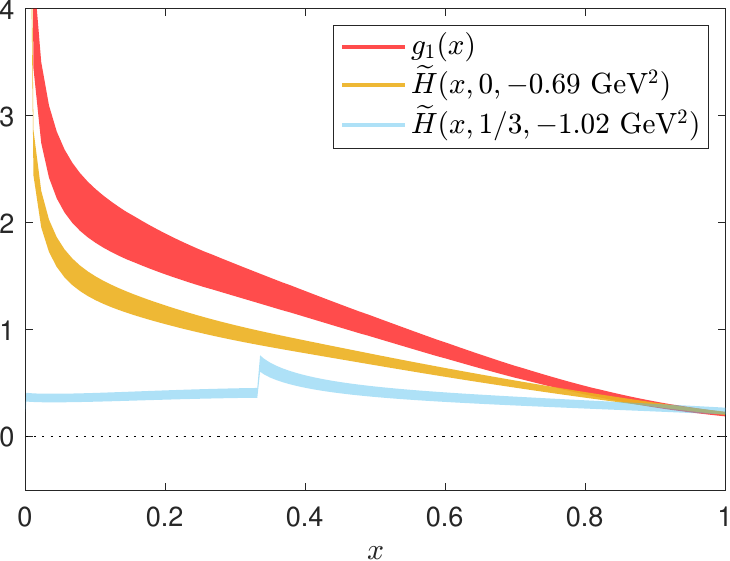}\hspace*{0.25cm} \includegraphics[scale=0.475]{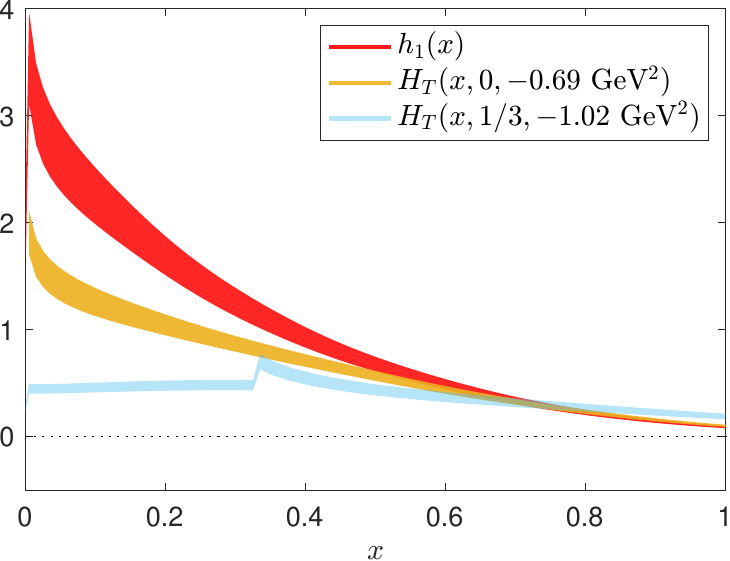}
    \caption{The non-polarized $H$ and $E$, helicity $\widetilde{H}$ and transversity $H_T$ GPDs at $\{t,|\xi|\}=\{0,0\}$,\,$\{-0.69\text{ GeV}^2,0\}$,
    \,$\{-1.02\text{ GeV}^2,1/3\}$ extracted from the 260-MeV pion mass lattice calculations of Ref.~\cite{Alexandrou:2020zbe,Alexandrou:2021bbo}.}
    \label{fig:latt_GPDs}
\end{figure}

\begin{figure}[tb]
\includegraphics[width=0.3\textwidth]{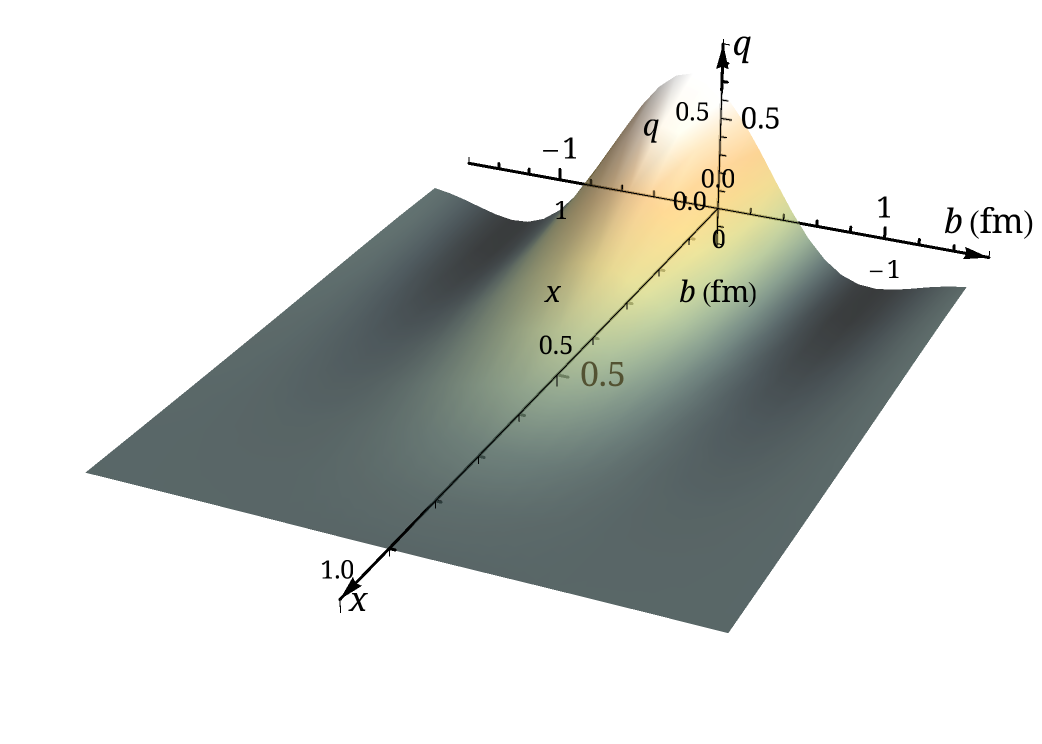}
\includegraphics[width=0.65\textwidth]{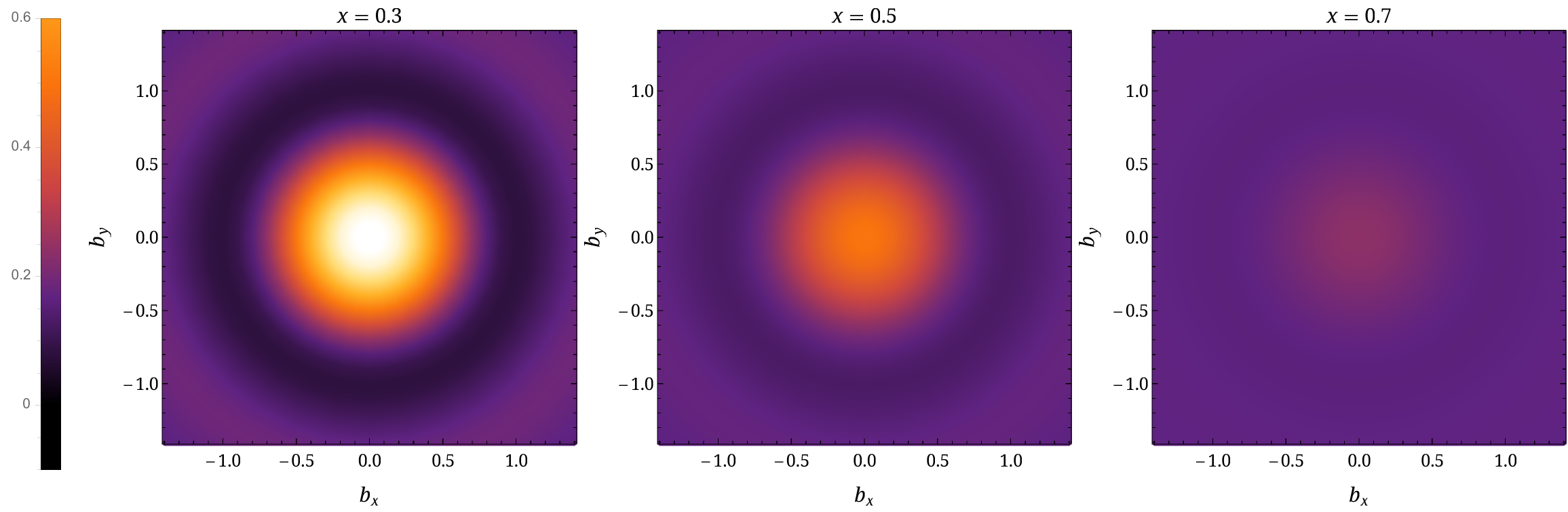}
\caption{
(left) Nucleon tomography: three-dimensional impact parameter--dependent parton distribution as a function of $x$ and $b$ using lattice $H$ at physical pion mass.
(right) Two-dimensional impact-parameter--dependent isovector nucleon GPDs for $x=0.3$, 0.5 and 0.7 from the lattice at physical pion mass. 
Source: Ref.~\cite{Lin:2020rxa}.
}
\label{fig:latt-tomography}
\end{figure}

Methods such as large momentum factorization (quasi-distributions) and short distance factorization (pseudo-distributions) are very promising in extracting the $x$-dependence of GPDs~\cite{Ji:2015qla,Liu:2019urm,Radyushkin:2019owq,Hou:2022sdf} avoiding the challenges associated with renormalization that are present in the calculation of GFFs mentioned above. However, the calculations are very taxing because, unlike FFs and GFFs, GPDs are frame dependent objects and are defined in a symmetric (Breit) frame. This increases significantly the computational cost, as a separate calculation is needed for each value of $t$.
The $x$-dependence of nucleon GPDs has already been explored, in the Breit frame, for the unpolarized ($H,\,E$), helicity ($\widetilde{H},\,\widetilde{E}$) and transversity ($H_T,\,E_T,\,\widetilde{H}_T,\,\widetilde{E}_T$) GPDs~\cite{Alexandrou:2020zbe,Alexandrou:2021bbo}. Such calculations are very timely, since the EIC will measure the DVCS process with polarized electrons and longitudinal and transverse polarized protons to extract the CFFs of $H,\,E$ and $\widetilde{H}$. It should be noted that, to date, lattice calculations of GPDs are exploratory and are available for only a few values of $t$ for zero and nonzero skewness, $\xi$. Nevertheless, lattice results are useful for a qualitative understanding of GPDs. For instance, one can find characteristics for the $t$ dependence for each operator under study. 
For instance, the lattice results of Fig.~\ref{fig:latt_GPDs} indicate that the decay of the GPD with $t$ is fastest in $H$, followed by $H_T$, and then $\widetilde{H}$. Also, one can compare the hierarchy of GPDs at each value of $t$. On this aspect, it is found that at $t=0$, $f_1\equiv H(t=0)$ is dominant, followed by $h_1\equiv H_T(t=0)$ and $g_1\equiv \widetilde{H}(t=0)$. As $-t$ increases, $H$ remains dominant, while the hierarchy of $H_T$, and then $\widetilde{H}$ interchanges. Finally, lattice results can be used to check sum rules. For more details we refer the reader to Refs.~\cite{Bhattacharya:2019cme,Alexandrou:2021bbo}. We emphasize that lattice calculations on GPDs are at the proof-of-concept stage, but results are promising. Once the lattice data can access a wide range of $t$, their $t$-dependence can be parameterized. This is very useful because the parameterizations can be used to extract the GPDs in the impact-parameter space as done in Refs.~\cite{Lin:2020rxa,Lin:2021brq} at physical pion mass. The green bands in Fig.~\ref{fig:latt_moments_GPDs} show the moments of lattice $x$-dependent GPD results at zero skewness; they are in nice agreement with the traditional local-operator methods, which shows there will be a promising future for lattice QCD contributions in GPD tomography. 
Figure~\ref{fig:latt-tomography} shows the first LQCD results of impact-parameter--dependent 2D distributions at $x=0.3$, 0.5 and 0.7~\cite{Lin:2020rxa}. Similar tomography results for helicity GPD, $\tilde{H}(x,\xi=0,Q^2)$ can be found in  Ref.~\cite{Lin:2021brq}.

The progress in the field of $x$-dependent GPDs from lattice QCD is being also extended to twist-3 GPDs~\cite{Dodson_Lattice:2021}. We anticipate that, in the near future, lattice results will be incorporated in phenomenological analysis of GPDs at both the twist-2 and twist-3 level.  Lattice-computed twist-3 GPDs can have advantages with regards to extracting twist-2 GPDs at kinematics where twist-3 contributions aren't negligible. In fact, this may even be a required step before one attempts to extract twist-2 GPDs from DVEP data.

\subsubsection{Transverse momentum dependent distributions}
    
In contrast to GPDs, TMDs describe the three-dimensional structure in terms of the longitudinal momentum-fraction $x$, and the transverse momentum of the partons. One of the additional challenges that arise in TMD calculations is the presence of the rapidity divergences that need an additional regulator. Such divergences can be factorized into the so-called soft function, which can be separated into a rapidity-independent and a rapidity-dependent part. The latter defines the Collins-Soper (CS) kernel, which depicts the rapidity evolution. One of the challenges is that the soft function is non-perturbative for small transverse momenta.

The TMDs involve the matrix elements of staple-like Wilson lines that extend along the light cone, imposing analogous restrictions on their calculation within lattice QCD as encountered for the case of PDFs and GPDs described above.  The first efforts at overcoming these restrictions employed space-like-separated staples that approached the light-cone as the length of the staple increased~\cite{Musch:2010ka}, in particular focusing on the time-odd Boer-Mulders and Sivers functions~\cite{Musch:2011er,Yoon:2017qzo} and their relation to the corresponding processes in Drell-Yan and SIDIS, including calculations for the pion~\cite{Engelhardt:2015xja}. 
    
\begin{figure}[htb!]
	\centering
	\includegraphics[scale=1]{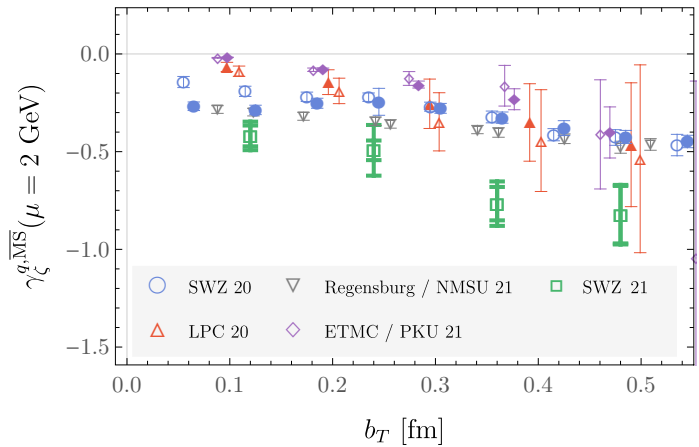}
	\caption{The Collins-Soper kernel as a function of $b_T$ as extracted from various lattice QCD calculations. We show results from SWZ~\cite{Shanahan:2020zxr,Shanahan:2021tst}, LPC~\cite{LatticeParton:2020uhz}, Regensburg/NMSU~\cite{Schlemmer:2021aij}, and ETMC/PKU~\cite{Li:2021wvl}. Open and filled symbols of the same shape and color correspond to results from the same lattice group. Source: Ref.~\cite{Shanahan:2021tst}.}
\label{fig:CS_latt_comp}
\end{figure}

More recently, there has been extensive work on exploring TMDs within the quasi-PDF approach~\cite{Ji:2014hxa,Ji:2018hvs,Ebert:2018gzl}, as well as the soft function~\cite{Ji:2019sxk,Ji:2019ewn}. The Collins-Soper kernel has been studied by a few collaborations~\cite{Shanahan:2020zxr,LatticeParton:2020uhz,Schlemmer:2021aij,Li:2021wvl,Shanahan:2021tst} and a comparison is shown in Fig.~\ref{fig:CS_latt_comp}. Presently, such a comparison is qualitative, as systematic uncertainties are not fully quantified. Nevertheless, the agreement is very good and encouraging. 
    
\subsubsection{Gluon and flavor-singlet structure}
The calculation of the flavor-singlet structure of hadrons is considerably more challenging than those for the flavor-non-singlet quantities that have been the focus of the most precise studies.  The challenges are primarily related to the degrading signal-to-noise ratio that impacts calculations both of the gluon distributions, and of the flavor-singlet quark distributions with which they mix.  Recently, the first calculations of the unpolarized $x$-dependent gluon distributions in the nucleon have been performed using quasi-PDF~\cite{Fan:2018dxu} and pseudo-PDF~\cite{Fan:2020cpa,HadStruc:2021wmh,Fan:2021bcr} methods, as well as the first lattice gluon helicity study~\cite{HadStruc:2022yaw}. Within the present statistical precision and through a qualitative comparison with global analyses of the gluon helicity distribution, the lattice calculation hinted at a positive gluon polarization contribution to the nucleon spin budget.

\begin{figure}[htb!]
    \centering
\includegraphics[width=0.45\textwidth]{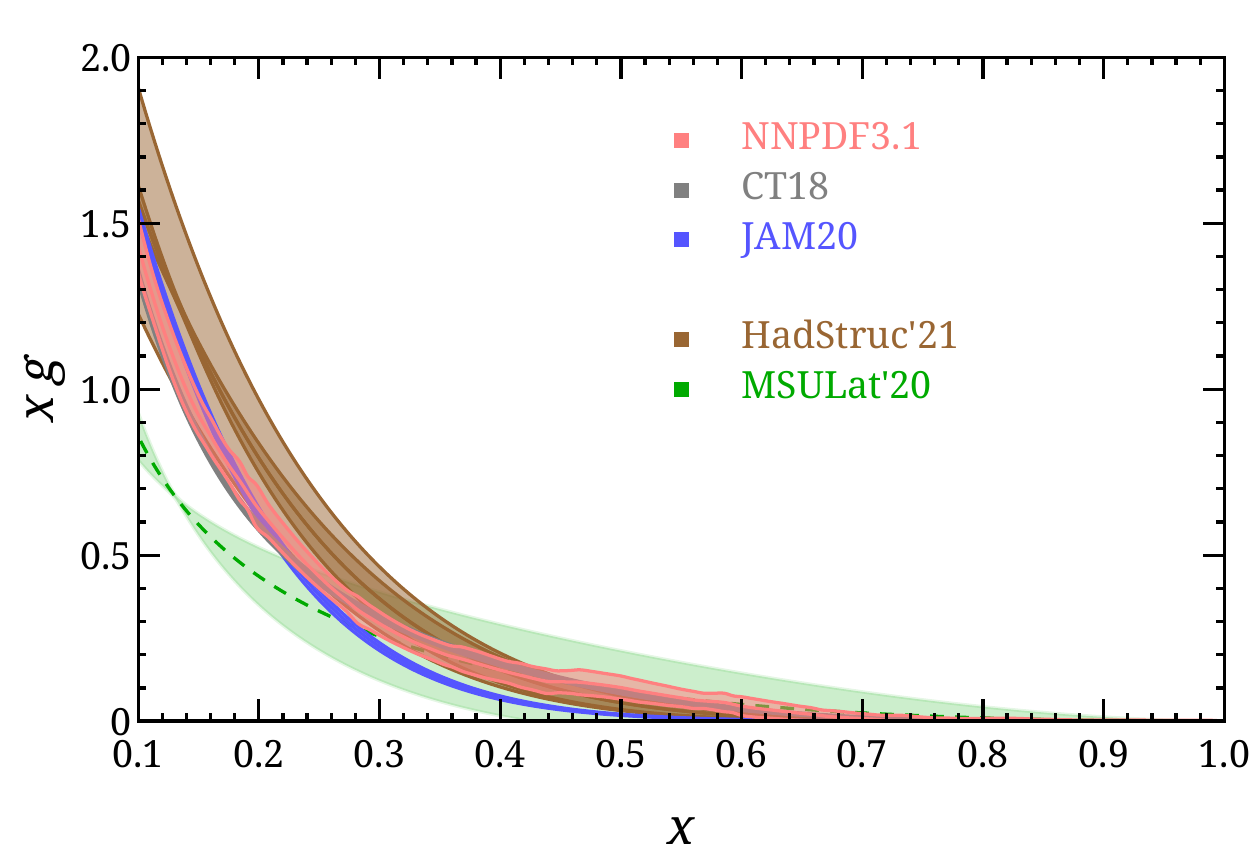}
    \includegraphics[width=3in]{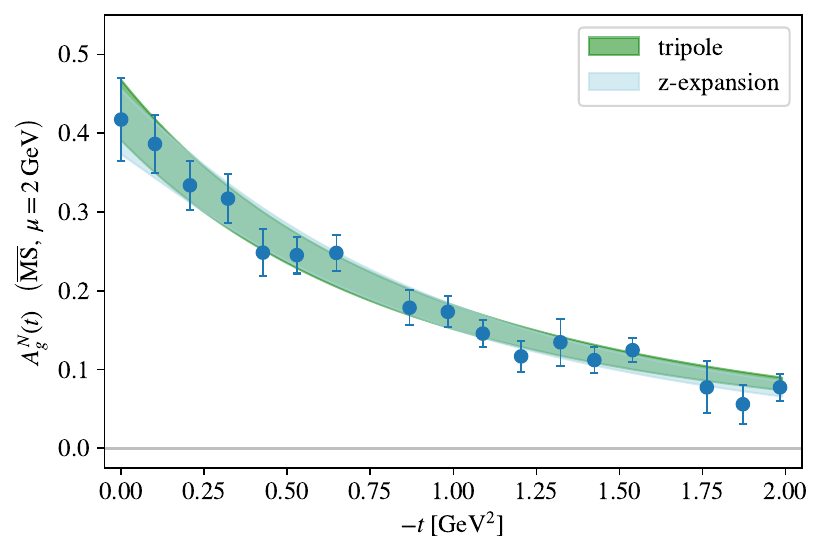}
    \vspace*{-0.25cm}
    \caption{Left: lattice results on the unpolarized nucleon gluon PDF using a two-parameter parametrization $x g(x) =N x^\alpha (1-x)^\beta$ by MSULat'20~\cite{Fan:2020cpa} and  HadStruc'21~\cite{HadStruc:2021wmh}  Also shown are the unpolarized gluon PDFs extracted from global fits to experimental data: CT18~\cite{Hou:2019efy}, NNPDF3.1~\cite{NNPDF:2017mvq}, and JAM20~\cite{Moffat:2021dji}. 
    Right: the gluon nucleon GFF in a lattice calculation corresponding to $M_\pi = 450 (5)~{\rm MeV}$; the bands show a multipole fit with $n = 3$ (green), and a model-independent $z$ expansion (blue). Source: Ref.~\cite{Pefkou:2021fni}.}
    \label{fig:latt_x_gluon}
\end{figure}
A comparison of the calculation with phenomenological parametrizations is shown as the left-hand panel in Fig.~\ref{fig:latt_x_gluon}. While this calculation is at unphysically large pion masses, with limited understanding of the systematic uncertainties, it demonstrates the potential of lattice QCD to complement and augment insights into hadron structure from experiment, notably at large $x$.

The calculation of the gluon contributions to three-dimensional structure of hadrons proceeds as in the case of that of the valence quarks described above.  In particular, the gluonic contribution to the GFF has been computed~\cite{Shanahan:2018pib,Shanahan:2018nnv,Pefkou:2021fni} thereby enabling, when combined with the corresponding quark contributions, the pressure and shear forces within a nucleon to be computed, shown as the right-hand panel of Fig.~\ref{fig:latt_x_gluon}.

\subsection{LQCD and Spectroscopy}
The ability to study multi-hadron states and resonances from lattice QCD calculations was transformed by the realization that, for the case of two-body elastic scattering, infinite-volume, momentum-dependent phase shifts could be related to energy shifts at finite volume on a Euclidean lattice~\cite{Luscher:1985dn,Luscher:1986pf,Luscher:1990ux}.  The formalism for elastic scattering has now been extended to coupled-channel scattering, and to multi-hadron final states facilitating a range of calculations that impact our understanding of the spectroscopy of QCD.  Notably, there are now calculations of coupled-channel scattering describing the nature of the isoscalar $a_0, f_0~{\rm and}~f_2$ resonances~\cite{Briceno:2017qmb}, and the first calculation of the decays of the exotic $1^{-+}$ meson~\cite{Woss:2020ayi}.

\begin{figure}[htbp!]
    \centering
    \includegraphics[width=4in]{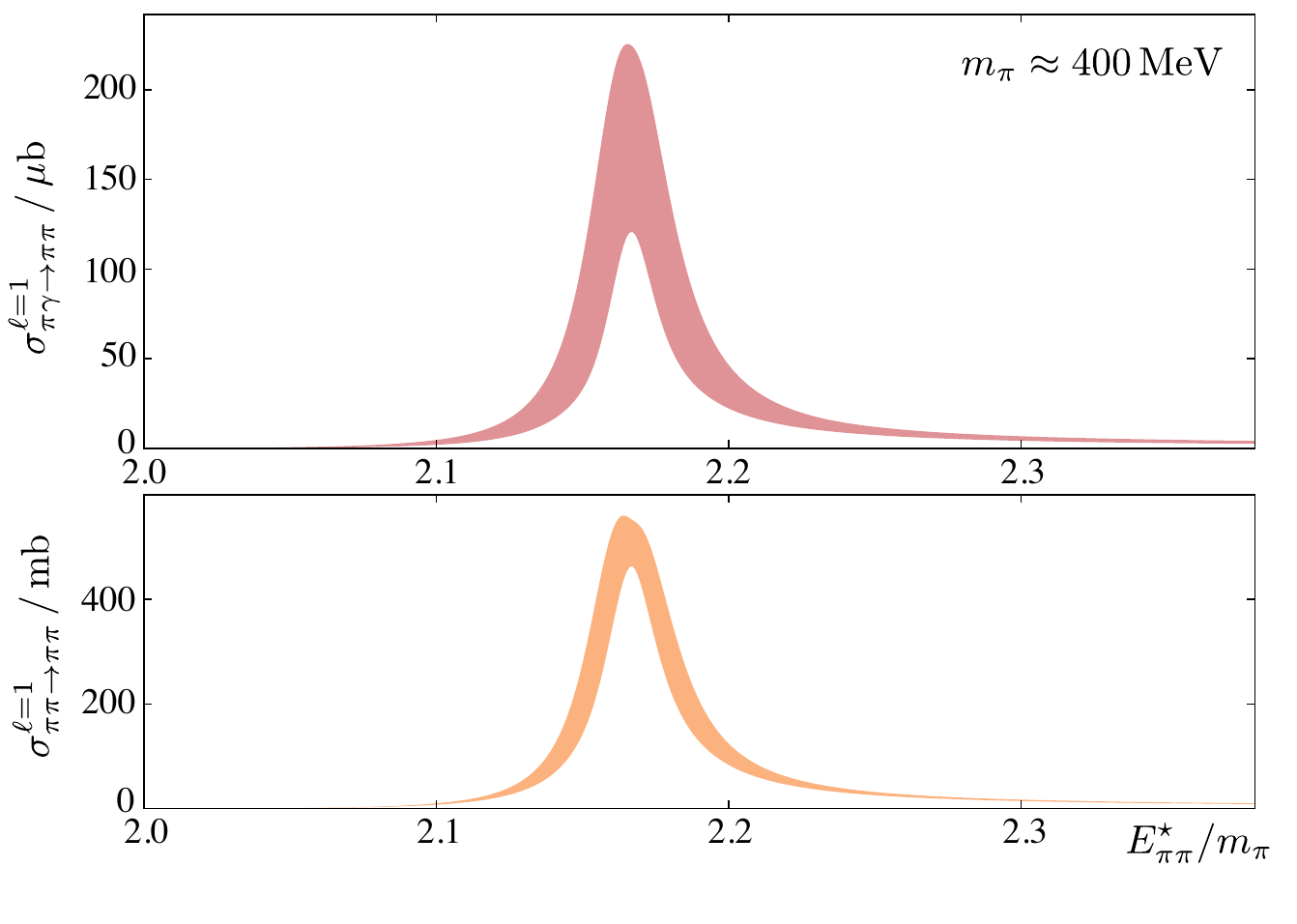}
    \vspace*{-0.4cm}
    \caption{The upper panel shows the $\pi^+ \gamma \rightarrow \pi^+ \pi^0$ cross section as a function of the $\pi\pi$  center-of-mass energy in a calculation with a pion mass $m_\pi \simeq 400~{\rm MeV}$.  The lower panel shows the $l=1$ elastic $\pi\pi$ scattering cross section, with the $\rho$ resonance visible in both cases. Source: Ref.~\cite{Briceno:2016kkp}.}
    \label{fig:lattice_trans}
\end{figure}
Beyond the challenge of computing the spectrum of resonances and their decays, an important development has been that of a formalism for the photo- and electro-production of two-hadron final states, an example of the so-called $1 + {\cal J} \rightarrow 2$ processes~\cite{Briceno:2014uqa,Briceno:2015csa}.  The formalism has been applied to the case of $\pi^+ \gamma \longrightarrow \pi^+ \pi^0$, shown in Figure~\ref{fig:lattice_trans}. Recently, this has been extended to the case of coupled-channel, multi-hadron final states~\cite{Briceno:2021xlc} thereby providing an essential framework underpinning the spectroscopy opportunities through photoproduction at the EIC.

 The calculation of the spectrum of the exotic charmonium and bottomonium states anticipated at the EIC poses several additional challenges beyond those encountered in the light-quark sector. Firstly, a precise understanding of light- and strange-quark spectroscopy is a precursor to precision calculations in the heavy-quark sector since the $c\bar{c}$ can mix with such states in many of the most interesting channels. Secondly, with increasing mass of the quark constituents, the splitting between the different energies on the lattice is compressed, with many $J^{P}$ states at similar energies requiring additional constraints to identify the states from the lattice data.  Finally, there are the many open channels that must be included.  The work so far is largely exploratory~\cite{CLQCD:2019npr,Prelovsek:2020eiw}, with the inclusion of only a limited number of coupled channels.  However, controlled calculations of many of the exotic states anticipated at the EIC are now computationally feasible, with studies both of the $\chi_{c1}(3872)$ and the $X(6900)$ most easily attainable.

\subsection{Outlook}
Many of the "no-go" theorems that until recently have imposed limitations on the range of quantities accessible to first-principles calculation in lattice QCD have now been circumvented through a progression of theoretical advances, with demonstrations of the ability of lattice QCD calculations to add to our understanding of the internal structure and spectroscopy of hadrons.  The advent of the era of exascale computing will enable the precision calculations needed to exploit the opportunities afforded by the EIC~\cite{Detmold:2019ghl,Joo:2019byq}.  Notably, in addition to the emerging precision computations of the isovector quantities, such calculations will be extended to the isoscalar sector.  Precise computations within lattice QCD of the three-dimensional measures of hadron structure, combined with the two-dimensional Generalized Form Factors accessible through exclusive processes at the EIC, will constrain the model dependence in global analysis of experimental data, and will facilitate a more precise three-dimensional imaging of hadrons that either experiment or first-principles calculation can achieve alone.

Despite these advances, there remain physical processes that elude current lattice QCD calculations, notably the direct calculation of real-time scattering cross sections, fragmentation functions, and nuclear response functions.  The rapid advance of Quantum Information Science, and its role as a high-priority research area, will play an increasingly important role in addressing many of these key problems, recognised in the report of the NSAC subcommmittee~\cite{NSAC_QIS}.  Thus far, the investigation of quantum field theory on quantum computers has been restricted to far simpler systems than that of QCD, but the role of QIS both in advancing lattice gauge theory is reviewed in ref.~\cite{Banuls:2019bmf}.  Further, strategies for exploiting quantum computing to directly address processes relevant to the EIC, such as Compton Scattering, are now being formulated~\cite{Briceno:2020rar}.

\section{Science of far forward particle detection}
\label{sec:forward}

\subsection{Far-forward detection overview}

\begin{figure}[htb]
    \centering
    \includegraphics[width =0.9\textwidth]{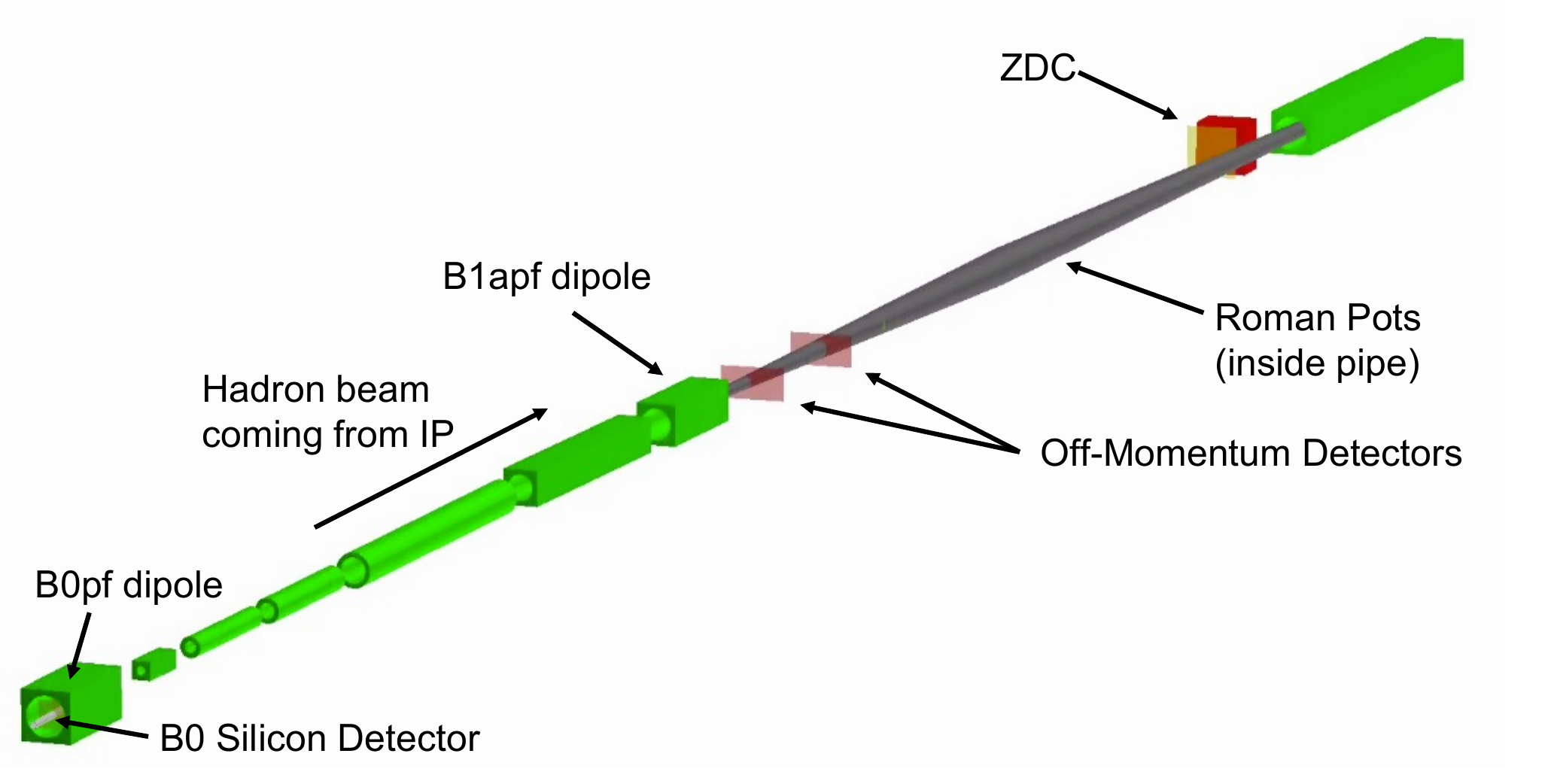}
    \caption{Layout of the IP6 Far-Forward region generated with the EICROOT simulation package~\cite{EICroot} including the dipole magnets (rectangular boxes), quadrupole magnets (cylinders), and the four detector subsystems currently proposed to cover the geometric acceptance.}
    \label{fig:IP6_FF_layout}
\end{figure}

In contrast to colliders that are mainly built to study particles produced at central rapidity, much of the EIC physics critically relies on excellent detection of the target and target fragments moving along, and often within, the outgoing ion beam. Consequently, EIC detectors are from the outset designed with an elaborate far-forward detection system that is closely integrated with the interaction region of the accelerator. The forward detection has several stages: the endcap of the central detector, trackers within a large-bore dipole magnet in front of the accelerator quadrupole (quad) magnets, two sets of Roman pots (one for charged particles at lower rigidity, so-called ``off-momentum detectors"; the other for tagging protons or light ions near the beam momentum) after a larger dipole behind the quads as seen in 
Fig.~\ref{fig:IP6_FF_layout} which shows the layout of IP6 during the time of the Yellow Report, which is largely unchanged. Additionally, a zero-degree calorimeter is employed for tagging neutrons and photons at very small ($<$5 mrad) polar angles. 

This arrangement allows for high-$p_{T}$ cutoffs to be determined by the magnet apertures, such as is the case for the neutron/photon cone going toward the zero-degree calorimeter (which must traverse the full hadron lattice), and for charged particles and photons being tagged in the first, large-bore dipole magnet after the IP, which contains a detector for far-forward particles at polar angles roughly between 5.5 and 20\,mrad. The bore of the first dipole (called B0pf in IP6) has a radius of 20\,cm (while the pre-conceptual design for IP8 has an equivalent dipole magnet with a slightly larger radius), which in principle allows for larger acceptance than 20 mrad, but support structure and services for the detectors will limit how much of the bore can be filled with active detector material. As designs progress, it may be possible to achieve a larger acceptance in the dipole spectrometer at both IP6 and IP8.

On the other hand, for lower-energy proton beams, unavoidable inefficiencies will occur in the transition regions. There is a low-$p_T$ cutoff due to the beam itself, which is most severe for the detection of recoil protons from mid- to high-energy beams (which provide the highest luminosity), for light ions at all energies, and for heavy ion fragments with A/Z close to that of the original beam. For ions, where the $p_T$ \textit{per nucleon} is usually small, acceptance at very low-$p_T$ is extremely important. With a traditional IR layout, low-$p_T$ acceptance can be improved by reducing the angular spread of the beam via reduced beam focusing. However, this has the drawback that it also reduces luminosity and still does not make it possible to reach $p_T$=0.

The kinematics of the EIC are uniquely suited to a more sophisticated forward detection concept than previous colliders. In DIS, the typical longitudinal momentum loss $dp/p \sim x$. At the same time, the intrinsic momentum spread of the particles in the beam is a few $\times 10^{-4}$. With a 10$\sigma$ margin, \textit{all} recoil protons with $x>0.01$ will thus separate out from the beam even at $p_T$=0, and at much lower $x$ for non-zero $p_T$. Since this method only relies on a fractional longitudinal momentum loss (magnetic rigidity), it is independent of the beam energy. For heavy ions, which typically only experience small changes in momentum, rigidity ($\sim A/Z$) can change through emission of nucleons. In particular, emission of a single neutron from an $A\sim100$ nucleus corresponds to a change in rigidity at the 1\% level, which in principle also allows the EIC to detect most nuclear fragments.

\begin{figure}[htb]
    \centering
    \includegraphics[width =0.99\textwidth, trim={51mm 0mm 0mm 0mm}, clip]{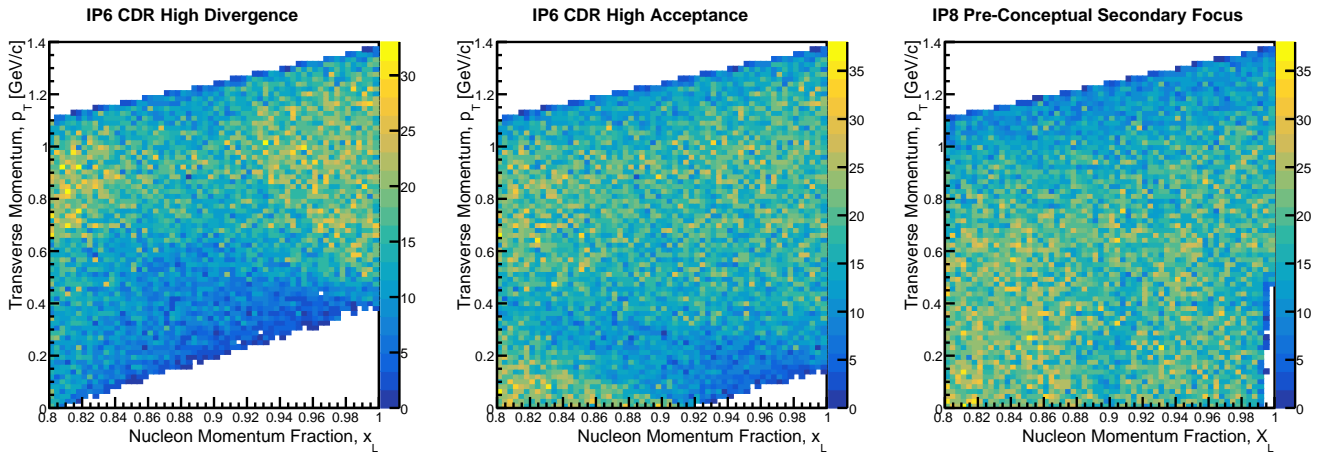}
    \caption{Two-dimensional plots of proton acceptance in transverse momentum, $p_{T}$ the nucleon momentum fraction. The acceptance is shown for three configurations: accepted protons in the IP6 Roman pots with the CDR high divergence optics (left), for accepted protons in the IP6 Roman pots with the CDR high acceptance optics (middle), and for accepted protons in the IP8 Roman pots at the secondary focus with the pre-conceptual optics configuration (right). All samples were generated for 18 GeV on 275 GeV protons with an $x_{L} > 0.8$ and with $0 < \theta < 5$ mrad; the cutoff at the top of the plots is due to the event generation region, while the acceptance in the bottom right varies with different configurations.}
    \label{fig:FF_layouts}
\end{figure}

To take full advantage of the EIC kinematics, the forward detection requires two elements: dispersion and focusing. The former is generated by dipole magnets and translates a momentum (rigidity) change into a transverse position offset: $dr = D dp/p$ (\textit{e.g.}, with $D=40 cm$, the transverse displacement for a particle with $dp/p=0.01$ and $p_T=0$ will be 4 mm). This value has to be compared with the (10$\sigma$) beam size at the detection point (Roman pot). Without focusing, this is typically a few cm, but with a secondary focus it can be reduced to 2-3 mm (depending on the beam momentum spread). The beam size on the Roman pot does in principle not depend on the focusing of the beam at the collision point ($\beta^*$), but in a practical implementation the same magnets are used to generate both the primary and secondary focus. However, in contrast to the unfocused case, this means that with a secondary focus the best low-$p_T$ acceptance is achieved at the highest luminosity.
A secondary focus could in principle be used at either IP6 or IP8 of the EIC. However, while the current IP6 layout has some dispersion (17 cm), it does not have a secondary focus. In contrast, IP8 is  designed for a much larger dispersion and incorporates a secondary focus – making it complementary to IP6 and opening up unique physics capabilities, as can be seen in Fig.~\ref{fig:FF_layouts}.

\subsection{Detection of recoil baryons and light ions}\label{sec:ff:recoilions}

As discussed in sections 7.2.2 and 7.3.8 of the Yellow Report~\cite{AbdulKhalek:2021gbh} and earlier in this paper, exclusive reactions on the proton and light nuclei form an essential part of the EIC physics program. The wide kinematic reach of the EIC makes it ideal for probing different parts of the nuclear wave function, revealing how the internal landscape of nucleons and nuclei changes with $x$. Measurements of exclusive processes require high luminosity, a range of collision energies, and excellent far-forward detection. Key issues are detector acceptance for the recoil proton or light ion and optimized reconstruction resolution of the momentum transfer, $t$.

\paragraph{Proton detection:} Detecting the recoiling nucleons is important to cleanly establish
the exclusivity of the reaction. It also makes it possible to reconstruct $t$ directly from the
proton. Since the EIC reaches its highest luminosity with the most asymmetric beam energies
(\textit{i.e.}, 5-10 GeV electrons colliding with hadrons at maximum energy), it is essential 
that the far-forward detection works optimally for high-energy protons. Here, the greatest
challenge is to detect low-$p_T$ protons which stay within the beam envelope. This capability 
can be improved by using a secondary focus, which essentially provides full acceptance for $x>10^{-2}$, and significantly improves the low-$p_T$ acceptance for lower $x$. For lower 
proton beam energies, a secondary focus is still useful, although less crucial. However, 
at lower energies, high-$p_T$ protons will start experiencing losses in the apertures of 
the accelerator quadrupole magnets, leading to a reduced acceptance for detectors downstream 
of these magnets. Embedding a tracking detector, such as is envisioned with the B0 tracker 
in IP6, provides increased coverage of high-$p_{T}$ protons at the lower beam energies. This 
issue can be alleviated by using magnet technologies that allow for higher peak fields, which 
makes it possible to increase the apertures, but there are other technical constraints that 
could make this challenging, especially at IP6, and more study is needed to determine what 
level of improvement is possible. In conjunction with a secondary focus, this would further 
enhance the capabilities of the EIC to do transverse proton imaging.

\paragraph{Determination of transverse momentum in exclusive reactions:} Another important consideration for
exclusive reactions is reconstruction of $t$. In principle it can be done either by using the
recoiling system detected in the far-forward detectors, or from the scattered electron and 
produced particle (charged meson or DVCS photon) detected in the central detector. There are
advantages to both methods. For example, the former method is very straightforward, but 
requires a good understanding of the beam effects (e.g. angular divergence). Ideally one 
would want to be able to apply both, but this requires that the central detector can provide 
a sufficiently good $p_T$-resolution. This is a challenge for a tracker, but even more so 
for the EM calorimetry if one wants to be able to determine $t$ ($\Delta_\perp$) from the 
DVCS photon (or the photons from $\pi^0$ production). However, while such a dual capability 
is useful for protons, it becomes essential for ions, where the ability to determine $t$ 
from the ion is more limited and vanishes entirely when the ion is not detected (at high A 
and low $p_T$). Being able to determine $t$ from the DVCS photon would thus greatly 
enhance the ability to do transverse imaging of ions.
One should note, however, that even if $t$ is reconstructed from the DVCS photon, the proton
still needs to be tagged in the far forward detectors for exclusivity, and low-$p_T$
acceptance would still remain very important.

\paragraph{Light ion detection:} Coherent exclusive scattering on light ions differs from protons
in that
scattered ions travel much closer to the beam, making low-$p_T$ acceptance very challenging (and conversely, the high-$p_T$ acceptance much less so, even for the high-$t$ tails). This is the combined result of two effects: cross sections for ions peak at lower $t$, and a given $t$ corresponds to a lower $p_T$ per nucleon. The former means that in contrast to the proton, clean imaging of light ions \textit{requires} an acceptance down to $p_T \sim 0$, and the latter that implementing such an acceptance is particularly difficult. A secondary focus is thus essential for high-quality measurements of coherent scattering on light ions.
However, if the central detector has the ability to reconstruct the $p_T$ from the produced photon or meson as discussed above, a secondary focus would also allow for a hybrid method where ions with higher $p_T$ (where the incoherent background is larger) are detected, while the low-$p_T$ part is reconstructed by vetoing the breakup.
Detecting the recoiling ion is always preferable, and a
hybrid measurement would not be as clean as one where all recoiling ions are detected, but it would make it possible to reach even lower $x$ and higher A, fully utilizing the capabilities of second focus to extend the discovery potential of the EIC.

Another consideration is that the EIC ion beam energies are restricted to a range between 100 GeV/A and $275 \times Z/A$ GeV/A (where $Z=A=1$ for protons) and a discrete energy at 41 GeV/A, where the upper limit comes from the ability of the arc dipoles to bend more rigid beams, while the lower limit (and 41 GeV/A value) arise from the need to synchronize collisions between the electrons and ions. For light ions the variation in $Z/A$ is considerable, and He-3 will thus be measured all the way up to 183 GeV/A, putting even greater emphasis on low-$p_T$ detection and a second focus.

\subsection{Spectator detection}

Detection of nuclear breakup is essential for a broad range of EIC physics topics. From a detection perspective, these broadly fall into two categories: spectator nucleons and nuclear fragments. In the first case, the spectator nucleon typically experiences a very small change in momentum, but its magnetic rigidity (A/Z) is very different from that of the original beam. A proton spectator will thus initially continue moving with the beam, but will separate quickly from it after passing the first dipole magnet. The detection challenge here thus lies primarily in providing adequate magnet apertures. An key example of spectator proton tagging are measurements of neutron structure in deuterium and $^3He$. 

In the case of nuclear fragments, they may be detected as a way of vetoing breakup or part of the direct measurement. The former case was discussed above (Sec.~\ref{sec:ff:recoilions} in the context of light and medium nuclei, but coherent processes on heavy ions are different in that even with a secondary focus, the high-$p_T$ tails cannot be measured directly as the ion always stays inside the beam envelope. A secondary focus can, however, make it possible to detect residual ions that have lost a single nucleon (A-1 tagging). Adding such a capability will significantly improve the efficiency for vetoing the large incoherent backgrounds, making a reasonably clean measurement possible. 

Finally, there are several measurements that rely on detection of the spectator nucleons, the residual nucleus, or nuclear fragments in the final state. One example is the case when the struck nucleon and its partner are in a short-range correlation with a high relative momentum. In this case, the spectator nucleon will not only have a different A/Z compared with the original ion, but also a large $p_T$. The breakup kinematics can then be best constrained if the residual A-2 nucleus can be detected, which is facilitated by a forward spectrometer with a secondary focus. A related topic is detection of rare isotopes produced in the interaction, which is discussed in Sec.~\ref{sec:ff:isotopes}. Additional detail, including discussion of the theoretical framework for several of the tagged measurements can be found in Ref.~\cite{Strikman:2017koc, Cosyn:2020kwu}

\paragraph{Neutron structure through spectator tagging}

Light ion beams can be used as an effective free neutron target via spectator tagging.
Deuterium is the simplest system, while $^3He$ can be polarized (70\%) and thus give 
access to the neutron spin structure. Spectator tagging can be applied to any primary 
measurement ($F_2$, DVCS, etc), but a key common challenge is to account for final-state interactions (FSI). However, recent studies~\cite{Jentsch:2021qdp} have shown that free
neutron structure can be extracted by on-shell extrapolation to the non-physical pole, 
where the neutron is by definition free and unaffected by FSI. In contrast to the pion, 
this approach is much more robust for the heavier nucleon where the extrapolation takes 
place over a shorter interval. The extrapolation is done by fitting the measured $t$ 
distribution, but focuses on the low-to-modest values of $t$ part, where the 
extrapolation has minimal model dependence.

Experimentally, this measurement relies on a high-resolution determination of the $p_T$ distribution, and of having sufficiently large magnet apertures to tag a spectator 
proton with low $p_T$~\cite{Jentsch:2021qdp}. As a cross check, it is also possible to 
apply the same method to the bound proton by tagging the neutron from deuterium in the ZDC.

\paragraph{Proton and neutron spectators from deuteron beams}

Deuteron beams can be used as an effective free neutron target via spectator tagging, where the undisturbed proton is measured to isolate scattering from the proton. To isolate nearly on-shell neutrons, the goal is to tag protons which had low initial momenta (corresponding to low $-t$) in the deuteron rest frame. Measurements will be made over a range of $t$, so that the extrapolation to the on-shell neutron can be performed over different ranges of $t$ to ensure stability of the extrapolation. As noted above, detection of these protons in the Off-Momentum Detector and Roman Pots is relatively straightforward and the key issue is minimizing the loss of acceptance in the apertures of the accelerator magnets. 

Similar studies of the proton structure of the proton structure can be performed with neutron tagging used to isolate scattering from a low-momentum proton. In this case, the results can be compared to the known proton structure, and these studies can be used to study the $t$-dependence and test the extrapolation to the on-shell proton. For the low $t$ values required for these measurements, the neutrons have $x_L$ near unity and small $P_T$ and are detected in the ZDC.

\paragraph{Double tagging from ${^3}$He breakup}

While the deuteron is the most common target used to study unpolarized neutron structure, polarized $^3$He serves as the most effective target for measuring neutron spin structure, as the neutron carries most of the spin in $^3$He. Inclusive measurements can provide some information, with the protons acting mainly as a dilution to the asymmetries associated with scattering from the polarized neutron. But double tagging of the two spectator protons in $^3$He can be used to isolate scattering from the neutron without dilution or corrections for the proton contributions~\cite{Friscic:2021oti}. In this case, the goal is to measure spectators with low momenta in the $^3$He rest frame which have momenta close to the beam momentum per nucleon and small $P_T$, but lower mass and therefore roughly 2/3 of the rigidity of the $^3$He beam. One can also examine events with one large-momentum proton to identify high-momentum neutrons in the initial state to look at the spin structure as a function of initial neutron momentum, which is relevant for understanding the spin EMC effect.

\paragraph {Tagged Pion structure - nucleon spectators from proton beams}

Measurement of the $\pi^+$ electromagnetic form factor can be accomplished at the EIC by the detection of the neutron spectator in coincidence with the scattered electron and $\pi^+$, i.e. an exclusive reaction with $e'-\pi^+-n$ triple coincidence. The neutron is emitted with 80-98\% of the proton beam momentum, and is detected in the ZDC.  The pion form factor measurement only requires $-t$ measurements up to about 0.4~GeV$^2$, so a moderate acceptance ZDC is sufficient to catch the events of interest.  Very good ZDC angular resolution is required for two reasons. First, to separate the small exclusive $\pi^+$ cross section from dominant inclusive backgrounds, a cut may be placed on the detected neutron angle in comparison to the reconstructed neutron angle (from $e'$ and $\pi^+$ using momentum conservation). Second, a $t$ reconstruction resolution better than $\sim$0.02~GeV$^2$ is necessary for a quality form factor measurement and such resolution is only possible when reconstructed from the initial proton and final neutron momenta.  The ZDC is thus of crucial importance to the feasibility of a pion form factor measurement at the EIC.

\paragraph{Tagged Kaon structure - hyperon spectators from proton beam}

As introduced in Sec.~\ref{sec:mass}, the Standard Model has two mechanisms for mass generation.  One is connected with the Higgs Boson (HB), while the other emerges as a consequence of strong interactions within QCD, particularly Dynamical Chiral Symmetry Breaking (DCSB).  DCSB is responsible for 98\% of the mass of the visible universe, and the properties of the pion and kaon are central to unraveling the mysteries of this mechanism~\cite{Arrington:2021biu}.
Measurements of the $K^+$ electromagnetic form factor at high $Q^2$ via the Sullivan process would yield valuable information towards this goal.  
The reaction of interest is $e + p \rightarrow e'+K^+ +\Lambda$, where the $\Lambda$ is emitted with $>70$\% of the proton beam momentum.  We expect that lower beam energies are optimal, to ensure a high $\Lambda$ decay fraction, as non-decayed $\Lambda$ will be impossible to distinguish from neutron hits.

The $\Lambda$ needs to be identified from its decay products to ensure the clean identification of the exclusive events from inclusive backgrounds, and to reconstruct $t=(p_p-p_{\Lambda})^2$ with sufficiently high resolution.  One complication is that the $\pi^-$ from the dominant $\Lambda\rightarrow p\pi^-$ decay channel cannot be detected in the far forward detectors for decays occurring at or after the B0 magnet (Fig. \ref{fig:IP6_FF_layout}). Such measurements would require dedicated detectors for negative particles or be limited to decays occurring sufficiently before B0.  The neutral $\Lambda\rightarrow n\pi^0\rightarrow n\gamma\gamma$ decay seems a better choice.  For the measurement to be feasible, three hit events need to be reliably identified in the ZDC with sufficiently good energy and angle resolution for $t$ reconstruction.  Even more challenging is confirming that the Sullivan process dominates at low  $-t$, which requires a measurement of the $\Lambda/\Sigma^0$ ratio. This entails the reliable detection of four neutral hits in the ZDC, from $\Sigma^0\rightarrow\Lambda\gamma\rightarrow n\pi^0\gamma$. Thus, this is a measurement that is significantly more challenging than that of the pion form factor, although if it is feasible, it would be an important addition to the EIC scientific program. The acceptance for neutral decay products could potentially be increased significantly if calorimetry were included in the B0 magnet. This option was mentioned as a possibility in the Yellow Report, but including both tracking and calorimetry is technically challenging due to spatial constraints inside the magnet and further design work is needed to know what is be possible.

\subsection{Tagging of active nucleons - high spectator momenta}

While the previous sections focused on tagging of relatively low-momentum spectators, other key studies are focused on isolating high-momentum nucleons and/or mapping out tagged nucleon structure over a wide range of initial virtualities. Studies of Short-Range Correlations between pairs of bound nucleons require tagging of final state nucleons at both high and low values of $p_T$ to fully exploit the measurement capability. This provides a unique challenge for the detector acceptances, as multiple far-forward subsystems play a role in covering the phase space. In general, the active nucleon in a reaction will be scattered with relatively large polar angles ($\theta > 5$~mrad), while the recoil nucleons and spectator nuclear fragments (for A $>$ 2) are usually at smaller values. Additionally, in the case of the recoil protons, there is a magnetic rigidity change with respect to the ion beam which further complicates detection. It is in principle also possible to tag an A-2 spectator nucleus, in the final state, but this is uniquely challenging to do the small scattering angles for the spectator nucleus, and the small rigidity change, dependent on the struck SRC pair. Tagging of A-2 nuclei can be enhanced with Roman Pots at a secondary focus.

In cases where both final-state nucleons from an SRC pair are measured, the spectator nucleon is detected in the far-forward region while the active (struck) nucleon is measured in the main or far-forward detectors. At higher energies, the acceptance is more complete when measuring a spectator neutron and active proton, since the polar angle coverage for struck protons is extended to $\sim20$~mrad in the B0 tracking detector, while the neutron acceptance is limited to $\sim5$~mrad by the magnet aperture. For active neutrons, the lower beam energy configurations (e.g. 5x41 GeV/n) are more beneficial since the larger active neutron scattering angle can place them in the acceptance of the main detector endcap hadronic calorimeter (i.e. $\theta > \approx30$~mrad). Additionally, if more of the open bore space in the dipole spectrometer can be used for active detector material, it would further enhance the capabilities for active proton tagging beyond the current 20 mrad assumption.

Having some capability for tagging in the higher-$p_T$ regime allows simultaneous study of both free nucleon structure and nuclear modifications with the same experimental setup. Studies of Short-Range Correlations and nuclear modifications enable the EIC to provide insight into the EMC effect and other physics at higher-x.

\subsection{Vetoing of breakup}

Separation of coherent and incoherent photoproduction of photons (Deeply Virtual Compton Scattering) and vector mesons is critical to many aspects of the EIC physics program.  In the Good-Walker paradigm, one can relate the coherent cross-section to the average nuclear configuration, while the incoherent cross-section is sensitive to event-by-event fluctuations of the nuclear configuration, including gluonic hot-spots~\cite{Klein:2019qfb}.  One can do a two-dimensional Fourier transform of $d\sigma_{\rm coherent}/dt$ to determine the transverse distribution of gluons in the nuclear target - the nuclear equivalent of the GPD.  By studying different mesons with different masses, and using photons with different $Q^2$, one can map out nuclear shadowing as a function of position within the nucleus.   

The challenge in these measurements is in adequately separating coherent and incoherent production, by detecting the products of nuclear breakup~\cite{Chang:2021jnu}. To determine the transverse gluon distributions, it is necessary to measure $d\sigma_{\rm coherent}/dt$ out to the third diffractive minimum~\cite{AbdulKhalek:2021gbh}, to avoid windowing artifacts in the Fourier transform.  At this minimum, a rejection factor of 500:1 is needed to adequately remove the incoherent background.   

In most cases, nuclear dissociation leads to neutron (or, less frequently, proton) emission from the target. These are relatively straightforward to detect, although very high efficiency is required. However, some soft excitations will produce excited nuclear states that decay by photon emission.  These photons typically have energies of a few MeV (or less) in the nuclear rest frame.  Gold (planned as the main EIC heavy nuclear target), is particularly bad.  It has a 77 keV excited state with a 1.9 nsec lifetime.  Because of the lifetime, this state is almost impossible to observe in an EIC detector.  Its next states have energies of 269 and 279 keV respectively.  The lab-frame energies depend on the EIC beam energies, but for 110 GeV/n gold beams, the maximum energy is 65 MeV.  For photon emission away from the far-forward direction, the energy will be lower.  This is likely beyond the reach of the planned EIC detectors, but could be accessible in an upgrade. Because the energy transfer to the target (and hence the energy spectrum of the excitations) depends on $t$, is it critical to be able to detect emission of protons, neutrons, and soft photons over the full phase space. As noted earlier, the addition of calorimetry in the B0 magnet would improve the acceptance, but is technically challenging.

Since the knockout of a single neutron (and possibly evaporation of another) is an important contribution to the incoherent background, the ability to tag and veto on A-1 nuclei (\textit{e.g.}, Zr-89 from a Zr-90 beam) is also very important for a clean measurement. High-resolution photon detection is also synergistic with a potential rare isotopes program discussed below.

It is also possible to mistake coherent production for incoherent, if a second collision in the same beam crossing dissociates a nucleus~\cite{Klein:2014xoa}.  This could affect the measurement of the incoherent cross-section at small $|t|$.  Although the background rate can be subtracted, statistical uncertainties will remain.  However, most of these events can be removed if the ZDC has very good timing.

\subsection{Backward (u-channel) photoproduction}

In backward (u-channel) photoproduction, the produced meson takes most of the energy of the incident proton, and so goes in the forward direction, while the proton is shifted many units of rapidity, and, at the EIC, is visible in the central detector~\cite{Gayoso:2021rzj}.  Instead of having small Mandelstam $t$, as in conventional photoproduction, $t$ is large (near the kinematic maximum) and $u$ is small. This process may be modelled using Regge trajectories involving baryons, but it is not easy to see how such simple reactions can lead to nucleons being shifted many units of rapidity; there may be connections with baryon stopping in heavy-ion collisions.   A systematic exploration of production of different mesons at higher energies is needed to fully characterize this reaction, and test the Regge trajectory approach. 

Reconstruction of these events requires a forward detector that is able to reconstruct multi-particle final states.  For the full 18 $\times$ 275 GeV beam energy, the products of light meson ($\omega$, $\rho$ or $\pi^0$) mostly end up with $\eta >6.2$, in the zero degree calorimeter (ZDC).  At lower beam energies, or with heavier mesons, the products at are at smaller pseudorapidity. This requires a forward detector with as full an acceptance as possible, {\it i.e.} with no holes in the acceptance) for both charged and neutral particles.

\subsection{Rare isotopes (including photons for spectroscopy)}\label{sec:ff:isotopes}

As discussed in the recent EIC Yellow Report, simulation studies suggest that the EIC has the potential to produce and detect rare isotopes along with their gamma photon decays, allowing this new machine to complement the results from dedicated rare isotope facilities.

\begin{figure}[htb]	
        \includegraphics[keepaspectratio=true,width=3.4in,page=1]{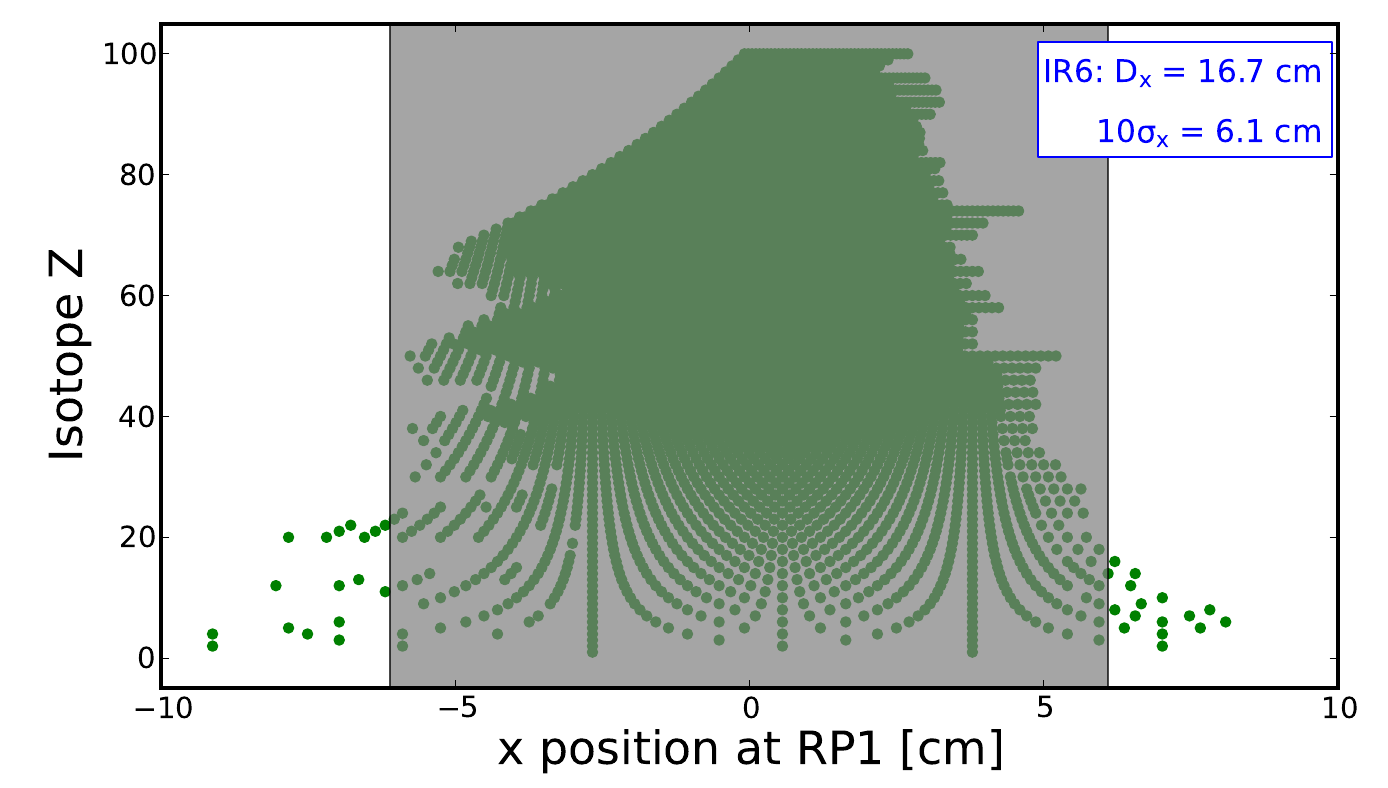}		
		\includegraphics[keepaspectratio=true,width=3.4in,page=1]{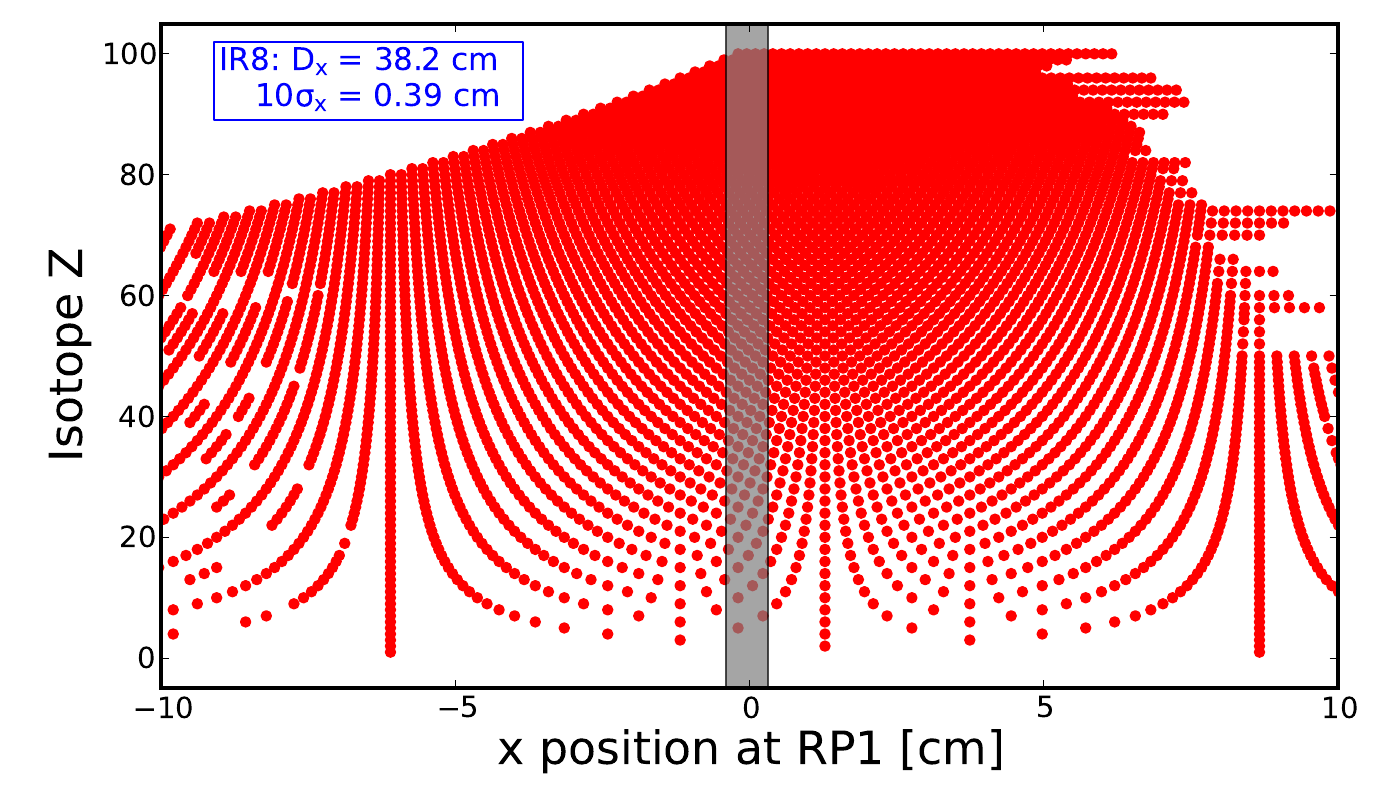}
	\caption{Left: Isotope Z vs. hit position in the first RP for the primary IR; right: Isotope Z vs. hit position in the first RP at the secondary focus for the second IR. The gray box on each plot shows the 10$\sigma$ beam exclusion area. The plots are made assuming a $^{238}$U beam.}
	\label{fig:rigidity_isotopes}
\end{figure}

Direct detection of the rare isotopes will use the Roman Pot (RP) detectors. At first approximation, the produced rare isotopes will have the same momentum-per-nucleon as the ion beam and no angle relative to the beam. The rigidity of an isotope is equal to its momentum divided by its charge (i.e. R $\propto$ p/Z). Under the above approximation, the rigidity of the isotope relative to the incoming ion beam ($R_{Rel}$) is directly related to the ratio of the isotope's mass number and atomic number as
\begin{equation}
R_{Rel} = \frac{R- R_{beam}}{R_{beam}} = \left(\frac{A}{Z}\right) / \left(\frac{A_{beam}}{Z_{beam}}\right) - 1 ~.
\end{equation}
In this approximation, the isotope's hit position in the dispersive direction at the RP gives a measurement of A/Z. Figure~\ref{fig:rigidity_isotopes} shows the expected hit positions for known and predicted isotopes both at the first RP for the primary IR and at the first RP located near the secondary focus in the second IR, assuming a $^{238}$U beam. Isotopes with the same Z and different A values are shown at the same vertical position in the plots. In addition, using the beam parameters from table~3.5 of the 2021 EIC CDR~\cite{EIC:CDR} for heavy nuclei at 110 GeV/A on electrons at 18 GeV, the 10$\sigma$ beam exclusion area is shown by the gray box. As can be clearly seen, none of the heavy rare isotopes can be detected in the primary IR, while the second IR has the potential to detect the majority of the isotopes.  At the RP in the second IR, isotopes with the same Z that differ by a single neutron are expected to be separated by 1.5 mm for Z = 100 and 5 mm for Z = 25.

For uniquely determining the isotope, a direct measurement of Z is needed. The simplest way to do this is by placing a Cherenkov detector behind the RPs at the secondary focus. The number of Cherenkov photons produced by the isotope will be proportional to Z$^2$.

Measuring gamma decay photons is also important as the level transitions reveal the structure of the final isotope. The photons are produced isotropically in the isotope's rest frame but can be Lorentz up-shifted significantly in the lab frame. This shift, as well as the requirement that these photons be detected in coincidence with an isotope, means that photon background will be small. LYSO crystals that do not require cryogenics can therefore be used for this measurement. In addition, while spectroscopy would benefit from a good photon acceptance, it would not be a critical requirement.
\section{Radiative effects and corrections} 
\label{sec:radiation}

\subsection{Introduction}
\label{intro}

QED radiative corrections (RC) are an integral part of the hadronic-structure studies with electron (or muon) scattering. In experiment, they can reach tens of per cent for unpolarized cross sections and several per cent for polarization asymmetries, while also altering dependence of observables on all kinamatic variables of DIS ($x, y, Q^2$) as well as altering dependence on azimuthal angles both in SIDIS and deep-exclusive reactions. Thus, they can become a significant source of systematics in a program of hadronic studies with EIC. 

Significance of electromagnetic RC  for analysis of scattering data should not be underestimated, as was clearly demonstrated by different outcomes of Rosenbluth and polarization methods for measurements of the proton electric form factor, see \cite{Afanasev:2017gsk} for an overview. Current and planned experiments probing 3D hadronic structure require precise measurements of GPD and TMD contributions to cross sections and spin asymmetries that may be possibly obscured or altered by radiative effects. For this reason, proper inclusion of RC is one of priority tasks in experiment planning and data analysis.

Historically, the approach developed by Mo and Tsai in 1960s \cite{Mo:1968cg} was successfully applied for both DIS and elastic electron scattering on protons and nuclei. In 1970s Bardin and Shumeiko developed a covariant approach to the infra-red problem in RC \cite{bardin1977exact} that was later applied to inclusive, semi-exclusive and exclusive reactions with polarized particles. 

Emission of multiple soft photons is conventionally included via exponentiation \cite{Yennie:1961ad}. A different approach for including higher-order corrections \cite{kuraev1985radiative} uses a method of electron structure functions based on Drell-Yan representation that allows RC resummation in all orders of QED. 

For high transferred momenta, such as in HERA or EIC, electroweak corrections have to be included. Corresponding formalism was developed for HERA \cite{Kwiatkowski:1990es,Charchula:1994kf}, while the codes presently used for JLab would have to be updated to include weak boson exchanges.

Higher precision of modern experiments presents new demands on the accuracy of RC. It is common to divide RC, in a gauge invariant way, into two categories, namely, model-independent and model-dependent. For model-independent RC, QED corrections do not involve extra photon coupling to a target hadron. Still, kinematics shifts due to extra photon emission require knowledge of hadronic response in off-set kinematics that can be handled either by iterative procedures, or existing data on the same reaction from other experiments, or input from theoretical models.  On the other hand, model-dependent corrections correspond to extra photon exchange or emission by a target hadron. They require knowledge of hadronic structure beyond what can be learned in a considered experiment from a given reaction.

\subsection{Monte Carlo generators for radiative events}
Classically, radiative corrections are applied to measured data post-hoc, i.e.\ a correction factor is calculated using analytical formulas and then multiplied onto the measured result, effectively mapping the measured radiative rate to an ideal Born-level rate (e.g.\ \cite{Mo:1968cg,Maximon:2000hm}). On the other hand, to calculate a cross section, the detector acceptance is also required, and either calculated analytically from geometry, or integrated numerically using Monte Carlo methods.

This post-hoc application of a---typically analytically integrated---correction has limited precision, as it must necessarily make simplifying assumptions about the detector acceptance, more so since radiative processes beyond a peaking approximation can radically shift the event kinematics.

Therefore, the Monte Carlo algorithms, classically used to calculate the acceptance, were extended to include full cross section and reaction models including radiative corrections. The MC result, together with the luminosity, is then not a calculation of the acceptance, but of the expected count rate, and results of experiments are often presented as the ratio of the observed to predicted count rates.  A proper implementation of this approach includes automatically all interactions between radiative corrections and other detector effects like bin-migration and detector acceptance, possibly even as a function of time. Such codes were developed for example for the HERA experiments H1 and ZEUS \cite{Kwiatkowski:1990es,Charchula:1994kf}.

Efficient MC simulations require a small variance of event weights. Radiative generators must overcome the fact that the radiative cross section varies by many orders of magnitude, with possibly multiple, unconnected regions of phase-space with high cross-section, for example for nearly collinear emission of photons along electron trajectories.

In these cases, naive rejection sampling methods show poor performance as only very few events are accepted. Automatic volume reweighting approaches like foams can in principle be effective, but suffer from the high derivatives near peaks. Efficient approaches therefore exploit the analytical structure of the underlying cross section to generate events efficiently. 

For fixed target electron scattering experiments, many suitable codes for QED radiative corrections exist, however mostly limited to first-order approximations, sometimes improved by approximate higher-level corrections (see e.g. \cite{Yennie:1961ad,akushevich1999radgen,Akushevich:2011zy,Gramolin:2014pva,Henderson:2016dea}).  Recently, true higher-order MC generators became available \cite{Banerjee:2020rww,Banerjee:2020rww}. The validity of such generators has been tested deep into the radiative tail, recently for lower energies in \cite{Mihovilovic:2016rkr}. 

The translation of these generators to collider kinematics is straight forward, with the caveat that numerical precision problems might crop up. 

Beyond DIS reactions, the mapping of the radiative process back to the Born-level base process becomes tedious. The QED radiative Feynman graphs resemble QCD higher-order graphs, opening the door to a unified approach that can handle both QCD and QED radiative effects, and corresponding algorithms are currently being implemented in HEP generators \cite{Buckley:2011ms}. Using the factorization theorem, the resummed leading logarithmic higher-order corrections can be described with distribution and fragmentation functions \cite{Liu:2020rvc,Liu:2021jfp}.  Higher-order corrections are resummed in the form of parton showers, treating partons and photons on equal footing \cite{Hoeche:2009xc}. The approach has to be extended
to include non-logarithmic higher order corrections.

\subsection{Opportunities to reduce model dependences}
\label{model-dep}

While QED radiative corrections seem straight forward to calculate, they often require external input and make model assumptions, for example about hadronic contributions. For example, recent experimental results on two-photon exchange, i.e.\ the next order of corrections for elastic scattering, are not particularly well predicted by current calculations (for an overview, see \cite{Afanasev:2017gsk}), and are an open research topic in theory and experiment.

Whether semi-analytical or Monte Carlo approaches are chosen for RC calculations, it is important that integration over the phase space of the radiated photon is done with a realistic hadronic tensor, as pointed out in Ref.\cite{PhysRevD.100.033005}. In particular, radiative tails from exclusive meson production can contribute to SIDIS or baryon resonance contributions would be enhanced due to kinematic shifts from the radiated photons. Uncertainties in large-x behavior of PDF may also affect RC calculations. In order to address these problems, the hadronic physics community needs to maintain a comprehensive database of exclusive and semi-inclusive reactions, whereby JLab/EIC data from lower energies and momenta transfers would be used for RC calculations for highest EIC energies. Artificial Intelligence approaches may also be instrumental in developing multi-dimensional iterative procedures, especially for SIDIS. In particular, SIDIS measurements at lower-energy Interaction Point at EIC may be used as an input for RC calculations for higher energies of the same machine, thereby providing necessary energy coverage for self-consistent RC approaches. Extension of conventional PDF analysis to large Bjorken $x$ values and studies of its impact on RC also have to be planned.

With an exception of elastic ep-scattering \cite{Afanasev:2017gsk}, most of approaches to exclusive electron scattering considered model-independent RC that include only coupling of the extra photon to lepton lines,  see, $e.g.$, Refs.\cite{PhysRevC.62.025501,PhysRevD.98.013005} for VCS and Ref.\cite{Afanasev:2002} for exclusive pion production.
Importance of model-dependent RC - still unaccounted for - is indicated both by experiment and theory. The JLab experiment \cite{PhysRevLett.113.022502}  measured DIS with a transversely polarized  3He target and revealed a few per cent spin asymmetry that only appears beyond Born approximation, and it is similar in magnitude to single-spin asymmetries due to T-odd effects arising from hadronic structure. Effects at a level of several per cent due to two-photon exchange were also predicted theoretically for exclusive electroproduction of pions \cite{Afanasev:2013,PhysRevC.101.055201}.

A collaborative effort between development of advanced models of hadronic structure, experimental data analyses and RC implementation will aim to minimize experimental systematics on one hand and provide access to hadronic PDFs, TMDs and GPDs in kinematics otherwise not accessible in direct measurements.  
In this respect,  dedicated workshops (e.g.\ \cite{Afanasev:2020hwg}) help bring together experts across several fields and facilitate such collaborations.

\section{Artificial Intelligence applications}%
\label{sec:ai}

Artificial Intelligence (AI) is defined as a ``machine-based system that can, for a given set of human-defined objectives, make predictions, recommendations or decisions influencing real or virtual environments”~\cite{NAIIA2020}. Among the topics that are grouped under the term AI, machine learning and autonomous systems are of particular importance for the EIC: 
\begin{itemize}
    \item Machine Learning (ML) represents the next generation of methods to build models from data and to use these models alone or in conjunction with simulation and scalable computing to advance research in nuclear physics. It describes how to learn and make predictions from data, and enable the extraction of key information about nuclear physics from large data sets. ML techniques have a long history in particle physics~\cite{Carleo2019, deiana2021}. With the advent of modern deep learning (DL) networks, their use expanded widely and is now ubiquitous to nuclear physics, as found promising for many different purposes like anomaly detection, event classification, simulations, or the design and operation of large-scale accelerator facilities and experiments~\cite{Bedaque:2021bja, Boehnlein:2021eym}. 
    \item Autonomous systems are of interest for monitoring and optimizing the performance of accelerator and detector systems without human control or intervention. This can include responsive systems that adjust their settings to background conditions as well as self-calibrating accelerator and detector systems. An ambitious goal is the usage of real-time simulations and AI over operational parameters to tune the accelerator for high luminosity and high degrees of polarization. 
\end{itemize}

The EIC community has started to incorporate AI into the work on the physics case, the resulting detector requirements, and the evolving detector concepts. Initiatives such as AI4EIC and the related AI working group in the EIC User Group will work with the community to systematically leverage these methodologies during all phases of the project. AI4EIC aims at identifying problems where AI can have an impact and at finding solutions that can be cross-cutting for the EIC community. The initiative will create a database with benchmark datasets and challenges to allow testing new AI approaches and methods and compare to previous ones. An overarching research theme of the EIC community is the work towards an autonomous experiment with intelligent decisions in the data processing from detector readout and control to analysis. 

AI will advance precision studies of QCD in both theory and experiment. An prominent examples is the applications of AI to the inverse problem of using measured observations to extract quantum correlation functions, e.g., with variational autoencoders (VAEs) that utilize a latent space principal component analysis to replicate lost information in the reconstruction of the posterior distribution \cite{9534012}. Other examples are AI methods to accelerate simulations for the design of experiments and for nuclear femtography to image quarks and gluons in nucleons and nuclei. 

\subsection{Accelerate Simulations with AI} 

Physics and detector simulations are being used to develop the physics case, the resulting detector requirements, and the evolving detector concepts for the experimental program at the EIC. The high-precision measurements envisioned for the EIC require simulations with high-precision and high accuracy. Achieving the statistical accuracy needed is often computationally intensive with the simulation of the shower evolution in calorimeters or the optical physics in Cherenkov detectors being prime examples. Fast simulations with parameterizations of detector response or other computationally efficient approximations that are pursued as alternative lack the accuracy required for high-precision measurements. Here, AI provides a promising alternative via fast generative models, e.g., generative adversarial networks (GANs) or VAEs. 

A promising approach is AI-driven detector design where the parameters of detector and its costs are being tuned using Bayesian optimization. AI-driven detector design has been used for detector components~\cite{Cisbani:2019xta} and recently for detector concepts~\cite{Fanelli:2022rdm}. 

\subsection{Nuclear Femtography and AI}
Tomographic images of the nucleon, referred to as nuclear femtography, are 
critical for understanding the origin of the mechanical properties of the nucleon such as mass and orbital angular momentum decompositions into contributions from quark and gluon dynamics. The development of the new imaging methodology, deeply-virtual exclusive processes in electron scattering, and their dedicated exploration through the future EIC's beam and detector technology, will make nuclear femtography a reality for the first time.

Efficiently constructing the images from future large complex experimental data sets along with first principles constraints from large-scale numerical lattice-QCD calculations requires the exploration of an ensemble of advanced AI and ML techniques. In the case of studies of GPDs, the data analytic strategy to go from
precisely understanding the performance of detectors in searching for high-energy diffractive events, through accurately extracting the Compton Form Factors as the key link between experimental data and the input for imaging construction, to generating the images through complex neural-network numerical regression that takes into account various physical constraints including direct lattice QCD results. To accomplish this, it is essential to assemble an interdisciplinary group of nuclear theorists and experimenters, along with computer scientists and applied mathematicians, to build the first AI/ML-based platform for the state-of-art nuclear sub-femto-scale imaging. The physical quantities connecting images and experimental data are 
Compton Form Factors (CFFs). To extract CFFs from data is complicated due to several CFF combinations corresponding to various quark-proton polarization configurations appearing simultaneously in the cross section terms for each beam and target polarization configuration. A neural-network (NN) approach, exploiting dispersion-relation constraints, was recently adopted to obtain the flavor-separated CFFs~\cite{Cuic:2020iwt}. 

Generally considered to offer the most robust and flexible method for multidimensional probability density estimation, Artificial Neural Networks (ANNs) represent a new paradigm to tackle this complex problem.
Initial ANN applications to CFF extraction were reported in~\cite{Moutarde:2019tqa,Kumericki:2019mgk,Grigsby:2020auv} using standard supervised NN architectures. 
The systematic application of AI to the extraction of multidimensional structure functions is currently in its initial stages.
A crucial aspect of these methods is the treatment of uncertainties and their propagation 
from direct experimental observables (such as cross sections and asymmetries) to the densities of physics interest (such as the distributions of electric charge or forces). With emerging JLab $12$~GeV data and beyond 
from various experimental sources, a suite of ML technologies will need to be explored to properly assess the optimal deep neural network architectures with proper treatment of uncertainty through robust uncertainty quantification (UQ) techniques such as Bayesian NN methods to quantify and separate model-dependent errors and ensembling techniques.
This ML strategy can also be 
systematically extended to extract the 
subleading CFFs once leading twist CFFs have been extracted with controlled uncertainties.
in the future which is, in part, made a more tangible goal once we have a better extraction of the leading ones. 
We will systematically compare performances and the influence of various choices, such as the detailed structure and depth of the ANN, prior assumptions of the local variation of the CFFs with respect to the kinematic variables, and prior assumptions of the full determination of the number and type of contributing CFFs.
A statistically rigorous analysis of the NN performance with respect to architecture (depth and width of the network), local variation of the CFFs with respect to the kinematic variables, and prior assumptions of the full determination of the number and type of contributing CFFs will need to be performed to fully quantify any systematic errors from using ANNs.
With the goal of extracting all eight of the leading CFFs from the chiral-even GPDs, one needs to develop eight independent ANNs, each with the goal of inputting experimental cross section data (e.g. DVCS asymmetries) with particular polarization configurations allowed by parity, and predicting a single CFF with minimal bias.

\subsection{Inverse problems of quarks and gluons with AI }

Since quarks and gluons are not directly observable states of nature due to confinement, understanding their emergent phenomena such as hadron structure and hadronization from experimental data is unavoidably an inverse problem. Traditionally, ML techniques have been mostly applied in the form of \emph{regression} that capitalize the model expressivity offered by ANN~\cite{NNPDF:2019ubu, NNPDF:2017mvq}. In recent years however, a number of machine learning  applications have been developed to tackle similar problems in nuclear physics, such as the reconstruction of neutron star equations of state from the observational astrophysical data~\cite{Fujimoto:2021zas,Fujimoto:2019hxv,Fujimoto:2017cdo,Soma:2022qnv}, the deconvolution problem of the Kaellen-Lehmann equation~\cite{Shi:2022yqw}, inverse Schroedinger equation solvers~\cite{Shi:2021qri}, inference on nuclear energy density functionals~\cite{Lasseri:2019ywk,Scamps:2020fyu}, and quantum many-body calculations~\cite{Molchanov:2021pjv} (see the recent review in Ref.~\cite{Boehnlein:2021eym}). The emerging features of these applications includes ML-theory emulators that mitigate large scale computational costs for parameter searches~\cite{Lasseri:2019ywk,Scamps:2020fyu}, generative models to improve Markovian sampling in lattice QCD~\cite{Albergo:2019eim},  design of explainable ML architectures for parton showers~\cite{Lai:2020byl} to mention few. Many of these applications are likely to cross pollinate the field of hadronic physics, and they will have a transformational impact for the scientific discoveries at the EIC.  

\section{The EIC interaction regions for a high impact science program with discovery potential} 
\label{EIC-IR}
\subsection{Introduction}

The compelling science program of the EIC focusing on the low to medium CM energies has been described in this document. Here we describe the two interaction regions (IRs) dedicated to the experimental programs, and some of the important differences between them. The overall layout of the EIC is shown in Fig.~\ref{EIC-concept}. 

\begin{figure}[ht!]
\centering{\includegraphics[width=0.95\columnwidth]{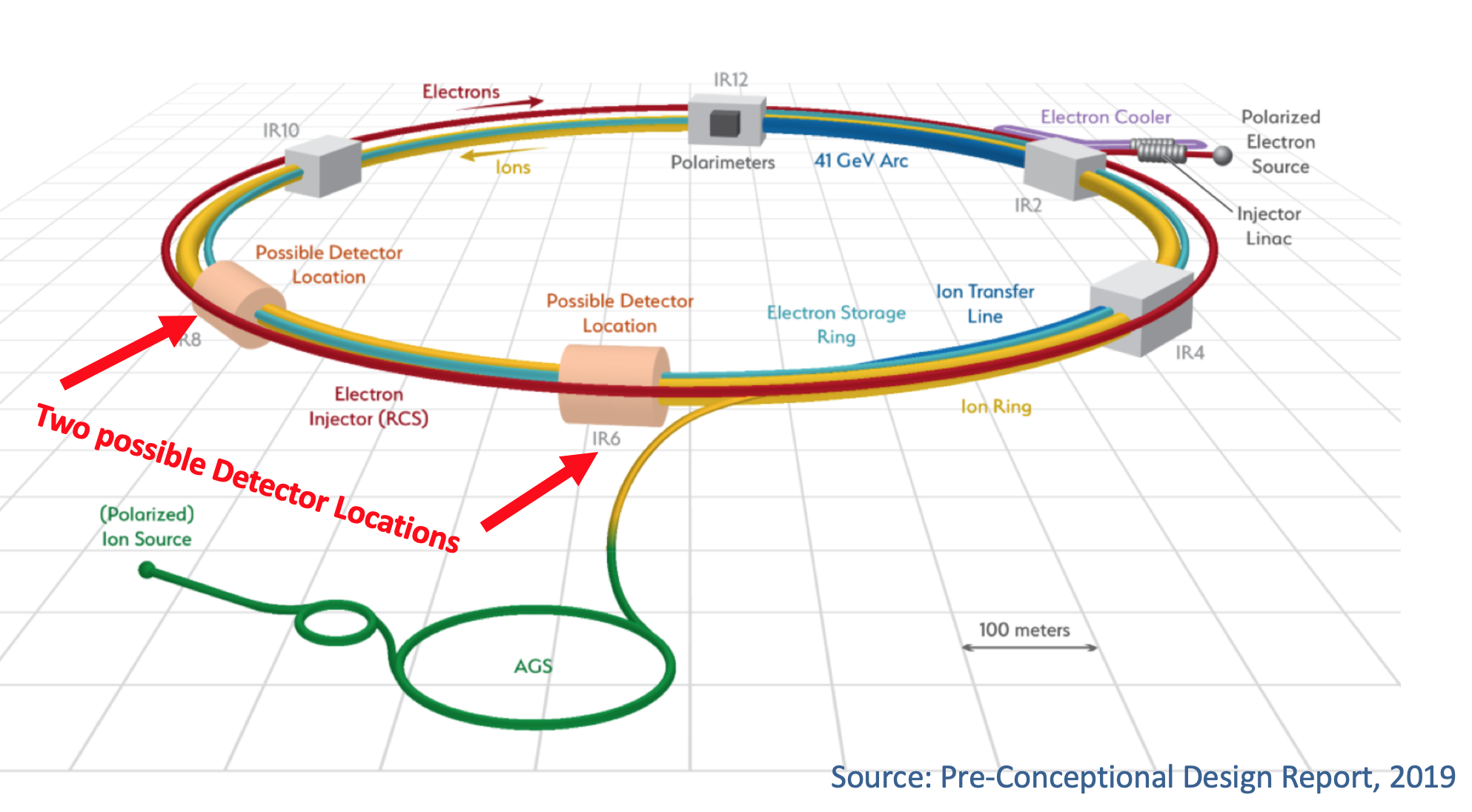}}
\caption{The EIC layout at Brookhaven National Laboratory. Electron and the ion beams directions are identified in the upper left. There are several beam intersection points (IPs); the 6 o'clock (IP6) and 8 o'clock (IP8) locations are suitable for the installation and operation of large-scale detector systems, with appropriate existing infrastructure. IP8 may be most suitable for high-luminosity optimization at low to intermediate CM energies as well as for the installation of a secondary focus for forward processes requiring high momentum resolution. Both beams will be highly polarized, with proton and electron beam polarizations over 70\%.}
\label{EIC-concept}
\end{figure} 

One of the EIC
design requirements is the capability of having two IRs. The EIC configuration therefore includes two IRs where collisions will occur, and where substantial
near-full-acceptance detectors may be installed. The two IRs are IR6
(for the primary IR at 6 o'clock) and IR8 (for the second
IR at 8 o'clock). Here the RHIC clock location
nomenclature is used, where the STAR detector is located in IR6 and
PHENIX/sPHENIX detector is located in IR8.

\begin{figure}[ht]
    \centering
    \vspace*{1mm}
    \includegraphics[width=0.95\textwidth]
    {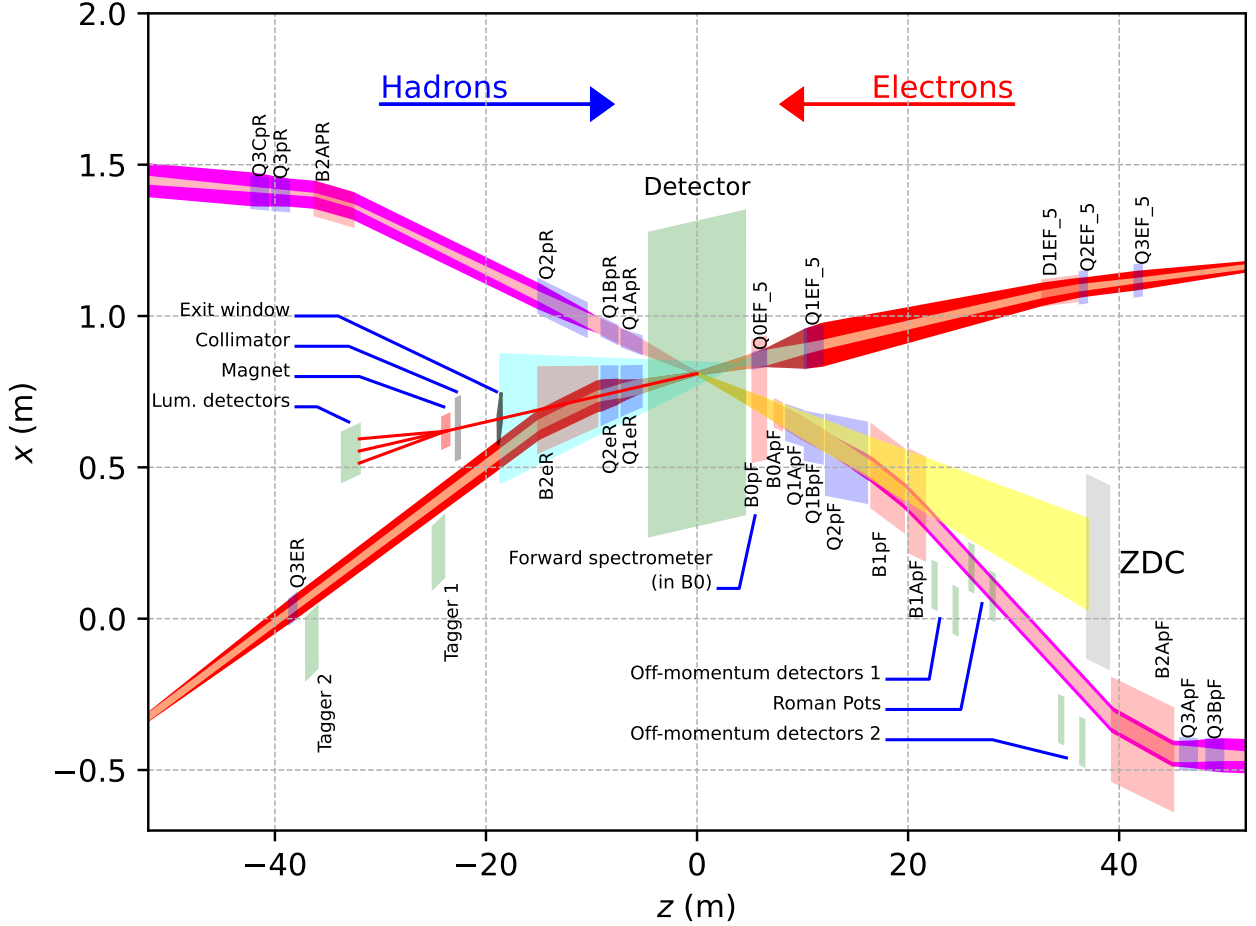}
    \caption{Schematic top view of the EIC IR6 primary IR, in the
      high divergence configuration. The
      y-axis positive direction points \textbf{inward} from the ring
      curvature; both beams cross from inner (positive y-axis) to
      outer (negative y-axis) beamlines. Beam envelopes and apertures are illustrated, as are quadrupoles (blue boxes), dipoles (red boxes), and detectors (green boxes).}
    \label{fig:IR6LayoutCDR}
      \vspace*{1mm}
  \end{figure}

IR6 and IR8 are not identical, nor are their existing experimental
halls. RHIC and EIC bring beams together horizontally for
collisions; in the arcs there is one ``inner'' beamline (closer to the arc center of curvature) and one
``outer'' beamline (further from the arc center of curvature). For the EIC, the IR6 crossing geometry is such
that both beams cross from inner to outer beamlines (illustrated in
Figure~\ref{fig:IR6LayoutCDR}), while the IR8
crossing geometry is from outer to inner beamlines. Hence the primary IR6 layout requires less
bending than the second IR layout at IR8. Other spatial layout and
RHIC experimental hall structural design differences exist that are
inherited by the EIC project.

The physical layout differences between IR6 and IR8, and their
separate implementation timelines, permit them to be developed to
enhance the overall facility science impact and discovery potential.
For example, IR6 might deliver the highest luminosities at highest CM
energies, while IR8 may be designed to provide higher luminosities at
mid-range CM energies. The former would emphasize discovery potential
such as gluon saturation, while the latter would emphasize rare
exclusive processes for 3D nuclear imaging and mechanical properties.

This section first briefly describes the primary IR design, as defined in the EIC Conceptual Design Report (CDR)~\cite{EIC:CDR}. This section then outlines the present implementation of the second EIC IR at
IR8, consistent with nuclear physics, accelerator, and engineering
requirements. The second IR may
also provide a different acceptance coverage than the first IR. We include discussion of the operation of both IRs over the entire
energy range of $\sim$20--140\,GeV center of mass, and include consideration of different modes of two-IR EIC
operations and their anticipated beam dynamics constraints.

\subsection{Primary IR design parameters} 
The luminosity and the design of the reference first EIC interaction
region is optimized emphasizing the discovery potential of the EIC by
providing the highest luminosity near the upper end of the CM energy
range, from $\sim$80--120\,GeV, while covering the entire range of
parameters required by the Nuclear Physics Long Range Plan. The
parameter set and design is based on 1160 colliding bunches in each
beam as described in the CDR~\cite{EIC:CDR}:
\begin{itemize}
\item{Peak luminosity of $L=10^{34}$\,cm$^{-2}$s$^{-1}$ at a
    CM energy of 105\,GeV;}
\item{Crossing angle $\theta_c=~$25\,mrad;}
\item{Maximum accelerator optical $\beta$-functions in the final focus quadrupole magnets,
    $\beta_{\rm max} \leq$1800\,m (for protons in the vertical
    direction) and acceptable nonlinear chromaticity resulting in
  sufficient dynamic aperture;}
\item{Intra-beam scattering (IBS) growth times in horizontal and longitudinal directions of
    $\tau_{\rm IBS} >$2\,hours.}
\end{itemize}

The design and layout of IR6 are reasonably mature, as illustrated in Figure~\ref{fig:IR6LayoutCDR}.

\subsection{Second IR design and downstream tagging acceptance} \label{sec:secondIrDownstreamTagging}
The EIC requirements include sufficient flexibility to permit alternative optimizations of the two experimental IRs. For example, the IRs may be optimized for highest luminosities at different CM
energies. Moreover, the two IRs and corresponding detectors may have acceptances and capabilities
optimized for different parts of the physics program as described in this white paper.

To first order, the luminosity at the IP is inversely proportional to
the distance between the last upstream and first downstream final
focus quadrupoles (FFQs). The statistical uncertainty of measurements
in the central detector scales as this distance. However, the closer
the beam elements are to the IP, the more they obstruct the acceptance
at shallow angles with respect to the beam axis and restrict the
acceptance for forward particles. The solenoidal field used in the
central detector region to measure the high $p_T$ particles in the
central detector is not effective in determining the momenta of
particles moving parallel to the beam direction, and additional fields
are needed.

From kinematics, the reaction products are biased towards small angles
around the original ion beam. In particular, the detection of
small-angle products requires acceptance to the recoiling target
baryon (3D structure of the nucleon), hadrons produced from its
breakup (target fragmentation), or all the possible remnants produced
when using nuclear targets (including the tagging of spectator protons
in polarized deuterium). The detection should be done over a wide
range of momenta and charge-to-mass ratios with respect to the
original ion beam. The second IR design should address these
measurement difficulties posed by the beam transport elements.

From machine design and luminosity considerations, it is not desirable
to leave a very large detector space free of beam focusing elements to
allow the small-angle products to accumulate sufficient transverse
separation from the incident beams. The solution is to let the
small-angle particles pass through the nearest elements of the machine
final-focusing system, which simultaneously perform the function of
angle and momentum analyzer for the small angle reaction products. Ideally, this forward detection system must be capable of accepting all reaction products that have not been captured by the central detector. In particular, similarly to the IR6 detector, this implies sufficiently large apertures of the forward ion final focusing quadrupoles to accommodate particle scattering angles from zero all the way up to the minimum acceptance angle of the central detector. Of course, detection of zero angle particles requires that they are outside of the beam stay-clear region in another dimension, namely, in the rigidity offset. The IR8 design is particularly optimized for separation of such particles from the beam and their detection as described below. A
significant challenge of this approach is to balance often
contradictory detector and machine optics requirements. For example, the choice of the apertures of the forward ion final focusing quadrupoles, and therefore the forward angular acceptance, are a balance of the detection requirements and engineering constraints. One would like to make the apertures sufficiently large without exceeding the technical limits on the maximum aperture-edge fields.

\begin{figure}[tbh]
    \centering
    \vspace*{1mm}
    \includegraphics[width=1.0\textwidth]
    {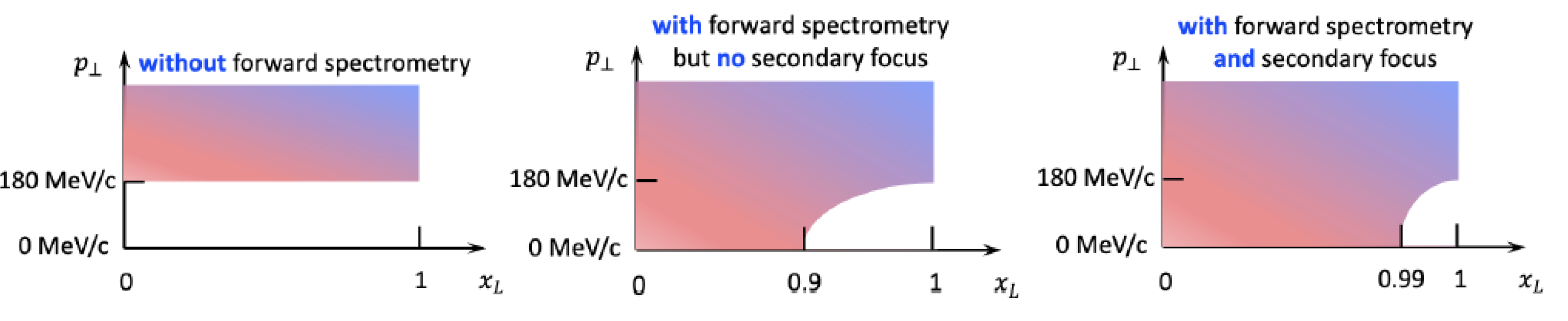}
    \caption{Illustration of forward spectrometry
      and secondary focus effects on detector acceptance (shaded) in the
      $\textit{x}_L - \textit{p}_T$ space for 275\,GeV protons.}
    \label{fig:xlpt}
      \vspace*{1mm}
  \end{figure}

Figure~\ref{fig:xlpt} illustrates $x_L-p_T$ acceptance with two
successive improvements to second IR acceptance. Without forward spectrometry
(left), the detection of low-angle scattered particles is limited by the beam
divergence at the IP. By introducing forward spectrometry (center), this limit can
be lowered, but particles with high ridigity
$x_L~=~1$ still escape detection. Adding a secondary focus point
with flat dispersion (right) improves the $x_L$ acceptance gap further.

The
maximum detectable $x_L$ at a point in the beam-line can be calculated
to first order using,
\begin{equation}
x_{L} < 1 -10\frac{\sqrt{\beta_x^{2nd}\epsilon_x + D_x^2 \sigma_{\delta}^2}}{D_x},
\end{equation}
where $\beta_x^{2nd}$ is the Twiss $\beta$-function at the second
focus, $\epsilon_x$ is the horizontal beam emittance, $D_x$ is the
horizontal dispersion at the second focus, and $\sigma_{\delta}$ is the beam
momentum spread. At a point in the lattice with low $\beta$ function and high
dispersion $D_x$, one can reach the fundamental limit for the maximum $x_L$ given by
\begin{equation}
x_{L} < 1 - 10\sigma_{\delta}\;.
 \end{equation}
The present EIC second IR secondary focus design is very close to this
theoretical limit. Further improvements are quite limited by space availability in the
experimental hall and magnetic field constraints.

The selection of crossing angle is an important design choice for the second IR. This crossing angle must not be too large ($>\sim$50\,mrad) for various reasons:
\begin{itemize}
    \item Constraints from the existing experimental hall geometry.
    \item The IP must be shifted towards the ring center to permit the Rapid Cycling Synchrotron (RCS) electron injector to bypass the detector.
    \item A large crossing angle requires more aggressive crabbing (or RF manipulation of both beams to compensate crossing angle and maximize luminosity); this aggressive crabbing in turn is limited by cost, impedance, and beam dynamics issues.
    \item Detector acceptance becomes unacceptably small at larger crossing angles.
    \item Limits proximity of final focus quads and overall IR luminosity.
\end{itemize}
The crossing angle must also not be too small ($<\sim$25 mrad), since the existing hall geometry requires spectrometer dipoles to bend towards the electron beam. Bending away as in the primary IR is not possible because of the second IR collision geometry. This pushes the second IR crossing angle away from the 25\,mrad used in the primary IR. The second IR design choice of crossing angle is presently 35\,mrad.

\begin{figure}[ht]
    \centering
    \vspace*{1mm}
    \includegraphics[width=0.95\textwidth]
    {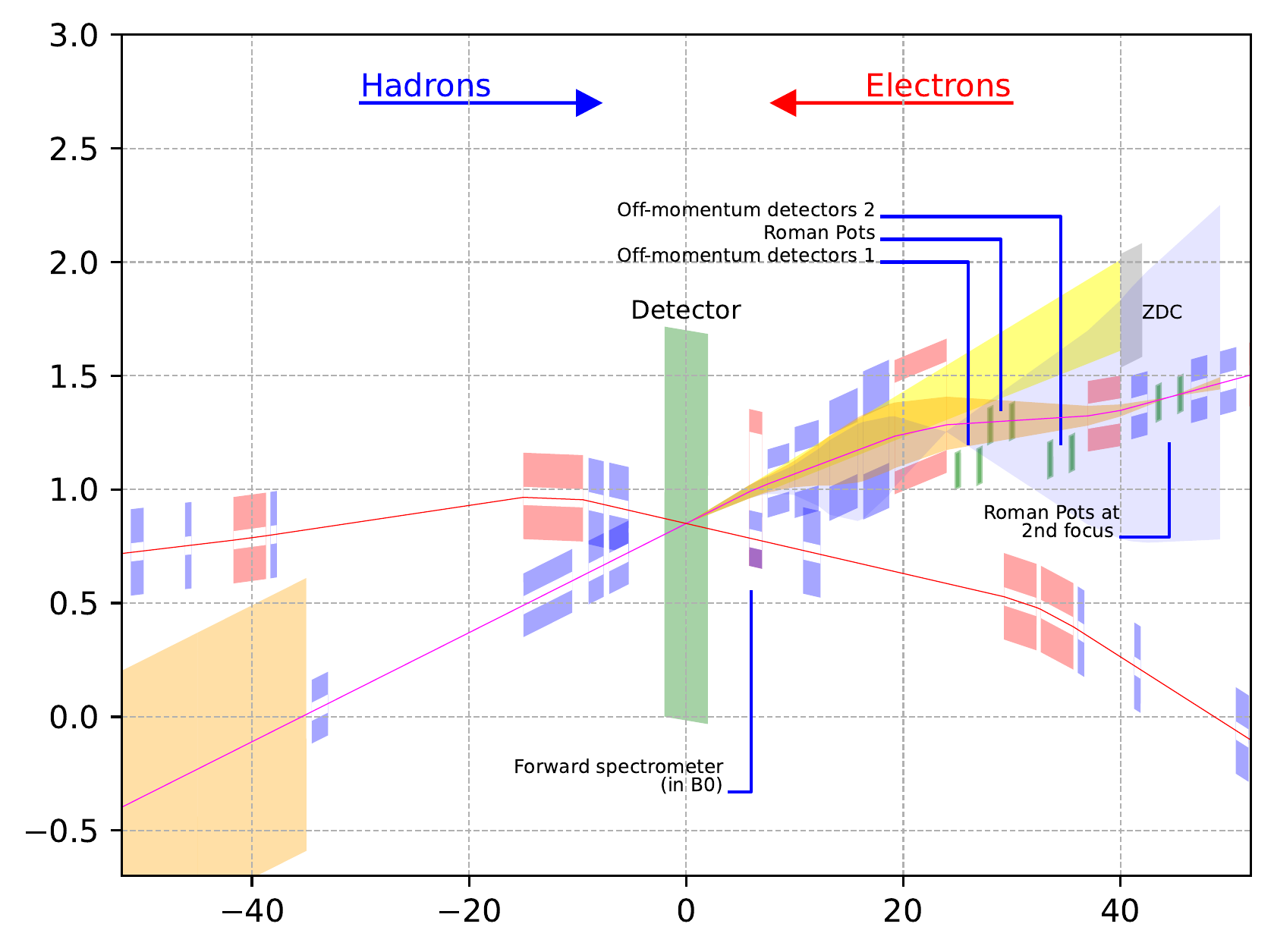}
    \caption{Layout of the second IR with a 35~mrad crossing angle
      indicating locations of the main forward and auxiliary detector
      component. The color shaded areas shows the $\pm$ 5~mrad~$p_{T}$
      acceptance for particles with yellow representing neutrons while
      orange and blue represent protons with $x_L=$ 1 and 0.5
      respectively. Magnets with their horizontal apertures are represented by pink (dipole) and blue (quadrupole) boxes.}
    \label{fig:ir8}
      \vspace*{1mm}
  \end{figure}

Figure~\ref{fig:ir8} shows the layout of the second IR with the
proposed detector component placements. The ancillary detectors in
the downstream hadron beam side have been integrated, while space is
available for luminosity monitor, low $Q^2$ tagger and local hadron
polarimetry.

\subsection{Technical design of an optimized low energy and high
  luminosity interaction region}

The above detection requirements make the detector and machine designs
intertwined and closely integrated. There is no longer a clear
separation between the detector and machine components. Several
detection parameters directly impact the design choices for the second
IR and vice versa. The major parameters critical to both detector and
machine aspects of the design are summarized in
Table~\ref{tab:2ndIRreqs}. This table also provides a
comparison of primary and second IR parameters.
 One of the important design differences is the inclusion of 
 a secondary focus in the second IR to provide improved downstream
 tagging resolution.
 
\begin{table}[htb]
\begin{center}
\vspace*{1mm}
\caption{Summary of second IR design requirements and their comparison
  to the first IR.
  \label{tab:2ndIRreqs}}
\begin{tabular}{c|l|c|c|l}
\# & Parameter & EIC IR \#1 & EIC IR \#2 & Impact \\ \hline
1 & Energy range & & & Facility operation \\
  & ~~~electrons [GeV] & 5--18 & 5--18 & \\
  & ~~~protons [GeV] & 41, 100--275 & 41, 100--275 & \\ \hline
2 & Crossing angle [mrad] & 25 & 35 	& $p_T$ resolution, \\
 & 					& 	& 		& acceptance, geometry \\ \hline
3 & Detector space  	& -4.5/+4.5 	& -5/5.5 & Forward/rear  \\ 
   &symmetry [m] 	& 			&  & acceptance balance \\ \hline
4 & Forward angular & 20 & 20--30 & Spectrometer dipole aperture	\\ 
 & acceptance [mrad] & 	&		 & 						\\ \hline
5 & Far-forward angular & 4.5 & 5 & Neutron cone, Max. $p_T$ \\ 
  &  acceptance [mrad] &        &           &   						\\ \hline
6 & Minimum $\Delta(B\rho)/(B\rho)$  & & & Beam focus with dispersion, \\
  & allowing for detection  & 0.1 & 0.003--0.01 & reach in $x_L$ and $p_T$ resolution, \\
  &  of $p_T=0$ fragments & & & reach in $x_B$ for exclusive proc. \\ \hline
7 & RMS angular beam diver-   & 0.1/0.2  	& $<$0.2 	& Min. $p_T$, $p_T$ resolution \\
  & gence at IP, h/v [mrad] 	& 			&	 	& 				\\ \hline
8 & Low $Q^2$ electron acceptance & $<$0.1 & $<$0.1 & Not a hard requirement \\ 
\end{tabular}
\end{center}
\end{table}

\subsubsection{Design constraints}

The design constraints for the second IR include:
\begin{itemize}
\item The second IR must transport both
  beams over their entire energy ranges with required path lengths. All second IR dipole
  magnets must have sufficient field integrals to
  provide the necessary bending angles keeping the IR footprint fixed
  from the lowest to the highest energy, while respecting geometric
  constraints of the existing infrastructure. The quadrupoles must also
  provide sufficient focusing to properly transport
  the beams over the entire energy range. Use of NbTi superconducting
  magnets implies that none of
  the second IR magnets can have aperture-edge fields higher than
  4.6~T at highest beam energies; more complicated magnets, such as
  the B0 spectrometer, may be limited to significantly lower fields~\cite{EIC:CDR}.
  For collisions, the second IR magnets
  must have sufficient strengths to focus the beams
  at the IP while having sufficiently large apertures to meet the
  detection requirements discussed below. Simultaneous operation of
  the two IRs is also subject to the beam dynamics constraints
  discussed later.

\item Consistent with the two detector complementarity approach, the 
  second IR could be designed to provide a near flat luminosity above $\approx$45\,GeV. This supports leveling of
  the EIC luminosity curve at a higher level over a
  wider energy range, as can be seen in Fig.~\ref{IR:EIC-lumi}. The second IR may also be designed to provide a different acceptance coverage than the first IR.

\begin{figure}[!ht]
\centering{\includegraphics[width=0.85\columnwidth]{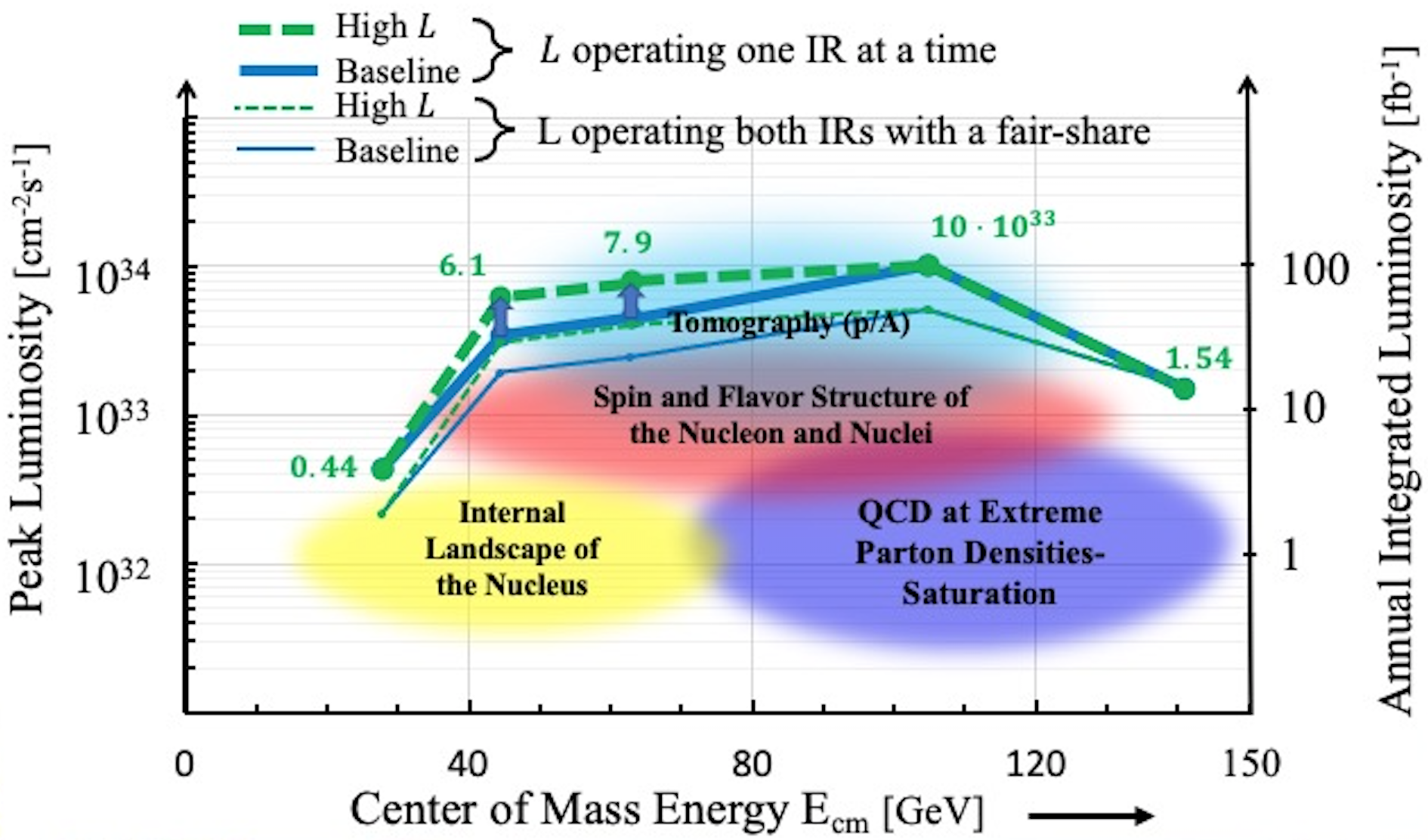}}
\caption{Estimated luminosity versus CM energies for the operation 
of one (thick lines) or two (thin lines) interaction regions. The blue lines 
show estimates of the reference luminosity. The green lines show the high luminosity 
operation with potentially improved beam optics and cooling at lower CM energies. 
(As shown in \cite{Gamage:2021summerEICUG})}
\label{IR:EIC-lumi}
\end{figure}

\item The ion and electron beams cross at a relatively large angle of
  35~mrad at the IP. High luminosity is preserved through the use of
  crab cavities. This angle moves the ion beam away from the electron
  beam elements and makes room for dipoles located just downstream of
  the central detector area. The dipoles serve two purposes. First,
  they shape the beam orbits providing their geometric match, making
  the IR footprint fit in the available detector hall and tunnel
  space, and creating room for detectors. Second, the dipole systems
  allow momentum analysis of the particles with small transverse
  momentum with respect to the beams. Particles with large
  transverse momenta are analyzed using the solenoidal field and the B0 magnet in the
  central detector.
\end{itemize}

\begin{figure}[ht]
    \centering
    \vspace*{1mm}
    \includegraphics[width=0.95\textwidth]{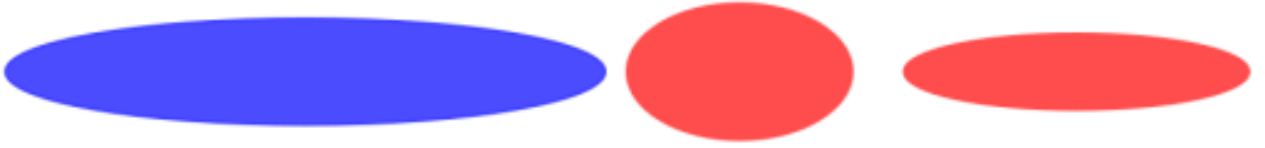}
\caption{Apparent horizontal broadening of the beam spot size at the IP due to the crab tilt. Blue left: RMS hadron bunch length $\sim 10$cm, red middle: Looking along the beam with no crabbing, and red right: What the RP sees $\sim$1.25mm.}
    \label{fig:crabtilt}
      \vspace*{2mm}
  \end{figure}

\subsubsection{Effect of horizontal crabbing in secondary focus}

Since the secondary focus is within the region where the hadron beam is crabbed, hadron crabbing effectively broadens the horizontal beam
spot size seen by the Roman Pot (RP) detectors in the secondary focus, as illustrated in
Figure~\ref{fig:crabtilt}. This beam spot size is one of
the sources of uncertainty in a $p_T$ measurement. Ignoring for the
moment other sources such as the beam angular spread at the IP, the
transverse position of a scattered particle at an RP $x_{RP}$ is
related to $p_T$ as
\begin{equation}
x_{RP} = M_{11} x_{IP} + M_{12} p_T /p ,
 \end{equation}
where $x_{IP}$ is the scattered particle's transverse position at the
IP and $p$ is the beam momentum. $M_{11}$ and $M_{12}$ are elements
of the linear beam transfer matrix from the IP to the RP known from
the magnetic optics design:
 \begin{eqnarray}
M_{11} & = & \sqrt{\beta_{RP}/\beta_{IP}}\cos\Delta\Psi , \\ \nonumber
M_{12} & = & \sqrt{\beta_{RP}\beta_{IP}}\sin\Delta\Psi ,
\end{eqnarray}
where $\beta_{RP}$ and $\beta_{IP}$ are the Twiss $\beta$-functions at
the RP and IP, respectively, and $\Delta\Psi$ is the betatron phase
advance from the IP to the RP. The measured $p_T$ can be expressed as
\begin{equation} 
p_T = p \frac{x_{RP}}{\sqrt{\beta_{RP}\beta_{IP}}\sin\Delta\Psi}-p\frac{1}{\beta_{IP}}\frac{\cos\Delta\Psi}{\sin\Delta\Psi}x_{IP}.
 \end{equation}
Since it is challenging to measure $x_{IP}$ precisely, the second
term on the right-hand side of the above equation represents a
measurement uncertainty
 \begin{equation} \label{eq:ptuncert}
\Delta p_T = \left| p\frac{1}{\beta_{IP}}\frac{\cos\Delta\Psi}{\sin\Delta\Psi}x_{IP} \right |.
\end{equation}
$x_{IP}$ consists of a random betatron component $x_{\beta}$ and a
longitudinal-position-correlated component $z\,\theta/2$:
\begin{equation}\label{eq:xip}
x_{IP}=x_{\beta}+z\,\theta/2, 
 \end{equation}
where $z$ is the particles longitudinal position from the center of
the bunch and $\theta$ is the total beam crossing angle.

The second term in Equation~\ref{eq:xip} describes the beam spot size
smear. It is typically much greater than the first term. Therefore,
the uncertainty term in Equation~\ref{eq:ptuncert} can be greatly
reduced by measuring the event's $z$ position. It has been suggested
that, with a feasible RP timing of $\sim 35$~ps, the $z$ position can
be resolved down to $\sim 1$~cm.

Another factor in the uncertainty term of Equation~\ref{eq:ptuncert}
is $\cos\Delta\Psi$. By placing the RP at a position with $\Delta\Psi$
close to $\pi/2$, $\Delta p_T$ in Equation~\ref{eq:ptuncert} can in
principle be made arbitrarily small. There may be practical
considerations limiting the available choice of $\Delta\Psi$ such as
the requirement of placing the RP before the crab cavities, which have
small apertures and kick the particles. In the presented design of the
second IR, $\Delta\Psi$ is adjusted as close to $\pi/2$ at the RP as
allowed by other constraints to minimize $\Delta p_T$.
  
Physics simulations set a requirement on the contribution of the
crabbing tilt to $\Delta p_T$ of
\begin{equation}\label{eq:dptreq}
\Delta p_T < 20\,{\rm MeV}. 
 \end{equation}
\begin{figure}[ht]
    \centering
      \vspace*{1mm}
    \includegraphics[width=0.5\textwidth]
    {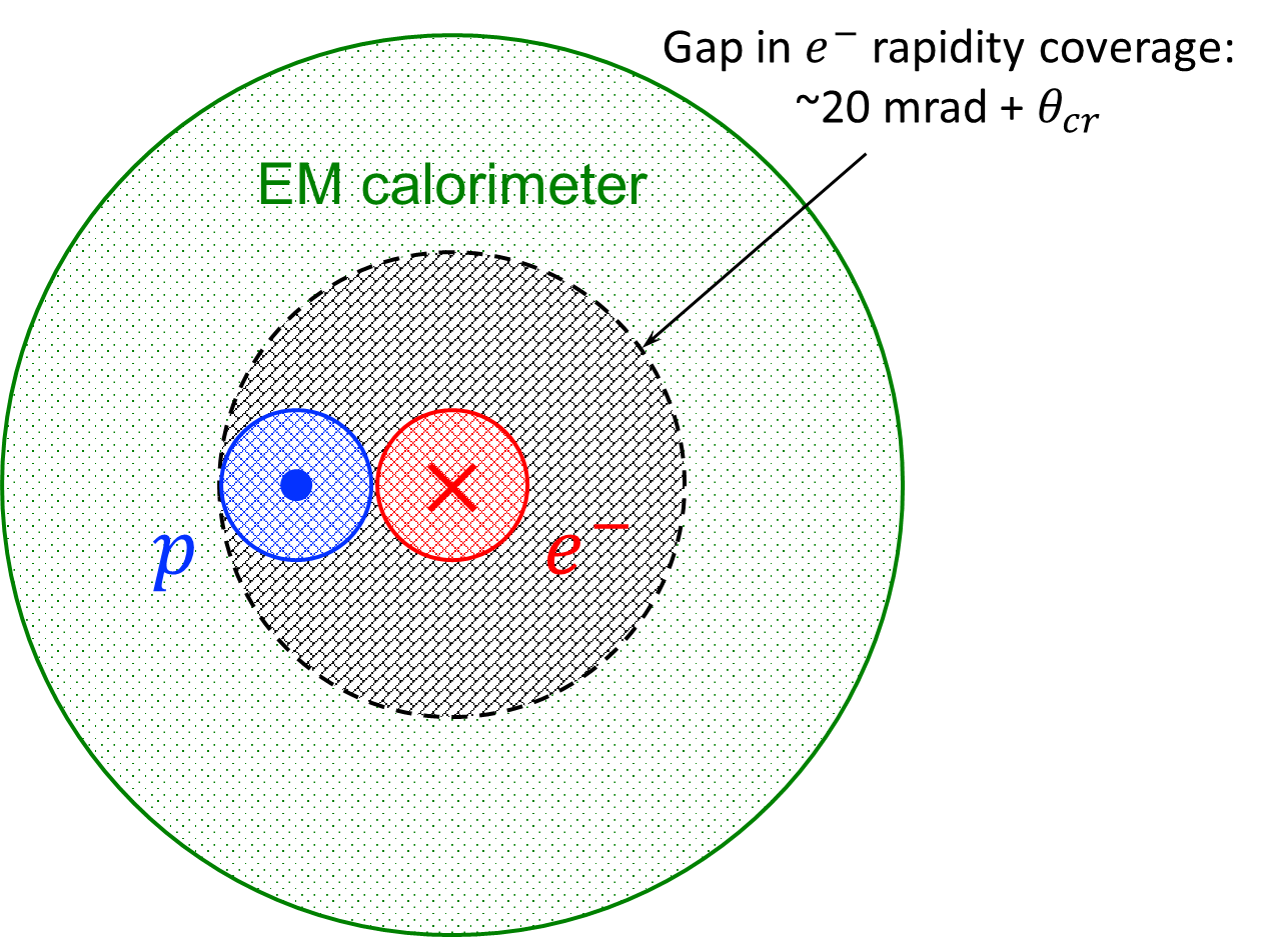}
    \caption{Gap in the electron rapidity coverage due to the crossing
      angle and the ion beam pipe. The blue and red circles represent
      the ion and electron beam pipes at the EM calorimeter location.
      The black dashed circle outlines the solid angle without full
      azimuthal detector acceptance.}
    \label{fig:rapiditygap}
      \vspace*{1mm}
  \end{figure}
Another issue with the size of the crossing angle is that it
contributes to the gap in the electron rapidity coverage in the rear
direction as illustrated in Figure~\ref{fig:rapiditygap}. There is
no full azimuthal coverage within an angle defined by the crossing
angle and the size of the ion beam pipe. Assuming 5~cm for the
radius of the ion beam pipe at a 2.5~m distance in the rear
direction from the IP, the total polar angle of the gap in the
rapidity coverage is about 20~mrad$+\theta_{cr}$.
There is also a subtle point worth mentioning regarding the impact of 
the crabbing on the RP resolution, and the advantage of measuring both 
the vertex z- and the time-coordinates. Z-coordinates will be measured by the MAPS Si vertex tracker.  However, if the vertex is measured e.g. z=+5cm, this does not determine if the collision happened at the leading edges of the two bunches (with mean x displaced in negative direction) or at the trailing edges of the collision (with mean x displaced in positive direction).  This is where the time measurement comes in, to determine where in the longitudinal profile of the crab the event happened.

\subsection{Operations with Two IRs}

At the time of this writing, the EIC construction project scope for IR8 has only nominal beam transport without magnets or optics support for collisions. Commissioning and operations will focus on beam parameter, luminosity, and polarization optimization for the single IR and detector within the project scope.

Later operations of the EIC with two IRs involves multiple scenarios, each with beam dynamics and design constraints that involve tradeoffs of available luminosity, operations time, and mode switching. The beam-beam force is the local nonlinear electromagnetic force colliding beams exert on each other; this force creates a nonlinear beam-beam tune shift that is a known limitation of many collider operations. This beam-beam tune shift is already optimized in the single-IR EIC design. Thus both IRs cannot operate simultaneously with full parameters necessary for maximum luminosity, as this would exceed the acceptable beam-beam tune shift limit. It is therefore infeasible to add net luminosity available to experiments by adding an IR in the EIC under optimized collider conditions where the beam-beam tune shifts limit integrated luminosity.

There are two alternatives to EIC operations with two IRs: EIC luminosity can be maximized separately for each detector in dedicated runs where only one IR is tuned for collisions; or EIC luminosity can be shared and optimized as much as possible between the two IRs in runs where both IRs are set up to share total facility luminosity.

The separate luminosity scenario is technically straightforward. The non-luminosity IR would be detuned to reduce chromatic effects, and beams would be steered to avoid collisions at that IR. For each run, the overall facility would then be optimized to maximize operational parameters necessary to optimize the science program for the given run time at the operating IR.

The shared luminosity scenario is technically more complicated. Section 4.6.4 of the EIC CDR~\cite{EIC:CDR} includes a section titled ``Beam-beam Effects with Two Experiments'' (pages 431--3) that describes one possibility for luminosity sharing. This involves design choices in the facility, and placement of the second IR and experiment in IR8, to enable an operating configuration that collides half the bunches at each of IR6 and IR8. Each individual bunch experiences only one collision per turn, so the total beam-beam tune shift limit for each bunch is respected. This CDR section also indicates that long-range beam-beam effects (present when beam timing is adjusted to share luminosity) may further limit the total luminosity available at both IRs.

The shared luminosity scenario may have other beam dynamics limitations (such as limitations of global chromatic correction) that would further limit the total available combined luminosity to both experiments. These beam dynamics considerations are being studied in the context of EIC second IR design and overall EIC lattice design optimization. Figure~\ref{IR:EIC-lumi} shows this best-case scenario as the ``fair-share'' curves, representing a 50\% sharing of total luminosity between the two IRs.

\newpage
\section{Acknowledgments} 
\noindent 
The PI's like to thank the Center for Frontiers in Nuclear Physics (CNFS) and the Center of Nuclear Femtography (CNF) for their support of this initiative and the dedicated discussions they devoted to its realization.

Special thanks go to  R. McKeown for the continuous support he gave to the initiative of developing this White Paper.

This work is supported in part by the U.S. Department of Energy, Office of Science, Office of Nuclear Physics, under contract numbers DE-AC02-05CH11231, DE-FG02-89ER40531, DE-SC0019230, DE-AC05-06OR23177. 
A. Afanasev is supported by National Science Foundation under Grant No. PHY-2111063. 
J. Bernauer is supported by National Science Foundation under Grant No. PHY-2012114.
A. Courtoy is supported by UNAM Grant No. DGAPA-PAPIIT IN111222 and CONACyT Ciencia de Frontera 2019 No. 51244 (FORDECYT-PRONACES). 
H.-W. Lin is partly supported by the US National Science Foundation under grant PHY 1653405 ``CAREER: Constraining Parton Distribution Functions for New-Physics Searches'' and PHY 2209424,  and by the  Research  Corporation  for  Science  Advancement through the Cottrell Scholar Award. 
K. Kumerički is supported by Croatian Science Foundation project IP-2019-04-9709.
Y. Oh was supported by the National Research Foundation under grants No. NRF-2020R1A2C1007597 and No. NRF-2018R1A6A1A06024970 (Basic Science Research Program). D. Glazier is supported by the UK Science and Technology Facilities Council under grants ST/P004458/1 and ST/V00106X/1. 
F. Ringer is supported by the Simons Foundation 815892, NSF 1915093. 
A. Signori acknowledges support from the European Commission through the Marie Sk\l{}odowska-Curie Action SQuHadron (grant agreement ID: 795475). 
D. Winney is supported by National Natural Science Foundation of China Grant No.~12035007
and the NSFC and the Deutsche Forschungsgemeinschaft (DFG, German Research Foundation) through the funds provided to the Sino-German Collaborative Research Center TRR110 ``Symmetries and the Emergence of Structure in QCD" (NSFC Grant No.~12070131001, DFG Project-ID~196253076-TRR~110). 
A. Vladimirov is supported by Deutsche Forschungsgemeinschaft (DFG) through the research Unit FOR 2926, “Next Generation pQCD for Hadron Structure: Preparing for the EIC”, project number 30824754". 
The work of Kirill M. Semenov-Tian-Shansky is supported by the Foundation for the Advancement of Theoretical Physics and Mathematics ``BASIS''. 
Krzysztof Cichy acknowledges support by the National Science Centre (Poland) grant SONATA BIS no. 2016/22/E/ST2/00013.
A. Vladimirov is funded by the \textit{Atracci\'on de Talento Investigador} program of the Comunidad de Madrid (Spain) No. 2020-T1/TIC-20204. This work was partially supported by DFG FOR 2926 ``Next Generation pQCD for  Hadron  Structure:  Preparing  for  the  EIC'',  project number 430824754.
C.R.~Ji is supported by DOE Contract No. DE-FG02-03ER41260. 
F.-K. Guo is supported by the Chinese Academy of Sciences under Grant No. XDB34030000, and by the National Natural Science Foundation of China under Grants No. 12125507, No. 11835015,
No. 12047503, No. 11961141012 and No. 12070131001. 
Xiaohui Liu acknowledges support by the National Natural Science Foundation of China under Grant No. 12175016.

\bibliography{main}
\clearpage
\appendix
\end{document}